\documentclass{tcibook}
\usepackage{fancyhea}
\usepackage{work}
\usepackage{bm}       
\usepackage{graphicx}
\usepackage{hyperref}      
\usepackage{xspace}
\usepackage{multirow}
\usepackage{amssymb}
\usepackage{cite}
\usepackage{adjustbox}

\newcommand{\nc}{\newcommand}  

\def\Acknowledgements{\bigskip  \bigskip \begin{center} \begin{large}
             \bf ACKNOWLEDGEMENTS \end{large}\end{center}}

\def\beq{\begin{equation}}
\def\eeq#1{\label{#1}\end{equation}}
\def\eeqn{\end{equation}}

\newenvironment{Eqnarray}%
   {\arraycolsep 0.14em\begin{eqnarray}}{\end{eqnarray}}
\def\beqa{\begin{Eqnarray}}
\def\eeqa#1{\label{#1}\end{Eqnarray}}
\def\eeqan{\end{Eqnarray}}

\nc{\ra}{\rightarrow}  
\nc{\slsh}{\slash\hspace*{-0.22cm}}
\def\Re{{\cal R \mskip-4mu \lower.1ex \hbox{\it e}\,}}
\def\Im{{\cal I \mskip-5mu \lower.1ex \hbox{\it m}\,}}

\nc{\vev}[1]{ \left\langle {#1} \right\rangle }
\nc{\bra}[1]{ \langle {#1} | }
\nc{\ket}[1]{ | {#1} \rangle }
\nc{\fb}{\,{\rm fb}^{-1}}
\nc{\ev}{{\rm eV}}
\nc{\kev}{{\rm keV}}
\nc{\Mev}{{\rm MeV}}
\nc{\gev}{{\rm GeV}}
\nc{\tev}{{\rm TeV}}
\nc{\mev}{{\rm MeV}}

\def\L{{\cal L}}

\def\W{{\cal W}}

\def\del{\partial}
\def\Dslash{\not{\hbox{\kern-4pt $D$}}}
\def\dslash{\not{\hbox{\kern-2pt $\del$}}}
\def\pslash{\not{\hbox{\kern-2pt $p$}}}
\def\ETmiss{ \not{\hbox{\kern-4pt $E$}}_T }

\def\BR{\mbox{\rm BR}}
\def\ee{e^+e^-}

\def\alphas{\alpha_s}
\def\msb{{\bar{\ssstyle M \kern -1pt S}}}

\begin{document}

\def\bibname{References}
\bibliographystyle{snowmass}

\raggedbottom

\pagenumbering{roman}

\parindent=0pt
\parskip=8pt
\setlength{\evensidemargin}{0pt}
\setlength{\oddsidemargin}{0pt}
\setlength{\marginparsep}{0.0in}
\setlength{\marginparwidth}{0.0in}
\marginparpush=0pt


\pagenumbering{arabic}

\renewcommand{\chapname}{chap:intro_}
\renewcommand{\chapterdir}{.}
\renewcommand{\arraystretch}{1.25}
\addtolength{\arraycolsep}{-3pt}



\newcommand{\draftnote}[1]{
\begin{center}\fbox{\begin{minipage}[c]{0.8\textwidth}
{\em #1} \end{minipage}}\end{center}}
\renewcommand{\draftnote}[1]{}


\renewcommand{\W}{\ensuremath{W}\xspace}
\newcommand{\Z}{\ensuremath{Z}\xspace}
\newcommand{\WW}{\ensuremath{WW}\xspace}
\renewcommand{\ee}{\ensuremath{e^+e^-}\xspace}
\newcommand{\ttbar}{\ensuremath{t\bar t}\xspace}
\newcommand{\pt}{\ensuremath{p_T}\xspace}
\newcommand{\as}{\ensuremath{\alpha_{s}}\xspace}
\newcommand{\amz}{\ensuremath{\as(M_{Z}^2)}\xspace}
\newcommand{\amt}{\ensuremath{\as(M_{\tau}^2)}\xspace}
\newcommand{\aqsq}{\ensuremath{\as(Q^2)}\xspace}
\newcommand{\NNNLO}{N$^{3}$LO\xspace}

\def\BR         {{\ensuremath{\cal B}}}
\def\invpb {\ensuremath{\mbox{\,pb}^{-1}}}
\def\invfb   {\ensuremath{\mbox{\,fb}^{-1}}}
\newcommand{\nsig}{\ensuremath{N_{sig}}}
\newcommand{\nback}{\ensuremath{N_{backg}}}
\newcommand{\nobs}{\ensuremath{N_{obs}}}
\def\L{{\ensuremath{\cal L}}}
\newcommand{\Pp}{\mathswitchr p}
\def\mathswitch#1{\relax\ifmmode#1\else$#1$\fi}
\def\mathswitchr#1{\relax\ifmmode{\mathrm{#1}}\else$\mathrm{#1}$\fi}
\def\mathswitchit#1{\relax\ifmmode{#1}\else$#1$\fi}
\newcommand{\TeV}{\unskip\,\mathrm{TeV}}
\newcommand{\GeV}{\unskip\,\mathrm{GeV}}
\newcommand{\MZ}{\mathswitch {M_\PZ}}
\newcommand{\pba}{\unskip\,\mathrm{pb}}
\newcommand{\nba}{\unskip\,\mathrm{nb}}
\newcommand{\born}{{\mathrm{Born}}}
\newcommand{\virt}{{\mathrm{virt}}}
\def\ga{\gamma}
\def\de{\delta}
\def\veps{\varepsilon}
\def\la{\lambda}
\def\si{\sigma}
\def\Ga{\Gamma}
\def\De{\Delta}
\def\La{\Lambda}
\newcommand{\EW}{{\mathrm{EW}}}
\newcommand{\QCD}{{\mathrm{QCD}}}
\newcommand{\QED}{{\mathrm{QED}}}
\newcommand{\veto}{{\mathrm{veto}}}
\newcommand{\LO}{{\mathrm{LO}}}
\newcommand{\NLO}{{\mathrm{NLO}}}
\newcommand{\Int}{{\mathrm{IF}}}
\newcommand{\full}{\mathrm{full}}
\newcommand{\rT}{{\mathrm{T}}}
\newcommand{\rM}{{\mathrm{M}}}

\newcommand{\mht}{\ensuremath{H_{T}\hspace{-1.3em}/\kern0.65em}\xspace}
\newcommand{\MHT}{\mht}
\newcommand\defMHT{\ensuremath{\mht = | - \sum_{i} \vec{p}_\mathrm{T} (\text{jet}_{i})|}\xspace}

\chapter{Working group report: QCD}
\label{chap:qcd}

\vspace{-2.9cm}
\begin{flushright}
\vbox{
\begin{tabular}{r}
ANL-HEP-CP-13-48 \\
FERMILAB-FN-0967-CMS-T
\end{tabular}
}
\end{flushright}

\begin{center}\begin{boldmath}

\vspace{1.75cm}



\begin{center}

\begin{large} {\bf Conveners: J. M. Campbell, K. Hatakeyama, J. Huston, F. Petriello} \end{large}

J.~Andersen,
L.~Barz\`e,
H.~Beauchemin,
T.~Becher,
M.~Begel,
A.~Blondel,
G.~Bodwin,
R.~Boughezal,
S.~Carrazza,
M.~Chiesa,
G.~Dissertori,
S.~Dittmaier,
G.~Ferrera,
S.~Forte,
N.~Glover,
T.~Hapola,
A.~Huss,
X.Garcia~i~Tormo,
M.~Grazzini,
S.~H\"oche,
P.~Janot,
T.~Kasprzik,
M.~Klein,
U.~Klein,
D.~Kosower,
Y.~Li,
X.~Liu,
P.~Mackenzie,
D.~Maitre,
E.~Meoni,
K.~Mishra,
G.~Montagna,
M.~Moretti,
P.~Nadolsky,
O.~Nicrosini,
F.~Piccinini,
L.~Reina,
V.~Radescu,
J.~Rojo,
J.~Russ,
S.~Sapeta,
A.~Schwartzman,
P.~Skands,
J.~Smillie,
I.~W.~Stewart,
F.~J.~Tackmann,
F.~Tramontano,
R.~Van de Water,
J.~R.~Walsh,
S.~Zuberi

\end{center}



\em\today

\end{boldmath}\end{center}


\section{Executive summary}
\label{sec:qcd-summary}

A quantitative description of Nature requires a detailed understanding of quantum chromodynamics (QCD) phenomenology.  The
success of Run 1 of the LHC relied upon advanced QCD simulation tools to support and guide experimental analyses, and the
discovery of the Higgs boson illustrated the indispensable role of the QCD community in enabling discovery science.    From
parton distribution functions with robust errors, through calculations to the next-to-next-to-leading order and beyond in
perturbative QCD, to the development of sophisticated Monte Carlo tools more faithful to the underlying hard dynamics, every
advance from over a decade of research was needed to make this historic discovery possible.  Run 2 of the LHC marks the
beginning of the precision phase in our study of the mechanism of electroweak symmetry breaking.  Quantitative QCD analyses
will become ever more indispensable in unraveling the origin of what we have found.  

In the nearly one decade since the previous Snowmass workshop, what was once considered impossible in the field of QCD has become commonplace.  The
technology in the area of next-to-leading order QCD computations has undergone an advancement so rapid as to invite comparison
to the Industrial Revolution.  We stand poised to enter a similar era for next-to-next-to-leading order calculations.  Parton
distribution function errors have become quantitative rather than qualitative, and parton showers matched with exact
next-to-leading order matrix elements have become the default simulation tools for experimental studies.  Any of these
breakthroughs would have been considered unlikely at best if predicted at the time of the previous Snowmass workshop.  It is
therefore our difficult task to summarize the level of the tools that exist now, to extrapolate their advancement into the
medium and long-term future, and to present a priority list as to the direction that the development of these tools should
take. 

Most of the efforts of the QCD working group concentrate on proton-proton colliders, at 14 TeV as planned for the next run of
the LHC, and for 33 and 100 TeV, possible energies of the colliders that will be necessary to carry on the physics program
started at 14 TeV. We also examine QCD predictions and measurements at lepton-lepton and lepton-hadron colliders, and in
particular their ability to improve our knowledge of $\alpha_s(M_Z^2)$ (at both types of colliders) and our knowledge of parton distribution
functions (PDFs) (at lepton-hadron colliders).  Since the current world average of strong coupling measurements is dominated by
the determinations made using lattice gauge theory we also explore possible improvements to our knowledge of $\alpha_s(M_Z^2)$
from such extractions. 

\draftnote{A detailed understanding of quantum chromodynamics (QCD) phenomenology, both perturbative and non-perturbative, is crucial for a detailed
understanding of physics at  hadron-hadron, lepton-hadron, and lepton-lepton colliders. The QCD sub-group is somewhat different from most of the
other sub-groups in the Snowmass workshop in that the emphasis is not on observables per se, but on the tools needed to understand the observables,
in physics processes at all of the colliders mentioned above. There has been a great deal of progress in the last 5-10 years on QCD-related tools for
calculation, simulation and analysis, at a level that would have been considered unlikely at best, if predicted at the time of the previous Snowmass
workshop. Thus, it is our difficult task to summarize the level of the tools that exist now, to perform this extrapolation into the medium and
long-term future, and to present a priority list as to the direction that the development of these tools should take. Most of our efforts concentrate
on proton-proton colliders, at 14 TeV as planned for the next run of the LHC, and for 33 and 100 TeV, possible energies of the colliders that will be
necessary to carry on the physics program started at 14 TeV. We also examine QCD predictions and measurements at lepton-lepton and lepton-hadron
colliders, and in particular their ability to improve our knowledge of $\alpha_s(M_Z^2)$ (both) and our knowledge of parton distribution functions (PDFs) (lepton-hadron colliders).
Since the current world average of strong coupling measurements is dominated by the determinations made using lattice gauge theory we
also explore possible improvements to our knowledge of $\alpha_s(M_Z^2)$ from such extractions.}

We summarize the main conclusions of this report below.  A more detailed listing of the results of the QCD working group studies can be found in the conclusions.

\begin{itemize}

\item Improvement in our current understanding of PDFs is needed in both the `precision region' relevant for
Higgs boson studies and in the `discovery region' of multi-TeV masses.  The latter will be addressed
by future LHC studies while the former is more difficult due to already-strong current constraints.  A future electron-hadron collider, such as the LHeC, would be the ultimate machine to provide PDFs for precision proton-proton physics.

\item Higher precision calculations are needed to fully realize the potential of the LHC and future hadron colliders.  Further progress is required on fixed-order calculations to next-to-next-to leading order and beyond in the QCD coupling constant, and on the resummation of large logarithms that appear in observables where the available phase space is restricted.
The inclusion of electroweak corrections into theoretical simulation programs is mandatory for physics studies in future high energy proton-proton collisions.  Unravelling the identity of the Higgs boson requires further advances in our QCD
calculational abilities.

\item Theoretical predictions at the 1\% level in QCD will require an understanding of numerous subtle conceptual issues, including the validity of the standard factorization approach and the universality of PDFs, power-suppressed contributions to cross sections, and perturbation theory in extreme kinematic configurations.

\item The most recent lattice and continuum determinations of $\alpha_s$, $m_b$ and $m_c$ could be used
to immediately reduce the parametric uncertainties in predictions for Higgs boson decay rates.  Future
precision determinations from both the lattice and experiment could reduce the error on $\alpha_s$ to 0.1\%.

\end{itemize}


\section{Introduction}
\label{sec:qcd-intro}

It is useful to recall the basic structure of a parton-level hadron collider cross section computed in
perturbative QCD.  The cross section can be written schematically as,
\begin{eqnarray}
\sigma &=& \sum_{a,b} \int_0^1 {\rm d}x_1 \, f_{a/A}(x_1, \mu_F^2)  \int_0^1 {\rm d}x_2 \, f_{b/B}(x_2, \mu_F^2) \Biggl\{
\int {\rm d}\hat\sigma_{ab}^{LO}\left(\alpha_s\right) \, \Theta^{(m)}_{\rm obs}  \nonumber \\
&& + \alpha_s(\mu_R^2) \left[ \int \left(
 {\rm d}\hat\sigma_{ab}^{V}\left(\alpha_s,\mu_R^2\right) 
+{\rm d}\hat\sigma_{ab}^{C}\left(\alpha_s,\mu_F^2\right) \right) \, \Theta^{(m)}_{\rm obs}
+\int {\rm d}\hat\sigma_{ab}^{R}(\alpha_s) \, \Theta^{(m+1)}_{\rm obs} \right] \Biggr\}
+ \ldots
\label{eq:qcd-xsec-sketch}
\end{eqnarray}
where we have sketched the terms that contribute up to the next-to-leading order (NLO) level in QCD.  The first ingredients in the perturbative
description are the PDFs, defined for a given species of parton $a$, $b$ inside incoming hadrons $A$, $B$. The PDFs
are functions of the parton momentum fractions $x_1$, $x_2$ and the factorization scale $\mu_F$. The leading order prediction depends on the hard
matrix elements, contained in the factor ${\rm d}\hat\sigma_{ab}^{LO}$, which in turn depend on the value of the strong coupling, $\alpha_s$, for
strongly-interacting final states. In this equation the quantity $\alpha_s$ is a shorthand notation since it must be evaluated at the renormalization
scale $\mu_R$, $\alpha_s \equiv \alpha_s(\mu_R^2)$.  The hard matrix elements may also depend upon additional parameters in the Standard Model Lagrangian, such as the quark masses.  The predicted cross section depends on the cuts that are applied to the
$m$-parton  configuration in order to define a suitable observable, $\Theta^{(m)}_{\rm obs}$. In the case of cross sections for multi-jet processes
this factor accounts for the jet definition that is required for infrared safety.  Finally, there may be additional non-perturbative inputs required, such as fragmentation functions or the matrix elements needed for processes such as quarkonium production.  At NLO there are further contributions, as indicated on the second
line of the equation.  The virtual diagrams contain an explicit dependence on the renormalization scale, ${\rm
d}\hat\sigma_{ab}^{V}\left(\alpha_s,\mu_R^2\right)$ while the collinear counterterms that are necessary in order to provide an order-by-order
definition of the PDFs introduce explicit factorization scale dependence, ${\rm d}\hat\sigma_{ab}^{C}\left(\alpha_s,\mu_F^2\right)$.  The effects of
real radiation, ${\rm d}\hat\sigma_{ab}^{R}(\alpha_s)$ now include a cut on the $(m+1)$-parton configuration. They may therefore be sensitive to
kinematic effects that are not present in the $m$-parton case, for instance the effect of jet vetoes in electroweak processes.
At next-to-next-to-leading order (NNLO) we would include terms in Eq.~(\ref{eq:qcd-xsec-sketch}) that have an explicit factor of $\alpha_s^2$
in addition to those present at leading order.  In outline the extension is clear, with the introduction of configurations that contain
$m$, $m+1$ and $m+2$ parton.  As a result NNLO calculations may be even more sensitive to kinematic effects that are only approximately modeled,
if at all, in lower orders.  However, the interplay between soft and collinear divergences in each of these contributions greatly complicates the
calculation of NNLO effects.

From this guiding equation it is clear that detailed QCD predictions require knowledge of:

\begin{itemize}
\item $\alpha_s(M_Z^2)$ and its uncertainty;
\item PDFs and their uncertainties;
\item higher order corrections to cross sections;
\item the impacts of restrictions on phase space, such as jet vetoes;
\item quark masses and other Standard Model parameters;
\item possible additional non-perturbative inputs, such as fragmentation functions or quarkonium matrix elements.
\end{itemize}

Measurements at 14 TeV and higher will access a wide kinematic range, where PDF uncertainties and the impact of higher order corrections may be
large. At scales large compared to the W mass, electroweak (EWK) corrections can be as important as those from higher order QCD; mixed QCD-EWK
corrections also gain in importance.  Higher energies also imply higher luminosities, which require the ability to isolate the physics of
interest from the background of multiple interactions accompanying the higher luminosities. Much of the physics of interest will still involve
the production of leptons, jets, etc at relatively low scales. Obtaining theoretical predictions in the presence of strict kinematic cuts,
especially those involving high transverse momenta, masses, etc., can result in the creation of large logarithms of ratios of scales
involved in the processes. All of these effects mean that, as the center-of-mass energy increases, both the perturbative and
non-perturbative environments may make precision measurements of such objects more difficult. 

In this contribution, we cannot hope for a comprehensive treatment of all of the above, but will try to summarize the most important aspects of QCD
for future colliders.


\section{The strong coupling from colliders}
\label{sec:qcd-alphas}

\draftnote{1-2 pages: K. Hatakeyama to start based on talks from 
G. Dissertori and contributions from the TLEP community; \\
summary of expected precision from LHC, LHeC, ILC, TLEP}


The strong coupling constant \as is one of the fundamental parameters of QCD. 
The coupling constant itself is not a physical
observable, but a quantity defined in the context of perturbation theory, 
which enters predictions for experimentally-measurable observables.
The uncertainty on \as is currently considered to be one of the major contributors to the
parametric uncertainties on the prediction for the $H\to b\bar b$ partial width.  Reducing
this uncertainty would enable a more precise measurement
of both this partial width and the total width of the Higgs boson at future lepton colliders,
where the experimental uncertainty no longer dominates~\cite{SnowmassHiggs}.
The size of \as is not given by theory, but 
can be extracted from experimental measurements
at $e^+e^-$, $ep$, $pp$, and $p\bar p$ colliders, as well as
from lattice QCD calculations. 

\begin{figure}[htb]
\begin{center}
\includegraphics[width=0.4\hsize]{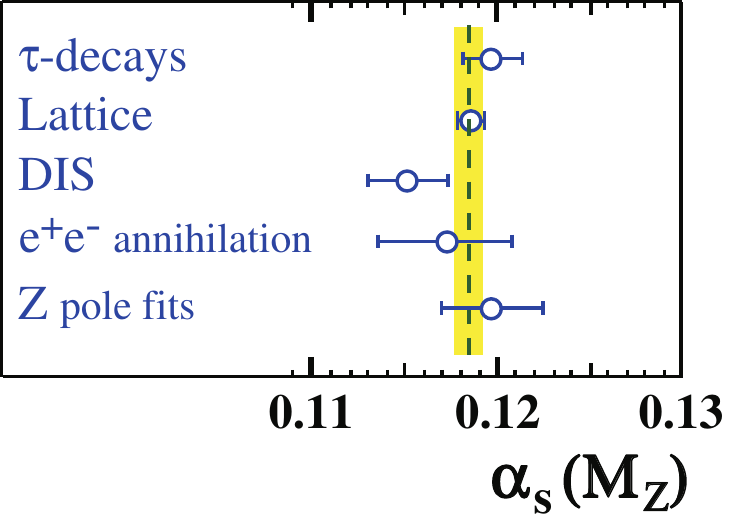}
\caption{Summary of values of $\amz$ obtained for
various sub-classes of measurements. 
The world average value of
$\amz = 0.1184 \pm 0.0007$ is indicated by the
dashed line and the shaded band.
Figure taken from \cite{Beringer:1900zz}.
}
\label{fig:alphas_average}
\end{center}
\end{figure}

A recent review on the determination of \as 
may be found in the 2012 PDG review~\cite{Beringer:1900zz}.
The current world average presented in the 2012 PDG review is:
$$
\amz = 0.1184 \pm 0.0007
$$
which has 0.6\% relative uncertainty and is summarized in
Fig.~\ref{fig:alphas_average}. The quoted uncertainty is 
a factor of 4 better than the value in the PDG review in 1992~\cite{Hikasa:1992je},
showing a significant progress on the determination of \as over
the last two decades.
As demonstrated in \cite{Beringer:1900zz}, the central value of
the world average of $\amz$ is rather stable against different inputs
to this average.
The result from lattice calculations, 
which has the smallest assigned uncertainty, agrees well with
the exclusive average of the other results; however,
it largely determines the size of the overall uncertainty.

Below we discuss various approaches to determine \as and
future possibilities to further improve
the determination of $\as$ with measurements at the LHC and
future accelerator facilities and with lattice QCD calculations.
Given the current \amz uncertainty,
the main theme is to see if and/or how we can potentially
reduce its uncertainty to the level of 0.1\% relative or
0.0001 absolute~\cite{Dissertori2013}.


\subsection{Strong coupling from \ee colliders}
\label{subsec:alphas_ee}
%
%


Various studies on \as have been performed using \ee annihilation data.
They include the determination of \as from hadronic $\tau$ decays,
heavy quarkonia decays, event shapes, jet rates, and the hadronic $Z$ decay
rate. Future prospects with some of these approaches are discussed below.


\subsubsection{Hadronic final states of \ee annihilations}

\draftnote{Summary of the talk by G. Dissertori at the Seattle meeting.}

Jet rates and hadronic event shapes have a strong sensitivity to \as,
and they have been studied extensively in the past.
For these observables, the theoretical predictions are
calculated up to NNLO and the resummation is achieved
up to NNLL or N${}^{3}$LL.

The typical experimental uncertainty for these measurements
is about 1\%, and it is considered that improvements should be possible.
The hadronization uncertainty estimated based on the difference
between various hadronization models is typically around 1--2\%.
And, the theoretical uncertainties for the scale choice and variation,
matching with resummed calculations, and quark mass effects
are typically 1--3\%.  For a review of these determinations see
for instance Ref.~\cite{Bethke:2012jm}.
By the time the next generation of \ee colliders come online,
it is conceivable that both these theoretical and hadronization uncertainties 
have been reduced;
however, going well below 1\% appears to be quite challenging, if not
impossible.


\subsubsection{Hadronic $W/Z$ decay widths from \ee annihilations}

An accurate determination of \as may be obtained from the precise
experimental measurement of hadronic \Z decays.
An advantage of using such inclusive observables is that the
theoretical predictions are known to \NNNLO and
non-perturbative effects are strongly suppressed.
The ratio $R_Z$ of the partial width of the $Z$ into hadrons to
that to one massless charged lepton flavour may be written as:
\begin{equation}
R_Z
\equiv
R_l^0
\equiv \frac{\displaystyle{\Gamma(Z\to\mathrm{hadrons})}}
  {\displaystyle{\Gamma(Z\to\mathrm{leptons})}}
= R_Z^\mathrm{EW}N_C(1 
  + \delta_\mathrm{QCD} 
  + \delta_\mathrm{m} 
  + \delta_\mathrm{np}),
\label{eqn:rz}
\end{equation}
$$
\delta_\mathrm{QCD} 
= \sum_{n=1}^{4} c_n \left( \frac{\as}{\pi} \right)^n
+ {\cal O}(\as^5),~~~c_1=1,~c_2=0.76264,~c_3=-15.490, c_4=-68.241,
$$
$$
\delta_\mathrm{m} \sim {\cal O}\left( \frac{m_q^2}{M_Z^2} \right),~~~~~
\delta_\mathrm{np} \sim {\cal O}\left( \frac{\Lambda^4}{M_Z^4} \right),
$$
where the $\delta_\mathrm{QCD}$, $\delta_\mathrm{m}$, and $\delta_\mathrm{np}$
terms are for the QCD, mass, and non-perturbative corrections, respectively.

The latest result from the LEP electroweak working group on $R_l^0$ is
$R_l^0=20.767\pm0.025$~\cite{ALEPH:2005ab}, whose relative uncertainty
is 0.12\%.
Up to a few years ago, when only NNLO predictions were available, and the Higgs
mass was still unknown, this measurement was translated to a value of \amz as (see eg.~\cite{Bethke:2004uy}
and references therein):
\begin{eqnarray*}
\amz &=& 0.1226 \pm 0.0038 (\mathrm{exp}) 
    \, {}^{+\, 0.0028}_{-\, 0.0005} (\mu=\, {}^{2}_{0.25}M_Z)
    \,\,\, {}^{+\, 0.0033}_{-\, 0.0} (M_H=\, {}^{900}_{100} \mathrm{~GeV}) \\
 && \phantom{0.1226} \pm 0.0002 (M_\mathrm{top}=\pm 5 \mathrm{~GeV}) \pm 0.0002 (\mbox{renormal. schemes}) \\
 &=& 0.1226 \, {}^{+\, 0.0058}_{-\, 0.0038}.
\end{eqnarray*}
Since the uncertainty due to the Higgs mass dependence is no longer relevant,
the top quark mass dependence is negligible, and
the pQCD scale uncertainty from latest \NNNLO calculations~\cite{Baikov:2008jh,Baikov:2012er}
is only 0.0002 on \amz, the question comes down to if 
a future $Z$ factory can measure  $R_l^0$ with an absolute precision of $\sim 0.001$. Such a reduction in the
experimental uncertainty, which dominates this \amz determination, will be necessary, since
the relative uncertainty on \amz is about 25 times larger than the relative uncertainty on $R_l^0$,
because $c_1 \amz / \pi \sim 0.04 = {\cal O}(1/25)$ in Eq.~(\ref{eqn:rz}).

The LEP measurement of $R_l^0=20.767\pm0.025$~\cite{ALEPH:2005ab}
is mainly limited by lepton statistics.
With $\sim10^{12}$ \Z event statistics expected from TLEP~\cite{Gomez-Ceballos:2013zzn} 
and assuming the selection efficiency uncertainties scale with statistics,
one might expect a reduction of the uncertainty by a factor of $\sim 200$.
At this level of precision, we will have to consider many subtle systematic uncertainties
and a detailed analysis would be necessary.
The $R_l^0$ measurement is sensitive to the electroweak vertex correction
$\delta_b$ on the $Zb\bar b$ vertex which may be sensitive to possible 
new physics effects; however, it can be constrained by the direct extraction
of $R_b=\Gamma(Z\to b\bar b)/\Gamma(Z\to\mathrm{hadrons})$, so 
this is not expected to be a limitation.
As discussed above, in order to achieve the absolute 0.0001 uncertainty on \amz,
about a factor of 30 reduction in the uncertainty is necessary.
This is challenging  and will require a lot of work, but 
the TLEP target goal of the measurement of $R_l^0$ with a relative
precision of $<10^{-5}$ meets this requirement.
This would be an interesting possibility to explore~\cite{Gomez-Ceballos:2013zzn}.
With the GigaZ running of the ILC, the \Z event statistical error improves by a factor of about 7
compared to LEP.  This may lead to an \amz determination with an
absolute uncertainty of $\sim0.0004-0.0006$~\cite{Flacher:2008zq}, which is an interesting possibility
of ILC.

Another interesting possibility suggested is to use the \W hadronic width,
i.e. $B_h \equiv (\Gamma_\mathrm{had}/\Gamma_\mathrm{tot})_W$,
which we can extract by measuring the branching fractions
of \WW events to the $l\nu\,l\nu$, $l\nu\,qq$, and $qq\,qq$ final states.
The previous LEP measurement of
$B_h =67.41 \pm 0.27$~\cite{Schael:2013ita} was limited by \WW event statistics of about
$4\times 10^4$ events.
With $0.5\times 10^8$ \W pairs expected at TLEP and assuming that selection
efficiency uncertainties scale with statistics, it may be possible
to reduce the uncertainty on $B_h$ by a factor $\sim 70$ and thus the absolute uncertainty
on \amz to  $\sim 0.0002$. 
This is an interesting possibility of the proposed TLEP facility.


\subsubsection{Hadronic $\tau$ decay width from \ee annihilations}

The $\tau$ lepton is the only lepton that can decay hadronically.  It provides an excellent testbed to study QCD at low energies
and determine \as.
This channel may be considered even more inclusive than $R_Z$ at the
\Z pole, as we are integrating over the hadronic invariant mass spectrum.
One interesting advantage of the \as determination from hadronic $\tau$ decays is the
fact that the uncertainty $\delta$ on a measurement of \aqsq
translates to an uncertainty $\delta'=(\as^2(M_Z^2)/\as^2(Q^2))\cdot\delta$ on \amz.
Therefore, the relative uncertainty on \amz is about a factor 3 smaller
than the relative uncertainty on \amt, when the \amt measurement at the $\tau$
mass scale is transported to the \Z scale using the QCD beta-function. 

The ratio $R_\tau$ of the $\tau$ decay width into hadrons to that
into an electron channel may be expressed by:
\begin{equation}
R_\tau
\equiv \frac{\displaystyle{\Gamma(\tau^-\to\nu_{\tau}+\mathrm{hadrons})}}
  {\displaystyle{\Gamma(\tau^-\to\nu_{\tau}e^-\bar{\nu}_e)}}
= S_\mathrm{EW}N_C(1 
  + \delta_\mathrm{QCD} 
  + \delta_\mathrm{np}),
\label{eqn:rtau}
\end{equation}
$$
\delta_\mathrm{QCD} 
= \sum_{n=1}^{4} c_n \left( \frac{\as}{\pi} \right)^n
+ {\cal O}(\as^5),~~~c_1=1,~c_2=5.202,~c_3=26.37.490, c_4=127.1,
$$
$$
\delta_\mathrm{np} = \frac{\mathrm{ZERO}}{m_{\tau}^2}
 + c_4 \cdot \frac{\langle O_4\rangle}{m_\tau^4} 
 + c_6 \cdot \frac{\langle O_6\rangle}{m_\tau^6} + \cdot\cdot\cdot,
$$
where the $\delta_\mathrm{QCD}$ and $\delta_\mathrm{np}$
terms are for the perturbative and non-perturbative QCD corrections, respectively, and
the notation of ref.~\cite{Altarelli:2013bpa} for  $\delta_\mathrm{np}$ is adopted.
The approach used at LEP was to fit simultaneously \as and the non-perturbative
coefficients by measuring various moments of the $\tau$ spectral function.
Interestingly, the non-perturbative contributions
turn out to be rather small, e.g. $\delta_\mathrm{np}=-0.0059\pm0.0014$~\cite{Davier:2008sk}.
It would be interesting to measure such moments again, with
much improved precision, e.g.\ with an uncertainty on $\delta_\mathrm{np}$ of $<0.0005$.
The challenge for a more precise \amz determination appears to be the fact that various methods of estimating the
impacts of higher-order QCD terms lead to differences in \amt of  $\gtrsim 5$\%,
which translates to $\gtrsim1$\% for \amz (see e.g. Refs.~\cite{Altarelli:2013bpa,Pich:2013sqa}).
Furthermore, also the importance of the
``ZERO'' term in the expansion for $\delta_\mathrm{np}$ is subject of discussion \cite{Altarelli:2013bpa}.
Thus, it seems unlikely that the relative uncertainty on \amz using $\tau$ decays will be
reduced well below the 1\% level in the near future.


\subsection{Strong coupling from hadron colliders}

%
%

The strong coupling constant has been extracted from measurements at
hadron colliders as well.
The results on the strong coupling constant from hadron collider
jet data are based on the
$\pt$-dependence of the inclusive jet cross section measurements by the 
CDF~\cite{Affolder:2001hn}, 
D0~\cite{Abazov:2009nc}, and 
ATLAS~\cite{Malaescu:2012ts} Collaborations,
the jet angular correlations by the D0 Collaboration~\cite{Abazov:2012lua},
and the ratio of the inclusive 3-jet and 2-jet cross sections
by the CMS Collaboration~\cite{Chatrchyan:2013txa}.
One of the complexities of the hadron-hadron environment is that it is hard to
disentangle the extraction of the value of the strong
coupling constant from the uncertainties in the gluon distribution, which can
be significant in the relevant $x$-range.
The most precise measurement to date, 
$\amz = 0.1161\, {}^{+\, 0.0041}_{-\, 0.0048}$,
is from the inclusive jet cross section measurement from D0.  The uncertainty is dominated by the 
experimental uncertainties from the jet energy calibration,
the \pt resolution, and the integrated luminosity, and 
as well as the uncertainties on non-perturbative corrections
and the renormalization and factorization scales.

The recent result on \amz from CMS using the \ttbar
cross section~\cite{Chatrchyan:2013haa} yielded
$\amz = 0.1178\, {}^{+\, 0.0033}_{-\, 0.0032}$ based on a full
NNLO QCD calculation for the inclusive \ttbar cross section.
This is the first determination of the strong coupling constant
from top-quark production and shows the best precision among hadron
collider measurements. 

%
%

There have been significant developments in measurements at hadron
colliders in recent years, and we expect that future LHC measurements will
improve the precision on \as further.
The \as extraction from the jet data is also likely to benefit from recent progress towards
a full NNLO QCD prediction for inclusive jet and dijet production~\cite{Ridder:2013mf} by the time
of the next LHC running.
However, given the currently-quoted scale uncertainty and
the experimental systematic uncertainty,
it will be challenging to achieve $<1$\% relative uncertainty on \amz{}.
The improved precision from hadron collider data at
relatively high-$Q^2$ is still important for the robustness of \as determinations,
and testing the running of \as and asymptotic freedom,
as the current world average of \amz is driven by low-$Q^2$ measurements.

%
%


\subsection{Strong coupling constant from $ep$ colliders}
\label{sec:alphas_lhec}

\draftnote{Text below is from M. Klein and V. Radescu}

\draftnote{KH: If we keep the current section order, the general LHeC introduction
will need to move from the PDF section to here. Need some smoothing with other sections.}

Studies of deep-inelastic scattering by HERA experiments
have led to a number of precise determinations of \as.
PDG quotes the average value of $\as=0.1151\pm0.0022$ from deep-inelastic scattering (DIS) measurements~\cite{Beringer:1900zz}.



One proposed future $ep$ collider is the Large Hadron Electron Collider
(LHeC)~\cite{AbelleiraFernandez:2012cc},
which can provide improvements
not only on PDFs but also on the determination of \as. 
Using the intense,
high energy hadron beams of the LHC, the LHeC would add a new
electron beam of typically $60$\,GeV energy to construct a first TeV energy scale
electron-proton and electron-ion collider.\footnote{As
HERA never accelerated ions, nor 
deuterons, the kinematic range in lepton-nucleus ($eA$)
DIS is extended with the LHeC by nearly
four orders of magnitude in four-momentum squared $Q^2$ and
towards low Bjorken $x$. This leads to a determination of 
the proton but also the neutron and nuclear
PDFs in a hugely extended range and with unprecedented diversity,
as is described in~\cite{AbelleiraFernandez:2012cc}.} As the first application of energy recovery 
techniques for high energy particle physics,
the LHeC is designed to achieve a luminosity in 
excess~\cite{AbelleiraFernandez:2012ty,Bruening:2013bga} of
$10^{33}$\,cm$^{-2}$s$^{-1}$. Very high integrated
$ep$ luminosities of several hundreds of fb$^{-1}$, i.e.
around $1000$ times more than at HERA,
can be collected by operating the new electron machine synchronously
with the LHC. Such a huge luminosity enables measurements 
close to $x=1$ and the exploitation of the full $Q^2$ range,
up to $Q^2 \simeq 10^6$\,GeV$^2$, exceeding the kinematic
range of HERA by a factor of $20$. 


Two independent approaches have been undertaken
in order to verify the potential of the LHeC to determine $\as$.
These analyses used a complete simulation of the experimental
systematic errors of the neutral currents (NC) and charged currents (CC)
pseudo-data with higher order QCD fit analysis techniques (see the LHeC conceptual design report (CDR)~\cite{AbelleiraFernandez:2012cc} for details).
The total experimental uncertainty on $\as$ is estimated to be $0.2$\,\%
from the LHeC alone and $0.1$\,\% when combined with HERA.  
The theoretical errors on this determination of \as from LHeC will be of order $0.5$\,\%, and will require a complete calculation of DIS to order N$^3$LO.  Relying solely on inclusive DIS $ep$ data at high $Q^2$,
this determination is free of higher twist, hadronic and nuclear corrections,
unlike any of the recent global QCD fit analyses.
There are known further parametric uncertainties in DIS determinations
of $\as$. These will be much reduced
with the LHeC as it resolves the full set of parton distributions,
$u_v,~d_v,~\overline{u},~\overline{d},~s,~\overline{s},~c,~b$
and $xg$ for the first time,
providing $x$ and $Q^2$ dependent constraints that do not arise from the assumed functional forms of the PDFs.
%





\subsection{Summary}

We have reviewed the current status of \as determinations using
collider data and discussed future prospects with various approaches.
These measurements are complementary and
sensitive to \as at different $Q^2$ values, thus providing a test of
asymptotic freedom, the driving principle of QCD.
Table~\ref{tab:alphas} shows a summary of \as determinations
from both collider data and lattice QCD, together with target precisions for
the next generations of experiments and calculations. The lattice QCD determination
of \as as well as quark masses are discussed in detail in section~\ref{sec:lattice}.
The current world average of $\amz=0.1184 \pm 0.0007$~\cite{Beringer:1900zz}
is dominated by lattice QCD results.  The lattice determination
of \amz may improve to the level of an absolute uncertainty of 0.0004
in the next 5 years or so. The GigaZ running of ILC may also reduce
the \amz uncertainty to a similar level~\cite{Flacher:2008zq},
and the improvements from LHeC and TLEP could reduce the error on \amz to 0.1\%.


We note that the \as uncertainty used by the Higgs cross section working group~\cite{Denner:2011mq}
is three times larger than that of the world average from PDG which is dominated by the lattice
determination.  Given its important role in future Higgs coupling extractions, 
this uncertainty should be revisited in order to fully benefit from on-going and future
improvements in \as determinations.
This is discussed in more detail in the following section.

{\footnotesize
\begin{table}[t]
\begin{center}
\begin{tabular}{|l|lr|l|} \hline
Method & Current relative precision & & Future relative precision \\
\hline
\multirow{2}{*}{$e^+e^-$ evt shapes} 
 & expt $\sim 1\%$ (LEP) & & $<1$\% possible (ILC/TLEP) \\
 & thry $\sim 1$--$3\%$ (NNLO+up to N$^3$LL, n.p. signif.)$\!\!\!\!\!\!\!\!\!\!\!\!\!\!\!\!\!\!\!\!\!\!$ & \cite{Dissertori:2009ik} & $\sim 1\%$ (control n.p. via $Q^2$-dep.)\\
\hline
\multirow{2}{*}{$e^+e^-$ jet rates} 
 & expt $\sim 2\%$ (LEP) & & $<1$\% possible (ILC/TLEP) \\
 & thry $\sim 1\%$ (NNLO, n.p. moderate) &  \cite{Dissertori:2009qa} & $\sim 0.5\%$ (NLL missing) \\
\hline
\multirow{2}{*}{precision EW} 
 & expt $\sim 3\%$ ($R_Z$, LEP) & & $0.1\%$ (TLEP~\cite{Gomez-Ceballos:2013zzn}), $0.5$\% (ILC~\cite{Flacher:2008zq})  \\
 & thry $\sim 0.5\%$ (N$^3$LO, n.p. small) & \cite{Baak:2012kk,Baikov:2012er} & $\sim 0.3\%$ (N$^4$LO feasible, $\sim 10$ yrs) \\
\hline
\multirow{2}{*}{$\tau$ decays} 
 & expt $\sim 0.5\%$ (LEP, B-factories) & & $<0.2$\% possible (ILC/TLEP) \\
 & thry $\sim 2\%$ (N$^3$LO, n.p. small) & \cite{Baikov:2008jh} & $\sim 1\%$ (N$^4$LO feasible, $\sim 10$ yrs) \\
\hline
\multirow{2}{*}{$ep$ colliders} 
 & $\sim 1$--$2\%$ (pdf fit dependent) & \cite{Gao:2013xoa,Alekhin:2012ig}, & $0.1\%$ (LHeC + HERA~\cite{AbelleiraFernandez:2012cc}) \\
 & (mostly theory, NNLO) & \cite{Ball:2011us,Martin:2009bu} & $\sim0.5$\% (at least N$^3$LO required) \\
\hline
\multirow{2}{*}{hadron colliders} 
 & $\sim 4\%$ (Tev. jets), $\sim 3 \%$ (LHC $t\bar t$) & & $<1\%$ challenging \\ 
 & (NLO jets, NNLO $t\bar t$, gluon uncert.) & \cite{Abazov:2009nc,Chatrchyan:2013haa,Czakon:2013goa}
 & (NNLO jets imminent~\cite{Ridder:2013mf}) \\
\hline
\multirow{2}{*}{lattice} 
 & $\sim 0.5\%$ (Wilson loops, correlators, ...) & & $\sim 0.3\%$ \\
 & (limited by accuracy of pert. th.) & \cite{McNeile:2010ji,Shintani:2010ph,Maltman:2008bx} & ($\sim5$ yrs~\cite{Mackenzie2013}) \\
\hline
\end{tabular}
\end{center}
  \caption{Summary of current uncertainties in extractions of
  $\amz$ and targets for future ($5-25$ years) determinations.
  For the cases where theory uncertainties are considered separately, 
  the theory uncertainties for future targets 
  reflect a reduction by a factor of about two.}
\label{tab:alphas}
\end{table}
}



\section{Quark masses and strong coupling from lattice QCD}
\label{sec:lattice}

\draftnote{Text from P. Mackenzie}

The single largest source of error in the theoretical calculation of the dominant Standard-Model Higgs decay mode $H\rightarrow b\overline{b}$ is  the parametric uncertainty in the $b$-quark mass.  Because this mode dominates the total Higgs width, this uncertainty is also significant for most of the other Higgs branching fractions.  Parametric uncertainties in \as and $m_c$ also play important roles in many of the Higgs decay channels, and are the largest sources of uncertainty in the partial widths $H\rightarrow gg$ and $H\rightarrow c\overline{c}$ , respectively.
In this section, the determination of these quantities with lattice-QCD calculations is discussed.
 
The most precise way of obtaining $m_c$ and $m_b$ from experiment using continuum perturbation theory
employs correlation functions of the quark's electromagnetic current~\cite{Chetyrkin:2009fv}.
Moments of these correlation functions have been calculated to third order in \as.
They can be determined experimental data in $e^+e^-$ annihilation. 
 
Such moments are also the most precise way known of determining the heavy-quark masses using lattice gauge theory,
as well as a good way of obtaining \as~\cite{Allison:2008xk,McNeile:2010ji}.
They can easily be calculated nonperturbatively in lattice simulations and then compared to the perturbative expressions to ${\mathcal O}(\as^3)$.
Lattice calculations of correlators of  quark bilinears offer several advantages over determination of these correlators from
experiment.  For example, the numerical lattice data for correlators are much cleaner than the experimental data.
Further, the lattice offers several choices of current operators and the most well-behaved one can be chosen
for the determinations; in practice, this turns out to be the pseudoscalar current.
The lattice calculations still need an input from experiment to set the overall energy scale, but this 
can be chosen in a way that also reduces final uncertainties.  For example, if $m_c$ is obtained 
from the pseudoscalar correlator, choosing $m_{\eta_c}$ to set the energy scale reduces sensitivity 
to the tuning of the bare charm-quark mass.
Using these methods, the HPQCD Collaboration obtains
$m_c(m_c,n_f=4)= 1.273(6)$~GeV in the $\overline{\mathrm{MS}}$ scheme \cite{McNeile:2010ji}.
By contrast, the Karlsruhe group obtains  $m_c(m_c,n_f=4)= 1.279(13)$~GeV
from $e^+e^-$ experimental data~\cite{Chetyrkin:2009fv}.
The most important reason for the greater precision of the lattice determination is that the
data for the lattice correlation functions is much cleaner than the $e^+e^-$ annihilation data.
The uncertainty is dominated by continuum perturbation theory,
and therefore may improve only modestly unless another order of perturbation theory is
calculated.
However,
these charm correlation functions are very easy to calculate with lattice QCD.  The 
lattice part of this determination will be checked by many lattice groups and should be very robust.

The $b$ quark mass can also be obtained in this way, with the result
$m_b(m_b, n_f=5) = 4.164(23)$~GeV~\cite{McNeile:2010ji}.
The sources of systematic uncertainty  are completely different than for $m_c$.
Perturbative uncertainties are tiny because $\as(m_b)^4  \ll \as(m_c)^4$.
However, the method requires treating the $b$ quark as a light quark, which is just barely working
at  lattice spacings used so far.  
Discretization errors dominate  the current uncertainty, followed by statistical errors.  The lattice result for $m_b$ is not currently as precise as the result from
 $e^+e^-$ experimental data, $m_b(m_b, n_f=5) = 4.163(16)$~GeV~\cite{Chetyrkin:2009fv}.
 Discretization and statistical errors should be straightforward to reduce by brute force computing
 power, and so are likely to come down by a factor of two in the next few years, 
 perhaps to 0.011~GeV or better.  Precisions of that order for $m_b$ have already been claimed from
 $e^+e^-$ data from
 reanalyses of the data and perturbation theory~\cite{Chetyrkin:2009fv}, and   coming lattice
 calculations will be able to check these using the computing power expected in the next few years.
 
 The strong coupling constant, $\as$, is also an output of these lattice calculations.  A very
precise value of $\as(M_Z, n_f=5) = 0.1183(7)$ has been obtained \cite{McNeile:2010ji}, 
with an uncertainty dominated by continuum perturbation theory.
Unlike the heavy-quark masses, for which the correlation function methods give the most precise
results at present, there are numerous good ways of obtaining $\as$ with both
continuum and lattice methods.
HPQCD has also obtained $\as$ from Wilson loops, obtaining
$\amz = 0.1184(6)$, comparable to their correlation function determination, but with
completely independent methods and uncertainties.  The Wilson loop determination makes heavy
use of lattice perturbation theory, while the correlation function determination makes none.
The precisions of both determinations are dominated by perturbation theory in one way
or another.
Several other quantities have been used to make good determinations $\as$ with lattice QCD, including
the Adler function~\cite{Shintani:2010ph},
the Schr{\"o}dinger functional~\cite{Aoki:2009tf},
and the ghost-gluon vertex~\cite{Blossier:2012ef}.
All of the lattice determinations are consistent, and each is individually more precise than the most
precise determination appearing in the 2012 world average~\cite{Beringer:1900zz} that does not use lattice QCD.
The  most precise current determination of $\as$ may improve only modestly over the next
few years, since the error is dominated by perturbation theory.
However, increasingly precise corroboration via an increasing number of quantities
should continue.

\begin{table}[t]\centering
\begin{tabular}{|c|c|c|c|c|c|c|}
\hline
	&  Higgs X-section 	& PDG\cite{Beringer:1900zz}	& Non-lattice 	& Lattice  & Lattice  & Targets of \\
	&   Working Group \cite{Denner:2011mq}	& 		&  			&  (2013) &  (2018)  & ILC/TLEP/LHeC \\
\hline
$\delta \as$ &0.002	&0.0007	&0.0012 \cite{Beringer:1900zz}		&0.0006 \cite{McNeile:2010ji}	& 0.0004	& 0.0001--0.0006 \cite{Gomez-Ceballos:2013zzn,Flacher:2008zq,AbelleiraFernandez:2012cc} \\
$\delta m_c$ (GeV) &0.03	&	0.025&	0.013 \cite{Chetyrkin:2009fv}	& 0.006 \cite{McNeile:2010ji}	& 0.004	& - \\
$\delta m_b$ (GeV) &0.06	&	0.03	&	0.016 \cite{Chetyrkin:2009fv}	& 0.023 \cite{McNeile:2010ji}	& 0.011	& -\\
\hline
\end{tabular}
\caption{Projected future uncertainties in $\as$, $m_c$, and $m_b$, compared with 
current uncertainties estimated from various sources.}
\draftnote{KH: added the lower one to illustrate the improvements from non-lattice sources. Need to check with Paul to see if it's ok with him.}
\label{tab:lattice}
\end{table}

Table~\ref{tab:lattice} gives the current uncertainties in $\as$, $m_c$, and $m_b$ from both lattice and non-lattice methods, along with projections for the lattice errors in the next five years.  For comparison, the uncertainties in these quantities estimated by the PDG and used by the LHC Higgs cross-section working group are also shown.
The current uncertainties in $\as$, $m_c$, and $m_b$ from lattice QCD are all around a half a per cent.
The lattice determinations of $m_c$ and $\as$ are currently the most precise in the world.
 The charm correlation functions used to determine $m_c$ are easy to calculate for many groups, and the uncertainty in $m_c$ will be dominated by estimates of the uncalculated fourth and higher orders of perturbation theory, which may improve modestly.
  The same is true of $\as$ determined using charm correlation functions.
For $\as$, there should be more corroboration from an increasing number of physical
quantities.
The most precise lattice determination of $m_b$ has uncertainties dominated by discretization and statistical errors, which should be reducible by perhaps a factor of two with the increasing amounts of computer power expected in the next few years.


\section{Parton distribution functions}
\label{sec:qcd-pdf}

\draftnote{Text below written by J. Rojo}

Parton distributions are an essential ingredient of present
and future phenomenology at hadron colliders~\cite{Forte:2013wc,DeRoeck:2011na,Perez:2012um}. They are one of the dominant theoretical uncertainties for
the characterization of the newly discovered Higgs-like boson at the
LHC, they substantially affect the reach of searches for new physics at
high final state masses and they limit the accuracy to which precision
electroweak observables, like the $W$ boson mass or the effective
lepton mixing angle, can be extracted from LHC data~\cite{Bozzi:2011ww}.


\subsection{Current knowledge and uncertainties}
\label{sec:qcd-pdf-current}

\draftnote{Text below written by J. Rojo;  J. Huston to add/edit}

The determination of the parton distribution functions of 
the proton from a wide variety of
experimental data has been the subject of intense activity in the last
years.
Various collaborations provide regular updates of their
PDF sets. The latest releases from each group are ABM11~\cite{Alekhin:2012ig}, 
CT10~\cite{Gao:2013xoa}, HERAPDF1.5~\cite{Radescu:2010zz,CooperSarkar:2011aa}, 
MSTW08~\cite{Martin:2009iq} and NNPDF2.3~\cite{Ball:2012cx}. 
A recent benchmark comparison of the most updated NNLO PDF sets
was performed in Ref.~\cite{Ball:2012wy}, where similarities and
differences between these five PDF sets above were discussed, and
where $W,Z$ and jet production data
 was used to quantify the level of agreement of the various PDF sets
with the Tevatron and LHC measurements.

\begin{figure}[h]
\begin{center}
 \includegraphics[width=0.48\hsize]{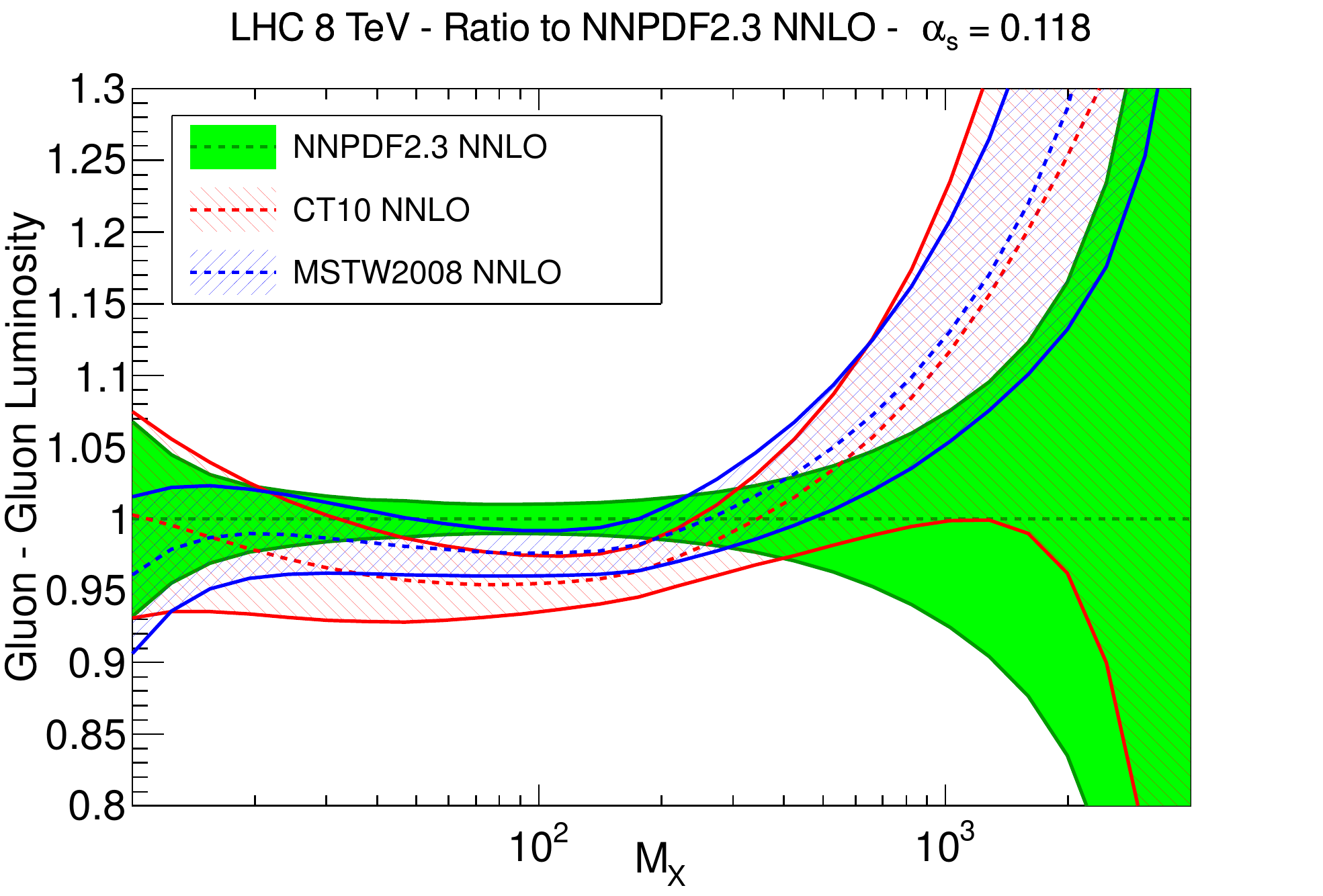}
 \includegraphics[width=0.48\hsize]{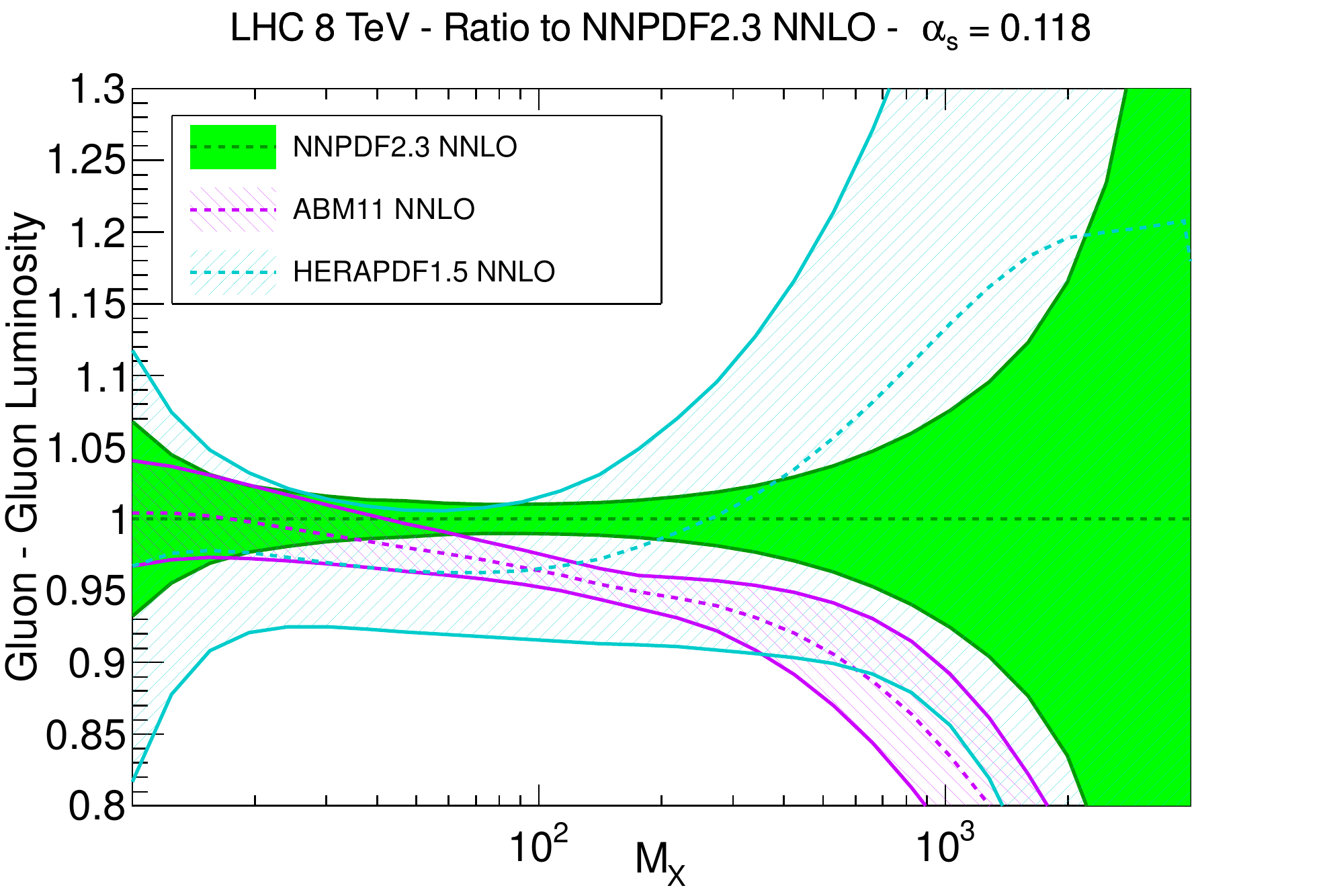}
 \includegraphics[width=0.48\hsize]{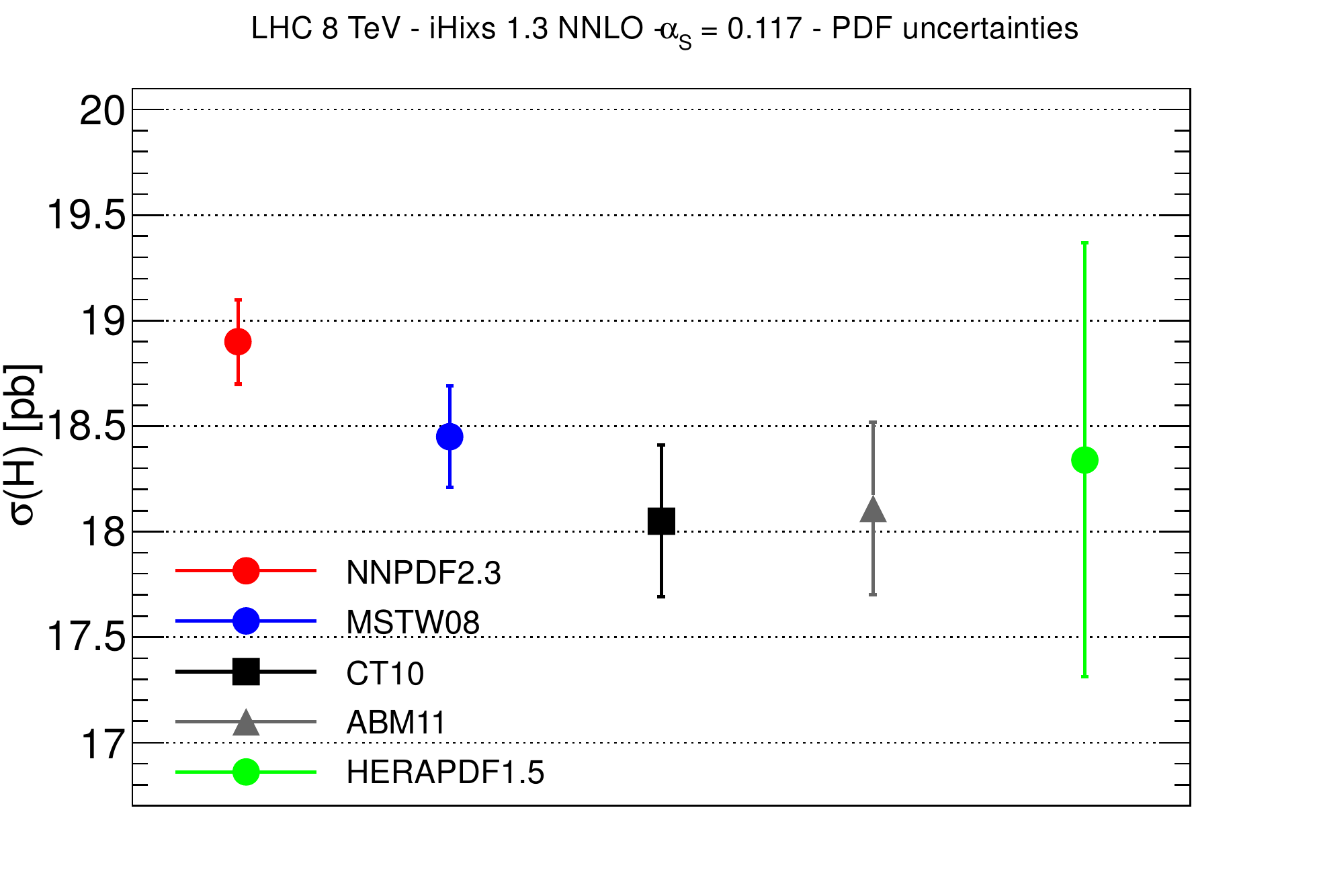}
 \includegraphics[width=0.48\hsize]{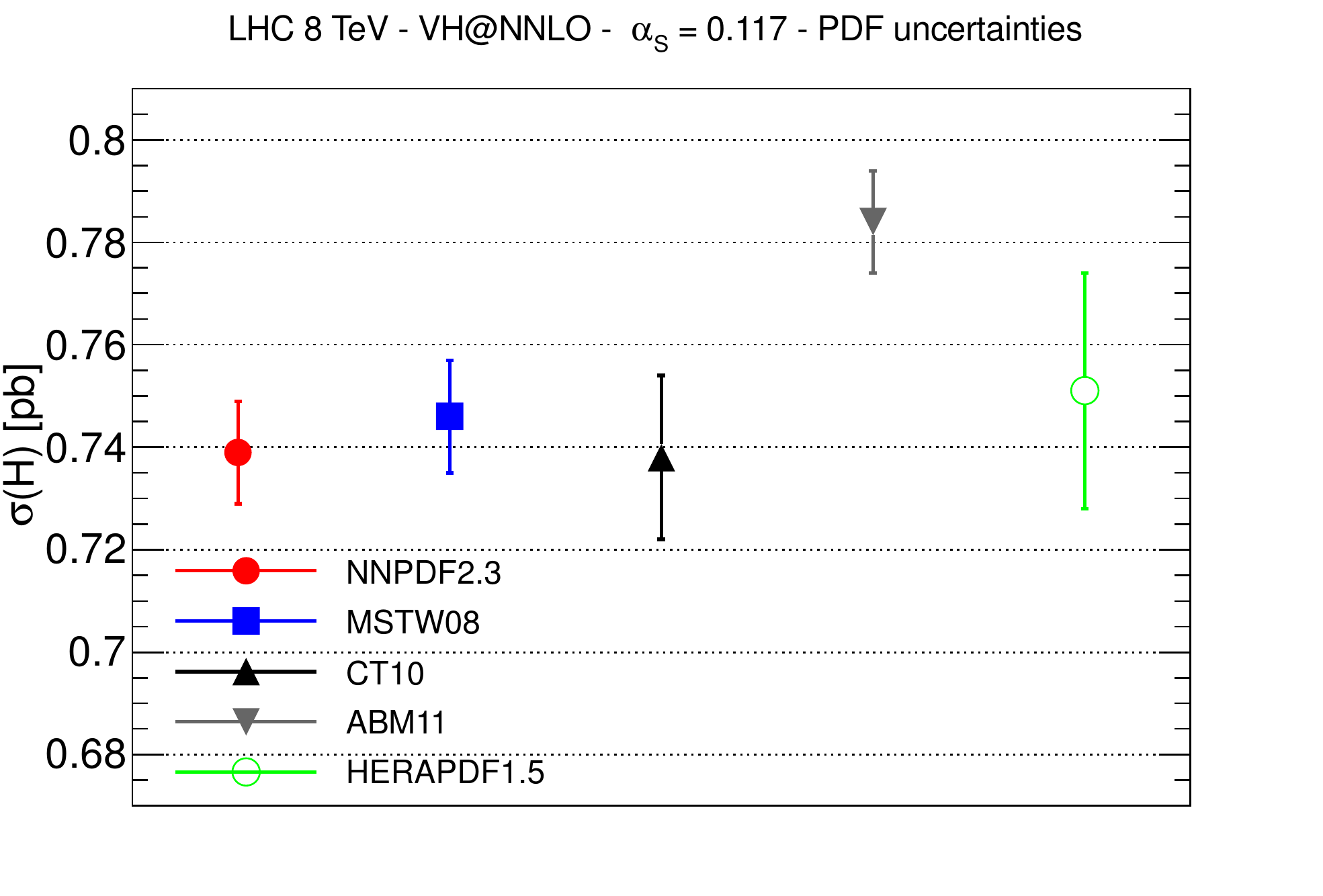}
\end{center}
\caption{Upper plots: Comparison of the gluon-gluon luminosity
at the LHC 8 TeV as a function of the final state
mass $M_X$ between the ABM11, CT10, HERAPDF1.5, 
MSTW and NNPDF2.3 NNLO PDF sets.
Lower plots: predictions for the Higgs production cross
sections at LHC 8 TeV for the same PDF sets, in the
gluon-fusion channel (left plot) and in the $WH$
channel (right plot).
\label{fig:rojo-bench}}
\end{figure}

A snapshot of the comparisons between recent NNLO PDFs at the level
of parton luminosities and cross section can be seen in 
Fig.~\ref{fig:rojo-bench}, where
we compare the gluon-gluon PDF luminosities between the five
sets.
We also show the predictions for the Higgs production cross section in the
gluon-fusion channel and in WH associated production.\footnote{
An extensive set of comparison plots for PDFs, parton luminosities,
LHC total cross sections and differential distributions at NLO and NNLO and
for different values of $\alpha_s(M_Z)$ can be found in~\url{https://nnpdf.hepforge.org/html/pdfbench/catalog/}} 
Results have been computed using the settings discussed in 
Ref.~\cite{Ball:2012wy}. 
%

In relation to previous comparisons,
one of the main conclusions of the benchmark study~\cite{Ball:2012wy}
was that the agreement between
the three global PDF sets, CT, MSTW and NNPDF has improved for
most PDFs and ranges of Bjorken-$x$. 
On the other hand, there are still important differences that need to
be understood, and that have substantial phenomenological impact.
To begin with, the gluon luminosity for the three PDF sets differs
maximally for $m_X \sim 125$ GeV, as can be seen from
Fig.~\ref{fig:rojo-bench}.  It would be important to
understand the source of these differences to improve the agreement
of the three sets for the gluon-induced Higgs production
cross sections. 
In addition, very large PDF uncertainties are present in the TeV range, affecting both the central values 
and the uncertainties for the production cross sections of massive particles.
These large uncertainties at large masses degrade the prospects
for eventual characterization of new BSM heavy particles.
On top of this, there are theoretical uncertainties due to
the choice of the heavy quark 
general-mass variable-flavor-number (GM-VFN) scheme, specific choices
in the fitted dataset and methodological differences that
still require further understanding to improve the agreement
between the various PDF sets.

In the next subsection we discuss the prospects to
obtain further constraints in PDFs from LHC data.


\subsection{Parton distributions with QED corrections}
\label{sec:qcd-pdf-qed}

\draftnote{Text below written by J. Rojo; note, we did not originally
have this as a separate section}

Precision predictions for electroweak processes at hadron colliders
require not only (N)NLO QCD corrections, but also the consistent inclusion
of QED corrections to parton distributions and photon-initiated
contributions.
QED and electroweak corrections for various relevant collider
processes have been computed in the last years in processes
like inclusive $W,Z$ production, vector boson pair production,
$t\bar{t}$ and dijet production among many others.
On the other hand, it is also known that a fully consistent treatment
of electroweak corrections requires the use of parton distributions
that also incorporate QED effects as well.
QED effects on parton distributions have two main implications: first
of all, the standard QCD DGLAP evolution equations are affected by
$\mathcal{O}(\alpha)$ corrections and the associated breaking of
isospin invariance, and, phenomenologically more important, the
photon PDF needs to be determined from experimental data
in parallel with the quark and gluon PDFs.

Until recently, only one PDF set with QED corrections was
available, the MRST2004QED set~\cite{Martin:2004dh}, where the
photon PDF was determined based on a model assumption.
However, now the NNPDF framework has also been extended to
provide PDF sets with QED corrections~\cite{Carrazza:2013bra,Carrazza:2013wua,Ball:2013hta}, and NNPDF2.3 QED
is available in the NNPDF {\tt HepForge} website.\footnote{ \url{https://nnpdf.hepforge.org/html/nnpdf23qed/nnpdf23qed.html}}
NNPDF2.3 QED avoids any model assumption on the photon
PDF and derives $\gamma(x,Q^2)$ and its associated uncertainties
from a global fit to DIS and LHC data,  where in the latter case
neutral current and charged current vector boson production
data provide stringent constraints on the shape and normalization
of $\gamma(x,Q^2)$.


Electroweak corrections to parton distribution functions
have important phenomenological implications, in particular
for the electroweak production of high invariant mass final states.
These include the measurement of the $W$ mass, searches for
$W'$ and $Z'$ resonances in the tails of the $W$ and $Z$
distributions and vector boson pair production among many
others.
The main effect is that the substantial uncertainties on
the large-$x$ photon PDF (that stem from the lack of
available experimental constraints) translates into very
large uncertainties from photon-initiated contributions
that can be as large as a factor 100\%.

\draftnote{Joey to add information on CT QED}

\begin{figure}
\begin{center}
 \includegraphics[width=0.48\hsize]{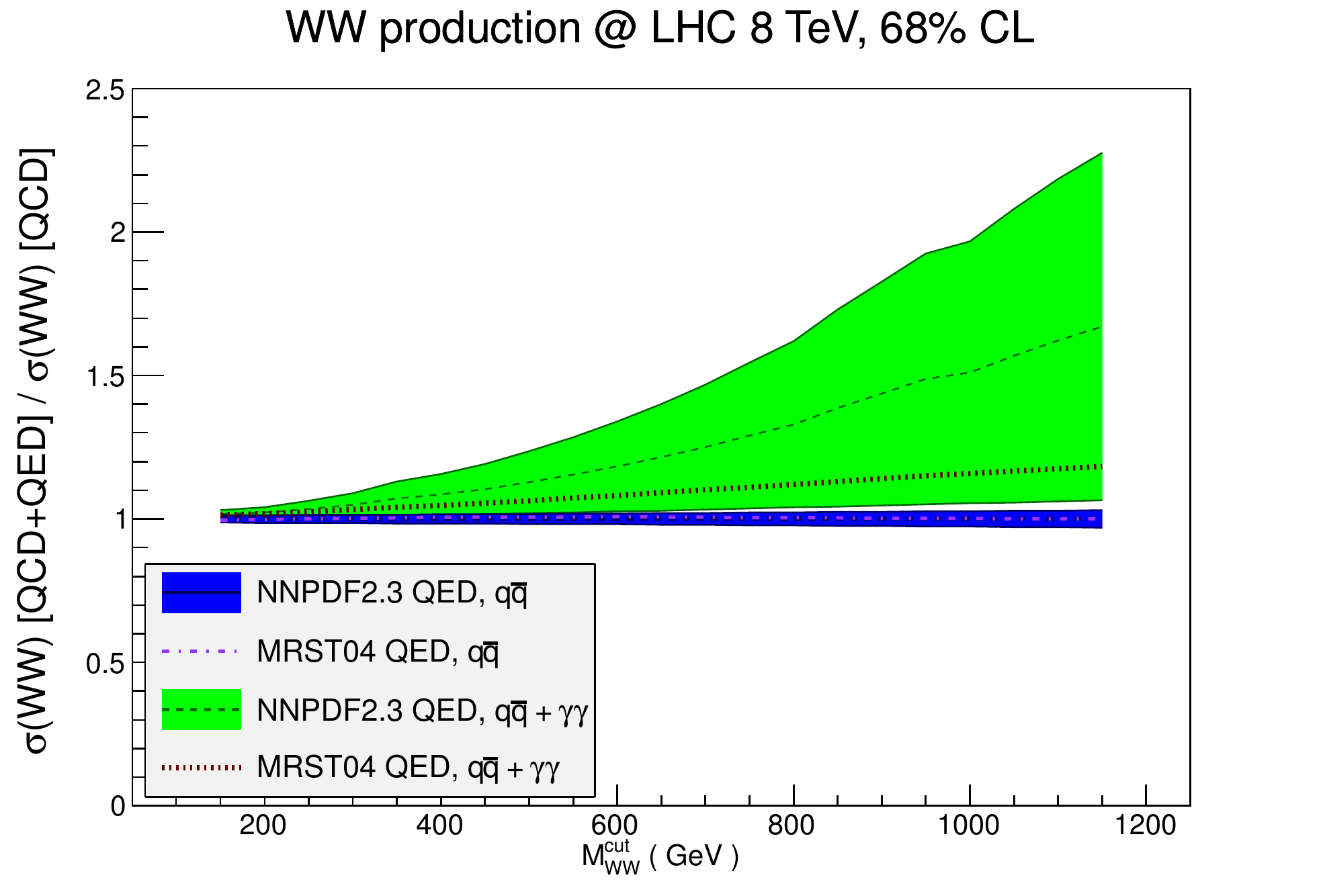}
 \includegraphics[width=0.48\hsize]{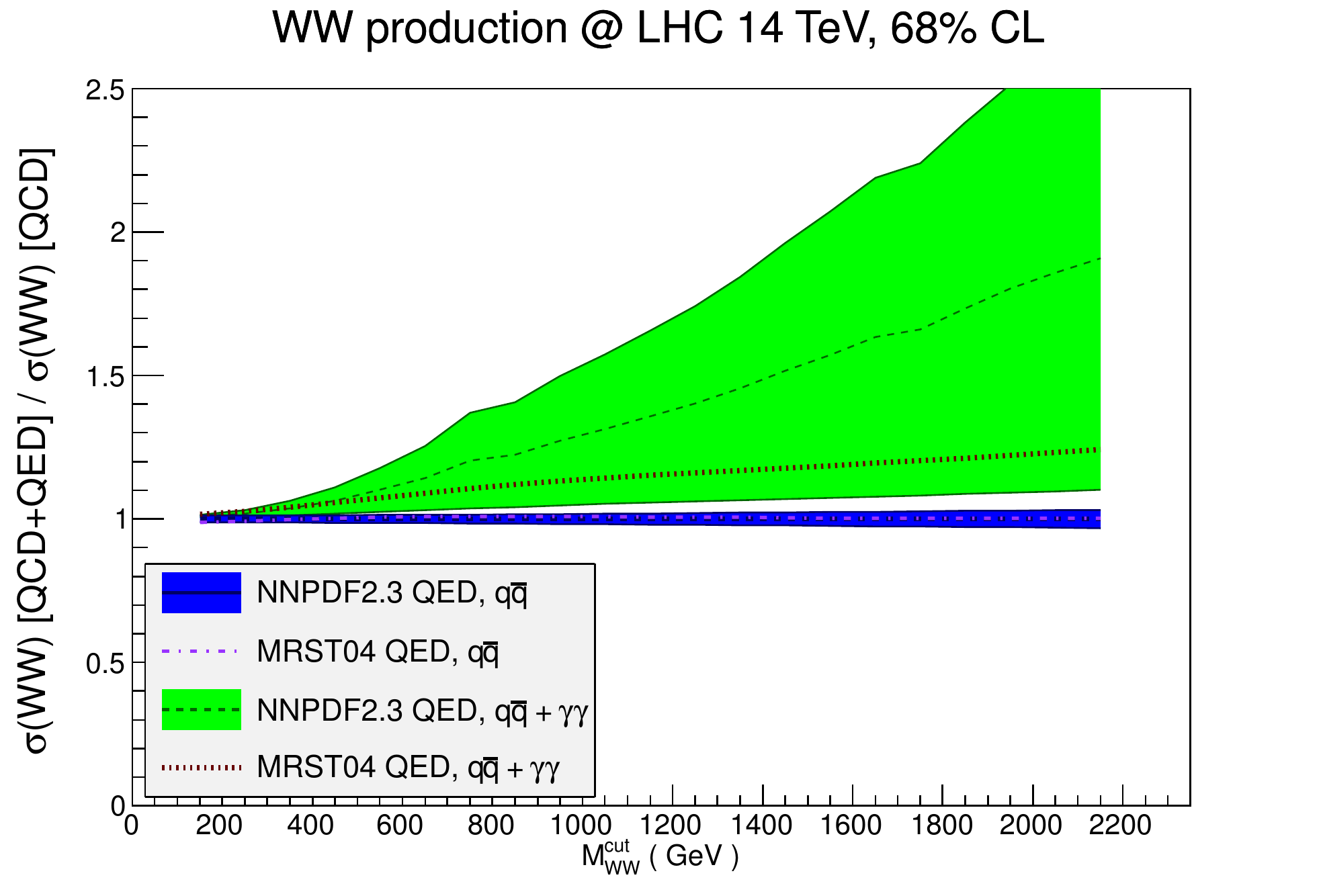}
 \includegraphics[width=0.48\hsize]{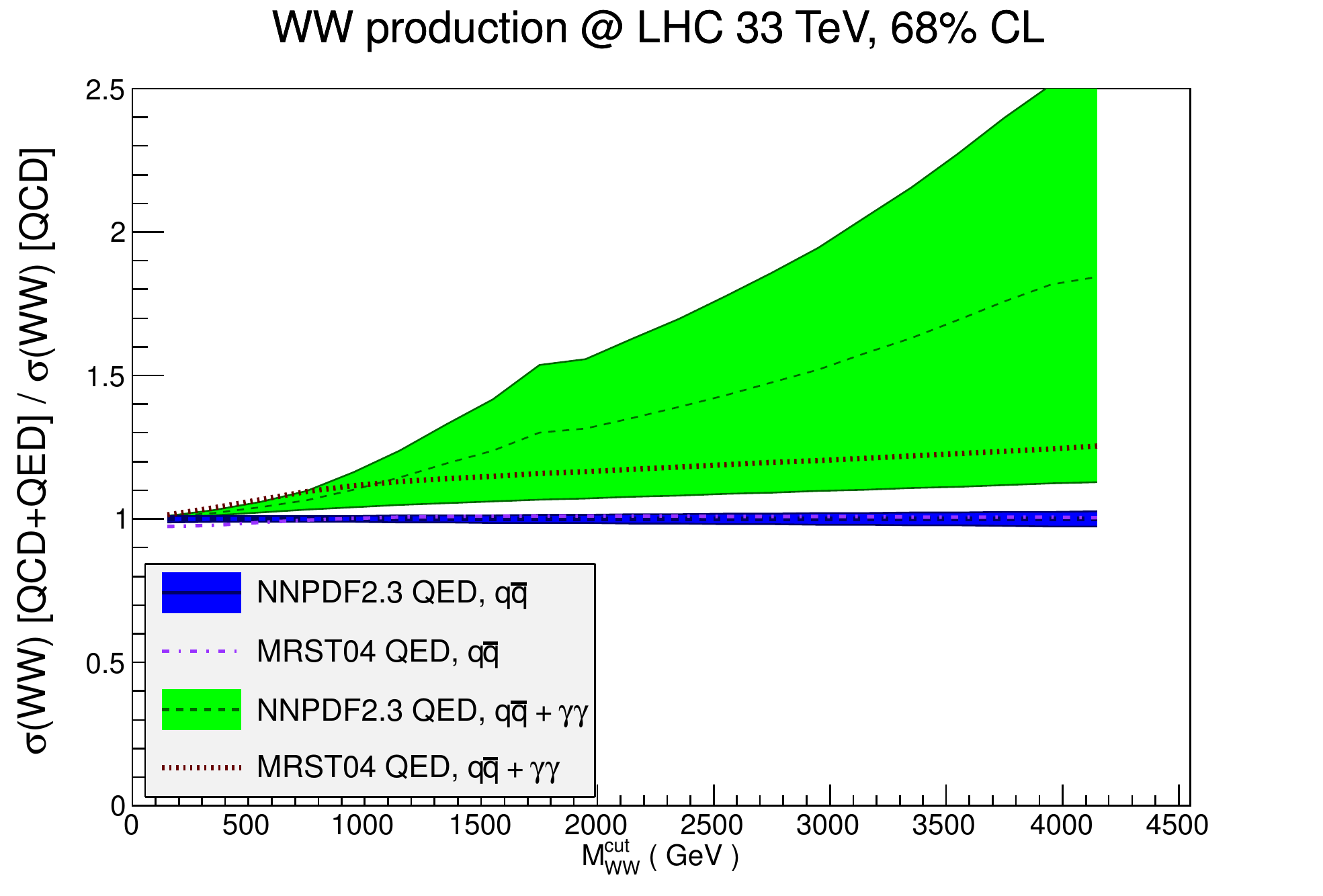}
 \includegraphics[width=0.48\hsize]{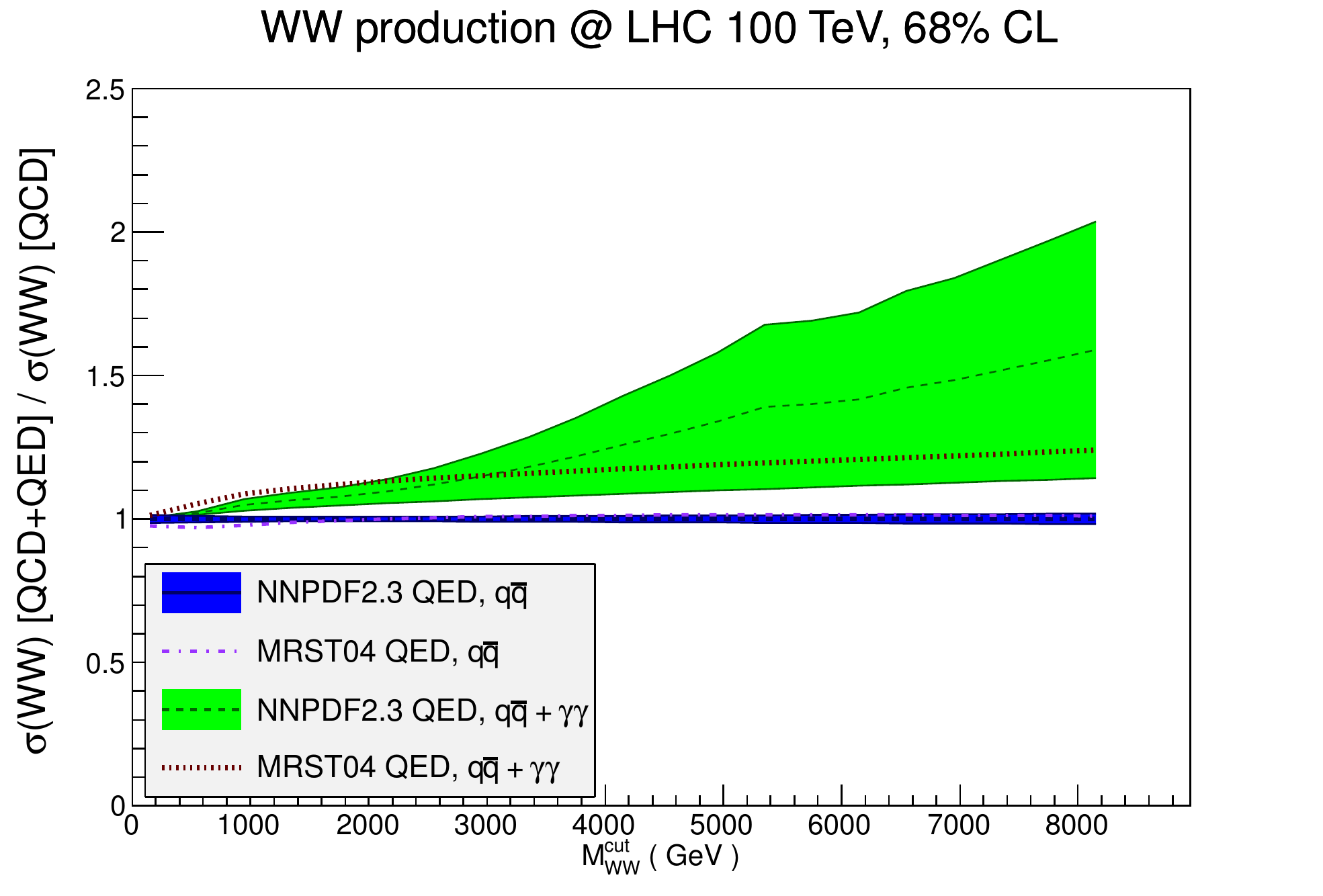}
\end{center}
\caption{The predictions of the NNPDF2.3 QED set for $WW$ production
at the LHC for various center of mass energies, compared with
the results of the reference NNPDF2.3 set, as well as with the predictions
from MRST2004QED. 
We show the total cross section as a function of the cut in the
$M_{\rm WW}$ invariant mass, for 8, 14, 33 and 100 TeV energies.
See text for more details.
\label{fig:rojo-ww}
}
\end{figure}

As an illustration of these phenomenological consequences, 
in Fig.~\ref{fig:rojo-ww} we have
computed the predictions of the NNPDF2.3 QED set for $WW$ production
at the LHC for various center of mass energies, compared with
the results of the reference NNPDF2.3 set, as well as with the predictions
from MRST2004QED. 
The computation has been done at leading order in the electroweak
coupling but including photon-initiated diagrams, using the
same settings as in Ref.~\cite{Bierweiler:2012kw}.
We show the total cross section as a function of the cut in the
$M_{\rm WW}$ invariant mass, for 8, 14, 33 and 100 TeV energies.\footnote{
We thank T. Kasprzick for providing us with these results.}
It is clear that the QED-induced theoretical uncertainties are
substantial and do degrade the constraining power
of BSM searches in this channel, for example, searches for heavy resonances
that decay into $WW$ pairs.  These effects are more severe
the higher the cut in the invariant mass of the $WW$ pair.
As the energy is increased, for a given value
of $M_{WW}$ the PDF uncertainties decrease, but they
are still very substantial at the highest available masses
in each case, a factor 2 at least.

In summary, the development of parton distributions with QED
corrections is an important ingredient of  fully consistent
theoretical predictions of hadron collider processes that include
both higher order corrections in the QCD and EW couplings.
On the other hand, the same analysis reveal the urgent need for
more experimental data to constrain the photon PDF $\gamma(x,Q^2)$,
and thus to reduce the currently large QED-induced uncertainties
that affect high mass electroweak production at the LHC.


\subsection{PDF constraints from future LHC data}
\label{sec:qcd-pdf-improvements}

\draftnote{Text below written by J. Rojo; also expect contributions
from U. Klein, M. Cooper-Sarkar}

The excellent performance of the Large Hadron Collider is substantially
increasing the range of processes that can be used to constrain PDFs
in a global analysis.
The traditional processes at hadron colliders 
that have been used for
PDF constraints are inclusive jet production and $W,Z$ production.
Inclusive jet and dijet data are now available up to the TeV region
from ATLAS and CMS~\cite{Aad:2011fc,Chatrchyan:2012bja}, 
and provide important constrains on the poorly known large-$x$ quarks
and gluons.
In addition, ATLAS has presented 
the measurement the ratio of inclusive jet cross sections
between 2.76 TeV and 7 TeV~\cite{Aad:2013lpa}.
Such ratios between different center of mass
energies~\cite{Mangano:2012mh} increase the PDF sensitivity of
data taken at a single energy, since on the one hand many experimental
uncertainties cancel in a dedicated measurement of a cross section
ratio, and on the other hand several theory systematics, like scale
variations, cancel as well. 
In the $x$-range currently dominated by HERA data ($x<0.1$ for quarks and $x<0.05$ for gluons), it will be difficult for hadron collider measurements to match the theoretical and experimental precision achieved at HERA. However, the HERA data provides little constraint
in the high mass regions relevant for BSM searches.  It is here where hadron collider measurements at high luminosity, at
14 TeV and at higher energies, can greatly contribute to improvements in the PDF uncertainties for both
quarks and gluons.
In this respect, recent progress towards the full NNLO QCD corrections
for inclusive and dijet production~\cite{Ridder:2013mf} 
should make it feasible to achieve
a few per-cent theoretical accuracy from scale dependence on these observables by the time of the 13 TeV
data taking. 
Although improving, it may be difficult for the expected experimental precision to match the expected
theoretical precision.

The precision measurements of $W$ and $Z$ boson production at hadron colliders
provide important information on quark flavor separation, 
and in addition reduce systematic uncertainties of important
observables like the $W$ boson mass~\cite{Bozzi:2011ww}.
While the $Z$ rapidity distribution and the $W$ lepton asymmetry from the
Tevatron and the LHC have been
by now available for some time, recently the range of available
processes has been extended by the measurement of the 
off-peak neutral current Drell-Yan process by ATLAS, CMS and LHCb.
High-mass measurements of the Drell-Yan process provide useful information
on the large-$x$ quarks and antiquarks.
Low-mass measurements provide information on the small-$x$ gluons and possible departures from linear DGLAP evolution.
Particularly striking signatures have been predicted for the LHCb kinematics.
Future measurements at higher energies will benefit from an increased coverage in the dilepton invariant mass, 
allowing  probes of very large-$x$ antiquarks, which are effected
by very large PDF uncertainties.  Measurements in the peak region will benefit from reduced systematics
and by negligible statistical uncertainties. To date, DGLAP evolution appears to
be sufficient to describe the LHC data, but runs at higher energies may demonstrate the 
presence of BFKL-like effects~\cite{Altarelli:2008aj}.  On the theoretical side, the NNLO QCD corrections to fully-differential $W$ and $Z$ production have been known for several years~\cite{Melnikov:2006kv,Gavin:2010az,Catani:2009sm}.  Recently, the NNLO QCD and NLO electroweak corrections to neutral-current dilepton production have been consistently combined~\cite{Li:2012wna}, and several combinations of NLO QCD, NLO electroweak and parton-shower effects for both $W$ and $Z$ production have appeared~\cite{Bernaciak:2012hj,Barze':2013yca}.

In addition to the traditional processes discussed above, many new collider observables
 have recently become available for the first time for use in PDF fits. 
The recent calculation of the full NNLO top quark production
cross section~\cite{Czakon:2013goa} makes possible for the first time the inclusion of top quark
data into a NNLO analysis to constrain the large-$x$
gluon PDF~\cite{Czakon:2013tha}.
This is an important result since top production is currently the only hadronic observable which is both directly
sensitive to the gluon and can be included in a  NNLO global fit
without any approximation.
In turn, the more accurate gluon PDF will translate into
an improvement of the theory predictions for various high-mass
BSM processes driven by the gluon luminosity.

%

Future precision measurements of differential distributions in top quark
pair production will allow  more precision constraints on the gluon distribution and an ability to enlarge the range of Bjorken-$x$ where the gluon
PDF is being probed, especially once the NNLO calculation of~\cite{Czakon:2013goa}
is extended to the fully differential case.
In addition to top quark data, the use of LHC isolated photon data and photon+jet
data has also been advocated in order
to pin down the gluon PDF~\cite{d'Enterria:2012yj,Carminati:2012mm}, though 
this process is affected by missing higher order 
and non-perturbative uncertainties.

Turning to the constraints on the quark distributions,
the production of $W$ and $Z$ bosons
in association with jets, for high $p_T$ values
of the electroweak boson, is a clean probe at
the LHC of both quark flavor 
separation and of the gluon PDF~\cite{Malik:2013kba}.
In particular, ratios of $W$ and $Z$ distributions at large $p_T$
provide constraints on quarks and antiquarks while benefiting from
substantial cancellations of experimental and theory uncertainties.
This is a good example of a process currently limited by statistics, and 
future LHC data will offer a much increased constraining power.
Another important source of information on the quark PDFs is
$W$ production in association with charm, which is directly sensitive to the strange
PDF~\cite{Stirling:2012vh}, the worst known of all the light
quark flavors. Measurements for this
important process have been recently reported by both ATLAS~\cite{ATLAS-CONF-2013-045} and CMS~\cite{Chatrchyan:2013uja}. 
Interestingly, the two measurements seem to pull the strange PDF in different
directions, with
 CMS showing
good agreement with the strangeness suppression of global
PDF fits derived from the neutrino DIS charm production data,
and ATLAS preferring a symmetric strange sea, as 
previously derived from their inclusive $W$ and $Z$ data.
Including all of these datasets into the 
global PDF fits is necessary to determine the optimal
strange PDF which accounts for all experimental constraints.

Putting everything together, it is clear that the LHC will provide in the next
years a plethora of new measurements that will be used to improve
our knowledge of parton distributions.
Quantitative projections in this respect are difficult since the precision
measurements used in PDF analyses are dominated by systematic errors,
which are notoriously difficult to predict.
In addition, correcting for pile-up in the high-luminosity phase of the LHC might
render such analyses even more complicated.
However, there are good prospects that in the next years PDFs
will be determined with increasingly better accuracy, in turn improving
the theory predictions needed for Higgs boson characterization and for new physics searches.


\subsection{Luminosities and uncertainties for $14$, $33$ and $100$ TeV}
\label{sec:qcd-pdf-higherenergy}

\draftnote{Text below written by J. Rojo; J. Huston to add/edit}

\begin{figure}[htb]
\begin{center}
 \includegraphics[width=0.32\hsize]{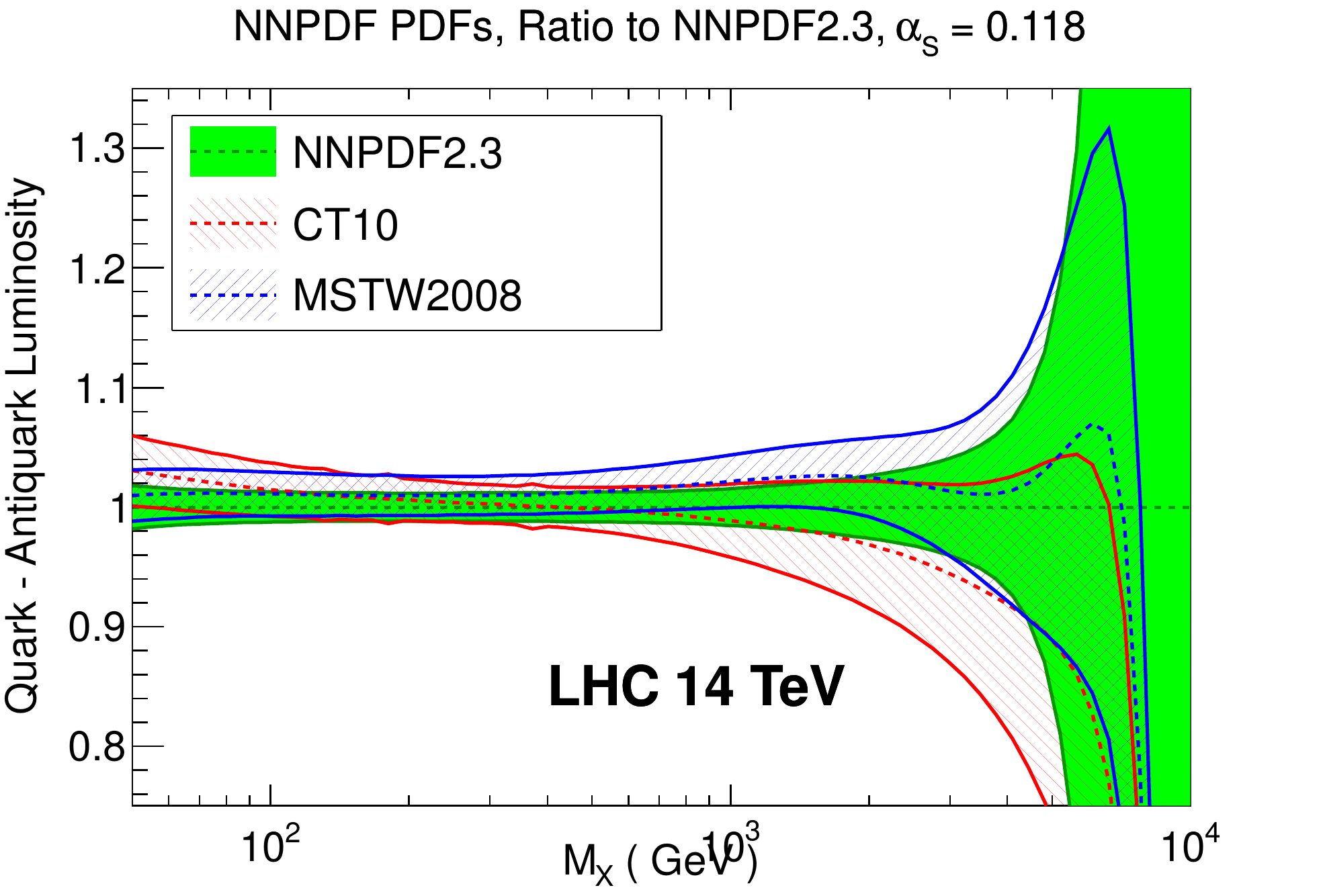}
 \includegraphics[width=0.32\hsize]{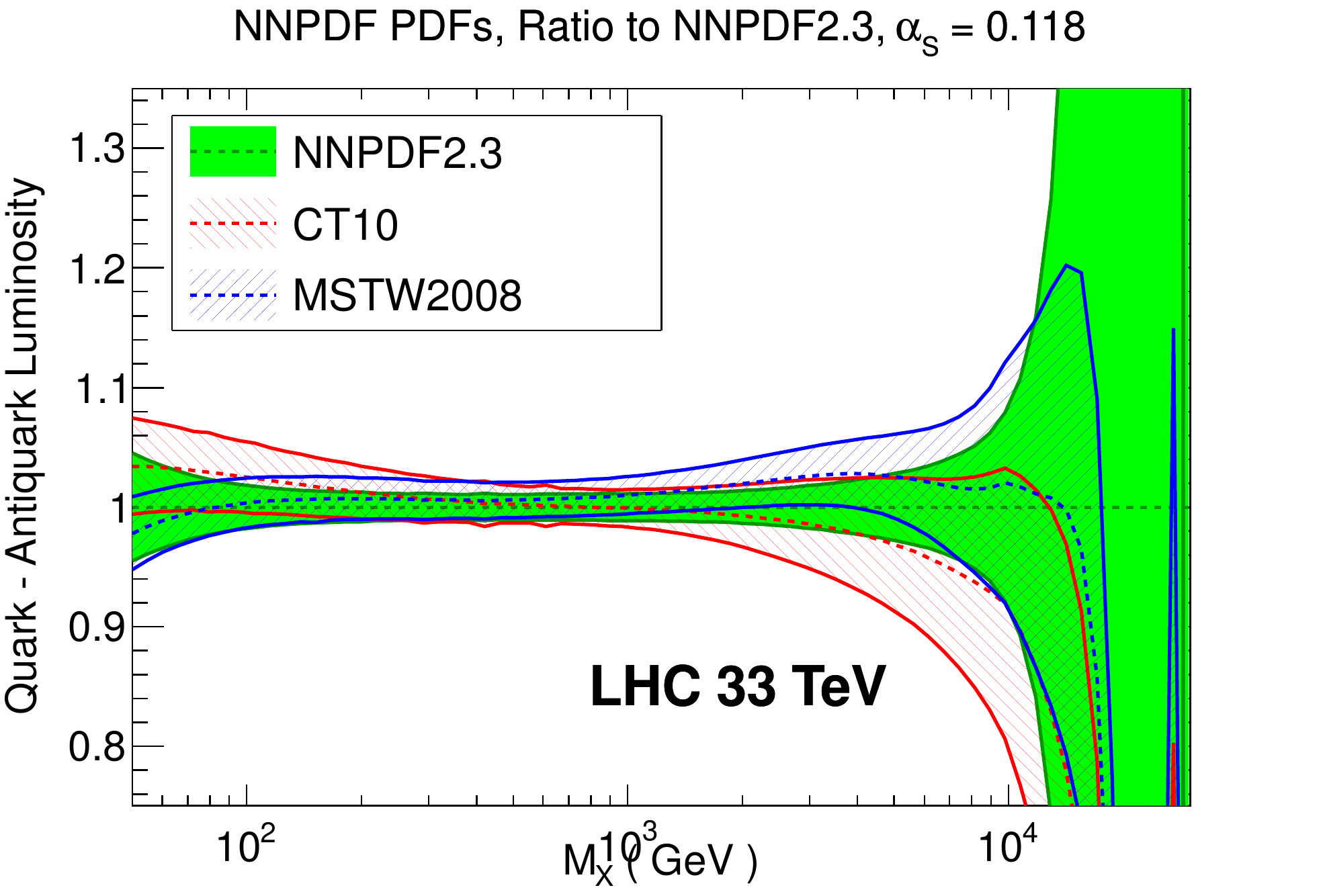}
 \includegraphics[width=0.32\hsize]{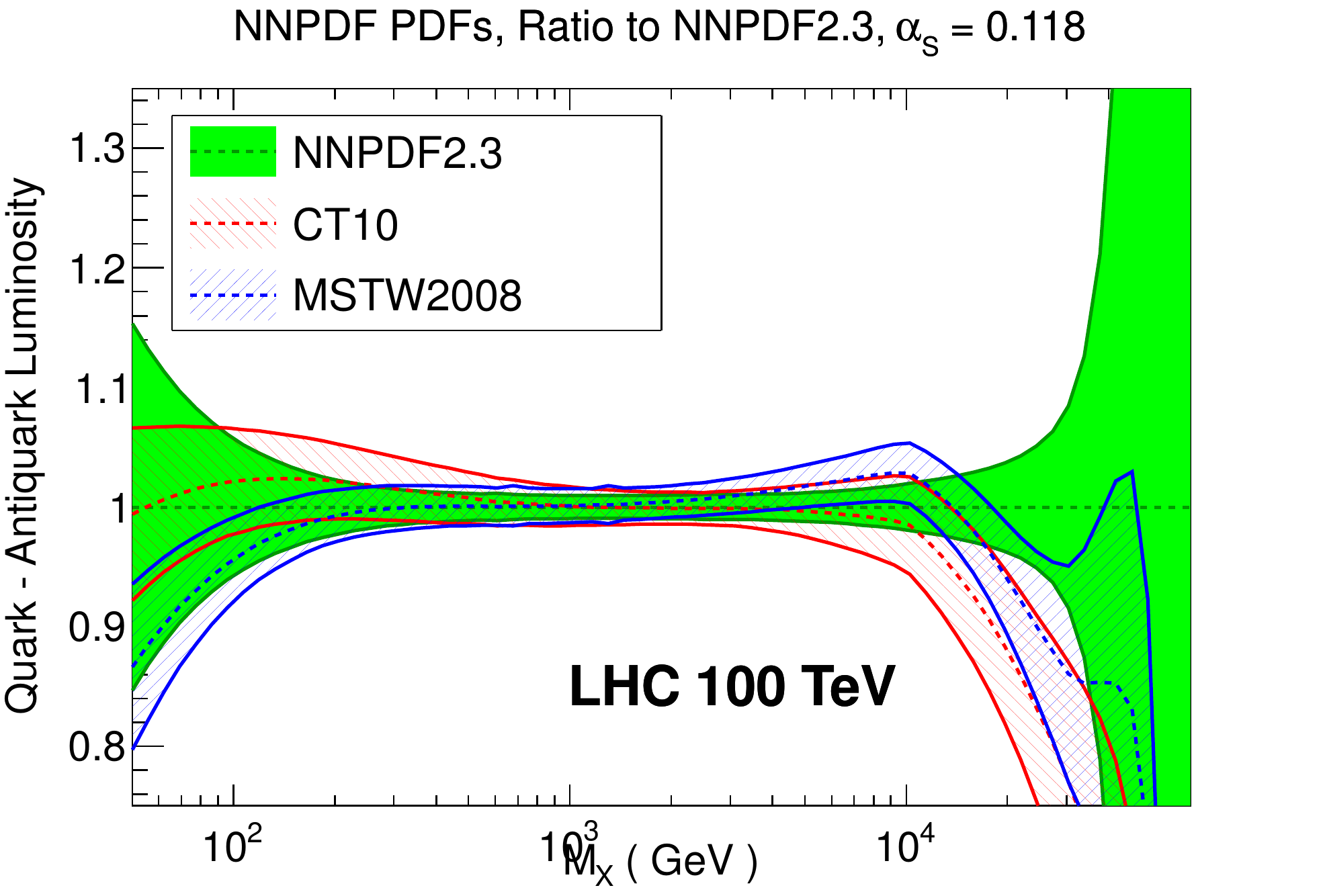}
 \includegraphics[width=0.32\hsize]{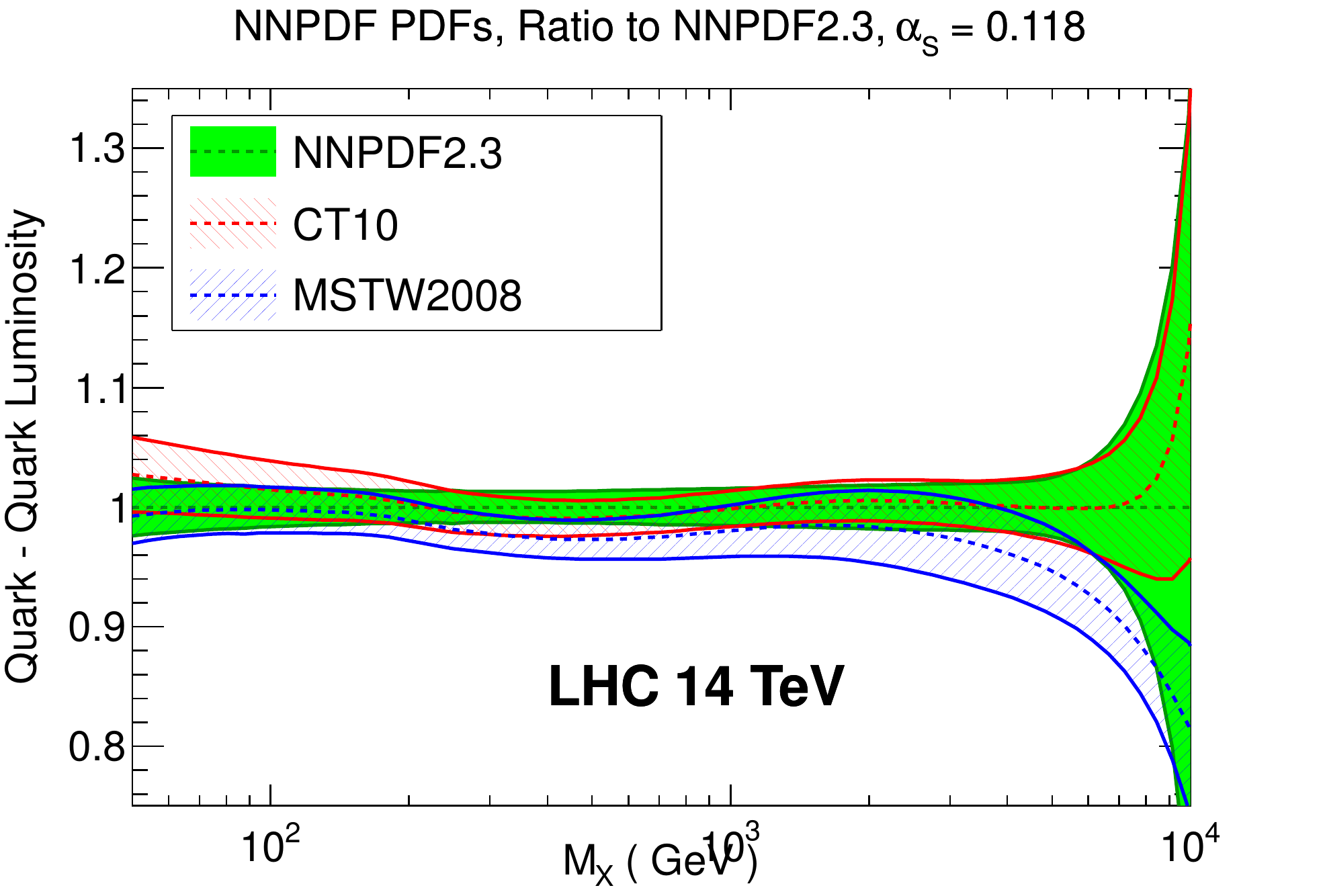}
 \includegraphics[width=0.32\hsize]{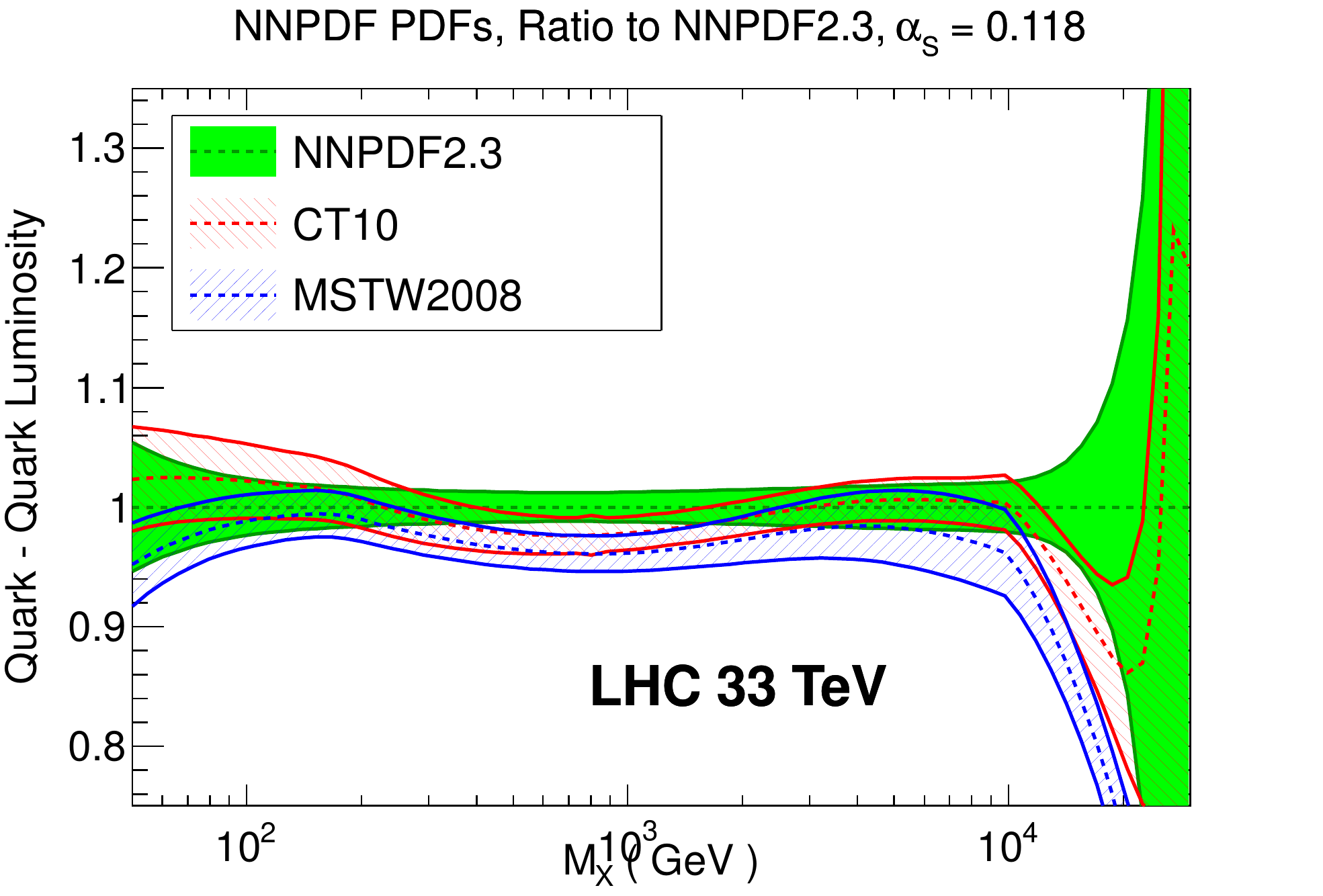}
 \includegraphics[width=0.32\hsize]{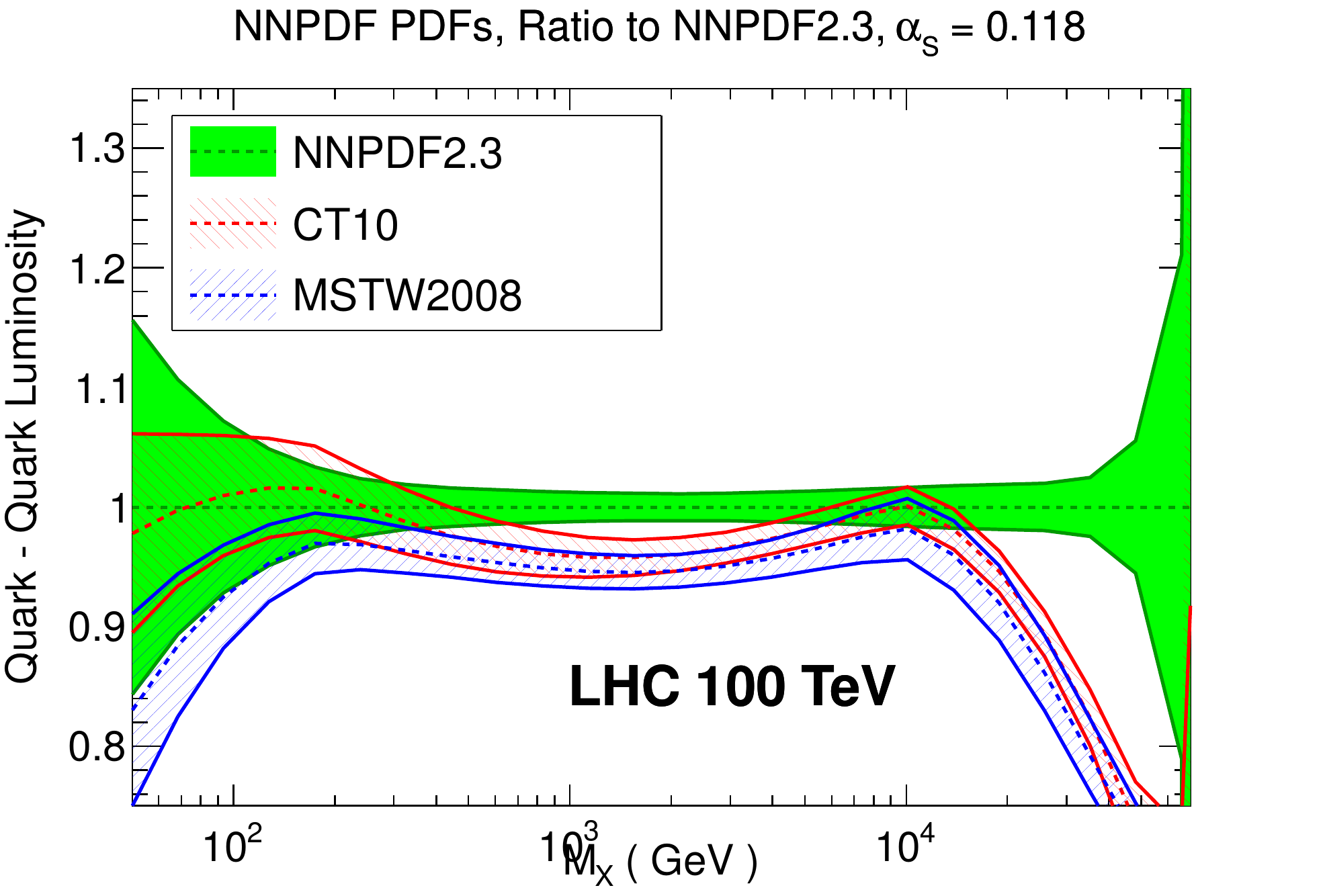}
\includegraphics[width=0.32\hsize]{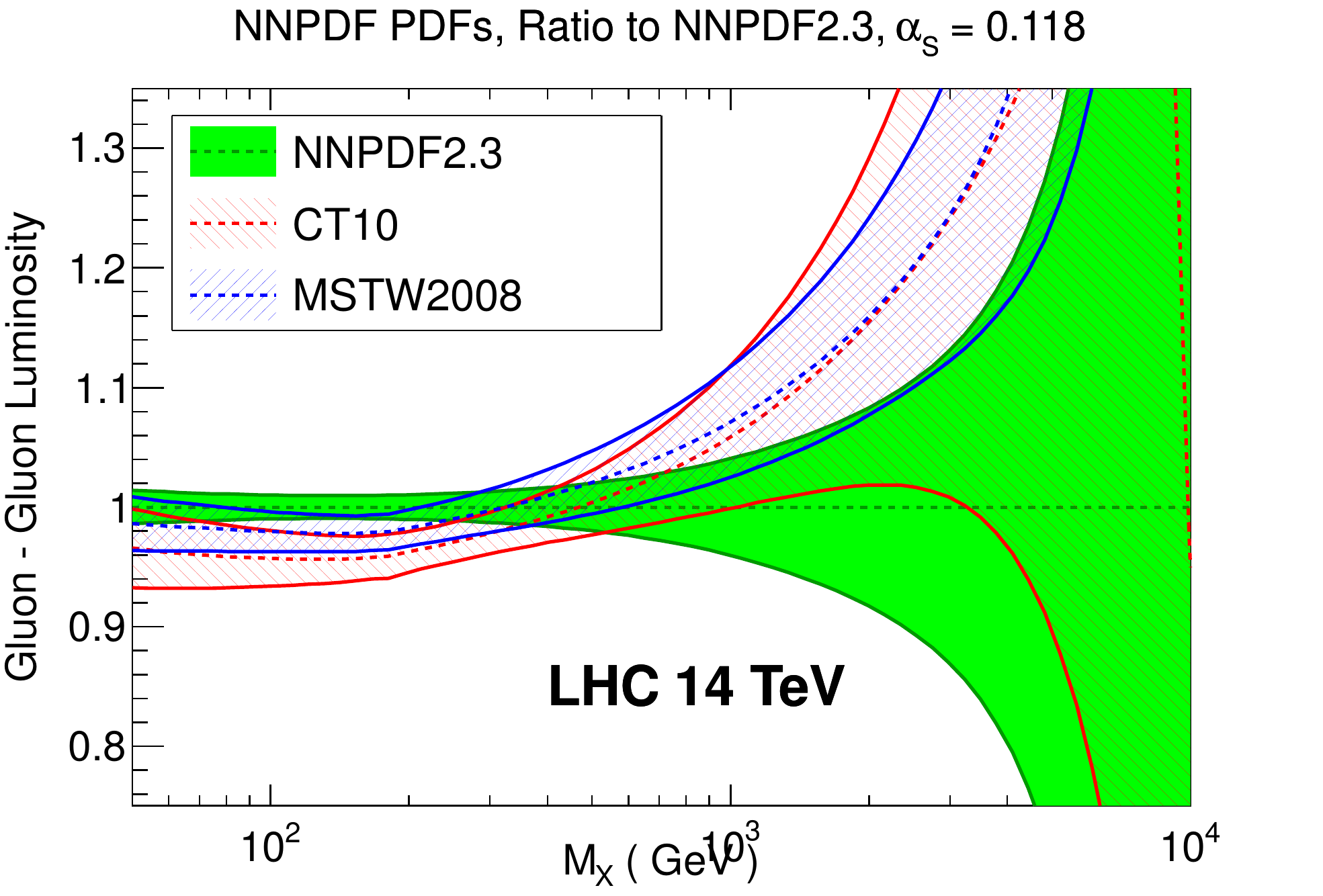}
 \includegraphics[width=0.32\hsize]{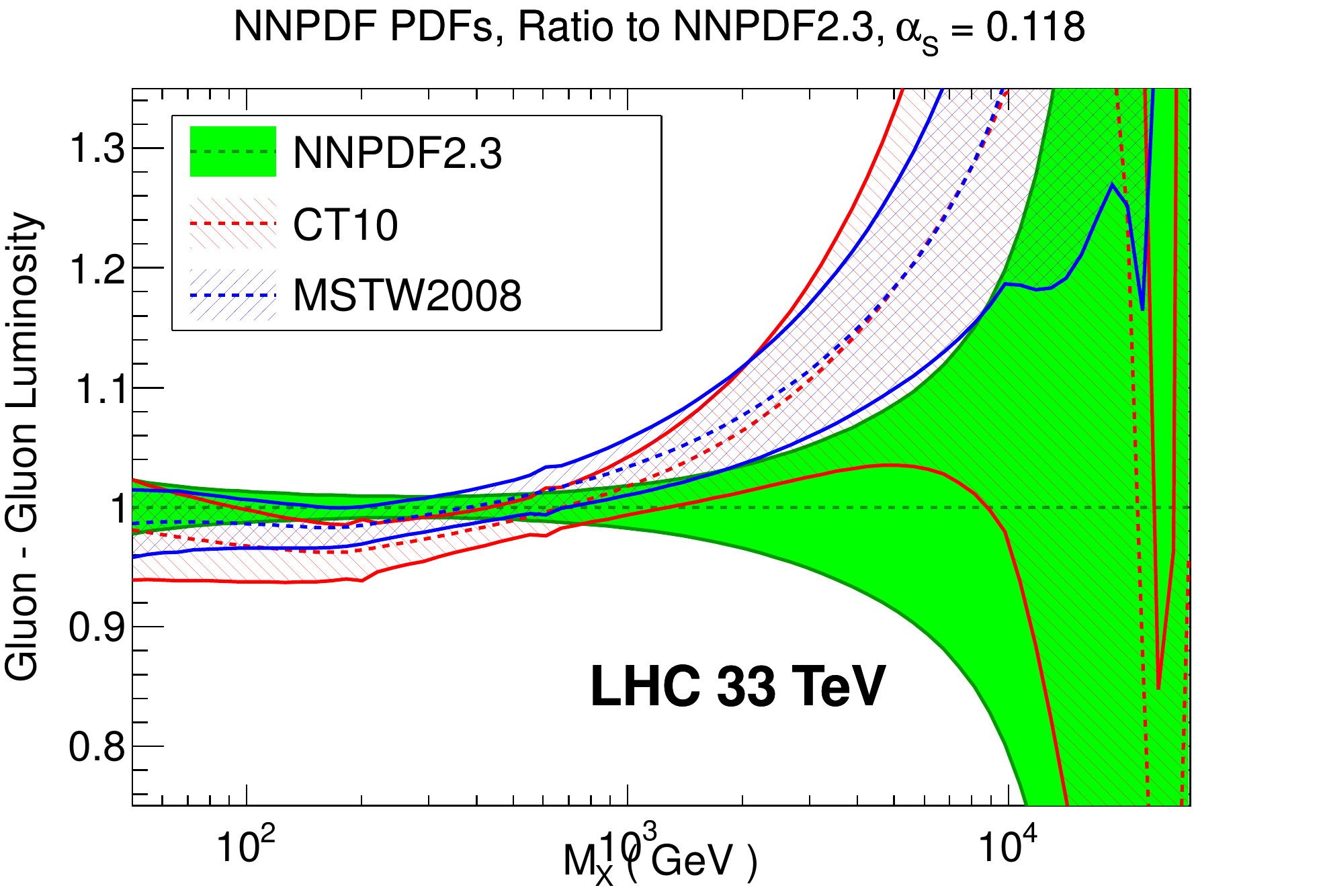}
 \includegraphics[width=0.32\hsize]{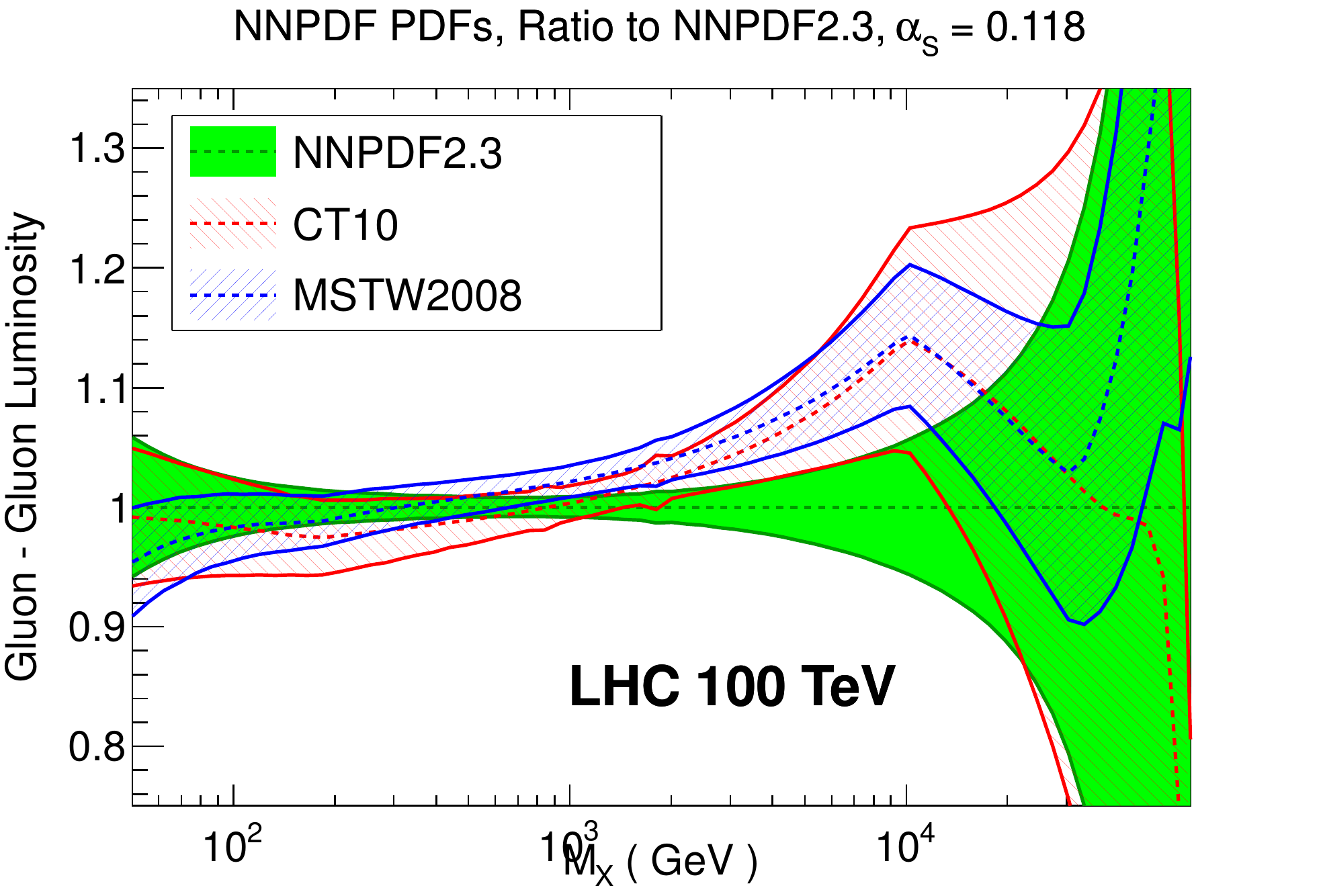}
\end{center}
\caption{ Comparison of the partonic luminosities at 14, 33 and
100 TeV between the CT10, MSTW and NNPDF2.3 NNLO PDF sets. From
top to bottom: quark-antiquark luminosity, quark-quark luminosity, and
the gluon-gluon luminosity. \label{fig:rojo1}}
\end{figure}

As discussed above, 
in order to assess similarities and differences between PDF sets, it
is useful to compare parton luminosities for different channels
as a function of the final state mass $M_X$ of the produced system,
for different values of the hadronic collider energy.
In the following we will redo the comparisons presented in Ref.~\cite{Ball:2012wy},
but this time for higher energy incarnations of the LHC: 14 TeV, 33 TeV and
100 TeV.
This comparison is shown in Fig.~\ref{fig:rojo1}, where we compare
the quark-quark, quark-antiquark and gluon-gluon luminosities between
the most updated CT, MSTW and NNPDF NNLO PDF sets at these three 
center of mass energies.

While in general we observe reasonable agreement between the three sets, there
are also some cases where the agreement is marginal, e.g. for the gluon-gluon
luminosity, or even non-existing, e.g. for the very high mass range for the quark-quark
luminosity at 100 TeV.
The quark-antiquark luminosities for the three PDFs in the $W/Z$ mass range agree well at 14 TeV, and even better at 7 and 8 TeV. For 33 and 100 TeV energies, the uncertainties in that mass range rapidly increase as smaller quark
$x$-values not well-constrained by HERA are probed. 



\subsection{Improvements from LHeC}
\label{sec:qcd-pdf-lhec}

\draftnote{1 page: M. Klein}

With the LHeC, very precise measurements of charged 
currents (CC) and the exploitation of $Z$ exchange in neutral
currents (NC) become possible, in addition to extending
photon exchange NC to extremely low $x$.
The question of gluon saturation
at low $x$ can be expected to be settled with precision measurements
of the structure functions $F_2$ and also $F_L$
down to $x \geq 10^{-6}$~\cite{AbelleiraFernandez:2012cc}
while the large $x$ determination of $xg$ is crucial for the LHC
Higgs and BSM 
program~\cite{AbelleiraFernandez:2012ty,Bruening:2013bga}.
The LHeC, combined with 
HERA to fill in the medium $Q^2$-larger $x$ 
region, provides a unique and complete DIS data set. 
With unprecedented precision 
there will be for the first time a determination possible of
$all$ PDFs, $u_v,~d_v,~\bar{u},~\bar{d},~s,~\bar{s},~c,~b$ and
even $t$, in furthermore a hugely extended kinematic range.
To explore this, a full set of NC and CC cross section
measurements has been simulated and a QCD fit analysis been
applied in order to study the potential for the determination
of the parton distribution functions in the proton.  

As detailed in~\cite{AbelleiraFernandez:2012cc},
the strange quark distribution will be measured 
in an accurate way with charm tagging of $W$ fusion in 
CC scattering. Precise measurements  of
the charm and the beauty quark distributions would be possible, from $Q^2$ values below
the quark masses squared up to $\sim ~10^5$\,GeV$^2$, based on 
the small beam spot size of $\sim 7$\,$\mu$m$^2$ and
a high resolution silicon detector of large acceptance. This, for
example, will determine the charm mass 
with the experimental error of $3$\,MeV~\cite{AbelleiraFernandez:2012cc},
an order of magnitude improved as compared to HERA, and similarly
for the bottom mass. Such high precision input will certainly
provide a new basis for higher order tests of the treatment of
heavy quarks in the $Q^2$ evolution, which currently is a significant
source of uncertainty in the understanding of PDFs and in
predictions for the LHC.

Monte Carlo data were simulated
for  NC and CC scattering assuming $e^{\pm}p$ luminosities of
$10$\,fb$^{-1}$ and a $40$\,\% polarization. Using the HERAFitter 
framework~\cite{Aaron:2009aa,Aaron:2009kv,James:1975dr}
with settings based on the HERAPDF NLO QCD fit analysis,
a set of PDFs has been generated (and is available on
 LHAPDF) for estimates of future measurement potentials. 
 An example  is the prediction of
the $gg \rightarrow H$ cross section at the LHC, which 
will have an uncertainty from PDFs and $\alpha_s$ of only about 
$0.4$\,\% and thus be sensitive to determinations of $M_H$
via the cross section~\cite{Bruening:2013bga}.
 Another example is the importance
of knowing $xg$ for high mass searches of SUSY particles
as has also been recently studied~\cite{AbelleiraFernandez:2012ty, posterEPS}.

The procedure used has been adopted from the HERA QCD fit
procedure~\cite{Aaron:2009aa} with a minimum $Q^2$ cut
of $3.5$\,GeV$^2$ and a
starting scale  $Q^2_0=1.9~ \rm GeV^2$, 
chosen to be below the charm mass threshold.
The fits have been extended to very low values of $x$ for systematic uncertainty
studies, even when at such low $x$ values non-linear
effects are expected to appear, eventually altering the evolution laws. 

The parameterized PDFs are the valence distributions
$xu_v$ and  $xd_v$,  the gluon distribution $xg$, and the
$x\bar{U}$ and $x\bar{D}$ distributions, where $x\bar{U} = x\bar{u}$,
$x\bar{D} = x\bar{d} +x\bar{s}$. This ansatz is natural to the
extent that the NC and CC inclusive cross sections determine
the sums of up and down quark distributions,
and their antiquark distributions, as the four independent
sets of PDFs, which may be transformed to the ones chosen
if one assumes $u_v = U -\overline{U}$ and $d_v = D - \overline{D}$,
i.e. the equality of anti- and sea quark distributions of given 
flavor.  The following standard functional form is used to parameterize the PDFs:
\begin{equation}
 xf(x) = A x^{B} (1-x)^{C} (1 + D x + E x^2),
\label{equ:pdf}
\end{equation}
where the normalization parameters ($A_{uv}, A_{dv}, A_g$)
are constrained by  quark counting and momentum  sum rules.
The parameters  $B_{\bar{U}}$ and $B_{\bar{D}}$ are set equal,
 $B_{\bar{U}}=B_{\bar{D}}$, such that
there is a single $B$ parameter for the sea distributions.
The strange quark distribution  at the starting scale
is  assumed to be a constant fraction of $\bar{D}$,
$x\bar{s}= f_s  x\bar{D}$,
chosen to be $f_s=0.5$ such that
$\overline{s}=\overline{d}$.
In addition, to ensure that $x\bar{u} \to x\bar{d}$
as $x \to 0$,
$A_{\bar{U}}=A_{\bar{D}} (1-f_s)$.
The $D$ and $E$ are introduced one-by-one until no further
 improvement in $\chi^2$ is found.
The best fit  resulted in a total of $12$ free parameters,
specifically fitting $B_g,~C_g,~D_g,~B_{uv},~C_{uv},
~E_{uv},~B_{dv},~C_{dv},~C_{\overline{U}},
~A_{\overline{D}},~B_{\overline{D}},~C_{\overline{D}}$.
While the LHeC NC, CC real data, and the inclusion of 
further information, as of $s,~c,~b$ and $F_L$, will certainly
lead to quite a different parameterization, it has been 
checked that with a more flexible set of $15$ parameters 
very similar results to the PDF uncertainties found here are obtained.

The PDFs are evolved using DGLAP evolution
equations at NLO in the $\overline{\mathrm{MS}}$ scheme with the
renormalization and factorization scales set to $Q^2$
using standard sets of parameters, such as for $\alpha_s(M_Z)$.
These, as well as the exact treatment of the heavy quark
thresholds, have no significant influence on the
estimates of the PDF uncertainties. The experimental uncertainties on the PDFs
are determined using the  $\Delta\chi^2=1$ criterion.
The LHeC Design Report~\cite{AbelleiraFernandez:2012cc}
contains a very detailed presentation
of the results of the present analysis for valence and sea quarks
with many remarkable features as the determination
of the $u/d$ ratio or the measurement of the valence quarks 
down to low $x \simeq 10^{-4}$. 
\subsubsection{Determination of the Gluon Distribution at the LHeC}
\begin{figure}[ht]
\centerline{\includegraphics[width=.8\textwidth]{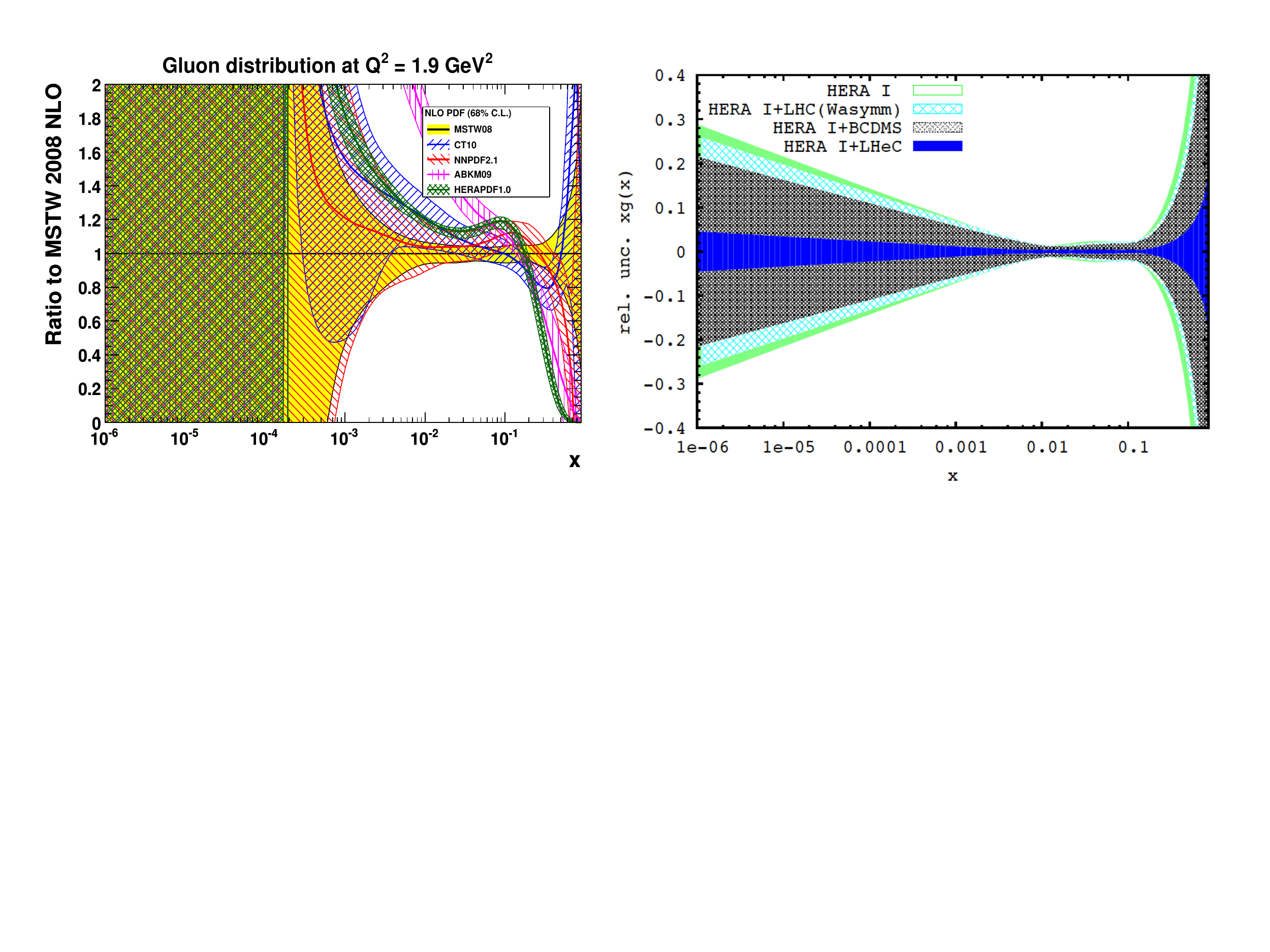}}
\vspace{-4.8cm}
\caption{Uncertainty of the gluon distribution at
$Q^2 = 1.9$\,GeV$^2$ as a function of Bjorken $x$, see text.
The LHeC PDF set, corresponding to the inner blue error band,
is available on LHAPDF.}
   \label{qcd-lhec-figxg}
\end{figure}

The result for the gluon distribution is presented in 
Fig.\,\ref{qcd-lhec-figxg}.  In the left panel,
recent gluon distribution determinations and their uncertainties
are shown plotted as a ratio to MSTW08. Below $x \simeq 10^{-3}$
the HERA data have vanishing constraining power due to
kinematic range limitations, and the gluon is just not determined
at low $x$.  At large $x \geq 0.3$ the gluon distribution
becomes very small and large variations differing by orders of magnitude appear in its
determination.  This
is related to uncertainties in jet data, theory uncertainties
and the fact that HERA had not enough luminosity to
cover the high $x$ region.  Moreover, the sensitivity
to $xg$ at HERA diminishes, as the valence quark evolution
is insensitive to it. The larger $x$ situation can be expected 
to improve with
LHC jet data and possibly top data and the HERA II data. The right panel shows
the experimental uncertainty of $xg$ based on the LHeC, on HERA alone
and in various combinations with further data;
see the LHeC design report~\cite{AbelleiraFernandez:2012cc} for more details.
At small $x$ a few per cent precision
becomes possible, as can be seen by comparing the right and left panels.
Note that the non-LHeC low $x$ uncertainty bands  
(right) remain narrow below $x \simeq 10^{-3}$,
as an artifact due to the parameterization
of $xg$. 

It is for the LHeC to discover whether
$xg$ saturates or not and whether indeed the DGLAP
equations need to be replaced by non-linear parton
evolution equations such as BFKL. This is important not only for QCD
but also for super high energy neutrino physics and
low $x$ physics at the LHC.
In the region of the Higgs data at the LHC, $x \sim 0.02$, the
LHeC will pin down the gluon extremely accurately and the 
$gg \rightarrow H$ cross section uncertainties will essentially be 
removed as has been discussed in~\cite{Bruening:2013bga}.
At large values of e.g. $x=0.6$ the LHeC can be expected to determine
$xg$ to $5-10$\,\% precision (inner blue band). This is crucial
for when the LHC operates at maximum luminosity 
and the searches approach the few TeV mass region, as in
$gg \rightarrow \tilde{g}\tilde{g}$~\cite{posterEPS}.  It is also important for testing QCD, as factorization and scales,
as well as electroweak effects at large $x$ in a future 
critical comparison of such $ep$ with LHC $pp$ data as
for jets, see also~\cite{AbelleiraFernandez:2012ty}.
 Similarly, surprises may result from comparisons with inclusive LHeC jet data, not considered here.
PDF physics  rests on controlling and testing 
the underlying theory.

\subsubsection{Final Remark}
It is important to emphasize that while the PDF analysis presented
here  serves as a valid starting point for comparison
with existing PDFs, the LHeC has a unique potential to release
the underlying simplifying assumptions 
and to provide a radically different and novel
way to determine the PDFs.  With the consideration of the direct
measurements of the strange, charm and beauty PDFs, and perhaps
even the top PDF, and with the addition of tagged $eD$ data, it will
be possible to analyze the behavior of not just $4$ suitable combinations
of PDFs but to determine the full set for the first time with crucial
direct input.  For example, the valence quarks at high $x$ will be determined by
high statistics CC data, and at low $x$ they will be measured with electroweak structure
functions. Light quarks can be determined independently of each other using 
$ep$ and $en$ and CC data. The LHeC the
will radically change the world of PDFs. The present study of uncertainties
to this extent is an illustration only and 
initially rather narrow in scope.  It yet becomes evident that
with the LHeC the development of QCD will hugely progress and 
the LH(e)C can be turned into a precision Higgs facility.
Electromagnetic substructure of the heaviest elementary particles may also be revealed.
Finally, the anticipated investment into the highest LHC luminosity
will be underpinned by the necessary precision QCD and PDF 
measurements by the LHeC without which highest mass limits must
remain weaker and interpretations of subtle new features
possibly  uncertain. The LHeC appears as 
an important upgrade to the LHC with which the symmetry between
$pp$, $ep$ and may be $e^+e^-$ can be restored at TeV energies.  
This fruitful symmetry allowed the Tevatron, HERA and LEP/SLC to eventually establish
the Standard Model of particle physics.


\section{Higher-order corrections}
\label{sec:qcd-xsecs}

The implementation of higher-order corrections in parton-level predictions
and Monte Carlo generators is essential for maximizing the potential
of future experiments.  This section presents a wide survey of both
current tools and directions of future development, with applications
to LHC operations at $14$~TeV and to proton-proton collisions at
$33$ and $100$~TeV.  The section concludes with
an overview of the highest-priority perturbative calculations, ones
that could feasibly be tackled in the next 5--10 years.


\subsection{NLO cross sections at  $14$, $33$ and $100$ TeV}
\label{sec:qcd-xsecs-mcfm}

\draftnote{Text below written by J. Campbell}

\begin{figure}[htb]
\begin{center}
\includegraphics[width=0.7\hsize,angle=90]{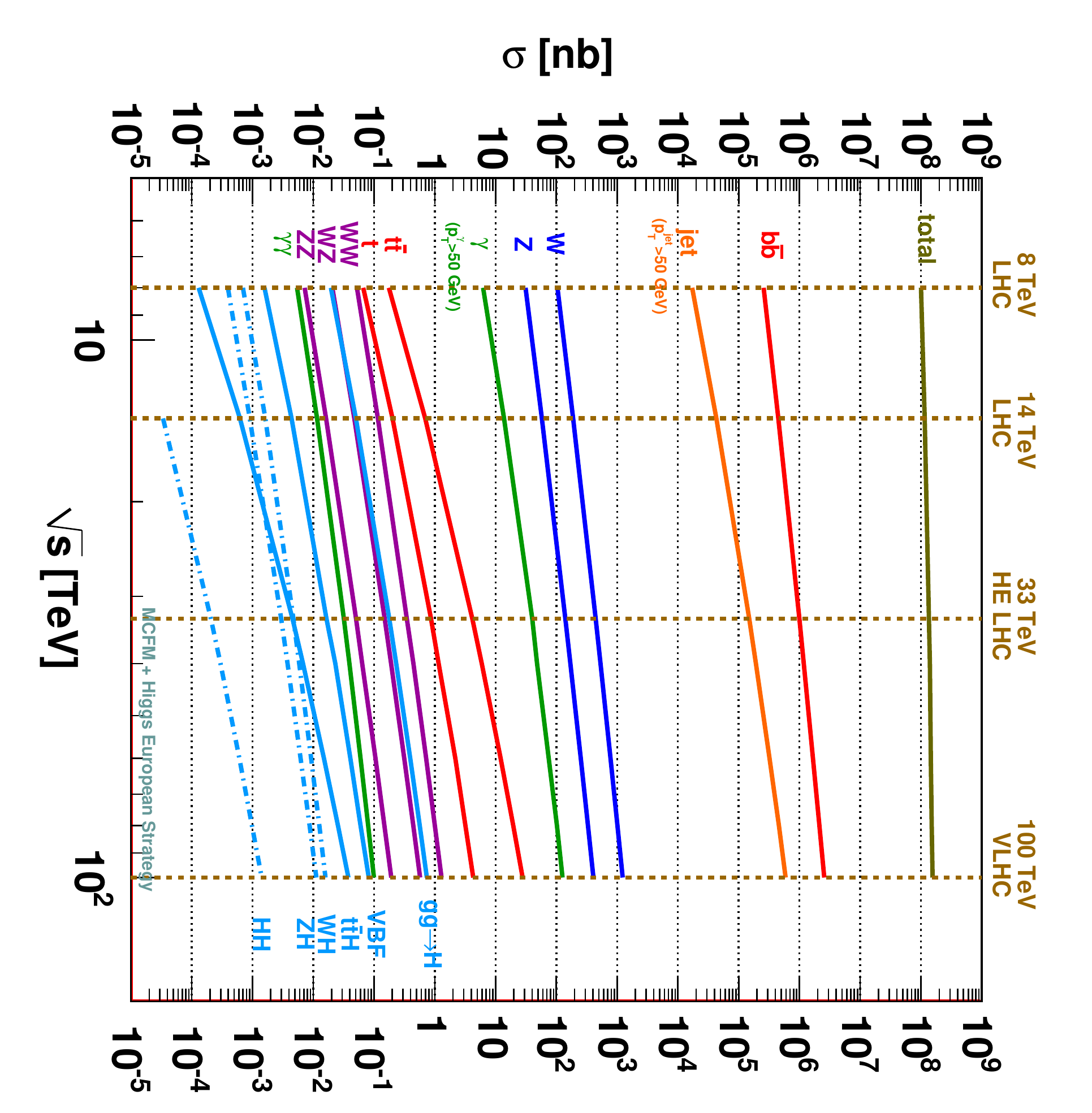}
\caption{Cross section predictions at proton-proton colliders
as a function of center-of-mass operating energy, $\sqrt s$.
}
\label{fig:qcd-xsecs-mcfm-Edep}
\end{center}
\end{figure}

As a first step towards investigating the physics potential of
future proton-proton colliders, 
it is interesting to investigate the center-of-mass energy dependence
of notable cross-sections at such machines.
Figure~\ref{fig:qcd-xsecs-mcfm-Edep} shows the predicted cross sections
for a selection of basic processes, ranging over twelve orders of magnitude
from the total inelastic proton-proton
cross section to Higgs boson pair-production.  For inclusive jet
and direct photon production, $50$~GeV transverse momentum cuts are
applied to the jet and the photon respectively.  The cross
sections presented in this figure have been calculated at next-to-leading
order in QCD using the MCFM program~\cite{Campbell:1999ah,Campbell:2011bn}, or taken from
the European Strategy report~\cite{HiggsEuropeanStrategy2012}
(in the case of Higgs cross sections).  

The growth of the cross-sections with $\sqrt{s}$ largely reflects
the behavior of the underlying partonic luminosities.  For instance,
the top pair cross section is dominated by the partonic process
$gg \to t\bar t$ and the gluon-gluon luminosity rises significantly
at higher values of $\sqrt{s}$.  The same holds true for the 
Higgs production channel $t\bar t H$ but, in contrast, the associated
production channels are dominated by quark-antiquark contributions and
rise much more slowly.  The different behavior means that, unlike at
current LHC operating energies, the $t\bar t H$ channel becomes the
third-largest Higgs production cross section at $33$~TeV and above.
As a figure of merit for estimating the difficulty of observing the
Higgs pair production process it is not unreasonable to consider the
ratio of its cross section to the top pair cross section.  In many of the possible Higgs boson decays the final states receive significant background contributions from the top pair
process.  The fact that both processes are predominantly gluon-gluon
induced means that this measure is approximately constant across
the range of energies considered.  It is therefore not clear that
the prospects for extracting essential information from the Higgs-pair
process would be significantly easier at a higher-energy
hadron-collider, even though the rates increase dramatically.

A different sort of contribution to event rates can also be estimated
from this figure.  The contribution of double parton scattering (DPS) --
where a single proton-proton collision is responsible for two
hard events -- can be estimated by,
\begin{equation}
\label{eq:qcd-dps}
\sigma_{XY}^{\rm DPS} \approx \frac{\sigma_X \sigma_Y}{15~{\rm mb}} \;.
\end{equation}
In this equation the DPS contribution for the final state $XY$
is related to the usual cross sections for individually producing
final states $X$ and $Y$ dividing by an effective DPS cross section.
This cross section appears to be approximately independent of
energy up to $8$~TeV and is approximately $15$~mb (for example, see
Ref.~\cite{Aad:2013bjm} for a recent measurement at $7$~TeV).
Of course the uncertainty on the effective cross section, and indeed
on the accuracy of Eq.~(\ref{eq:qcd-dps}) itself, is such that this
should be considered an order-of-magnitude estimate only.
A particularly
simple application of this is the estimation of the fraction of events
for a given final state in which there is an additional DPS contribution
containing a pair of $b$-quarks. This fraction is clearly given 
by the ratio, $\sigma_{b\bar b}/(15~{\rm mb})$.  From the figure this fraction
ranges from a manageably-small $2$\% effect at $8$~TeV to a much more
significant $20$\% at $100$~TeV.  More study would clearly be required
in order to obtain a true estimate of the impact of such events on the
physics that could be studied at higher energies, but these simplified
arguments can at least give some idea of the potentially troublesome
issues.

\begin{figure}[t!]
\begin{center}
\includegraphics[width=0.3\hsize,angle=90]{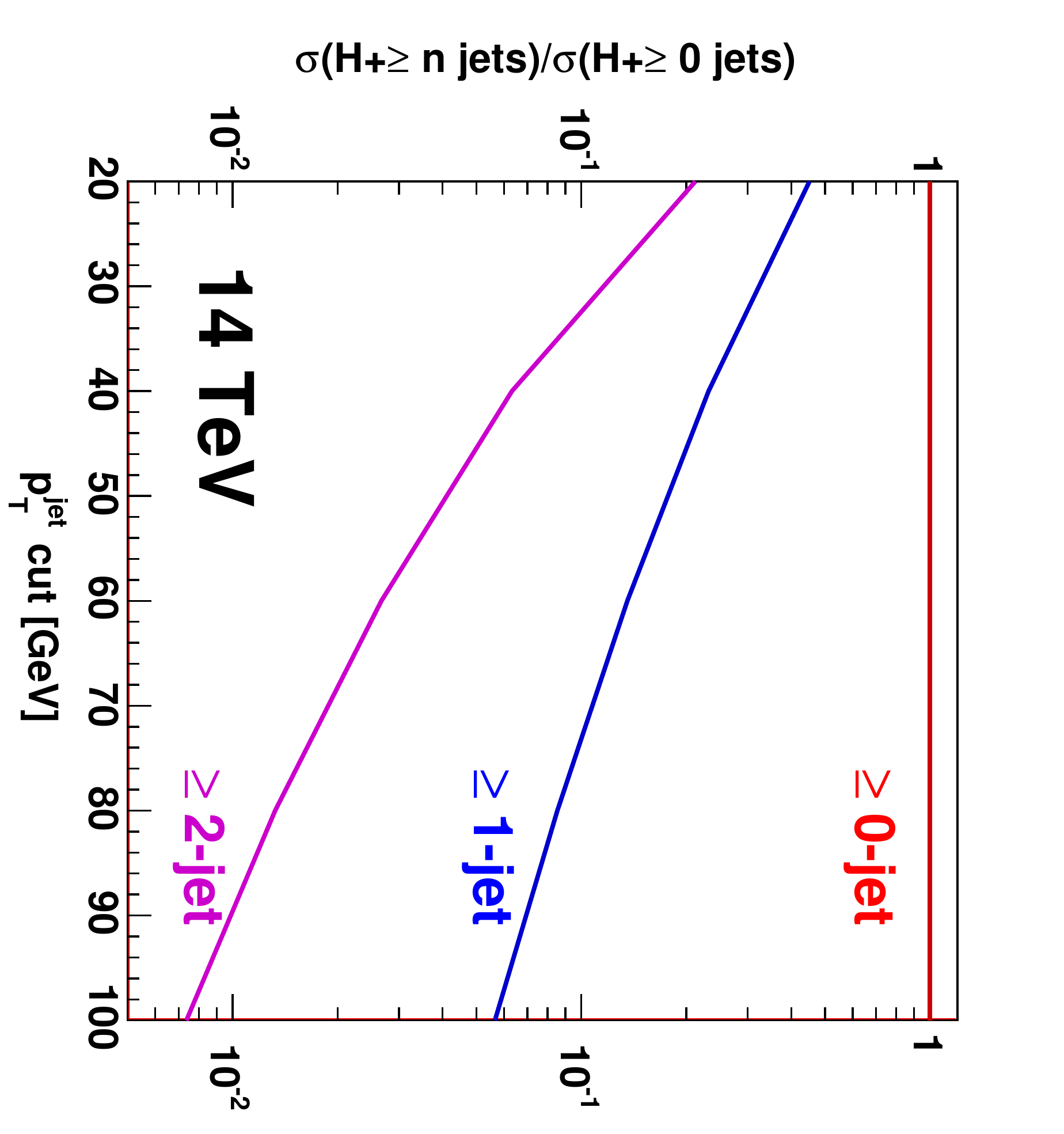}
\includegraphics[width=0.3\hsize,angle=90]{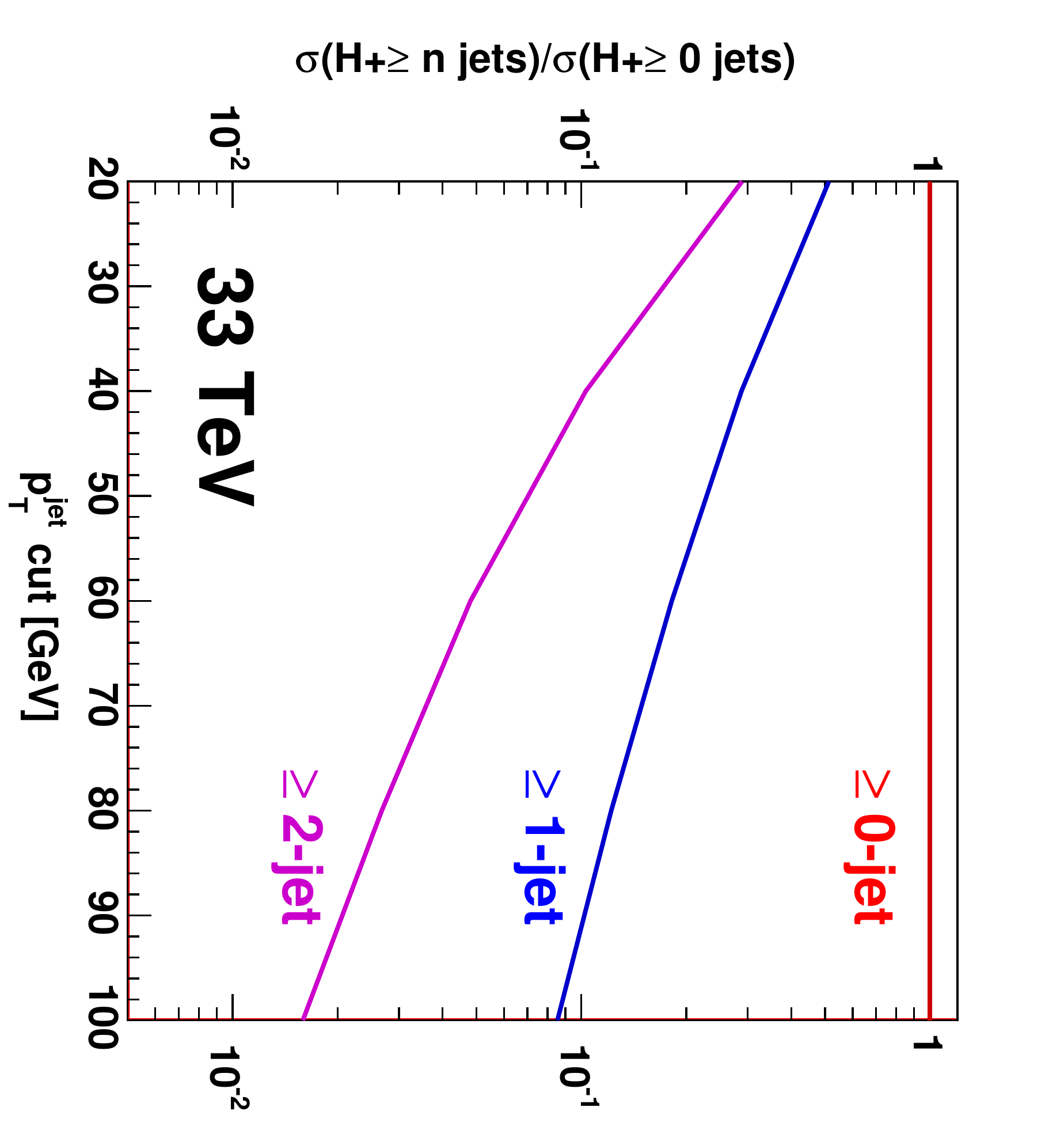}
\includegraphics[width=0.3\hsize,angle=90]{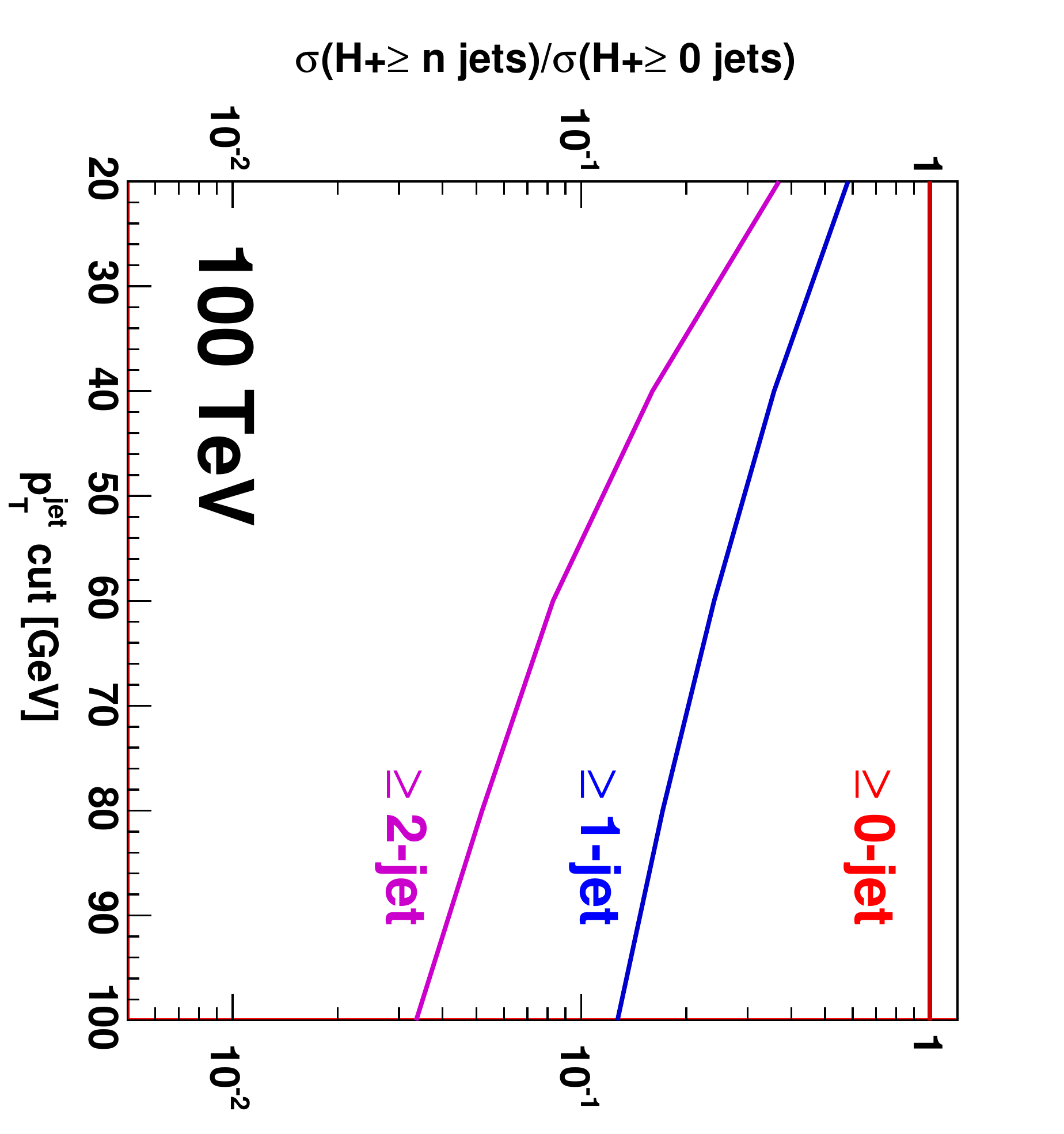}
\caption{Cross sections for the production of a Higgs boson produced
in association with $n$ or more jets, for $n=0,1,2$, normalized to
the inclusive Higgs cross section ($n=0$). Cross sections are
shown as a function of the minimum jet $p_T$ and are displayed
for a proton-proton collider operating at $14$~TeV (left),
$33$~TeV (center) and $100$~TeV (right).
}
\label{fig:qcd-xsecs-higgsjets}
\end{center}
\end{figure}

As an example of the behavior of less-inclusive cross sections at
higher energies, Fig.~\ref{fig:qcd-xsecs-higgsjets} shows predictions
for $H+n~{\rm jets}+X$ cross sections at various values of $\sqrt{s}$
and as a function of the minimum jet transverse momentum.  The
cross sections are all normalized to the inclusive Higgs production
cross section, so that the plots indicate the fraction of Higgs events
that contain at least the given number of jets.  The inclusive Higgs
cross section includes NNLO QCD corrections, while the $1$- and $2$-jet 
rates are computed at NLO in QCD.

The extent to which additional jets are expected in Higgs events
is strongly dependent on how the jet cuts must scale
with the machine operating energy.  For instance, consider a jet cut
of $40$~GeV at $14$~TeV, a value in line with current analysis projections.
For this cut, approximately $20$\% of all Higgs boson events produced through
gluon fusion should contain at least one jet.  The fraction with two or more
jets is expected to be around $5$\%.  To retain approximately the same
jet compositions at $33$ and $100$~TeV requires only a modest increase
in the jet cut to $60$ and $80$~GeV respectively.

At higher operating energies it is especially interesting to compare
predictions produced using the standard perturbative expansion, here at NLO,
with alternative formalisms that directly appeal to the high energy limit.
One such formalism is encoded in the program
HEJ (``High Energy Jets'')~\cite{Andersen:2009nu,Andersen:2011hs} that
implements a resummation scheme based on the factorization of scattering
amplitudes in the high energy limit.  For this study we investigate
predictions for $H+2$~jet events, with particular interest in the
region where two of the jets are separated
by a large rapidity span.  As well as being relevant
for separating the gluon fusion and vector boson fusion processes,
this region is expected to be particularly sensitive to differences between
the predictions of NLO QCD and HEJ~\cite{Binoth:2010ra}.  Jets
are reconstructed using the $k_T$ algorithm with $D=0.6$ and $|y|<5$. In
the first scenario we consider a minimum transverse momentum cut of $40$~GeV
for operating energies of $14$, $33$ and $100$~TeV.  In the second scenario
the jet cut is doubled to $80$~GeV at $33$~TeV and again to $160$~GeV at $100$~TeV.

\begin{figure}[htb]
\begin{center}
\includegraphics[width=0.3\hsize,angle=90]{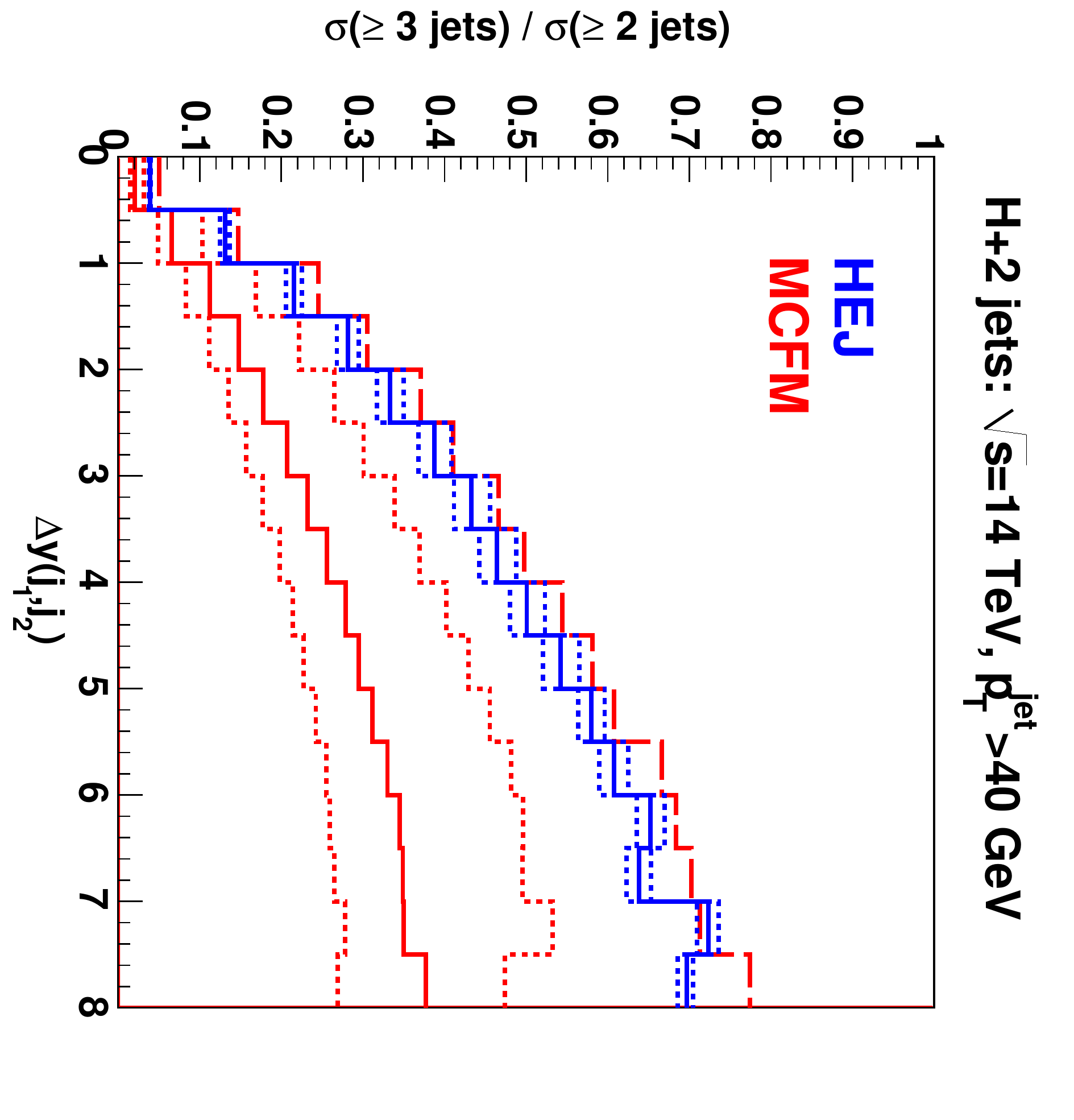}
\includegraphics[width=0.3\hsize,angle=90]{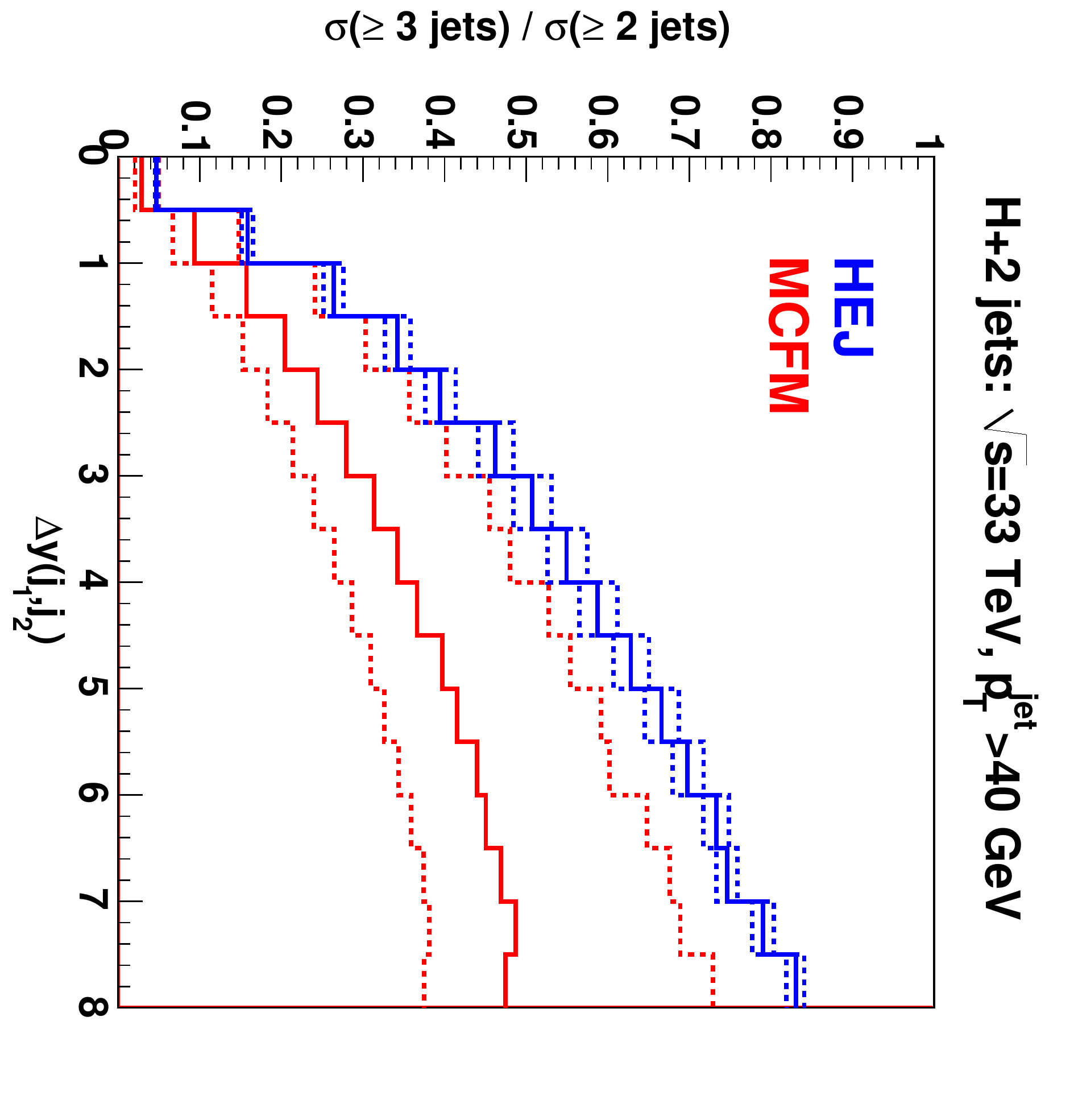}
\includegraphics[width=0.3\hsize,angle=90]{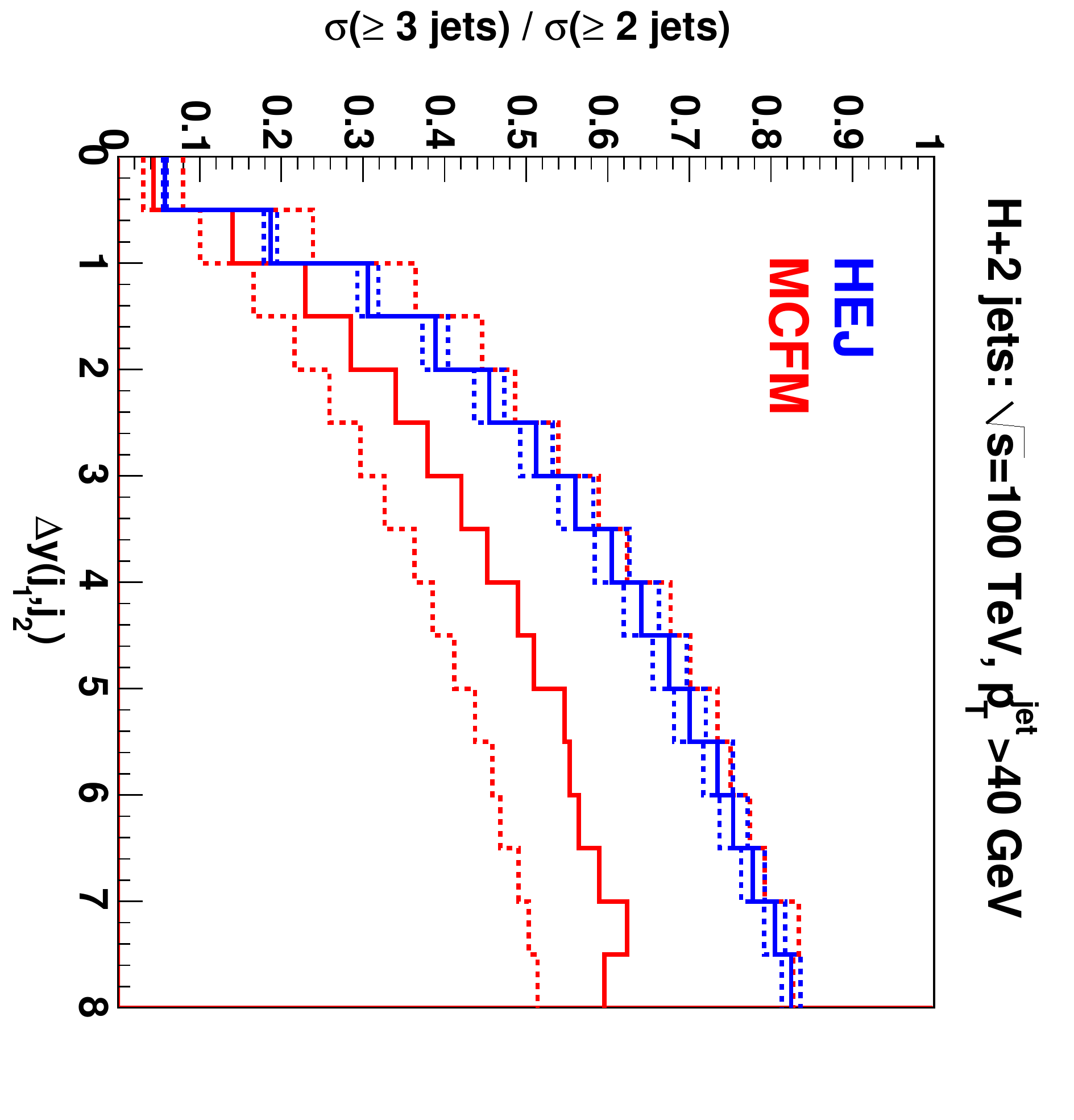} \\
\hspace*{0.3\hsize}
\includegraphics[width=0.3\hsize,angle=90]{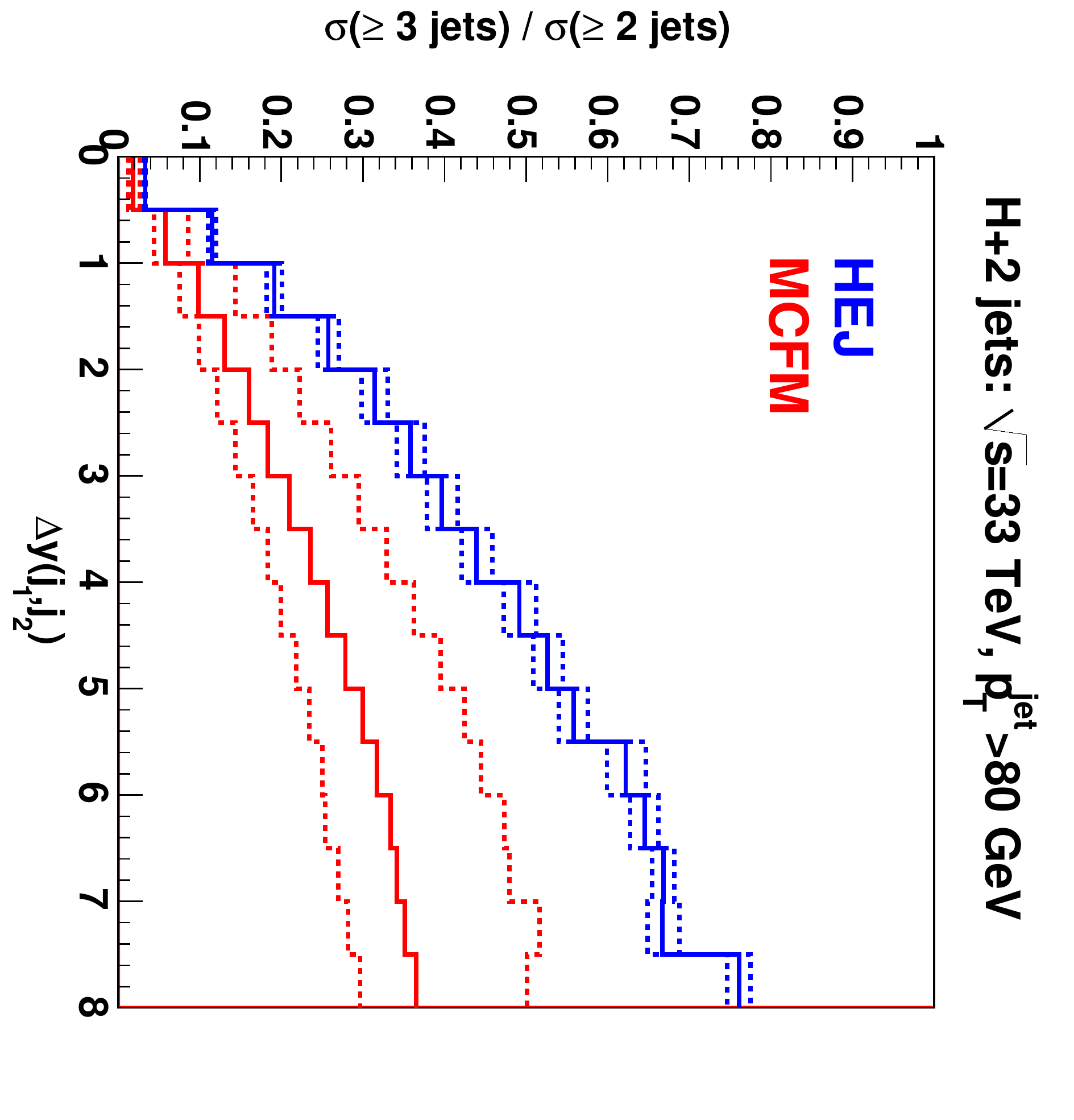}
\includegraphics[width=0.3\hsize,angle=90]{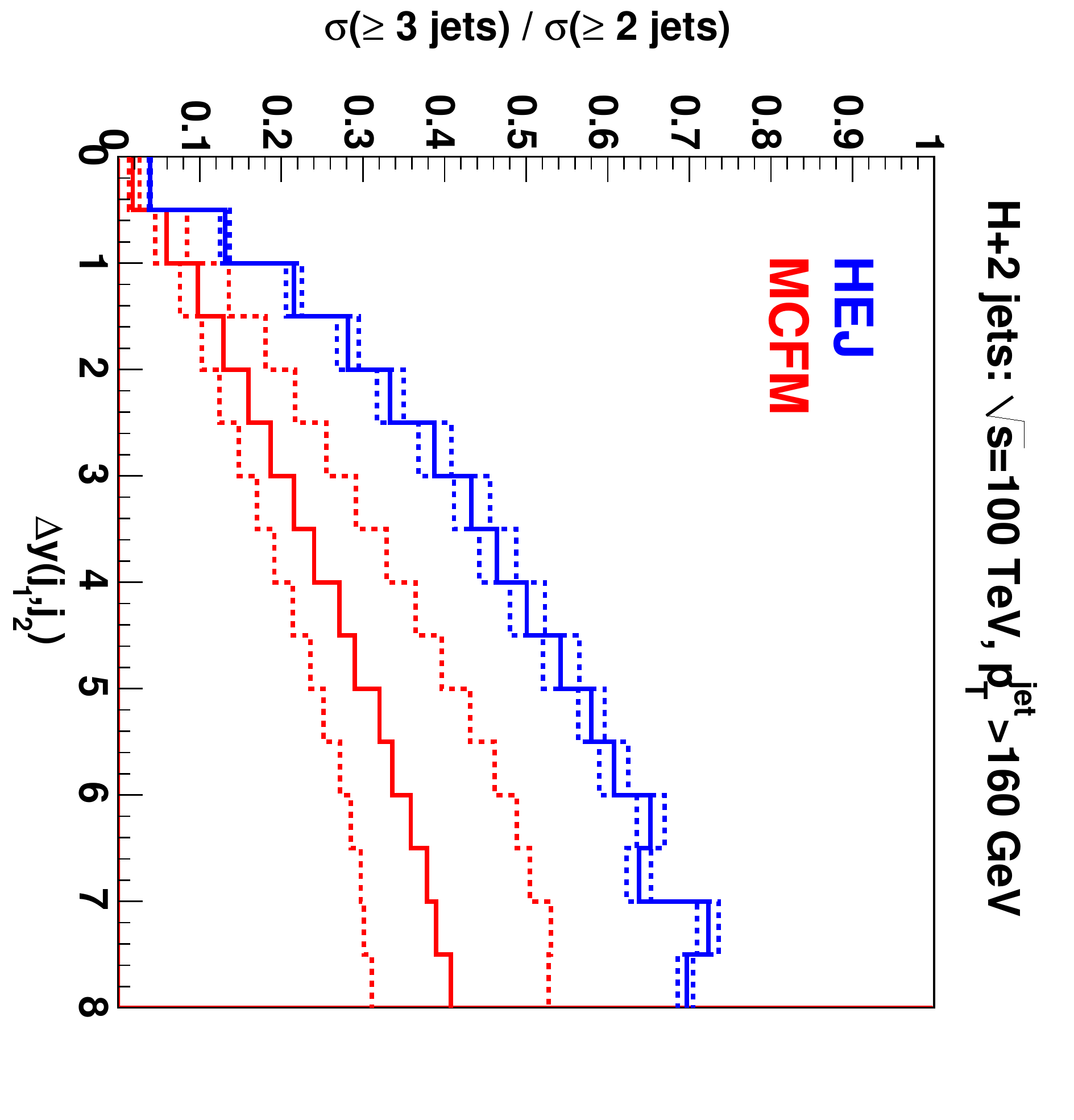}
\caption{The ratio of number of events that contain at least three jets to the number
that contain at least two jets, as a function of the rapidity difference between the two most
widely-separated jets.  Predictions are obtained using the NLO calculation of the
$H+2$~jet process and are shown at three operating energies.  The jet transverse
momentum cut is at $40$~GeV (top), or scales with the operating energy (bottom).
}
\label{fig:qcd-xsecs-mcfmvhej}
\end{center}
\end{figure}
The results of this study are shown in Fig.~\ref{fig:qcd-xsecs-mcfmvhej}.
Predictions are shown for the ratio of inclusive $3$-jet to inclusive $2$-jet
events, as a function of the rapidity difference between the two most
widely-separated jets.  The uncertainty band is obtained by varying the scale
choice by a factor of two about the central value ($H_T/2$, where $H_T$ is the
sum of the total transverse momentum of all objects in the final state).
The predictions for ratios of cross sections from HEJ are
very stable against variations in the factorisation and
renormalisation scales. This is because both numerator and
denominator are resummed predictions, with virtual
corrections counter-acting the renormalisation scale
dependence of the born level predictions.
The NLO MCFM results are much more
sensitive to changes in the scale, as can be seen from the wider uncertainty
bands.  This is further illustrated in the first plot (for collider energy
14~TeV and jet $p_T>40$~GeV) where the additional dashed line shows the MCFM
result for an even lower scale choice of $H_T/6$.  A typical $H_T$ value for
these events is around 250~GeV, so this choice is closer to the scale of the jet
$p_T$.   The prediction for this choice is very close to the HEJ prediction
throughout the range.

In all five scenarios, the predictions from HEJ for the ratio of $3$-jet to
$2$-jet events rise faster and reach a higher value than the predictions from
the NLO calculation as the rapidity span of the event increases.  This is the
region where the higher-order corrections included in HEJ become significant.  
As the collider energy is increased while the jet $p_T$ cut is kept constant the
phase space for the production of higher numbers of jets increases.  This is
illustrated in the top row of Fig.~\ref{fig:qcd-xsecs-mcfmvhej}, where it can
be seen that the value of the ratio increases in both descriptions.
As the energy of the collider increases, the difference between the HEJ and MCFM
descriptions is reduced.  As a results, at $\sqrt{s}=100$~TeV,
the HEJ curve practically coincides with the MCFM prediction using $H_T/4$.

In the cases where the jet $p_T$ cut is also increased (bottom row), the predictions
are rather similar across energies.  However the higher $p_T$ cuts mean that the
central scale is much larger in the MCFM calculation and, as a result, the scale
dependence band is smaller than in the case of a fixed cut.  This leads to
a larger difference in the predictions of HEJ and MCFM using our default scale
variation, although once again the HEJ curves could be mimicked by choosing a
slightly smaller scale of $H_T/6$ in MCFM.

\draftnote{FP: Two further aspects of these plots are confusing to me.
The HEJ scale uncertainties are extremely small.
Do we understand why the improvement is so dramatic upon moving away from MCFM,
and do we believe it?  - JC, added comments.
In the second row, the MCFM uncertainty decreases upon increasing the jet cut,
worsening the discrepancy with HEJ.  Should we add some editorial comment on this?}


\subsection{Extrapolation from existing NLO results}

The progress within the last 5 years in the calculation of NLO
corrections for complex final states  has been truly impressive, as
witnessed for example by the calculation of $W+5$~jets by the
Blackhat+Sherpa Collaboration~\cite{Bern:2013gka}. Of course, there is a limit,  as
increasing the number of partons in the final state by definition
increases the complexity of the calculation, while the physics reward
(typically) decreases. Within a matrix-element + parton shower
framework, additional jets can be added either at leading order or
through the parton shower. In addition, there are heuristic tools that
can be developed to extrapolate cross sections for higher jet
multiplicity based on the patterns observed at lower jet multiplicity.
For instance, known results for $W + 2$ through $W+5$~jet production
can be used to assess the scaling behavior of the $W+n$~jet
cross sections~\footnote{$W + 1$~jet behaves differently because of missing
production channels and kinematic differences.}
Defining the quantity $R^{\pm} = \sigma(W^\pm+n~{\rm jets})/\sigma(W^\pm+n-1~{\rm jets})$,
Blackhat+Sherpa have developed predictions for this ratio
for $n \geq 3$, in $pp$ collisions at $7$~TeV, for jets with $p_T>25$~GeV. The predictions
are:
\begin{eqnarray}
R^+_{\rm NLO}&=&0.263\pm0.009-(0.009\pm0.003) \, n, \nonumber \\
R^-_{\rm NLO}&=&0.248\pm0.008-(0.009\pm0.002) \, n. \nonumber
\end{eqnarray}
From these formulae, the cross sections at NLO for $W + 6$ or $7$ jets can
be predicted, without any actual calculation. Of course, such scaling
formulae are strongly dependent on the kinematic regions being
considered, but can easily be re-assessed for different cuts or
center-of-mass energies. 
However, there seems to be no a priori reasons why similar predictions would not be possible
for other final states for which $n$-jet cross sections are known. 

\begin{figure}[htb]
\begin{center}
 \includegraphics[width=0.48\hsize]{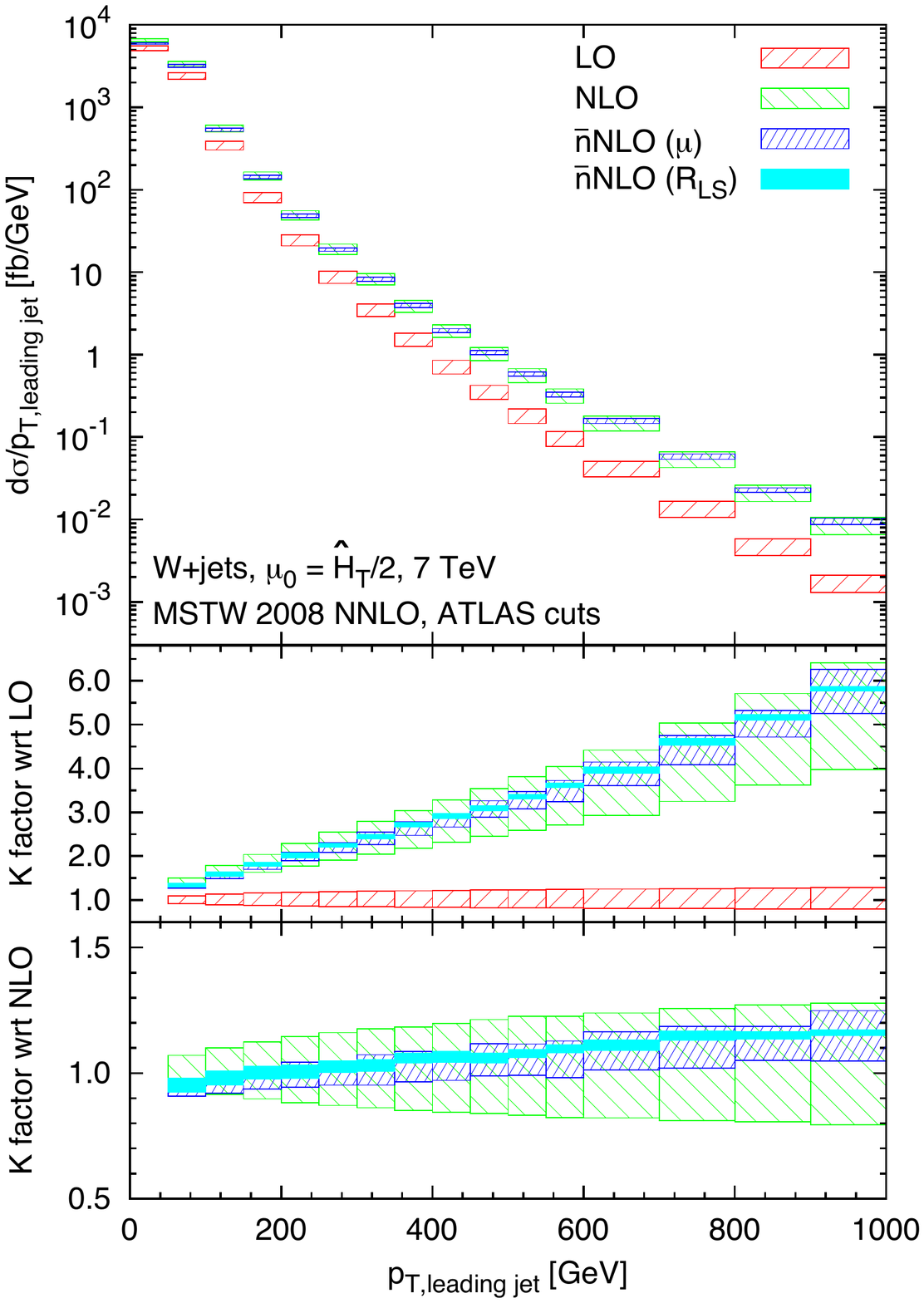}
 \includegraphics[width=0.48\hsize]{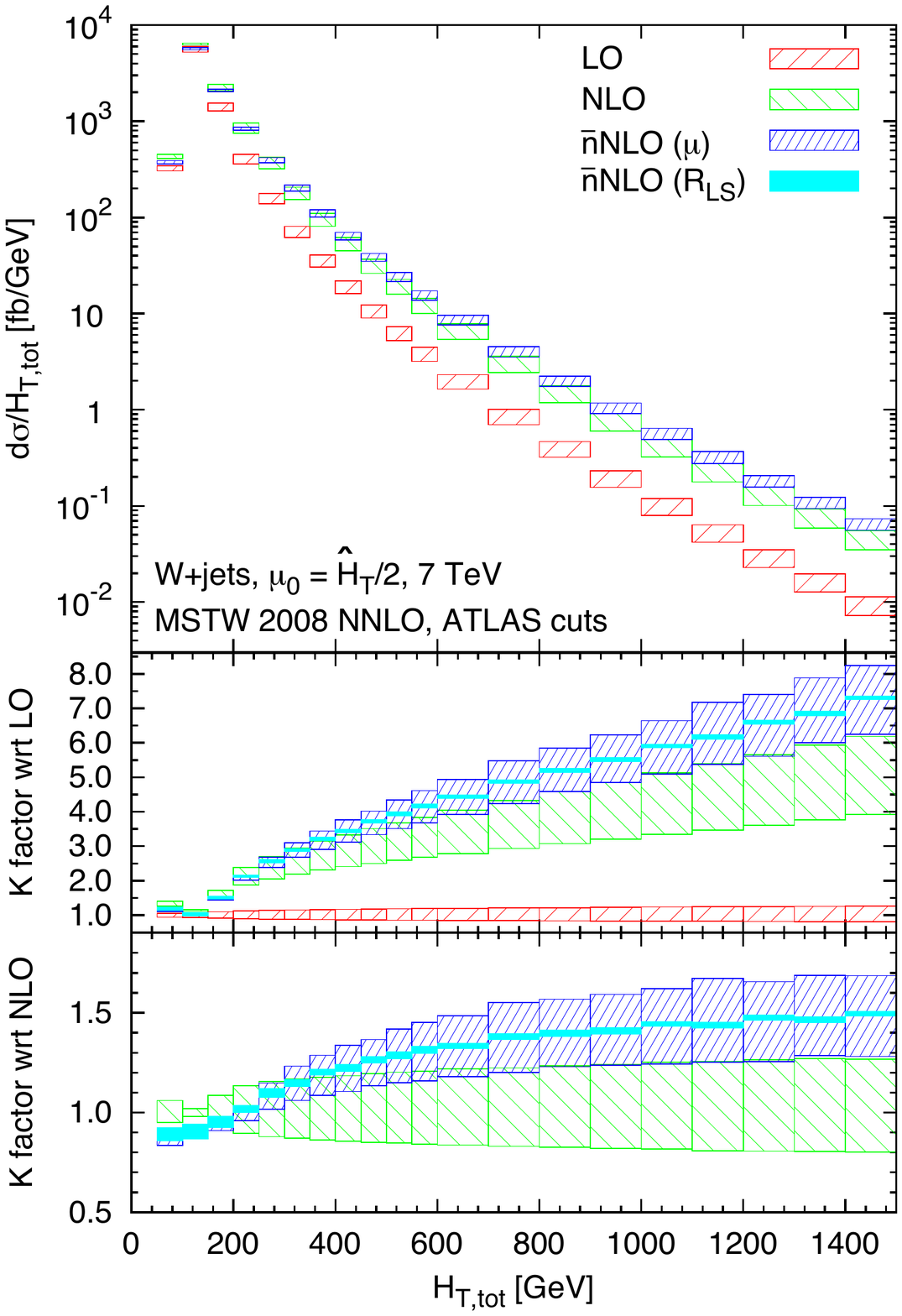}
\end{center}
\caption{The differential cross sections and $K$-factors for $p_{T,leading \ jet}$ and $H_{T,tot}$ at parton level at LO, NLO and
$\bar{n}$NLO. The bands correspond to varying $\mu_R=\mu_F$ by factors of 0.5 and 2.0 around the central value. 
\label{fig:loopsim}}
\end{figure}

There also exist other techniques for approximating NNLO (or higher) contributions to hard cross sections. One such technique is LoopSim~\cite{Rubin:2010xp}. The LoopSim method allows for the merging of NLO Monte
Carlo samples of different jet multiplicities in order to obtain approximate NNLO predictions, and has been applied to processes such as $W$ and $Z$ + jets~\cite{Maitre:2013wha} and $WZ$
production~\cite{Campanario:2012fk}. The LoopSim method makes use of any existing virtual matrix elements in the merged samples, and where these are not available, determines exactly the singular (or logarithmic) terms of
the loop diagrams, which, by construction, match precisely the corresponding singular terms of the real diagrams with one extra parton.  The approximate NNLO cross section thus constructed differs from the
complete NNLO cross section only by the constant terms. In some sense, this technique makes use of matrix element information from a number of multi-leg NLO calculations in a way similar to that carried out by a
matrix element-parton shower combination such as MEPS@NLO~\cite{Hoeche:2012yf}. This technique can potentially be  very powerful  when new sub-processes at higher jet multiplicities result in substantial additional contributions to
the cross section. One example is the production of a $W$ boson with  1  or more jets at the LHC (7 TeV). In Fig.~\ref{fig:loopsim}, the $p_T$ distribution for the leading jet (left) and for the
$H_{T,tot}$ distribution (the sum of the transverse momenta of all of the jets and $W$ boson decay products) (right) are shown, with predictions at LO, NLO and approximate NNLO
($\bar{n}$NLO)~\cite{Maitre:2013wha}. There are sizable NLO QCD corrections for the leading jet $p_T$ distribution, increasing with the jet transverse momentum,  and a considerable scale dependence even at NLO.
Both result from the addition of new topologies where the $W$ boson is soft or collinear. At $\bar{n}$NLO, there is a modest increase in the cross section  (within the NLO uncertainty band),  but a considerable
reduction in the scale dependence. In contrast, the $H_{T,tot}$ distribution receives large corrections at  $\bar{n}$NLO (outside of the NLO scale uncertainty band) with a smaller relative reduction in the scale
dependence. The increase results from the inclusion of a third jet in the $W+\ge1$ jet cross section calculation; even though the third jet will be relatively soft, its inclusion in a steeply falling distribution
results in sizable effects for the $H_{T,tot}$ cross section. This technique can also be applied to produce approximate NNNLO cross sections.

\subsection{Computational complexity}

The advances of NLO and NNLO multi-leg calculations have resulted in programs
that are (1) very complex  and (2) very time-consuming to run using the
resources available to a typical user. It can thus be difficult for a full
dissemination of the theoretical results to the experimental community. The
Blackhat+Sherpa Collaboration has partially addressed these issues by releasing
results for $W/Z + n$~jets in ROOT ntuple format, with all information needed for
calculation of physical cross sections with a variety of jet parameters and
cuts. In addition, the results can be re-weighted for different scales and
different PDFs. The cost is the presence of very large outputs, but subdivided
into files of a few GB that are ideal for parallel processing. Ntuples are
probably not practical, though, for some of the more complex NNLO calculations,
such as inclusive jet production, where the number of files needed would be
prohibitive. Making the resultant program user-friendly, though, can take a
great deal of time.

An alternative is to make such complex  programs available to the user on high
performance computing platforms, which inherently have much greater processing
and storage capabilities; in addition, the programs may be pre-compiled by the
authors insuring that the calculation is run correctly. Initial tests of HPC
platforms have been successfully carried out during the Snowmass workshop. We
have shown weak scaling up to 16000 cores on IBM BlueGene/Q, up to 8192 cores on
CRAY XK7, and strong scaling up to 1024 cores on CRAY XK7 using Sherpa. Another
possibility for large-scale production runs may be the Open Science Grid.  More details on 
the use of HPC resources in high energy theoretical research can be found in Ref.~\cite{Hoche:2013zja}.


\subsection{Improvements to matrix element and parton shower matching}

\draftnote{Text below written by S. H\"oche}

General-purpose event generators have been undergoing tremendous 
development over the past years. Both the fixed-order and logarithmic 
accuracy underpinning their simulations of perturbative QCD have been 
improved in order to match increased precision needs in the experiments.

The basis for the these developments was established by the MC@NLO 
technique~\cite{Frixione:2002ik}, which matches NLO QCD calculations 
to parton showers, such that fully differential events can be generated 
at the particle level. The POWHEG~\cite{Nason:2004rx,Frixione:2007vw} 
method was later introduced to eliminate negative weights from simulations.

Parton-shower matched predictions have been provided for some of 
the most challenging NLO calculations to date, including, for example, 
$pp\to W+3$~jets~\cite{Hoeche:2012ft} and $pp\to t\bar{t}+$~jet~\cite{Alioli:2011as}. 
The current limitation in matching to even higher multiplicity processes 
is not of algorithmic nature, but purely computational, and it is related 
to the memory consumption of executables. Fixed-order calculations 
benefit from the fact that they can be split into different parts, 
corresponding to Born, virtual correction, integrated infrared subtractions 
and real-emission correction minus real subtraction. Due to the intricate 
interplay between real and virtual corrections in MC@NLO and POWHEG, 
such a splitting is harder to achieve when matching to a parton shower.

Different proposals were made to combine MC@NLO simulations of varying 
jet multiplicity into inclusive event samples~\cite{Hoeche:2012yf,Frederix:2012ps}.
They are natural extensions of the CKKW(-L)~\cite{Catani:2001cc,Lonnblad:2001iq}
and MLM~\cite{Mangano:2001xp} leading-order merging schemes to the 
next-to-leading order, respectively. Another, entirely independent method 
was introduced earlier, which relies on a different subtraction 
scheme~\cite{Lavesson:2008ah,Lonnblad:2012ix}. 
The simulations provided by these new techniques can be used to obtain 
NLO-accurate predictions for different jet multiplicities at the same time. 

Achieving this level of fixed-order accuracy has been a priority in
the development of Monte-Carlo event generators for more than a decade. 
The current technology has undoubtedly benefited greatly from the
advances in computing fixed-order NLO QCD corrections at large jet
multiplicity in a fully automated manner. These calculations provide 
the parton-level input for the new merging methods.

NLO-merged predictions have been provided for $pp\to W/Z+$~jets with up to 
two jets described at NLO accuracy~\cite{Hoeche:2012yf}, $pp\to H+$~jets 
in gluon fusion, with up to one jet at NLO 
accuracy~\cite{Lonnblad:2012ix,Frederix:2012ps}
and $pp\to t\bar{t}+$jets with up to one jet at NLO 
accuracy~\cite{Frederix:2012ps,Hoeche:2013mua}. All related
implementations are fully automated and can, in principle, be used
for any type of reaction. However, because MC@NLO parton-level predictions 
are needed as an input, the current limitations are identical to those for 
MC@NLO. They are not due to algorithmic deficiencies, but mostly due to 
memory constraints on production systems and restrict the usage of the
methods to processes with fewer than four light jets in the final state
computed at NLO.

It is conceivable that working techniques for matching NNLO fixed-order 
calculations to parton-shower simulations will be constructed in the near 
to mid-term future. Such a matching, which could be dubbed `MC@NNLO', 
would further stabilize predictions for the differential cross section 
at lowest multiplicity, and eliminate the unitarity violation observed 
in most NLO merging methods in a natural way. 

An alternative technique already exists, which does not rely on a modified
subtraction scheme to construct counterterms for fixed-order calculations,
but on constructing counterterms for the parton shower, at the order at
which the shower is to be matched~\cite{Lavesson:2008ah}. This is very easy 
to achieve. The method has been used as proof of principle to provide NNLO 
matched predictions for $e^+e^-\to$~jets production~\cite{Lavesson:2008ah}.

At the same time that matching to fixed-order NNLO calculations is being
developed, the logarithmic accuracy of parton shower simulations must be 
improved systematically, in order not to degrade the precision of the 
fixed-order result after matching. This involves two developments.

Firstly, corrections which are sub-leading in the number of colors, $N_c$,
must be included. A proposal to do this was formulated some time 
ago~\cite{Nagy:2007ty}, but first steps to implementation were taken 
only recently~\cite{Hoeche:2011fd,Platzer:2012np}. The importance of 
sub-leading $N_c$ corrections in processes with non-trivial color 
structure at Born level was observed in an analysis of the $t\bar{t}$ 
forward-backward asymmetry~\cite{Hoeche:2013mua}.
Respecting the full color structure during parton evolution will allow
to include all next-to-leading logarithmic effects in the parton shower,
in a manner that is independent of the actual evolution variable, and
therefore does not rely on angular ordering.

Secondly, it will be beneficial to systematically extend parton-showers 
to higher logarithmic accuracy, for example by including higher-point 
splitting functions. An alternative approach, based on the matching of 
parton showers to analytic calculations at higher logarithmic accuracy 
also seems promising. Such an approach was used already to generate 
predictions for the thrust distribution and several event shapes in 
$e^+e^-\to$~jets~\cite{Alioli:2012fc}. First results for $pp\to e^+e^-$ 
have been reported~\cite{Alioli:2013vza}.

Logarithmic enhancements of the cross section at high energy, which are 
resummed in the HEJ framework~\cite{Andersen:2009nu,Andersen:2011hs}
could be crucial to understanding the structure of multi-jet events 
at the LHC. Including these contributions in event generators may become
important~\cite{Andersen:2011zd}.


\subsection{Beyond NNLO: Higgs boson production }
\label{sec:qcd-xsecs-approxn3lo}

\draftnote{ Text below written by S. Forte}

Small $x$ ``BFKL'' (or high-energy) resummation and large $x$ ``Sudakov'' (or
threshold/soft-gluon) resummation provide information on the
all-order behavior of a wide class of hadron collider observables in
two opposite kinematic limits. Because the Mellin transform of a
partonic cross section $\sigma(N,\alpha_s(M^2))$
is an analytic function of the variable $N$
which is conjugate to the longitudinal momentum scaling variable (usually
called $x$ or $\tau$), this information provides powerful constraints
on the unknown higher order perturbative corrections to the
cross-section. 

The use of resummation to determine approximate higher
order perturbative corrections has a long history, and, in particular,
approximate NNLO jet cross sections determined using results from
threshold resummation~\cite{Kidonakis:2000gi} are routinely used in PDF fits.
Recently, in Ref.~\cite{Ball:2013bra}, it was suggested that especially accurate
results can be obtained if maximal use is made of analyticity
constraints, by not only combining information from different kinds
of resummation, but also by making sure that the known all-order analytic
properties of the cross section are reproduced as much as
possible. So, for instance, while as $N\to\infty$
$\sigma(N,\alpha_s(M^2))\sim \sum_k(\alpha_s(Q^2) \ln^2 N)^k$, the
cross section is expected to have poles and not cuts when  $N=0$.  Indeed, a more detailed analysis reveals that the logarithmic
behaviour of the cross section only arises through functions such as
$\psi_0(N)$, which indeed has a simple pole at $N=0$ even though
$\psi_0(N)\sim \ln N$ as $N\to\infty$.

In Ref.~\cite{Ball:2013bra} it was shown that indeed this approach leads to a very good
approximation to the known NLO and NNLO expressions for the total
cross section for Higgs production in gluon fusion with finite $m_t$.
\draftnote{In particular, this approximation is rather better than it would be found by simply
expanding out the standard resummed result of
Ref.~\cite{deFlorian:2012yg}.} An approximate expression for the
N$^3$LO correction to the cross section was then constructed.

\begin{figure}[htb]
\begin{center}
 \includegraphics[width=0.49\hsize]{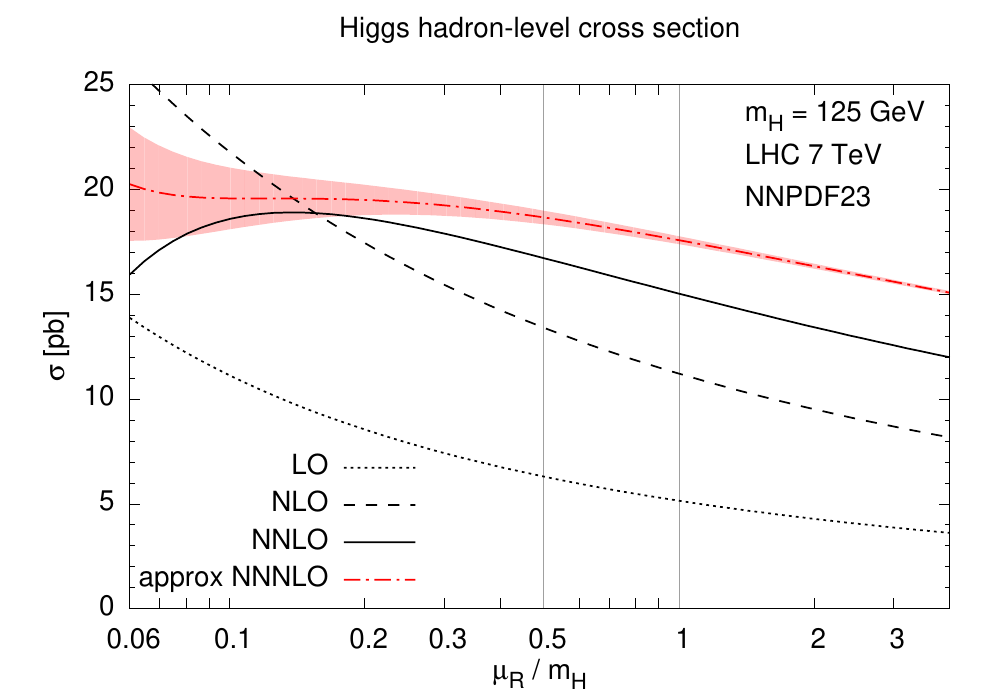}
 \includegraphics[width=0.49\hsize]{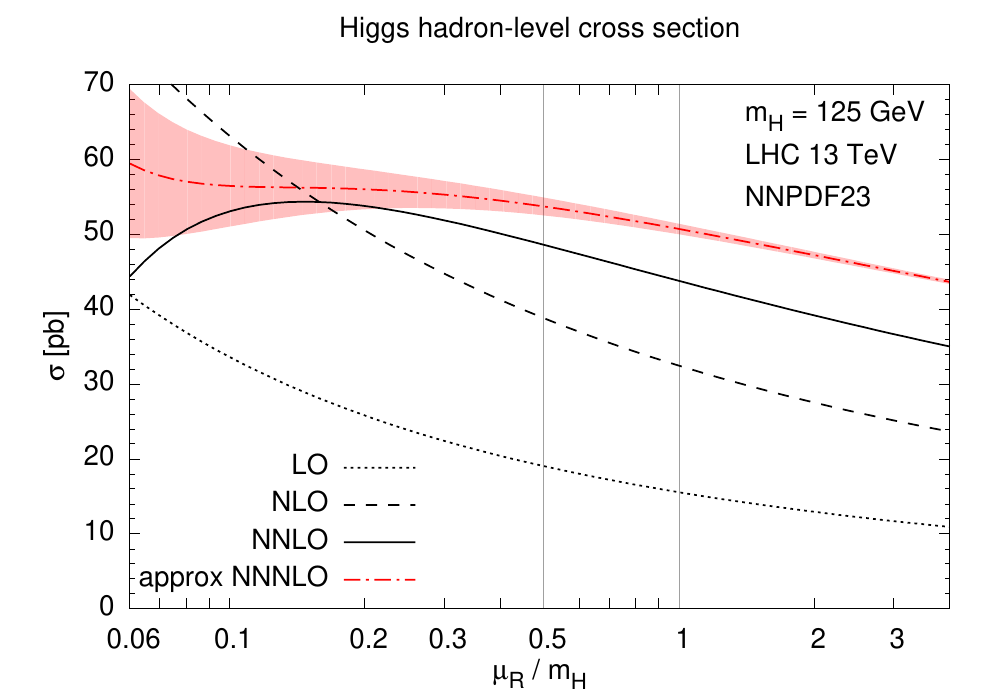}\\
 \includegraphics[width=0.49\hsize]{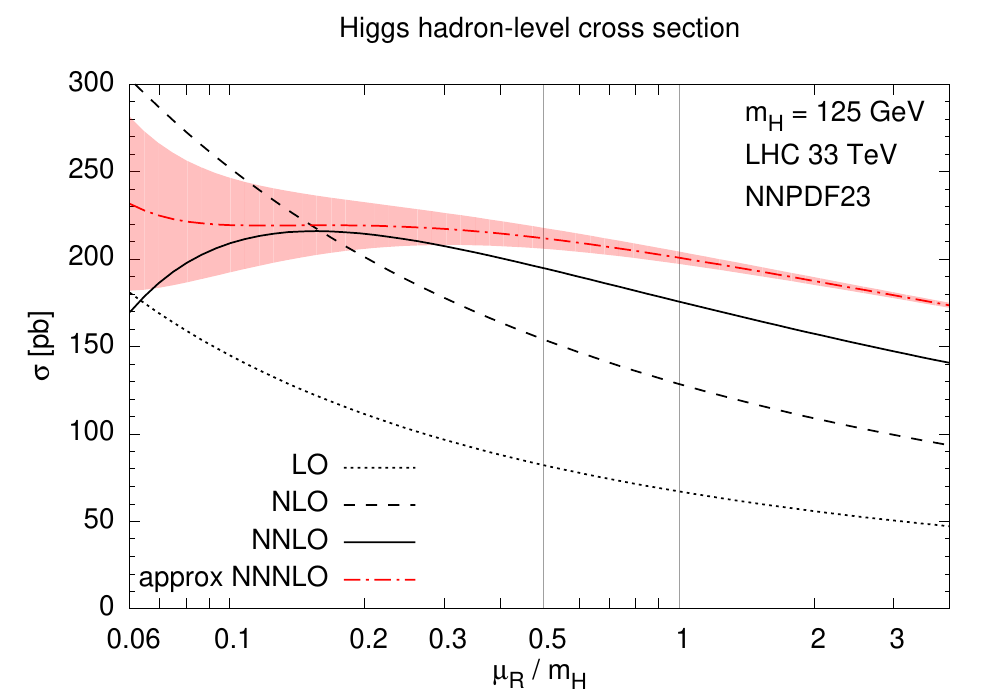}
 \includegraphics[width=0.49\hsize]{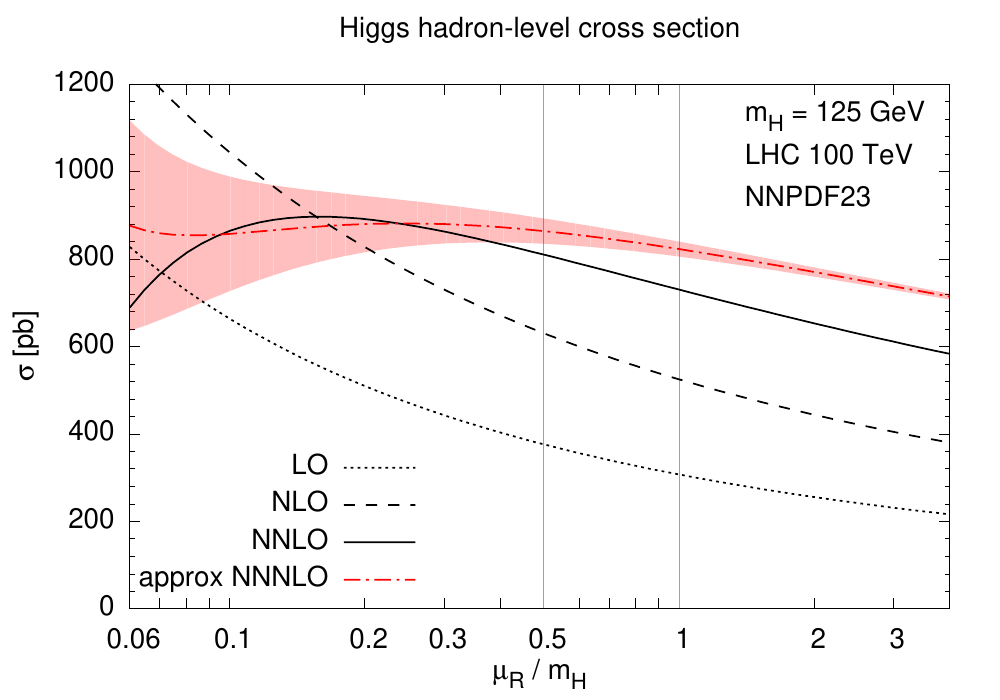}\\
\caption{Dependence on the renormalization scale
of the LO, NLO, NNLO and approximate N$^3$LO contribution from the
gluon-gluon channel to the total cross section for Higgs production at
a proton-proton collider with four different values of the collider
energy. The results shown are obtained using the NNPDF2.3 PDF set with
 $\amz=0.118$}
\label{fig:qcd-xsecs-approxn3lo-scale}
\end{center}
\end{figure}

The full N$^3$LO Higgs production cross section at the LHC
at $\sqrt{s}=8$~TeV, with $m_H=125$~GeV was found to be
\begin{eqnarray}
\sigma_{\rm approx}^{{\rm N}^3{\rm LO}}(\tau,m_H^2)&=&\sigma^{(0)}(\tau,m_H^2)
\left[ \sum_{ij}\left(\delta_{ig}\delta_{jg}+ \alpha_s K_{ij}^{(1)}+\alpha_s^2 K_{ij}^{(2)} \right)+\alpha_s^3 K_{gg,{\rm approx}}^{(3)} \right]
\nonumber\\
&=& \left( 22.61 \pm 0.27  +0.91\cdot 10^{-2} \bar g_{0,3} \right) {\rm ~pb} \qquad {\rm for}~\mu_R=m_H \label{eq:qcd-higgs-approxn3lo} \\
&=& \left( 24.03 \pm 0.45  +1.55\cdot 10^{-2} \bar g_{0,3} \right) {\rm ~pb} \qquad {\rm for}~\mu_R=m_H/2, \nonumber 
\end{eqnarray}
using the NNPDF2.1 PDF set with $\alpha_s(M_z)=0.119$, where the error  shown is an estimate of the uncertainty in the
approximation procedure, and the coefficient $\bar g_{0,3}$ is
unknown. The
known perturbative behaviour of the coefficients $g_{0,i}$, which
provide constant corrections to the cross section (i.e. neither
logarithmically enhanced nor power-suppressed as $N\to\infty$)
suggests that  $\bar g_{0,3}$ is possibly of order ten.
The renormalization scale dependence of the contribution from the
gluon-gluon channel to the cross section is shown in Fig.~\ref{fig:qcd-xsecs-approxn3lo-scale}
for various choices of the collider energy (red band), and compared to
the exact LO, NLO, and NNLO results, and also to the a different soft
approximation and its collinear improvement (see below). Note that the
factorization scale dependence of the result is known to be
essentially negligible even at LO, more so at NLO and NNLO.

\begin{figure}[htb]
\begin{center}
 \includegraphics[width=0.49\hsize]{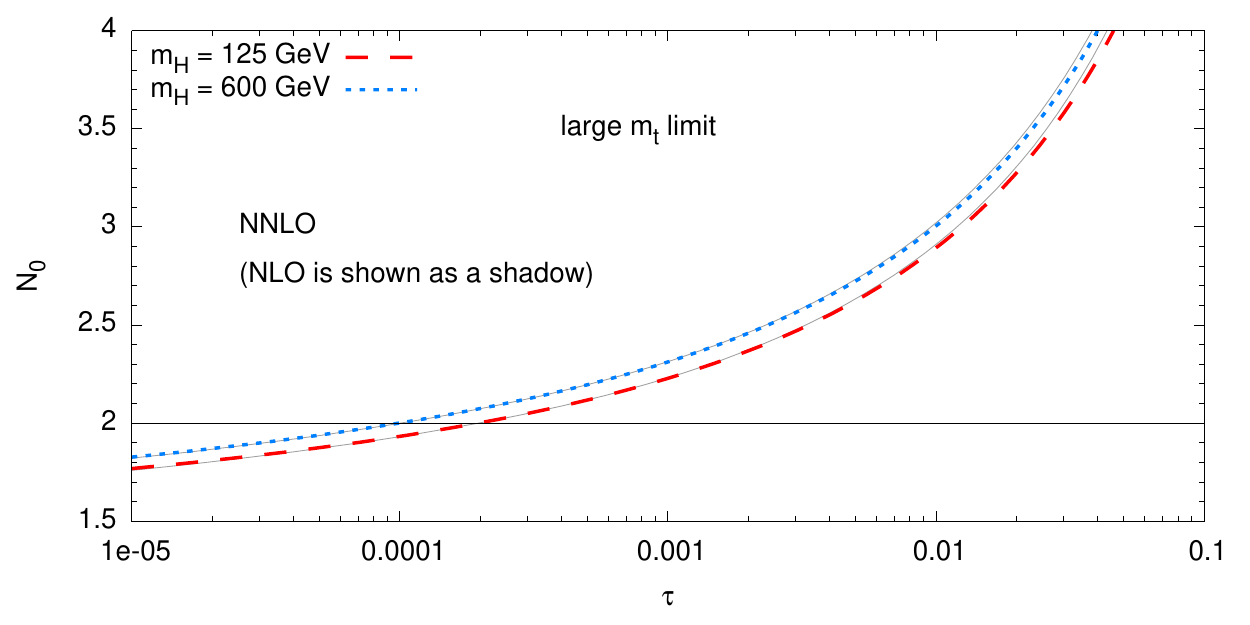}
 \includegraphics[width=0.445\hsize]{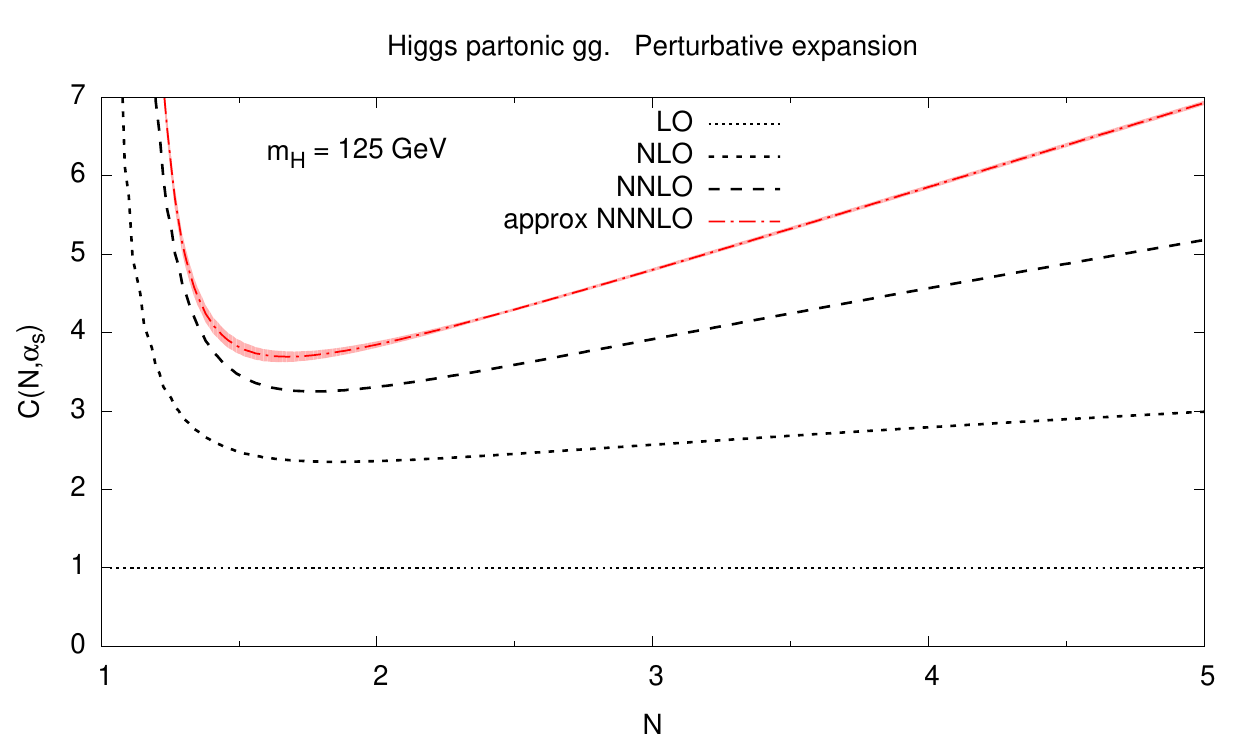}
 \caption{Position of the saddle-point value of $N$ which dominates
   the Mellin-space cross section for production of a 125~GeV Higgs
   boson in gluon fusion as a function of the collider energy (left),
   and perturbative expansion of the cross-section (right)}
\label{fig:qcd-xsecs-approxn3lo-saddle}
\end{center}
\end{figure}

The main features of this approximate result are the following:
\begin{itemize}
\item The perturbative expansion converges quite slowly: in
  particular, it is clear that at each order the next-order result is
  not contained within the range found varying the
  scale by a factor two about either $m_H$ or $m_H/2$.
\item The perturbative expansion converges better as the collider
  energy increases. The reason for this can be understood by computing the
  value of $N$ which dominates the cross-section~\cite{Bonvini:2012an},
  which is fully determined by the collider energy and the Higgs mass,
  and then studying the perturbative behavior of the cross section
  for the given value of $N$ (see
  Fig.~\ref{fig:qcd-xsecs-approxn3lo-saddle}). 
\item For all collider energies, the scale dependence is considerably
  reduced by the inclusion of 
  the N$^3$LO corrections.
\item The central prediction of Ref.~\cite{Ball:2013bra}, Eq.~(\ref{eq:qcd-higgs-approxn3lo}), 
amounts to a rather
  substantial correction, of order of 17\% for $\mu_R=m_H$ at LHC~8~TeV. \draftnote{FP: I would like to replace this 
  sentence by a comment on the shift from $\mu_R=m_H/2$, which I think is more representative of the 
  magnitude of this effect.}
\item The N$^3$LO truncation of the resummed result of
  Ref.~\cite{deFlorian:2012yg}, together with its collinear
  improvement according to 
  Ref.~\cite{Catani:2003zt} would predict a rather smaller  correction to the
  NNLO result, of order of 6\%  for $\mu_R=m_H$.
\item The whole NNLL
  correction to the NNLO result from  Ref.~\cite{deFlorian:2012yg}
  modifies the NNLO result by about 8\%, 6\% of which, as mentioned,
  comes from the N$^3$LO, and the remaining 2\% or so from higher
  orders. This means that the resummation is
  perturbative in this region.  It mostly amounts  to a prediction
  for the N$^3$LO correction.
\item The discrepancy between the prediction of a 6\% correction
  N$^3$LO 
  (expanding out the resummation of Ref.~\cite{deFlorian:2012yg}) and a a 17\% correction (using the
  approximation of Ref.~\cite{Ball:2013bra}) is partly due to the choice of the
  value for the constant $\bar g_{0,3}$. In fact, using the value of
  the constant which is implicit in the resummed result of
  Ref.~\cite{deFlorian:2012yg} reduces the N$^3$LO correction of
  Eq.~(\ref{eq:qcd-higgs-approxn3lo}) from about 17\% to about 12\%. The remaining
  difference \draftnote{between the approximations of Ref.~\cite{deFlorian:2012yg} and 
  Ref.~\cite{Ball:2013bra},} is due to the (allegedly more accurate) approximation of
  Ref.~\cite{Ball:2013bra}.
\item The difference between the  approximation of
  Ref.~\cite{Ball:2013bra} and the expansion of the resummed result is mostly due
  to the fact that the soft approximation in     Ref.~\cite{Ball:2013bra} is
  designed to
  preserve the small $N$ singularity structure. The explicit inclusion
  of the correct small $N$ terms from the ``BFKL'' resummation stabilized somewhat the scale dependence at the very
  lowest edge $\mu_R/m_H < 0.1$ of the scale variation range of
  Fig.~\ref{fig:qcd-xsecs-approxn3lo-scale}, but it otherwise has a small impact.
\item The scale dependence of the N$^3$LO result was also determined
  in Ref.~\cite{Buehler:2013fha} as a function of the value of the cross-section at
  the reference scale $\mu_R=m_H/2$. It
  was found that if this value is such that the scale dependence of
  the N$^3$LO is smaller than that of the NNLO, then the  N$^3$LO is
  in the same ballpark as found in Ref.~\cite{Ball:2013bra}.
\end{itemize}

There are ongoing efforts to complete a full N$^3$LO calculation of the
$gg \rightarrow H$ cross section, i.e. to provide the value of
the coefficient $\bar g_{0,3}$ in Eq.~\ref{eq:qcd-higgs-approxn3lo}.
Given the slow convergence of the
perturbative series for this process,  a full calculation to this order may
be necessary to achieve the needed theoretical precision for Higgs
production, both for $14$~TeV and for still higher energies. An extraction of PDFs at N$^3$LO 
is strictly needed in order to have a complete prediction for a hadron-collider cross section at N$^3$LO.  
In particular, the DGLAP kernels which control the $Q^2$ evolution of the PDFs would need to be computed to one higher order than currently known.  This calculation is unlikely to be performed soon.  However, it is easy to check that the shifts in predicted Higgs cross sections upon changing from NLO to NNLO PDFs is much smaller than the change induced by going from NLO to NNLO in the coefficient function, once the value of $\alpha_s$ is fixed~\cite{qcd:sforte}.  This mismatch in the order of perturbation theory and the PDF order is therefore likely not a major phenomenological obstacle.


\subsection{Wishlist for higher order QCD and EW corrections}
\label{sec:qcd-qcdew-wishlist}

\draftnote{3 pages: J. Huston to start, based on Les Houches wishlist \\
Wishlist and putting calculations together in one framework}

The Les Houches NLO wish list, consisting of calculations that were phenomenologically important for LHC physics, and
were feasible  but difficult to calculate at NLO in perturbative QCD, was started in 2005. After being incremented in
2007 and 2009 it was terminated in  2011. By 2011, every calculation on the wish list had been completed, and
technology had advanced far enough that any reasonable multi-parton calculation could be carried out at NLO using
semi-automated technology. In 2013, the NLO QCD wish list was replaced by one focussing on NNLO and the combination
of QCD and EW corrections, the new calculational frontier. The inclusion of electroweak corrections is
vital given the precision inherent at NNLO.  The new wish list~\footnote{To appear in the 2013 Les Houches proceedings.} is shown in Tables~\ref{tab:qcd-wishlist1},~\ref{tab:qcd-wishlist2}
and~\ref{tab:qcd-wishlist3}, giving the level to which the current calculation is known, and the level which is desired
for full exploitation of physics at the LHC and higher energy hadron-hadron colliders. For the Higgs boson processes
listed in Table~\ref{tab:qcd-wishlist1} the improved calculations will enable more accurate extractions of Higgs
couplings. The processes involving heavy quarks and jets, Table~\ref{tab:qcd-wishlist2}, will predominantly provide
better extractions of PDFs.  The vector boson processes given in Table~\ref{tab:qcd-wishlist3} will be essential in
investigating the precise nature of electroweak symmetry breaking, by providing more accurate predictions for channels
that are sensitive to vector boson scattering at high energy and to anomalous cubic and quartic gauge boson couplings.

Until recently, the state-of-the-art for NNLO was the calculation of $2\rightarrow1$ processes. Within the last few
years, several calculations of $2\rightarrow2$ processes have been completed. Indeed, the year 2013 has seen the
completion of a number of landmark calculations at NNLO, namely the total cross section for top pair
production~\cite{Czakon:2013goa} and first approximations of jet production~\cite{Ridder:2013mf} and the Higgs + jet
process~\cite{Boughezal:2013uia}. It is noteworthy that the wish list even contains $2\rightarrow3$ processes at NNLO.
Adding to the complexity is the need for the inclusion of decays for many of the massive
final state particles. Given the recent progress in the field, it is difficult to speculate as to what length of
time will be needed for the completion of this new list, but a period of 10 years may be a reasonable estimate. 

Note that for many processes the higher order QCD and the higher order EW
corrections are currently known separately, while the desire is to have combined
corrections, often at NNLO in QCD and NLO in EW. One of the ambiguities in
situations where the corrections are known separately is whether the two
corrections are multiplicative or additive, i.e., whether the EW corrections are
affected by the (often) large QCD corrections. The degree to which the
corrections are multiplicative or additive no doubt depends on the particular
process and even on the observable within that process. The joint calculations
posited here will resolve this ambiguity.

\def\mathswitchr#1{\relax\ifmmode{\mathrm{#1}}\else$\mathrm{#1}$\fi}
\newcommand{\rd}{{\mathrm{d}}}
\def\ga{\gamma}
\newcommand{\Pb}{\mathswitchr b}
\newcommand{\Pc}{\mathswitchr c}
\newcommand{\Pg}{\mathswitchr g}
\newcommand{\Pj}{\mathswitchr j}
\newcommand{\Pt}{\mathswitchr t}
\newcommand{\PH}{\mathswitchr H}
\newcommand{\PV}{\mathswitchr V}
\newcommand{\PW}{\mathswitchr W}
\newcommand{\PZ}{\mathswitchr Z}
\newcommand{\sigtot}{\sigma_{\mathrm{tot}}}
\newcommand{\dsig}{\rd\sigma}

\begin{table}[p]
\begin{center}
\begin{tabular}{|l|l|l|l|}
\hline
Process & known & desired & details
\\
\hline
$\PH    $    & $\dsig$ @ NNLO QCD & $\dsig$  @ NNNLO QCD + NLO EW           & H branching ratios   \\
             & $\dsig$ @ NLO EW           & MC@NNLO   & and couplings \\
             & finite quark mass effects @ NLO  & finite quark mass effects @ NNLO &   \\
\hline
$\PH+\Pj$    & $\dsig$ @ NNLO QCD ($\Pg$ only) & $\dsig$ @ NNLO QCD + NLO EW            & H $p_T$ \\
             & $\dsig$ @ NLO EW           &   finite quark mass effects @ NLO & \\
             & finite quark mass effects @ LO  &   & \\
\hline
$\PH+2\Pj$   & $\sigtot$(VBF) @ NNLO(DIS) QCD & $\dsig$ @ NNLO QCD + NLO EW            & H couplings \\
             & $\dsig(\Pg\Pg)$ @ NLO QCD      &                     & \\
             & $\dsig$(VBF) @ NLO EW          &                     & \\
\hline
$\PH+\PV$    & $\dsig$ @ NNLO QCD &   with $H \to \Pb\bar\Pb$ @ same accuracy                 & H couplings  \\
             & $\dsig$ @ NLO EW   &                     & \\
\hline
$\Pt\bar\Pt\PH$ & $\dsig$(stable tops) @ NLO QCD & $\dsig$(top decays) & top Yukawa coupling\\
                &                                & @ NLO QCD + NLO EW  & \\
\hline
$\PH\PH$ & $\dsig$ @ LO QCD (full $m_t$ dependence) & $\dsig$ @ NLO QCD (full $m_t$ dependence)& Higgs self coupling\\
         & $\dsig$ @ NLO QCD (infinite $m_t$ limit) & $\dsig$ @ NNLO QCD (infinite $m_t$ limit) & \\
\hline
\end{tabular}
\end{center}
\caption{Wishlist part 1 -- Higgs ($\PV=\PW,\PZ$).
\label{tab:qcd-wishlist1}}
\end{table}

\begin{table}[p]
\begin{center}
\begin{tabular}{|l|l|l|l|}
\hline
Process & known & desired & details
\\
\hline
$\Pt\bar\Pt$ & $\sigtot$ @ NNLO QCD            & $\dsig$(top decays) & precision top/QCD, \\
             & $\dsig$(top decays) @ NLO QCD   & @ NNLO QCD + NLO EW & gluon PDF, effect of extra \\
             & $\dsig$(stable tops) @ NLO EW   &                     & radiation at high rapidity,  \\
             &                                 &                     & top asymmetries               \\
\hline
$\Pt\bar\Pt+\Pj$ & $\dsig$(NWA top decays) @ NLO QCD & $\dsig$(NWA top decays)  & precision top/QCD   \\
                 &                                   & @ NNLO QCD + NLO EW      & top asymmetries  \\
\hline
single-top        & $\dsig$(NWA top decays)  @ NLO QCD & $\dsig$(NWA top decays) & precision top/QCD, $V_{tb}$ \\
                  &                                    & @ NNLO QCD (t channel)  & \\
\hline
dijet        & $\dsig$ @ NNLO QCD ($\Pg$ only) & $\dsig$             & Obs.: incl. jets, dijet mass\\
             & $\dsig$         @ NLO weak & @ NNLO QCD + NLO EW & $\to$ PDF fits (gluon at high x)\\
             &  &   & $\to$ $\as$\\
\hline
$3\Pj$       & $\dsig$ @ NLO QCD & $\dsig$             & Obs.: $R3/2$ or similar \\
             &                   & @ NNLO QCD + NLO EW & $\to$ $\as$ at high scales\\
             &  &   & dom. uncertainty: scales \\
\hline
$\ga+\Pj$ & $\dsig$ @ NLO QCD & $\dsig$ @ NNLO QCD & gluon PDF\\
          & $\dsig$ @ NLO EW & +NLO EW & $\ga+\Pb$ for bottom PDF\\
\hline
\end{tabular}
\end{center}
\caption{Wishlist part 2 -- jets and heavy quarks.
\label{tab:qcd-wishlist2}}
\end{table}

\begin{table}[p]
\begin{center}
\begin{tabular}{|l|l|l|l|}
\hline
Process & known & desired & details
\\
\hline
$\PV    $    & $\dsig$(lept. $\PV$ decay) @ NNLO QCD & $\dsig$(lept. $\PV$ decay)    & precision EW, PDFs  \\
             & $\dsig$(lept. $\PV$ decay) @ NLO EW   & @ NNNLO QCD + NLO EW          &  \\
             &                                       & MC@NNLO   &  \\
\hline
$\PV+\Pj$    & $\dsig$(lept. $\PV$ decay) @ NLO QCD & $\dsig$(lept. $\PV$ decay) & $\PZ+\Pj$ for gluon PDF \\
             & $\dsig$(lept. $\PV$ decay) @ NLO EW  & @ NNLO QCD + NLO EW  & $\PW+\Pc$ for strange PDF\\
\hline
$\PV+\Pj\Pj$ & $\dsig$(lept. $\PV$ decay) @ NLO QCD & $\dsig$(lept. $\PV$ decay) & study of systematics of  \\
             &                                      & @ NNLO QCD + NLO EW  & $\PH+\Pj\Pj$ final state\\
\hline
$\PV\PV'$    & $\dsig$($\PV$ decays) @ NLO QCD & $\dsig$($\PV$ decays) & off-shell leptonic decays\\
             & $\dsig$(stable $\PV$) @ NLO EW  & @ NNLO QCD + NLO EW  & TGCs\\
\hline
$\Pg\Pg \to \PV\PV$    & $\dsig$($\PV$ decays) @ LO QCD & $\dsig$($\PV$ decays) & bkg. to $H \to VV$\\
                       &                                & @ NLO QCD             & TGCs\\
\hline
$\PV\ga$    & $\dsig$($\PV$ decay) @ NLO QCD & $\dsig$($\PV$ decay) & TGCs\\
             & $\dsig$(PA, $\PV$ decay) @ NLO EW  & @ NNLO QCD + NLO EW  &  \\
\hline
$\PV\Pb\bar\Pb$ & $\dsig$(lept. $\PV$ decay) @ NLO QCD & $\dsig$(lept. $\PV$ decay) @ NNLO QCD & bkg. for $\PV\PH\to\Pb\bar\Pb$ \\
                & massive $\Pb$                        & massless $\Pb$                        &  \\
\hline
$\PV\PV'\ga$ & $\dsig$($\PV$ decays) @ NLO QCD & $\dsig$($\PV$ decays) & QGCs \\
                &                              &  @ NLO QCD + NLO EW        &  \\
\hline
$\PV\PV'\PV''$ & $\dsig$($\PV$ decays) @ NLO QCD & $\dsig$($\PV$ decays) & QGCs, EWSB\\
                &                              &  @ NLO QCD + NLO EW        &  \\
\hline
$\PV\PV'+\Pj$ & $\dsig$($\PV$ decays) @ NLO QCD & $\dsig$($\PV$ decays) & bkg. to H, BSM searches\\
                &                              &  @ NLO QCD + NLO EW        &  \\
\hline
$\PV\PV'+\Pj\Pj$ & $\dsig$($\PV$ decays) @ NLO QCD & $\dsig$($\PV$ decays) & QGCs, EWSB \\
                &                              &  @ NLO QCD + NLO EW        &  \\
\hline
$\ga\ga$     & $\dsig$ @ NNLO QCD & & bkg to $H\to \ga\ga$ \\
\hline
\end{tabular}
\end{center}
\caption{Wishlist part 3 -- EW gauge bosons ($\PV=\PW,\PZ$).
\label{tab:qcd-wishlist3}}
\end{table}


\section{Electroweak corrections and Sudakov logarithms}
\label{sec:qcd-qcdew-sudakov}

\draftnote{5 pages: solicit text from Dittmaier?;
 \\ photon-photon processes, electroweak corrections}

\newcommand{\ensuremathr}[1]{\ensuremath{\mathrm{#1}}}
\newcommand{\alphaw}{\ensuremath{\alpha_{\mathrm{w}}}}
\newcommand{\rw}{\ensuremathr{w}}
\newcommand{\order}[1]{\ensuremath{ {\mathcal{O}\left( #1 \right)} }}
\newcommand{\lvert}{\left|}
\newcommand{\rvert}{\right|}
\def\met {\ensuremath{{E\!\!\!/}_{\mathrm{T}}}\xspace}

At future high-energy collider experiments the electroweak (EW) radiative corrections are
generally expected to become more important due to the fact that they include terms of the form,
$\alphaw\ln^2\left(Q^2/M_\PW^2\right)$, where $Q$ denotes the energy scale of the hard-scattering process,
$M_\PW$ is the $W$-boson mass, and  $\alphaw=\alpha/\sin^2\theta_\rw=e^2/(4\pi\sin^2\theta_\rw)$ with
$\theta_\rw$ denoting the weak mixing angle.  The corrections that contain such terms, called Sudakov logs,
are generated by diagrams in which  virtual and real gauge bosons are radiated by external leg particles.  They 
correspond  to the soft and collinear singularities appearing in QED and QCD, i.e. when massless  gauge
bosons are involved.  At variance with this latter case, the weak boson  masses act as a physical cutoff on
these ``singularities'', so that virtual and real weak  bosons corrections can be considered separately.
Experimentally the radiation of real  weak bosons is in principle detectable and event selections can me made
such that it is not included.  In this case the physical effect of  virtual  corrections is singled out and
can amount to several tens of per cent, or more. The one-loop  Sudakov logs are naturally included in any
complete calculation of  NLO EW radiative corrections to a given process. 

Often, including only the effect of double and single Sudakov logs can be a reasonable
approximation to the full NLO electroweak corrections for a process. 
This approach misses finite contributions of order $\alpha$, but
can work very well in the Sudakov regime, where $s$ and $|t|$ are both large ($\gg m_W^2$). 
This type of approach is well-suited to implementation by process-independent methods in 
Monte Carlo event generators. 



For many of the interesting kinematic regions at both the LHC, and at higher
energy colliders, there can be simultaneously large QCD corrections and large
electroweak corrections. In the absence of a calculation that accounts for the
presence of both corrections simultaneously, one can provide an
interim solution by either multiplying or adding together
the separate QCD and EW corrections.  These two possibilities lead to 
cross sections of the schematic forms
\begin{eqnarray}
\sigma_{add} &\sim& \sigma_0 \left[ 1+{\cal O}(\alpha_s)+{\cal O}(\alpha_{EW}) \right], \nonumber \\
\sigma_{mult} &\sim& \sigma_0 \left[ 1+{\cal O}(\alpha_s)\right] \times \left[ 1+{\cal O}(\alpha_{EW})\right],
\end{eqnarray}
where the combinations are denoted by $\sigma_{add}$ and $\sigma_{mult}$, and the
Born-level cross section  by $\sigma_0$.  Expanding the product of brackets in
$\sigma_{mult}$ shows that these two prescriptions differ  by terms of order ${\cal
O}(\alpha_s \alpha_{EW})$.  Only a complete calculation of these mixed corrections  can
determine whether either of these prescriptions provides an accurate description of the
perturbative expansion.  Even in current data, this ambiguity can result in significant
uncertainties in the comparison of standard model predictions to the data.  
The relative ${\cal O}(\alpha_s \alpha_{EW})$ corrections have been calculated for inclusive Higgs production, 
in a tractable parametric limit~\cite{Anastasiou:2008tj}. And recently, the relative ${\cal O}(\alpha_s \alpha_{EW})$ corrections 
have also been calculated for W and Z production, using a resonance expansion around the W/Z pole~\cite{Dittmaier}. 
While both results suggest that the multiplicative combination is a good approximation, this conclusion 
is not necessarily applicable to other processes.

In the following each of these issues will be discussed in turn. Also, the
whitepaper by K.~Mishra {\it et al.}~\cite{Mishra:2013una} presents a nice summary of implications
of electroweak corrections at high energies.


\subsection{Importance of Sudakov logarithms for basic processes}
\label{sec:qcd-qcdew-mishra}

\draftnote{Text below from K. Mishra}

\begin{figure}[htb]
\begin{center}
\includegraphics[width=0.37\textwidth]{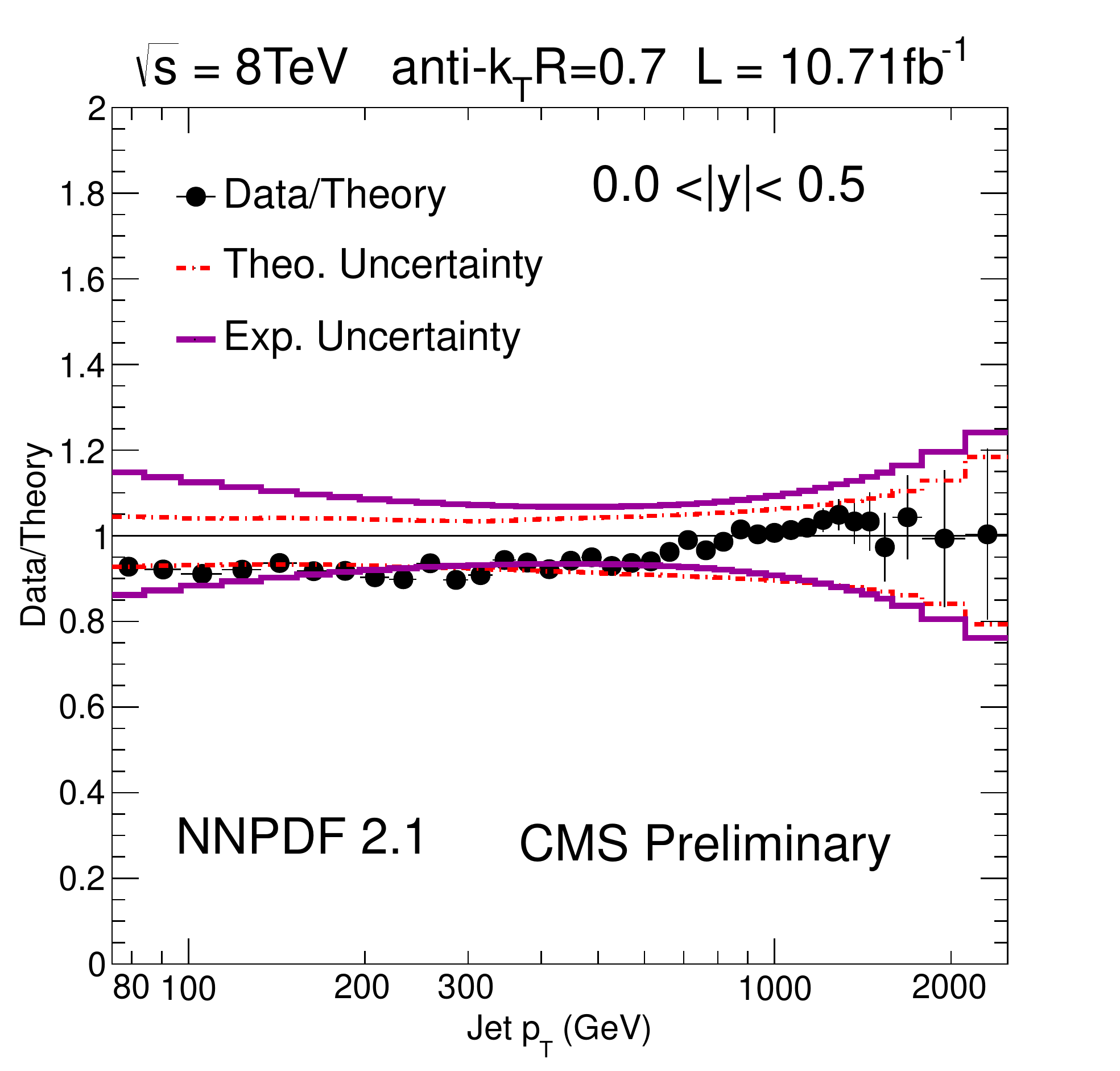}
\includegraphics[width=0.45\textwidth,bb=301 356 510 520,clip=]{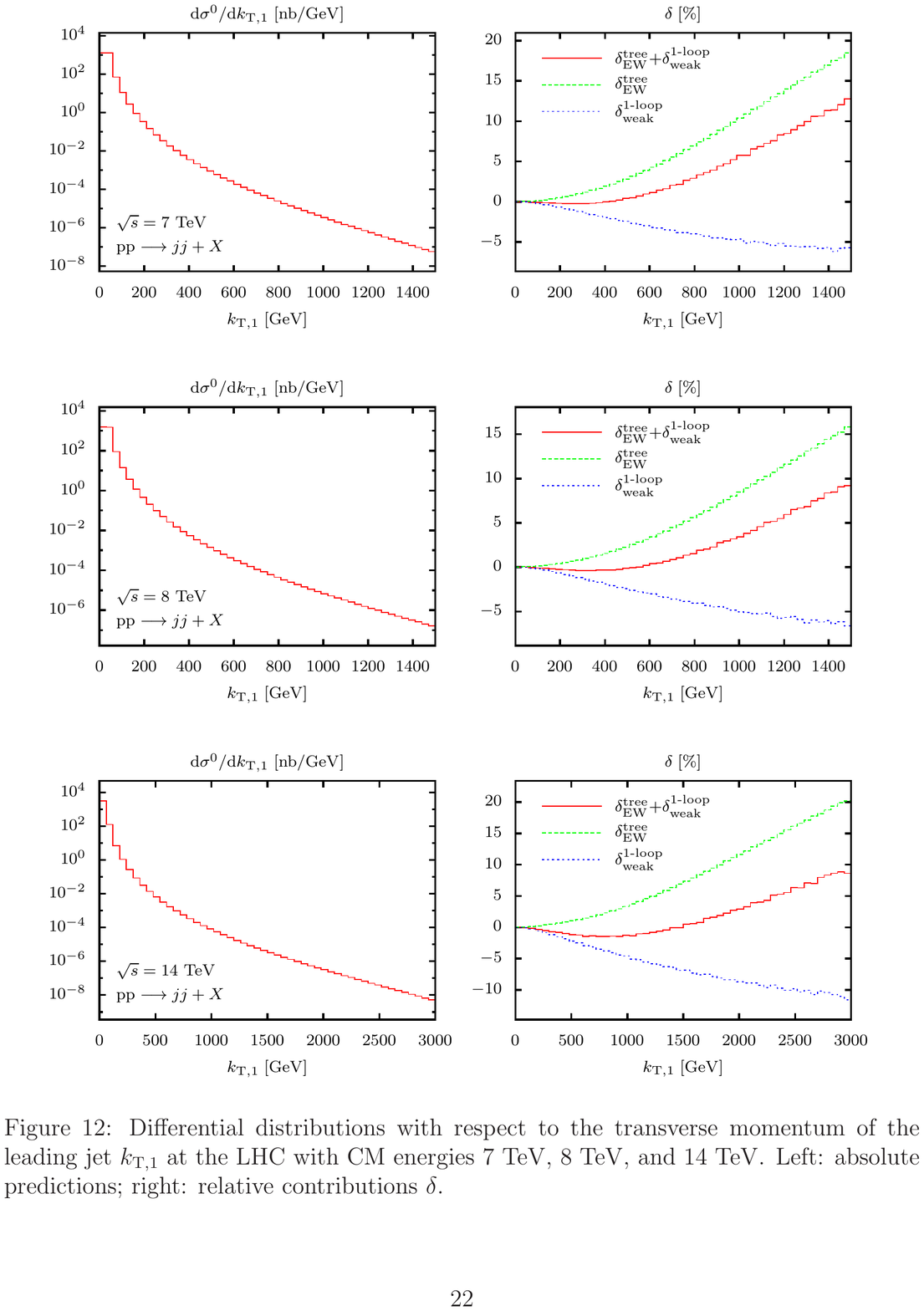}
\caption{The uncertainties in the jet transverse momentum measured in 8 TeV pp collisions~\cite{cms:dijet}
(left) and the relative magnitude of electroweak corrections in pp collisions at 8~TeV~\cite{Dittmaier:2012kx}
(right).}
\label{fig:dijet_exp}
\end{center}
\end{figure}

\begin{figure}[htb]
\begin{center}
\includegraphics[width=0.7\textwidth]{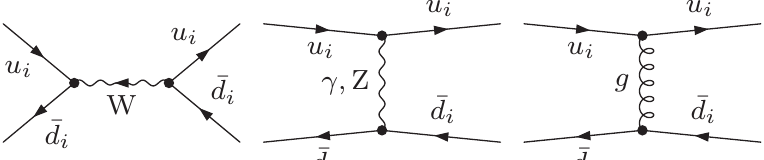} \\
($d_1$)\hspace*{0.2\textwidth}($d_2$)\hspace*{0.2\textwidth}($d_3$)
\\ \vspace*{0.1cm}
\caption{Representative tree-level diagrams that enter the calculation of EW
corrections to dijet production presented in Ref.~\cite{Dittmaier:2012kx}.
\label{fig:dijet-ewtree}}
\end{center}
\end{figure}

\paragraph{Dijet production}
The inclusive production of two jets (dijets) allows for a 
detailed study of QCD at TeV energies. It is also the main 
background for searches of new heavy particles from Beyond 
Standard Model (BSM) physics decaying into dijet signatures. 
Inclusive jet and dijet production have been analyzed by the
ATLAS~\cite{Aad:2011fc} and CMS~\cite{cms:dijet} 
Collaborations at $\sqrt{s} = 7$ and 8~TeV showing 
sensitivity to dijet invariant masses of up
to $5~\TeV$ and jet transverse momenta of up to $2~\TeV$
at the LHC. At the 
current level of experimental and theoretical accuracy,
the SM is able to describe data well. 
However, the size of the EW correction~\cite{Dittmaier:2012kx} is comparable 
to the experimental uncertainty for the highest \pt bins,
as shown in Fig.~\ref{fig:dijet_exp}.  Given this sensitivity
we thus say that the dijet measurement at $\sqrt{s} = 8~\TeV$ has already 
started probing the ``Sudakov zone''.  
The electroweak corrections in Ref.~\cite{Dittmaier:2012kx} comprise
NLO corrections through order $\alphas^2\alpha$ as well as additional
tree-level contributions of order $\order{\alphas\alpha,\,\alpha^2}$.
The tree-level contributions arise through diagrams such as the ones
shown in Fig.~\ref{fig:dijet-ewtree} and correspond to
$|d_1+d_2|^2$ and the interference $(d_1+d_2) d_3^*$.~\footnote{
The contribution $|d_3|^2$ is of course part of the leading order QCD result.}
The tree-level contributions are typically of the same 
size as the loop corrections at $\sqrt{s} = 8~\TeV$. 
The total correction to the integrated cross section 
is negligible, typically staying below the per-cent level. 
However, the Sudakov logarithms 
affect the tails of the distributions in the dijet
invariant mass and in the transverse momenta of the two jets.
The magnitude of the corrections at  $\sqrt{s} = 8~\TeV$ and 14~TeV 
were found to be similar in Ref.~\cite{Dittmaier:2012kx}. 

\begin{figure}[htb]
\begin{center}
\includegraphics[page=1,width=0.45\textwidth,bb=321 351 530 515,clip=]{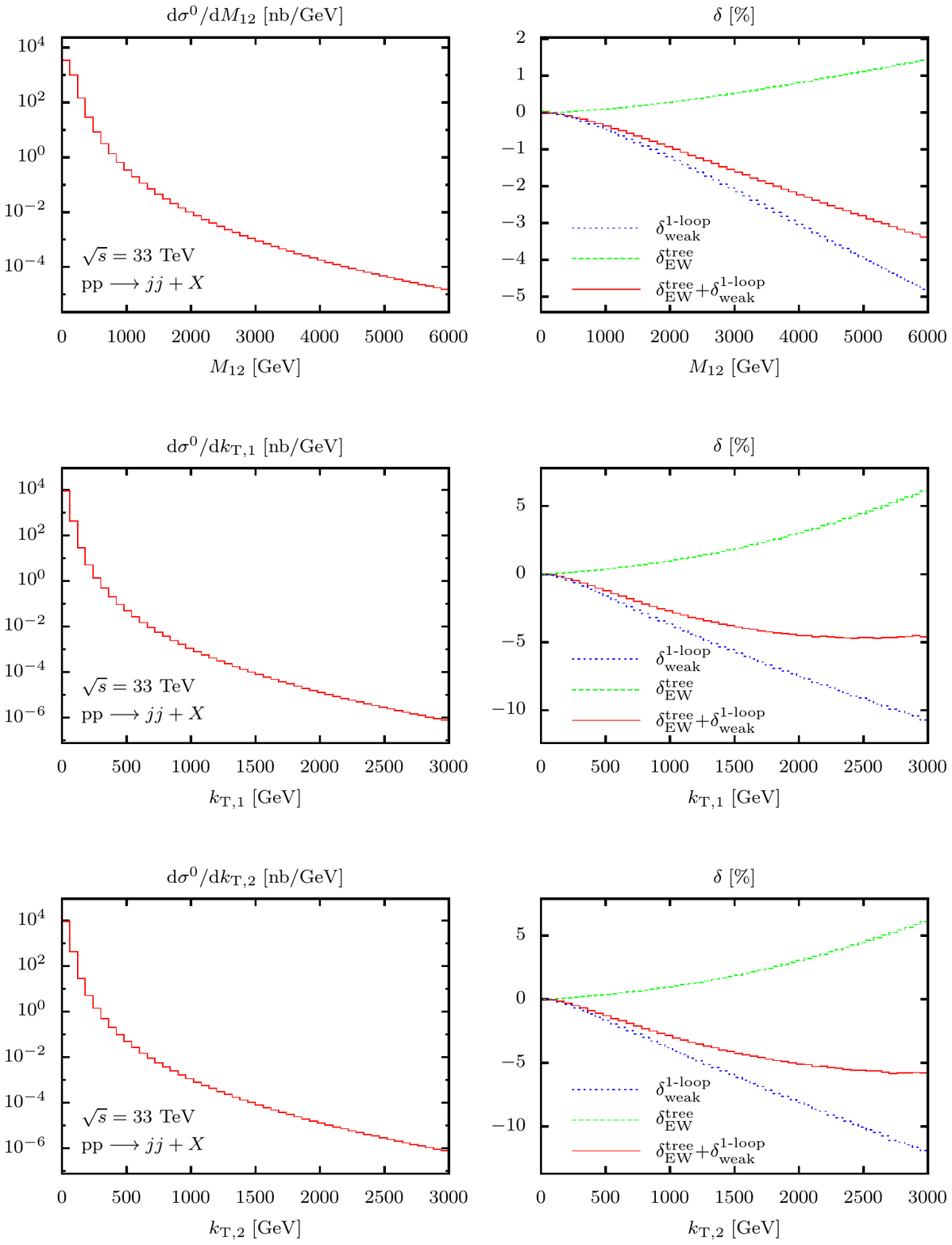}
\includegraphics[page=3,width=0.45\textwidth,bb=321 351 530 515,clip=]{QCD/Fig_dijet_corr.pdf}
\caption{The relative magnitude of electroweak corrections on the leading jet
transverse momentum~\cite{Dittmaier:2012kx} in pp collisions at $33$~TeV (left)
and $100$~TeV (right).}
\label{fig:dijet_corr_higherenergies}
\end{center}
\end{figure}

Results for the same same observable at  
$\sqrt{s} = 33$ and $100$~TeV are shown in Fig.~\ref{fig:dijet_corr_higherenergies}.
The 1-loop virtual corrections do not depend strongly
on the collider energy, while 
the tree-level corrections decrease with $\sqrt{s}$.  The latter effect can be
explained by the fact that the tree-level contributions do not depend on the gluon
distribution and are therefore relatively less important.
The cancellation between the loop and tree contributions is therefore
less perfect and, as a result, the virtual negative corrections dominate
in the kinematic tails  at large $\sqrt{s}$. Since the kinematic reach becomes larger 
as $\sqrt{s}$ increases, these corrections will become progressively  
more important.


\paragraph{Inclusive vector boson production}
The production of a single electroweak boson is one of the basic 
hard-scattering processes at the LHC and constitutes major 
background for BSM searches like $W'\to\ell\nu$ and $Z'\to\ell\ell$ 
and important SM measurements like $H\to ZZ{}^*\to 4\ell$.
Like in the diboson case, the virtual corrections due
to electroweak-boson exchanges can become quite significant. 
Since we are considering single electroweak-boson
production, without additional radiation of soft or collinear
$W$ or $Z$ bosons, the cross section will contain 
Sudakov logarithms that can be as large as 20\% for boson 
$\pt\sim 1~\TeV$ at the LHC. 
These effects need to be included for precise prediction 
of kinematic distribution in the region $\pt\gg M_\PV$. 
Inclusive $W$ and $Z$ spectra have been analyzed by the
ATLAS~\cite{atlas:Zll} and CMS~\cite{cms:Wlnu} 
collaborations at $\sqrt{s} = 8~\TeV$ showing 
sensitivity to invariant masses of up
to $2~\TeV$ and boson transverse momenta of up to $800~\GeV$.
The current experimental uncertainty for invariant masses above 
1~TeV is somewhat larger than the size of the EW corrections
computed in Ref.~\cite{Becher:2013zua}. However, the 
same measurements at 14 TeV will be sensitive to 
probing the Sudakov zone.  Studies of these effects
up to $100$~TeV indicate that the relative corrections
depend only on \pt and are are essentially independent of $\sqrt{s}$.
However, since the kinematic reach naturally increases,
these corrections will become progressively  more important~\cite{Mishra:2013una}.

\paragraph{Vector boson production in association with jets}
The production of a $W$ or $Z$ boson in association with jets 
has played a special role in collider physics. It was the 
dominant background to top-quark pair production at the Tevatron. 
At the LHC it remains an important background for processes 
involving lepton, missing energy, and jets. Prominent examples are 
measurements of top quarks, Higgs boson, and multi-boson production 
and BSM searches for supersymmetry signatures. 
Such measurements also permit stringent tests of the predictions
of the Standard Model.
Measurements of $W$ and $Z$ 
boson production in association with multiple jets have been made by
the ATLAS~\cite{Aad:2012en,Aad:2011qv,atlas:Zjets} and 
CMS~\cite{Chatrchyan:2011ne,Chatrchyan:2012vr,cms:Wbb} 
collaborations at $\sqrt{s} = 7$ and 8~TeV, showing 
sensitivity to boson and leading jet transverse momenta of up 
to about $500~\GeV$ at the LHC. 


The full NLO EW corrections for $W$ + 1-jet production have been
computed for the final state containing a charged
lepton, a neutrino, and a hard jet~\cite{Denner:2009gj}.
The full NLO EW corrections for $Z$ + 1-jet production have also been
computed for the final state containing two charged
leptons and a hard jet~\cite{Denner:2011vu}, and for the monojet scenario,
where the $Z$ decays into two undetected neutrinos~\cite{Denner:2012ts}.
The overall magnitude of these corrections as a function 
of boson \pt is similar to the inclusive $W/Z$ case.
As in that case, the current experimental uncertainty for the highest \pt bins 
is somewhat larger than the size of the EW corrections.  
The same measurements at 14 TeV will be sensitive to 
probing the Sudakov zone.

Results from repeating the same calculation as in 
Ref.~\cite{Denner:2011vu} for $Z$ + 1-jet events at  
$\sqrt{s} = 100~\TeV$ are listed in Table~\ref{tab:corr:Zlljetptj100} 
as a function of  the leading jet \pt.
The relative corrections show very weak dependence   
on $\sqrt{s}$ and depend more strongly on the leading
jet \pt.  A similar statement holds for the dependence of the
corrections on the invariant mass of the boson system. 
As the kinematic reach increases  
with increase in $\sqrt{s}$ these corrections will become progressively  
more important.  
\begin{table} 
                                    $$ \begin{array}{c|rrrrrr}
                                                                  \multicolumn{7}{c}{pp
                                                                    \to \ell^+\ell^-\; \mathrm{jet} + X \;\mbox{at} \;\sqrt{s} =100 \TeV} \\
              \hline p_{\rT,\mathrm{jet}} / \GeV & 100-\infty \;\;\; & 200-\infty \;\;\; & 400-\infty \;\; & 800-\infty \;\; & 2000-\infty \;\; & 4000-\infty \; \\ 
  \hline \hline 
                 \si_{\born}^{\mu = \MZ}/\pba \; & \;              114.29(1)                  \; & \;              23.772(3)                  \; & \;              3.5452(4)                  \; & \;             0.42003(4)                  \; & \;            0.017238(1)                  \; & \;          0.00094403(9)                  \\ 
              \si_{\born}^{\mathrm{var}}/\pba \; & \;              118.30(1)                  \; & \;              23.762(3)                  \; & \;              3.1922(3)                  \; & \;             0.31583(3)                  \; & \;           0.0091290(9)                  \; & \;          0.00035205(3)                  \\ 
  \hline \hline 
                   \de_{\EW}^{\mu = \MZ} / \% \; & \;               -5.62(1)                  \; & \;               -9.57(1)                  \; & \;              -16.86(2)                  \; & \;              -27.11(8)                  \; & \;               -43.5(1)                  \; & \;               -58.8(1)                  \\ 
    \de_{\EW}^{\mathrm{rec}\,,\mu = \MZ} / \% \; & \;               -4.65(3)                  \; & \;               -8.72(2)                  \; & \;              -16.08(2)                  \; & \;              -26.29(4)                  \; & \;              -43.15(7)                  \; & \;               -58.5(2)                  \\ 
 \hline 
                  \de_{\EW}^{\mathrm{var}}/\% \; & \;               -5.50(1)                  \; & \;               -9.29(1)                  \; & \;              -16.38(3)                  \; & \;              -26.36(4)                  \; & \;               -43.2(2)                  \; & \;               -57.5(1)                  \\ 
   \de_{\EW}^{\mathrm{rec}\,,\mathrm{var}}/\% \; & \;               -4.48(2)                  \; & \;               -8.52(2)                  \; & \;              -15.62(2)                  \; & \;              -25.64(4)                  \; & \;              -42.21(7)                  \; & \;               -56.8(1)                  \\ 
  \hline \hline 
                    \de_{\QCD}^{\mu = \MZ}/\% \; & \;                97.4(2)                  \; & \;               146.0(1)                  \; & \;               215.2(2)                  \; & \;               288.7(2)                  \; & \;               378.0(3)                  \; & \;               472.6(5)                  \\ 
                 \de_{\QCD}^{\mathrm{var}}/\% \; & \;                85.4(2)                  \; & \;               130.0(2)                  \; & \;               201.7(1)                  \; & \;               298.8(2)                  \; & \;               487.9(3)                  \; & \;               769.0(7)                  \\ 
 \hline 
              \de_{\QCD,\veto}^{\mu = \MZ}/\% \; & \;                35.7(2)                  \; & \;                54.2(1)                  \; & \;                66.7(1)                  \; & \;                61.3(1)                  \; & \;                13.2(2)                  \; & \;               -43.1(1)                  \\ 
           \de_{\QCD,\veto}^{\mathrm{var}}/\% \; & \;                29.6(2)                  \; & \;                47.4(1)                  \; & \;                65.5(1)                  \; & \;                76.4(3)                  \; & \;                65.6(2)                  \; & \;                51.5(1)                  \\ 
  \hline \hline 
               \de_{\ga,\born}^{\mu = \MZ}/\% \; & \;              0.1218(3)                  \; & \;              0.1400(4)                  \; & \;              0.1681(5)                  \; & \;              0.2114(7)                  \; & \;               0.291(1)                  \; & \;               0.382(1)                  \\ 
            \de_{\ga,\born}^{\mathrm{var}}/\% \; & \;              0.1407(3)                  \; & \;              0.1799(5)                  \; & \;              0.2482(7)                  \; & \;               0.365(1)                  \; & \;               0.630(2)                  \; & \;               1.006(5)                  \\ 
 \hline \hline 
     \si_{\full,\veto}^{\mathrm{var}}/\pba/\% \; & \;               147.0(2)                  \; & \;               32.86(4)                  \; & \;               4.767(4)                  \; & \;               0.475(1)                  \; & \;             0.01124(2)                  \; & \;           0.0003343(7)                  \\ 
\end{array} $$

  \caption{$Z$ + 1-jet production: 
    Integrated cross sections for different cuts on the \pt 
    of the leading jet (jet with highest \pt) at a proton-proton 
   collider with $\sqrt{s} = 100~\TeV$. 
   The LO results are shown both for a variable and for a constant scale. 
   The relative EW corrections $\de_{\EW}$ are given with and without
   lepton-photon recombination. The QCD corrections $\de_{\QCD}$ 
   are presented for a fixed as well
   as a for variable scale and with or without employing a veto on
   a second hard jet. The EW corrections and the corrections due 
   to photon-induced processes, $\de_{\ga}$, are presented for the
   variable scale. Finally, the last row shows the full NLO cross section
   $\si_{\full,\veto}^{\mathrm{var}}$. The error from the
   Monte Carlo integration for the last digit(s) is given in parenthesis as far as significant. 
   See Ref.~\cite{Denner:2011vu} for details.}
\label{tab:corr:Zlljetptj100}
\end{table}

\paragraph{Vector-boson pair production}
Vector-boson pair production is among the most important 
SM benchmark processes at the LHC, because of its 
connection to the electroweak symmetry breaking.  It is a probe of Higgs boson production in 
$gg/q\bar{q}\to H \to WW^*$, $ZZ^*$, $\gamma\gamma$ processes, 
and of gauge boson self interactions.  
Diboson production can also help to gain a deeper 
understanding of the electroweak interaction in 
general, and to test the validity of the SM at 
highest energies.

\begin{figure}[htb]
\begin{center}
\includegraphics[width=0.8\textwidth,bb=1 201 500 400, clip=]{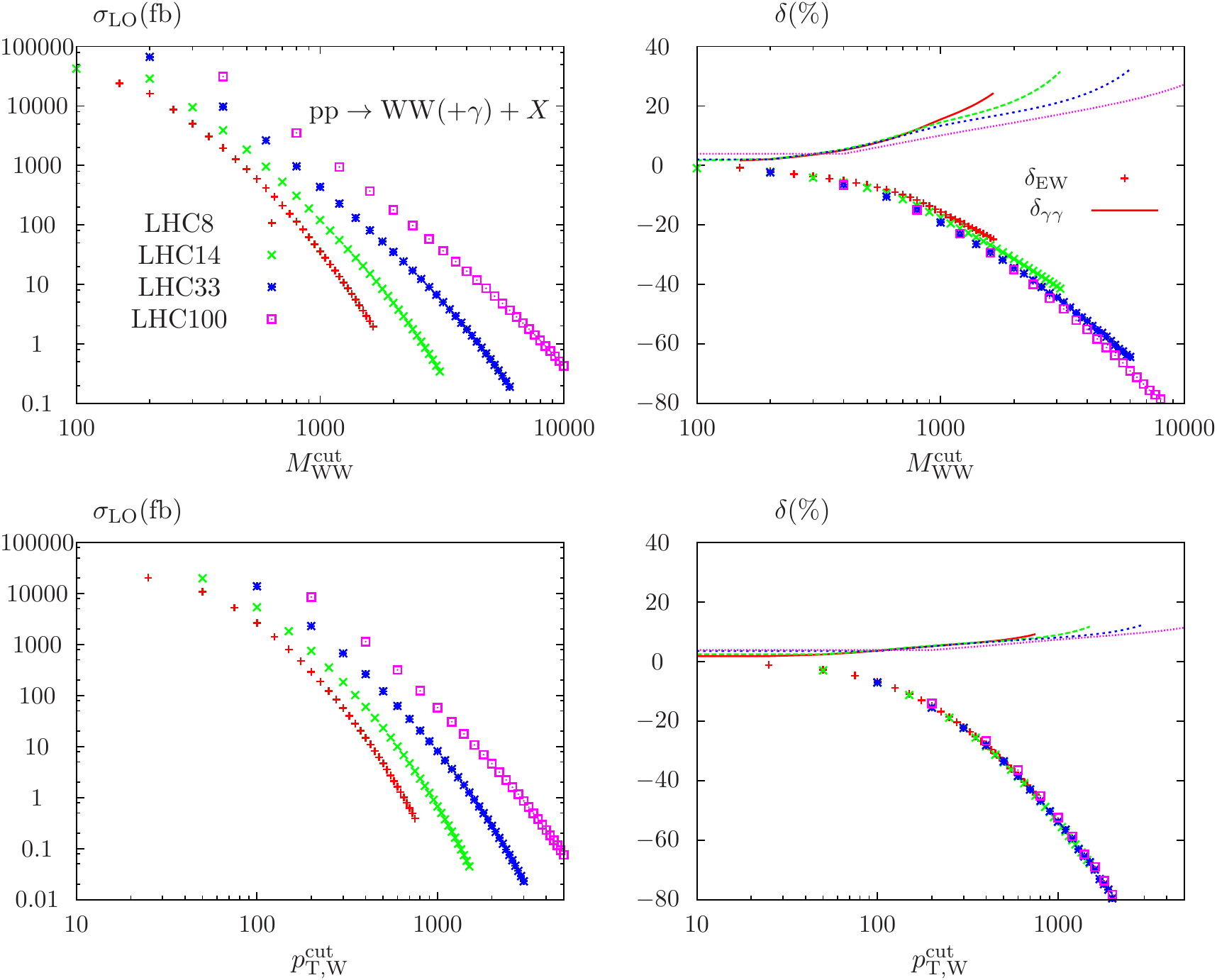}
\caption{$WW$ production: Total cross sections for $W$-pair 
  production for different cuts on $WW$ invariant mass evaluated at $pp$ collision energies 
  of 8, 14, 33, and 100~TeV. 
    Left: absolute predictions; right: relative electroweak corrections
    ($\delta_{\mathrm{EW}}$) and relative contributions from the $\gamma\gamma$-induced
    process ($\delta_{\gamma\gamma}$).}
\label{fig:WW_corr_rtsdep}
\end{center}
\end{figure}

The case of $WW$ production at 
large invariant masses or large $W$ \pt is the 
kinematic regime of high interest.  It can be subject to 
large EW corrections. 
This kinematic regime has recently been analyzed by the
ATLAS~\cite{atlas:WWllvv} and CMS~\cite{cms:WWlnuqq} 
collaborations at $\sqrt{s} = 8~\TeV$, showing 
sensitivity to invariant masses of up
to $1~\TeV$ and boson transverse momenta of up to $500~\GeV$.
The size of the experimental uncertainty for the highest kinematic end points
is comparable with the full one-loop EW corrections 
to on-shell $WW$ production~\cite{Bierweiler:2012kw,Baglio:2013toa}.
As noted in Ref.~\cite{Bierweiler:2012kw}, the corrections 
due to photon-induced channels can be large at high energies, 
while radiation of additional massive vector bosons does not 
influence the results significantly. 
Results from repeating the exact same calculation for  
$\sqrt{s} = 33~\TeV$ and 100~TeV are shown in Fig.~\ref{fig:WW_corr_rtsdep}.
While the relative NLO EW corrections hardly depend on the 
collider energy, the relative photon-induced contributions 
are suppressed at to larger values of $\sqrt{s}$. 
As a result, the overall corrections show very little dependence on $\sqrt{s}$. 
However, as in the case of other processes described earlier, 
the EW corrections will become progressively more important with   
increase in $\sqrt{s}$ due to the extended kinematic reach.  

\paragraph{Summary}
The above discussion represents a survey of the most abundant processes at LHC 
for sensitivity to electroweak corrections at various proton-proton 
collision energies relevant for LHC and future hadron colliders. 
The observations are summarized in Table~\ref{tab:summary}. 
For most processes, the overall electroweak 
corrections do not change much as the collider energy increases.  
However, the corrections become more important at high collider 
energies simply because of the increase in kinematic reach at 
high $\sqrt{s}$, where the corrections are inherently large. 
\begin{table}[htbp]
\begin{center}
\caption{Are we in the Sudakov zone yet?}
\label{tab:summary}
\begin{tabular}{l c  c  c} \hline \hline
Process & $\sqrt{s}=8~\TeV$ &  $\sqrt{s}=14~\TeV$  & $\sqrt{s}=33, 100~\TeV$ \\
\hline
Inclusive jet, dijet          &  Yes  & Yes  & Yes  \\  \hline
Inclusive $W/Z$ tail          &  $\sim$ Yes  & Yes  & Yes \\
W$\gamma$, Z$\gamma$ tail ($\ell\nu\gamma, \ell\ell\gamma$) & No &  $\sim$ Yes &  Yes \\  
$W/Z$ + jets tail             & $\sim$ Yes & Yes  & Yes  \\  \hline
$WW$ leptonic  & Close        & $\sim$ Yes  & Yes  \\  
$WZ$, $ZZ$ leptonic           & No & No  & Yes  \\ 
$WW$, $WZ$, $ZZ$ semileptonic & $\sim$ Yes & Yes  & Yes  \\  \hline
\end{tabular}
\end{center}
\end{table}


\subsection{Interplay of electroweak and QCD corrections in Drell-Yan production}

\draftnote{Text from Y. Li and F. Petriello}

\begin{figure}[h!]
\begin{center}
\includegraphics[width=3.5in,bb=51 201 600 600,clip=]{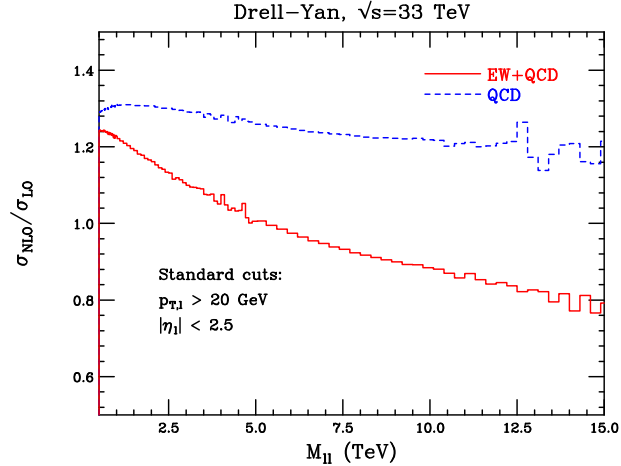}
\caption{QCD corrections and combined electroweak-QCD corrections to lepton-pair production as a function of the lepton-pair invariant mass, at a 33~TeV $pp$ collider.}
\label{fig:qcd-EW-Fmll}
\end{center}
\end{figure}

The effects of electroweak Sudakov logarithms on more differential quantities, and their interplay with higher-order QCD corrections, have been studied for the example case of lepton-pair production via the Drell-Yan mechanism at a 33~TeV $pp$ collider.  The results shown are obtained with the numerical program FEWZ~\cite{Melnikov:2006kv,Gavin:2010az,Li:2012wna}, which additively combines higher-order QCD and electroweak corrections.  MSTW PDFs at the appropriate order in QCD perturbation theory are used.  Shown first in Fig.~\ref{fig:qcd-EW-Fmll} is the lepton-pair invariant mass distribution, with minimal acceptance cuts imposed on the transverse momenta and pseudorapidities of the leptons.  The shift due to NLO QCD corrections alone is shown, as is the result of combining the full NLO electroweak correction with the QCD one.  Both shifts are normalized to the leading-order prediction.
Over a broad range of invariant masses, the QCD corrections increase the cross section by $20-30\%$.  The electroweak corrections grow in importance with invariant mass, and lead to a decrease in the cross section.  The electroweak corrections begin to overtake the QCD ones at $M_{ll} \approx 5$~TeV, and the reduction in the cross section from the combined corrections reaches 30\% at invariant masses of 15~TeV.

\begin{figure}[h!]
   \centering
   \includegraphics[width=0.49\textwidth,bb=51 201 600 600,clip=]{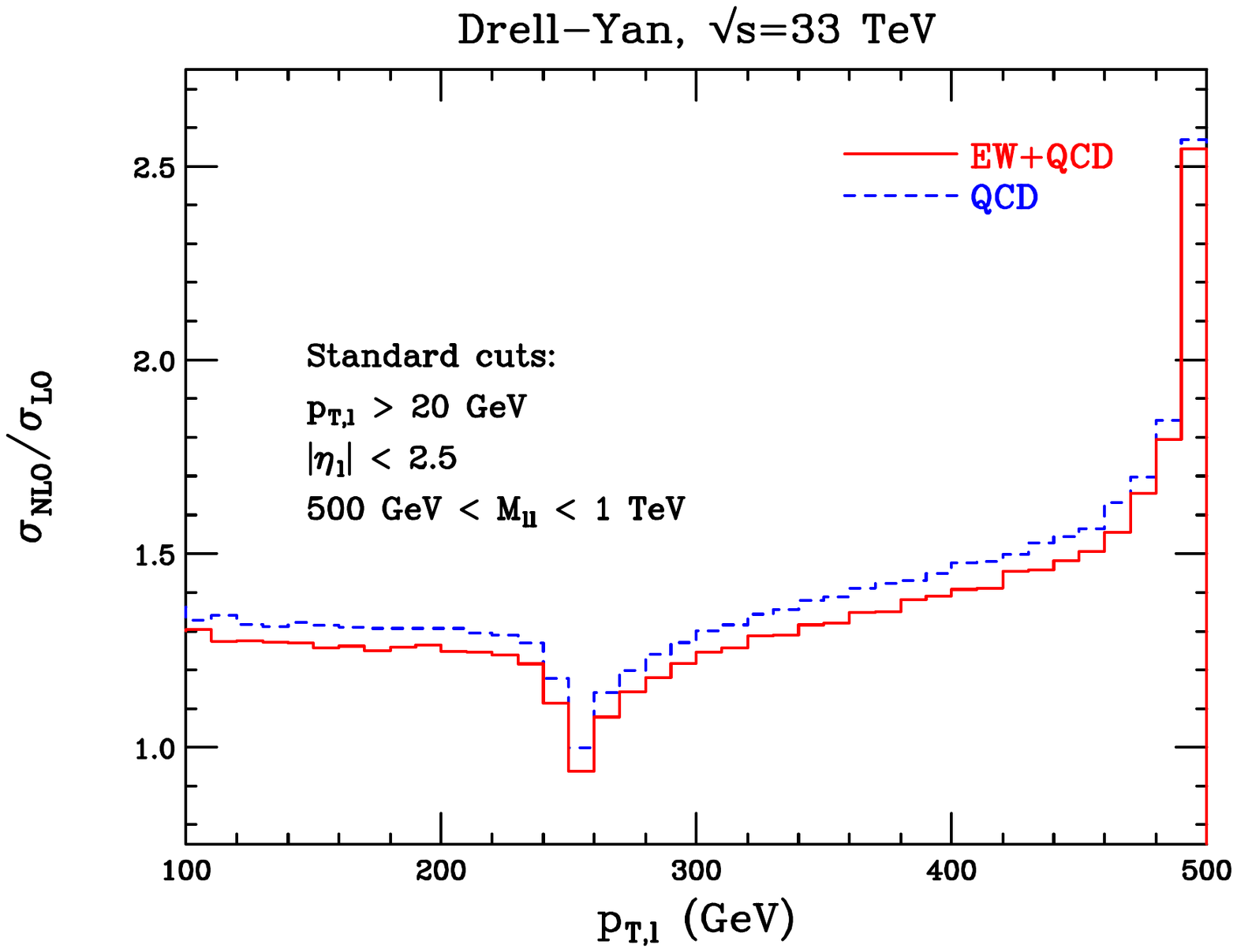}
   \includegraphics[width=0.49\textwidth,bb=51 201 600 600,clip=]{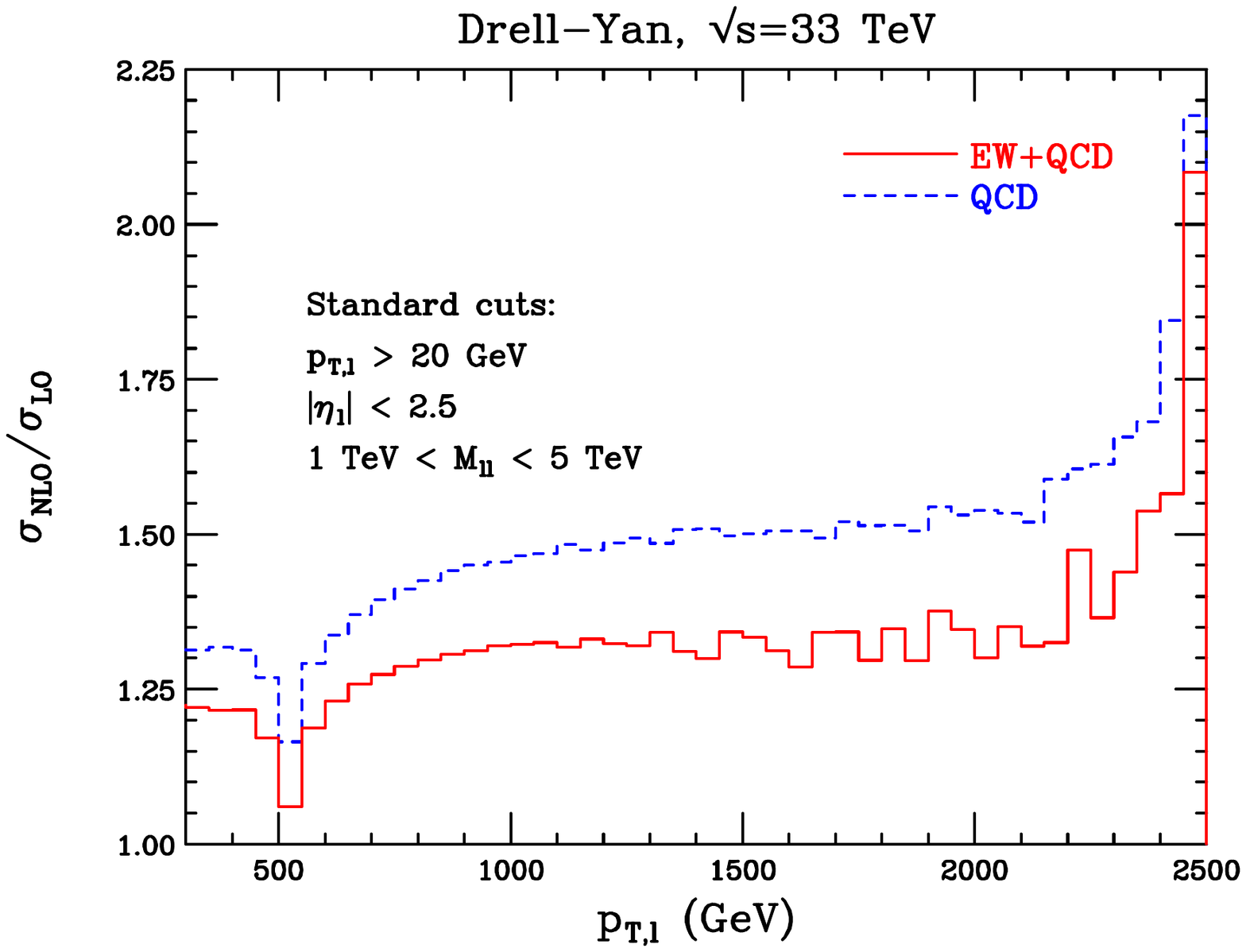}
   \caption{QCD corrections and combined electroweak-QCD corrections to lepton-pair production as a function of the lepton transverse momentum, at a 33~TeV $pp$ collider.  Results for two bins of lepton-pair invariant mass, $M_{ll} \in [500\, {\rm GeV}, 1 \, {\rm TeV}]$, and $M_{ll} \in [1\, {\rm TeV}, 5 \, {\rm TeV}]$, are shown.}
   \label{fig:qcd-EW-Flep1}
\end{figure}

\begin{figure}[h!]
   \centering
   \includegraphics[width=0.49\textwidth,bb=51 201 600 600,clip=]{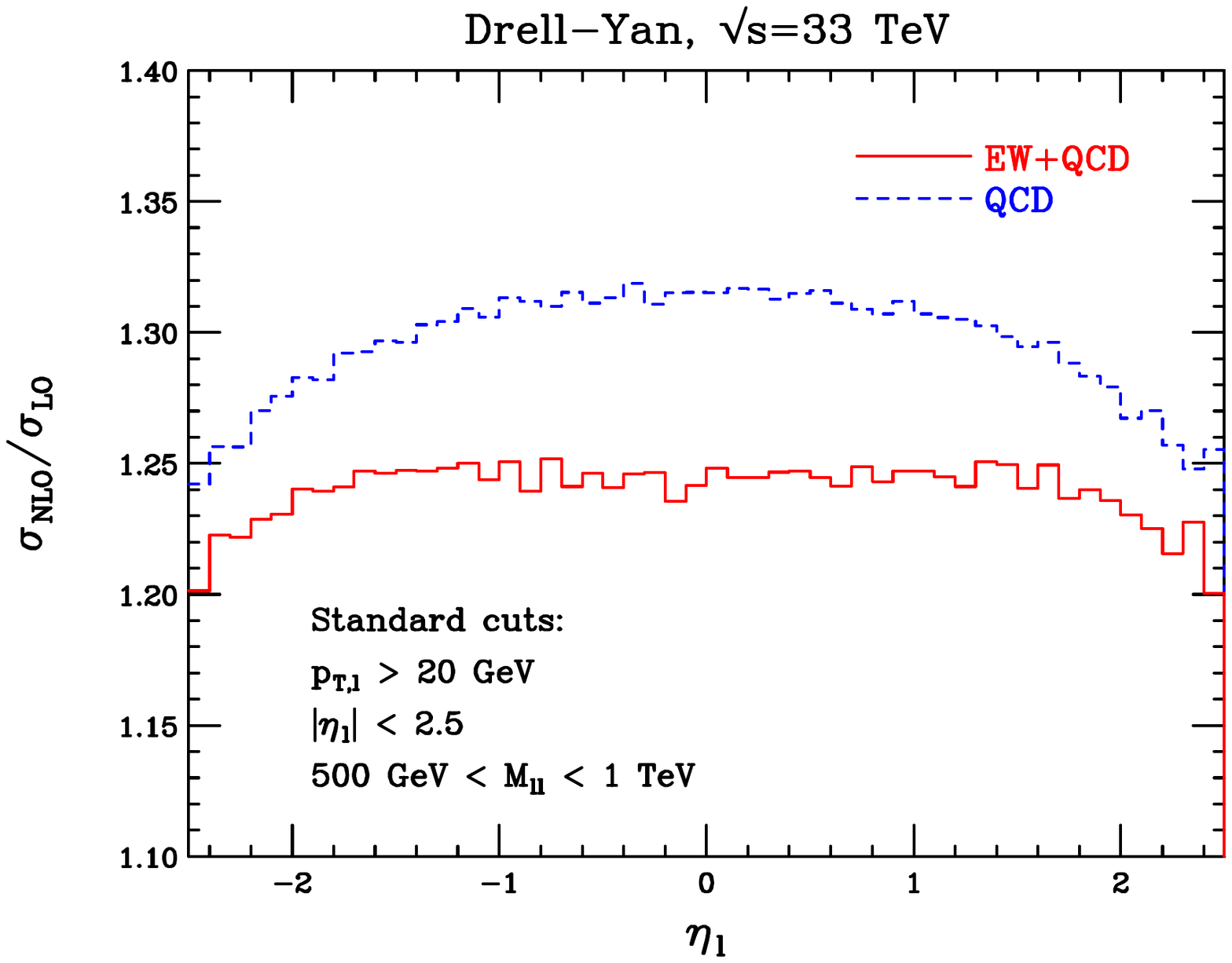}
   \includegraphics[width=0.49\textwidth,bb=51 201 600 600,clip=]{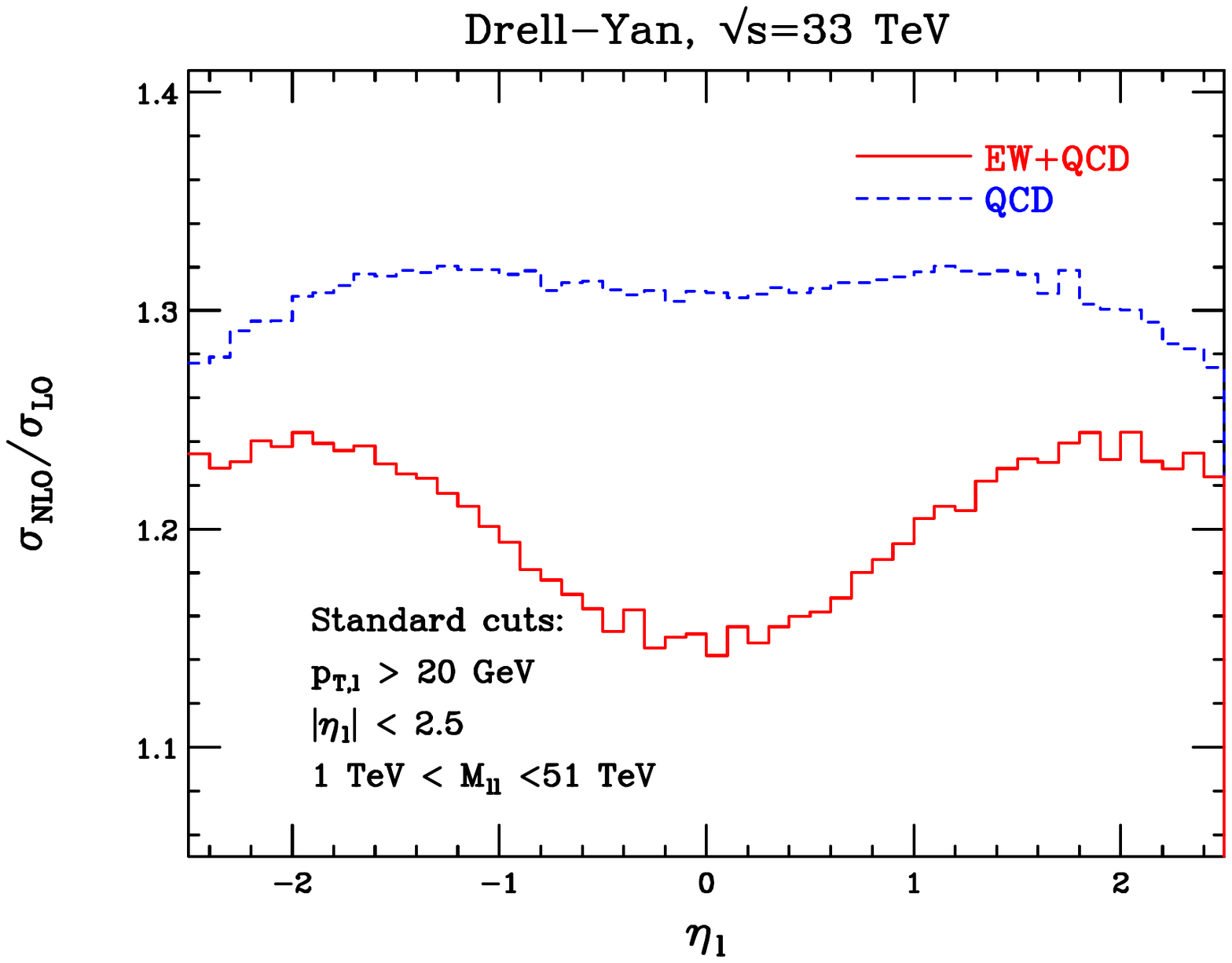}
   \caption{QCD corrections and combined electroweak-QCD corrections to lepton-pair production as a function of the lepton pseudorapidity, at a 33~TeV $pp$ collider.  Results for two bins of lepton-pair invariant mass $M_{ll} \in [500\, {\rm GeV}, 1 \, {\rm TeV}]$, and $M_{ll} \in [1\, {\rm TeV}, 5 \, {\rm TeV}]$, are shown.}
   \label{fig:qcd-EW-Flep2}
\end{figure}

\begin{figure}[h!]
   \centering
   \includegraphics[width=0.49\textwidth,bb=51 201 600 600,clip=]{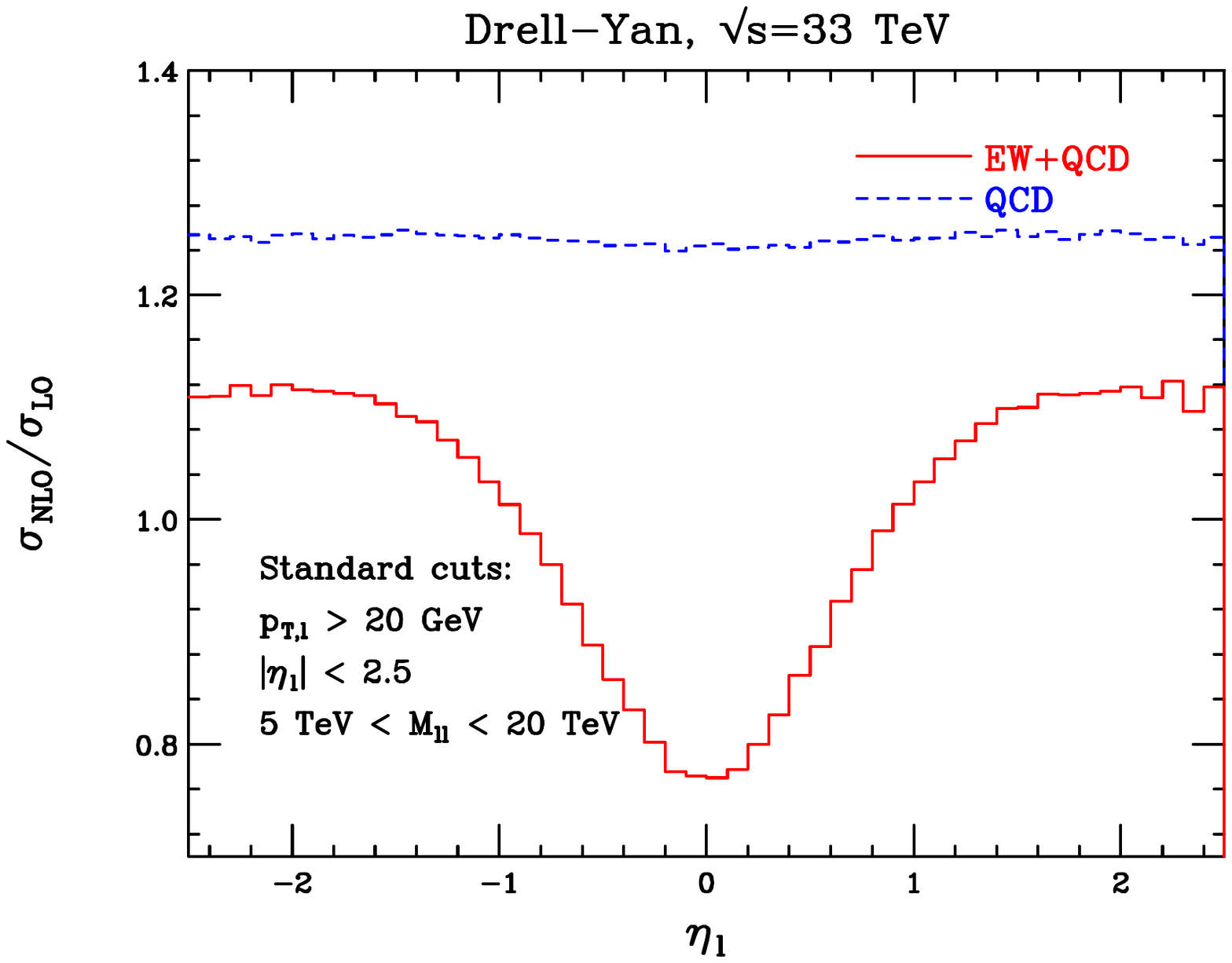}
   \includegraphics[width=0.49\textwidth,bb=51 201 600 600,clip=]{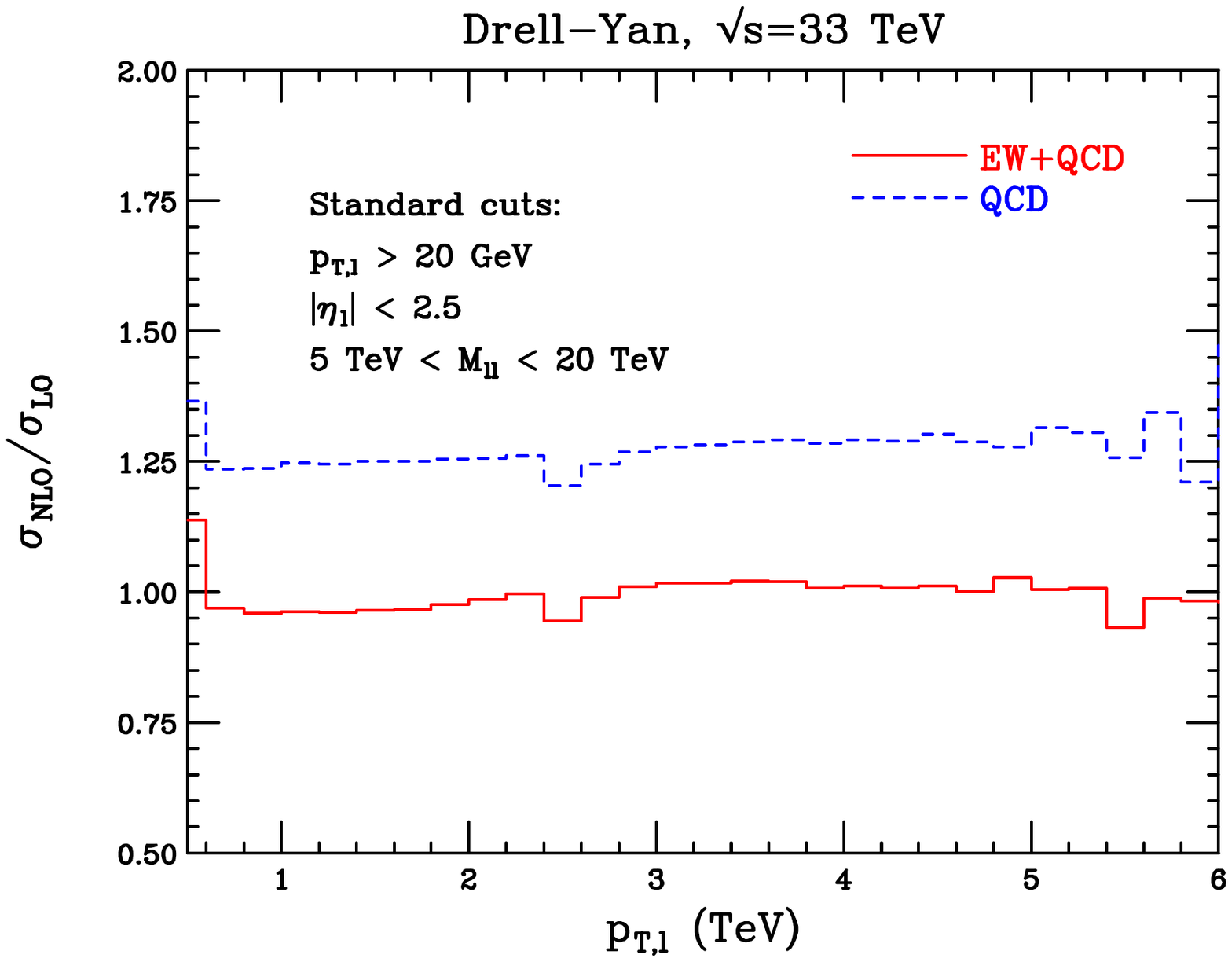}
   \caption{QCD corrections and combined electroweak-QCD corrections to lepton-pair production as a function of the lepton pseudorapidity and transverse momentum, at a 33~TeV $pp$ collider.  Results for the bin of lepton-pair invariant mass  $M_{ll} \in [5\, {\rm TeV}, 20 \, {\rm TeV}]$ are shown}
   \label{fig:qcd-EW-Flep3}
\end{figure}

The shifts induced by the combined QCD and electroweak corrections on the lepton differential distributions are considered next.  The cross section is first divided into the following invariant mass bins: $M_{ll} \in [500\, {\rm GeV}, 1 \, {\rm TeV}]$, $M_{ll} \in [1\, {\rm TeV}, 5 \, {\rm TeV}]$, and $M_{ll} \in [5\, {\rm TeV}, 20 \, {\rm TeV}]$.  The lepton transverse momenta and pseudorapidity distributions in each bin are then studied.  The results are shown in Figs.~\ref{fig:qcd-EW-Flep1},~\ref{fig:qcd-EW-Flep2},~and~\ref{fig:qcd-EW-Flep3}.  The QCD and electroweak corrections have the same shape as a function of lepton $p_T$.  The dips appearing in the corrections at half the lower bin edge, and the rise at the upper bin edge, are artifacts produced by the Jacobian peaks present in the leading-order result.  An interesting feature emerges in the lepton $\eta$ distributions at higher invariant masses.  The electroweak corrections act more strongly for central pseudorapidities, leading to a dip in the combined corrections that is quite large for the highest invariant mass bin.  

\subsection{Electroweak corrections to $Z + 2$ and $Z + 3$~jets 
hadroproduction in the Sudakov zone}

\draftnote{Text below from F. Piccinini}


Important searches for new physics (NP) beyond the Standard Model 
(SM) at present and future proton-proton ($pp$) colliders are based on the analysis of 
events with jets and missing transverse momentum ($\rlap\slash{\!p_T}$). 
The main SM backgrounds to this signature are given by the 
production of weak bosons accompanied by jets ($W/Z + n$~jets), 
pure QCD multiple jet events and $t \bar{t}$ production. Among these processes only
 $Z + n$~jets (in particular with $Z \to \nu \bar\nu$) 
constitutes an irreducible background and is particularly relevant for final states 
with 2 and 3 jets. 


The one-loop  Sudakov logs are naturally included in any complete calculation of 
NLO EW radiative corrections to a given process. 
Up to now, these calculations are available for a limited class of $2 \to 2$ and $2 \to 3$ 
processes, see for instance~\cite{Kuhn:2005az,Kuhn:2007cv,Maina:2003is,Maina:2004rb,
                                  Moretti:2006ea,Denner:2009gj,
                                  Denner:2011vu,Denner:2012ts,Dittmaier:2012kx,Actis:2012qn},
many of which have already been discussed in Section~\ref{sec:qcd-qcdew-mishra}.
On the other hand, by virtue of their universality, 
the Sudakov logarithmic contributions can be accounted for by means of process-independent 
methods, as shown in Ref.~\cite{Ciafaloni:1998xg,Beccaria:1999fk,Ciafaloni:2000df,Ciafaloni:2000rp,
                                Ciafaloni:2001vt,Denner:2000jv,Denner:2001gw,
                                Kuhn:2004em,Kuhn:2005gv,Kuhn:2007qc,Accomando:2006hq}. 
In particular, Denner and Pozzorini presented a method to single out the double and single  
Sudakov log contributions, as detailed in~\cite{Denner:2000jv,Denner:2001gw}.
The generality of this algorithm has been recently combined~\cite{Chiesa:2013yma} with the 
leading order matrix element event generator {\tt ALPGEN v2.1.4}~\cite{Mangano:2002ea}, 
to obtain a tool able to calculate NLO EW 
Sudakov corrections to processes involving multijet final 
states. Phenomenological results have been presented in 
Ref.~\cite{Chiesa:2013yma} for the 
production processes $Z(\to \nu \bar \nu) + 2(3)$~jets in $pp$ 
collisions at $\sqrt{s} = 7,14$~TeV.

The considered processes are,  
at the leading order $\as
^\mathrm{njets} \alpha$, of neutral current type, so 
the EW contributions can be separated into purely weak corrections 
(which contain the Sudakov logs and are the subject of the present contribution) 
and QED corrections. The latter can be treated separately together with their real counterparts 
and, for sufficiently inclusive event selections, they give rise to rather moderate corrections. 

In this section, the focus is on the virtual EW Sudakov corrections to
$Z + 2$ and $Z + 3$~jets hadroproduction, studying their scaling with the center-of-mass energy 
of proton-proton collisions from $14$~TeV to $33$ and $100$~TeV. The event selections 
and considered observables are those of Ref.~\cite{Chiesa:2013yma}. The 
parameters and PDF setting are the  {\tt ALPGEN} defaults. In particular, 
for $Z + 2$~jets, the observable/cuts are those presently adopted by ATLAS~\cite{Aad:2011ib}, namely 
\begin{eqnarray}
&& m_{\rm eff} > 1~{\rm TeV} \qquad \, \, \, \, \, 
\rlap\slash{\!E_T}/m_{\rm eff} > 0.3  \nonumber\\
&& p_T^{j_1} > 130~{\rm GeV} \qquad \, \, 
p_T^{j_2} > 40~{\rm GeV} \quad \, \, |\eta_{j}| < 2.8 \nonumber \\
&& \Delta\phi ({\vec p}_T^j,\rlap\slash{\!\vec{p}_T}) > 0.4 \quad 
\Delta R_{(j_1, j_2)} > 0.4  \, 
\label{eq:atlascut}
\end{eqnarray}
where $j_1$ and $j_2$ are the leading and next-to-leading $p_T$ jets. 
For the $Z+3$~jets final state the observables/cuts used by 
CMS~\cite{Collaboration:2011ida,Chatrchyan:2012lia} are considered, namely
\begin{eqnarray}
&& H_T > 500~{\rm GeV} \qquad \, \, \, \, \, \, 
|\rlap\slash{\!\vec{H}_T}| > 200~{\rm GeV}  \nonumber\\
&& p_T^j > 50~{\rm GeV} \qquad  \, \, |\eta_j| < 2.5 
\quad \Delta R_{(j_i, j_k)} > 0.5 \nonumber\\
&& \Delta\phi ({\vec p}_T^{j_1, j_2},\rlap\slash{\!{\vec H}_T}) > 0.5 
\qquad  \Delta\phi (\vec{p}_T^{j_3},\rlap\slash{\!{\vec H}_T}) > 0.3 \, ,
\label{eq:cmscut}
\end{eqnarray}
where $H_T = \sum_i p_{T,i}$ and 
$\vec{\rlap\slash{\!H_T}} = 
- \sum_i \vec{p}_{T,i}$.

\begin{figure}[t!]
\begin{center}
\includegraphics[scale=0.42]{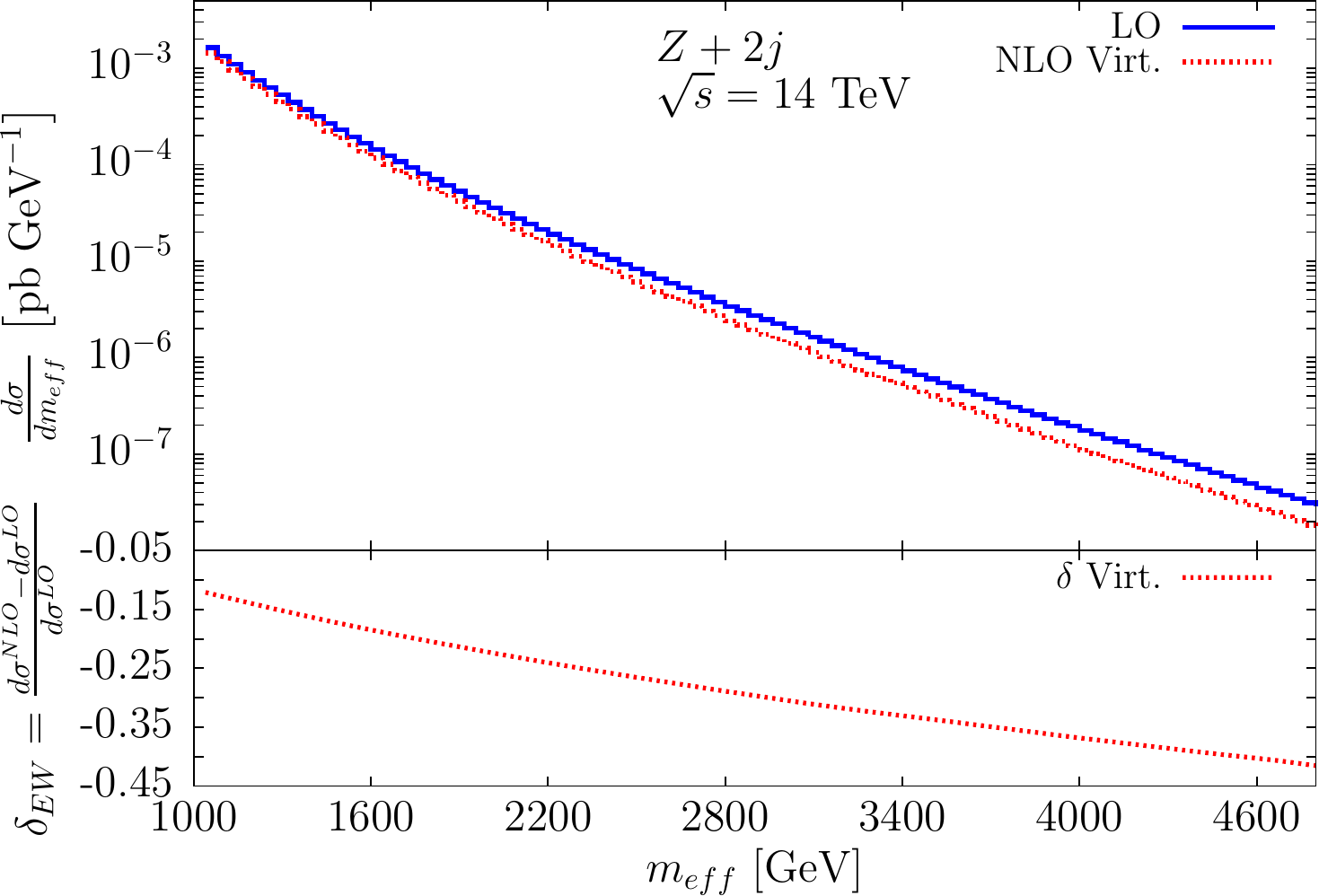}
\hspace{0.5cm}\includegraphics[scale=0.42]{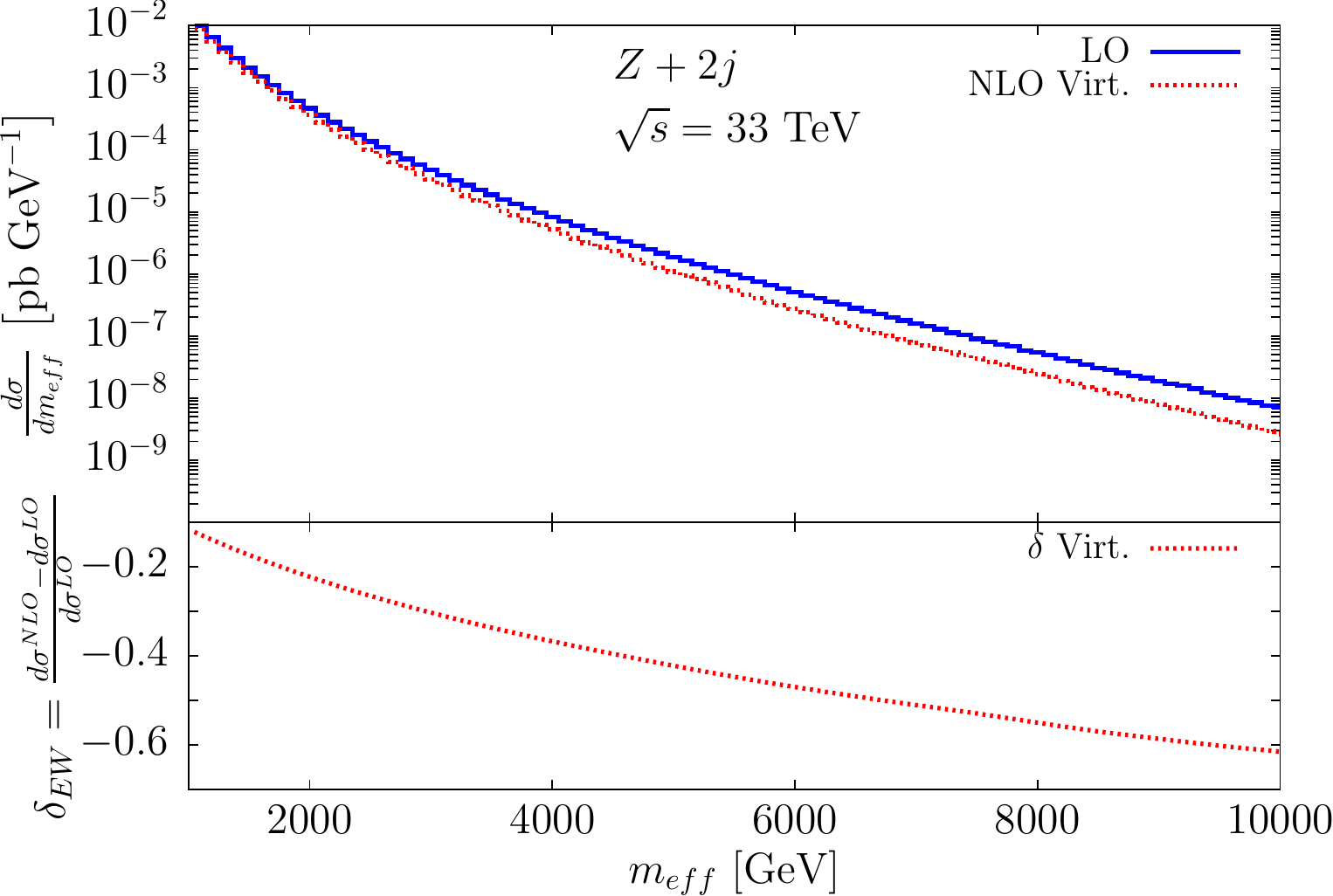}
\includegraphics[scale=0.42]{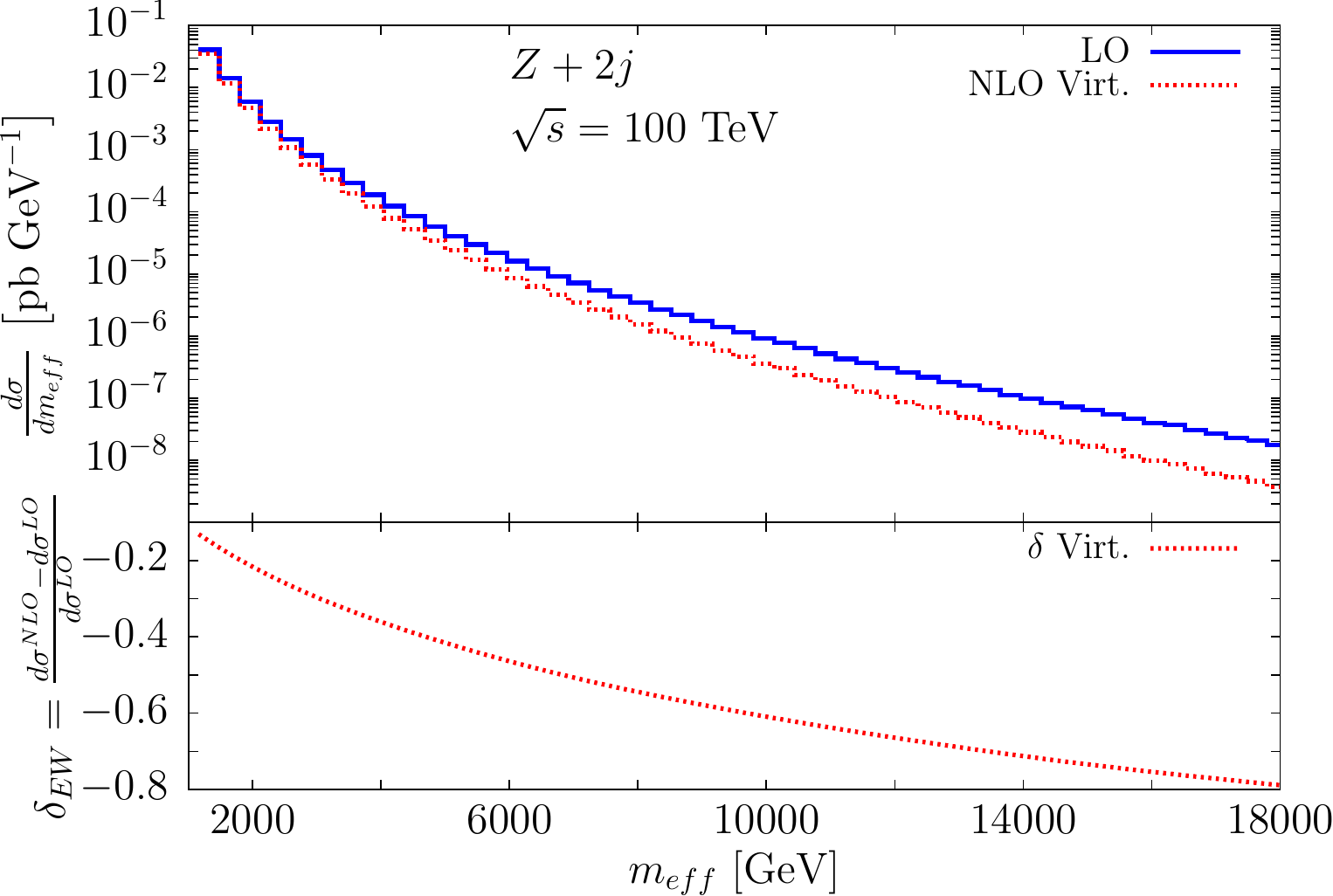}
\caption{\label{fig:meff2j_lhc} $Z+2$~jets: ATLAS $m_{\rm eff}$ and EW correction 
at $\sqrt{s} = 14$,~$33$ and $100$~TeV.}
\end{center}
\end{figure}

\begin{figure}[t!]
\begin{center}
\includegraphics[scale=0.42]{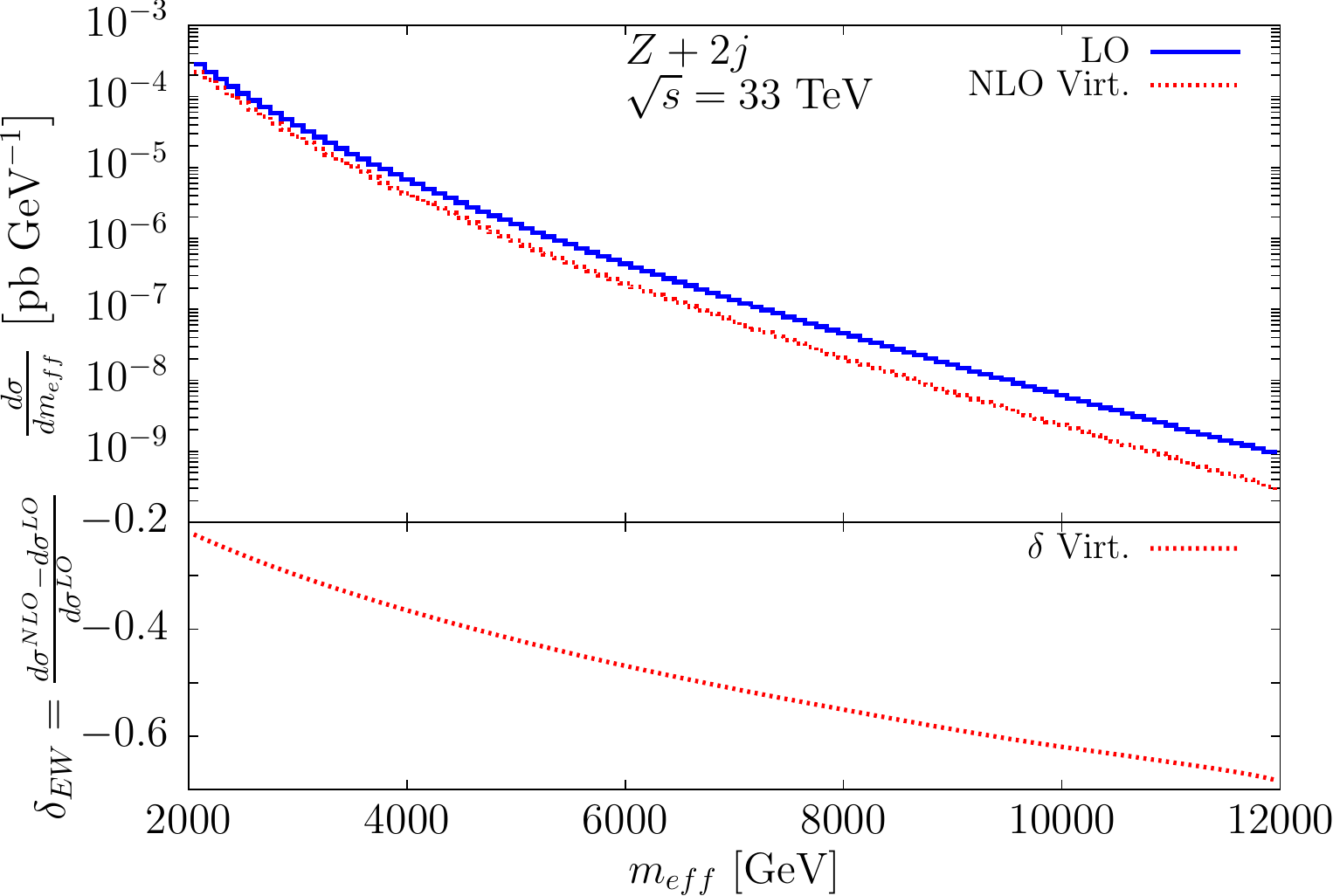}
\includegraphics[scale=0.42]{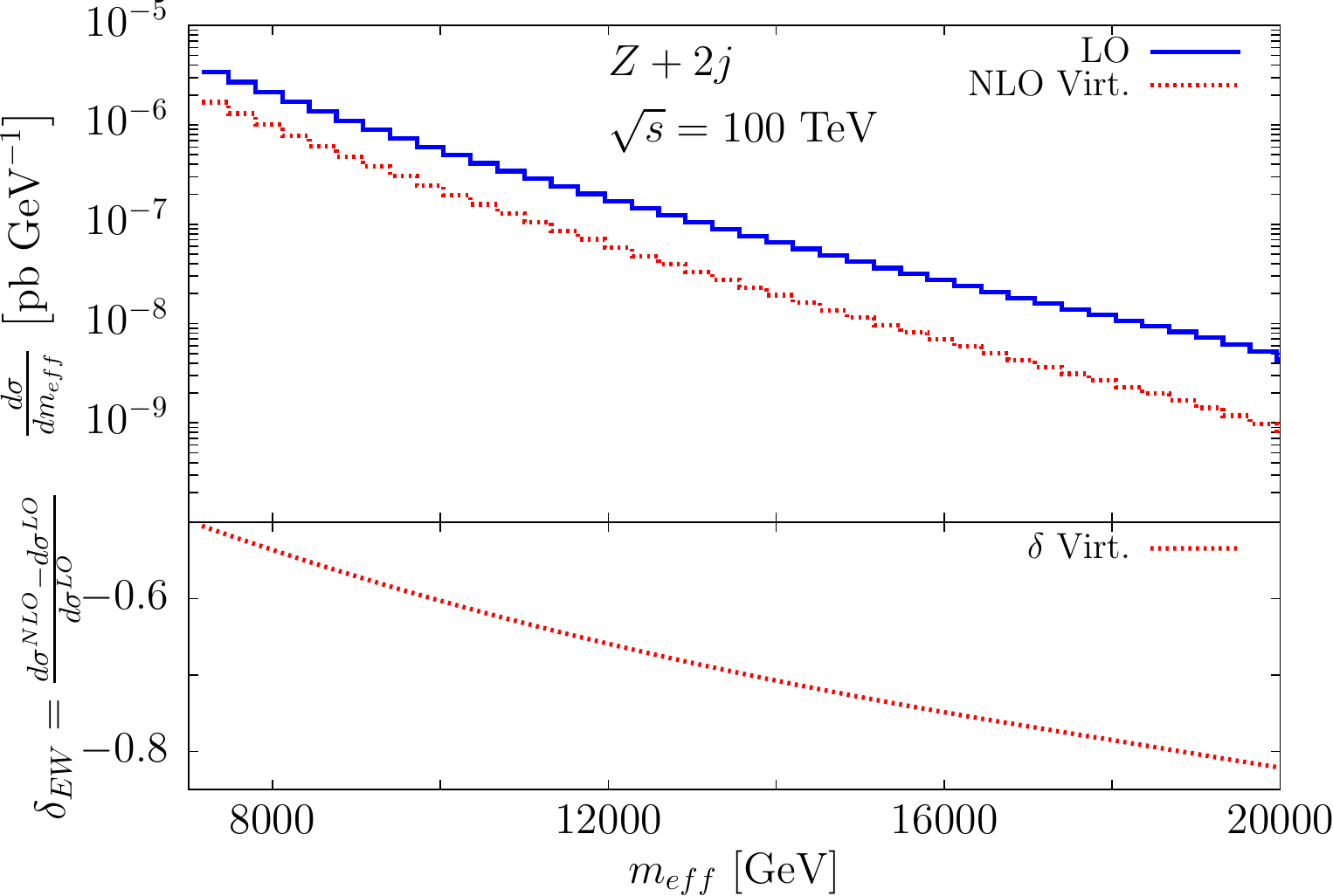}
\caption{\label{fig:meff2j} The same as Fig.~\ref{fig:meff2j_lhc} 
with rescaled cuts at $33$ and $100$~TeV, as described in the text.}
\end{center}
\end{figure}

In Ref.~\cite{Chiesa:2013yma} it has been shown that Sudakov virtual corrections 
to $Z + 2$ and $Z + 3$~jets production are negative and can be as large as 
about $-40$\% at $\sqrt{s} = 7, 14$~TeV. Here that analysis is extended to 
c.m. energies of future hadronic colliders, namely $33$ and $100$~TeV. For the sake 
of reference, we report here some partial results from Ref.~\cite{Chiesa:2013yma} 
corresponding to the c.m. energy of $14$~TeV. 
Fig.~\ref{fig:meff2j_lhc}  
shows the effect of the Sudakov logs on the effective mass 
distribution in the process $Z+2$~jets according to the event selection of Eq.~(\ref{eq:atlascut}). 
All plots start from  $m_{\rm eff} = 1$~TeV and have different upper limits: 
$5$~TeV for $\sqrt{s} = 14$~TeV, $10$~TeV for $\sqrt{s} = 33$~TeV 
and $18$~TeV for $\sqrt{s} = 100$~TeV. 
The upper panels display the effective mass distribution at LO 
(blue, solid) and including the approximate NLO virtual corrections (red, dotted) 
due to weak bosons in the Sudakov zone as calculated with {\tt ALPGEN}, according 
to Ref.~\cite{Chiesa:2013yma}, respectively. The lower panels show the relative 
corrections due to virtual weak corrections. 
As can be seen, the negative correction due to Sudakov logs is of the order of 
some tens of percent, increasing to about 40\% (60\%, 80\%) in the extreme regions at 
$\sqrt{s} = 14$ ($33$, $100$)~TeV, respectively. 
As can be naively expected, for a given bin of the $m_{\rm eff}$ distribution, the relative EW corrections 
is practically the same, independently of the collider c.m. energy.
It is interesting to study whether and how the effects on the $m_{\rm eff}$ distribution 
change when adopting acceptance cuts scaled with the c.m. energy w.r.t. the case $\sqrt{s} = $~14 TeV
(for simplicity the geometrical acceptance cuts are kept fixed). 
Figure~\ref{fig:meff2j} shows the predictions for 
\begin{eqnarray}
&& m_{\rm eff} > 2(7)~{\rm TeV}\, ,\, \,   (\sqrt{s}= 33(100)~{\rm TeV)}\,,  \nonumber \\
&& p_T^{j_1} > 260(910)~{\rm GeV}\, , \, \, (\sqrt{s}= 33(100)~{\rm TeV)}\, , \nonumber \\
&& p_T^{j_2} > 80(280)~{\rm GeV},\, \,  (\sqrt{s}= 33(100)~{\rm TeV)}\, , \nonumber
\label{eq:rescaledatlascuts}
\end{eqnarray}
with the rest of the cuts unchanged w.r.t. Eq.~(\ref{eq:atlascut}).
The effects of the tighter acceptance cuts are very mild, the leading effect 
being given by the $m_{\rm eff}$ cut. 

\begin{figure}[t!]
\begin{center}
\includegraphics[scale=0.42]{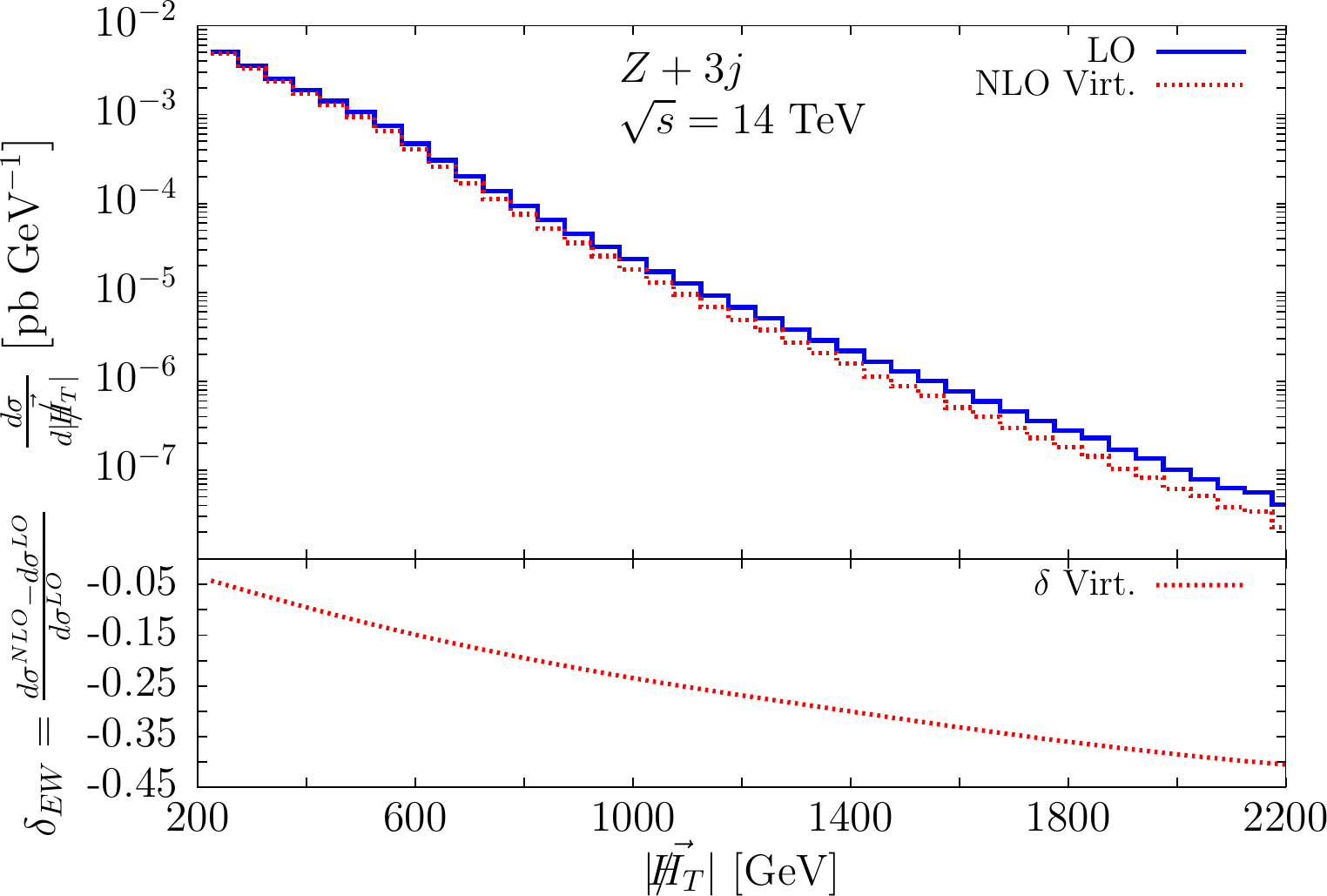}%
\includegraphics[scale=0.42]{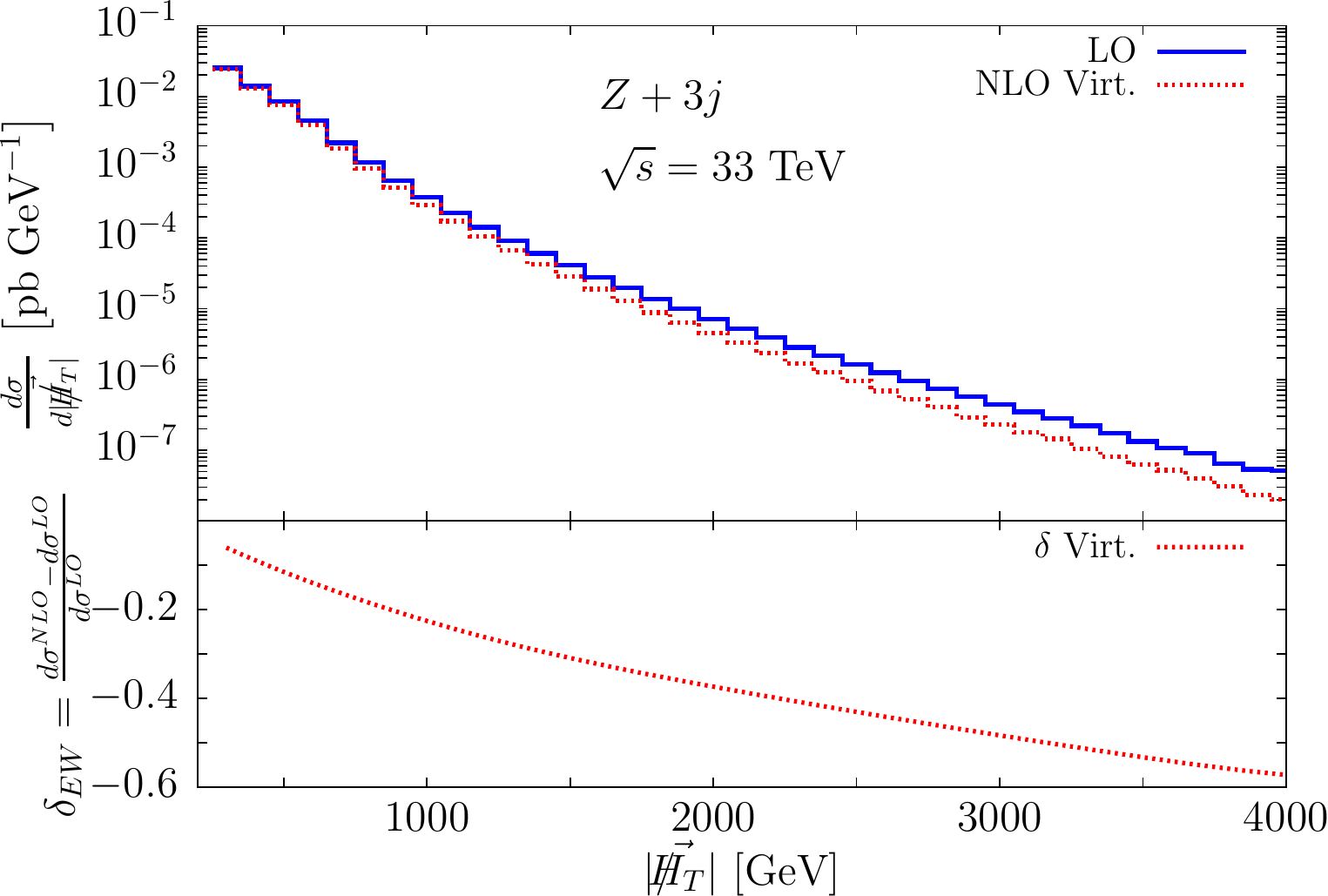}
\includegraphics[scale=0.42]{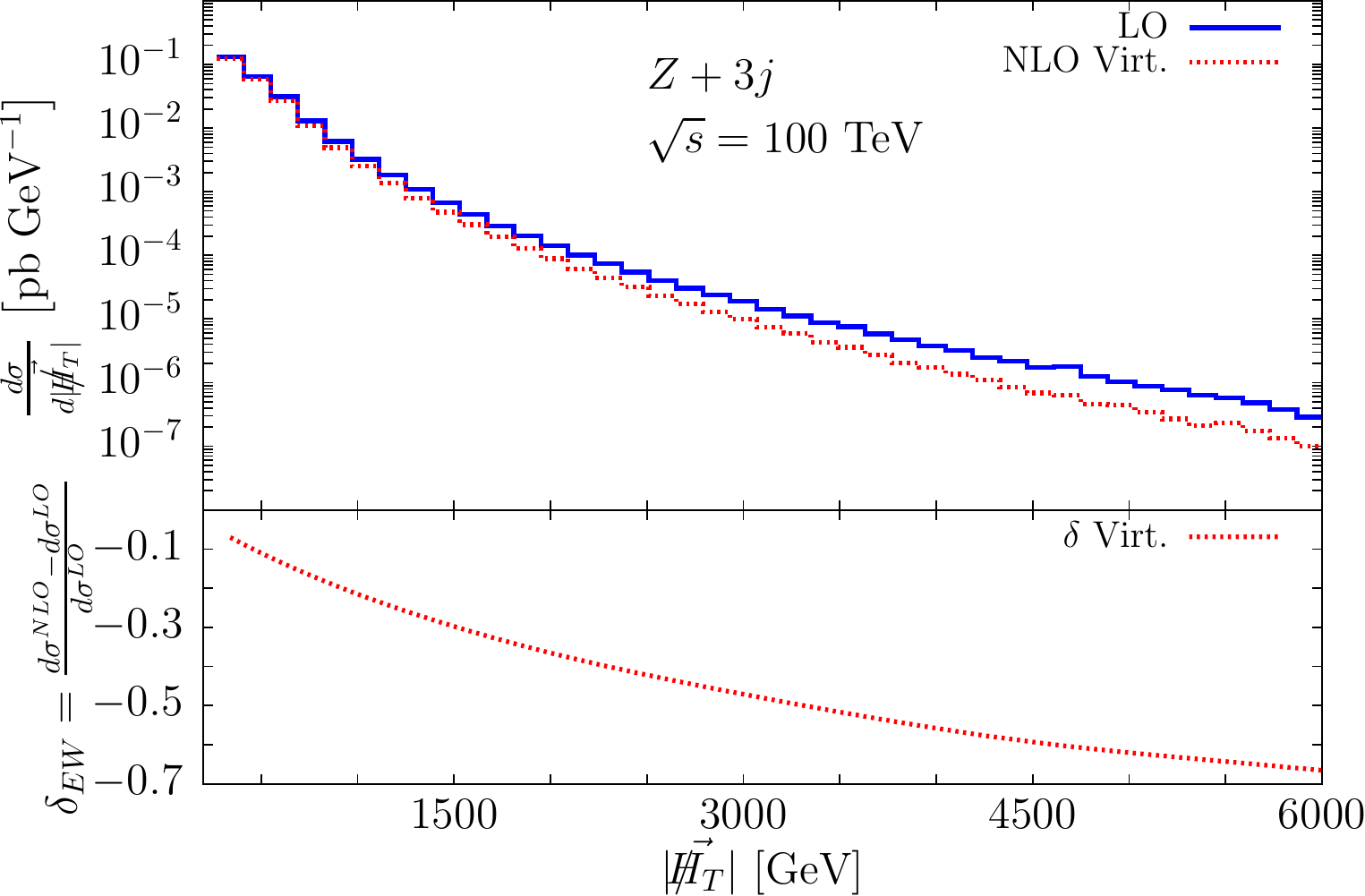}%
\caption{\label{fig:Z3j_lhc} $Z+3$~jets: CMS $|\rlap\slash{\!H_T}|$ 
and EW correction at $\sqrt{s} = 14$,~$33$ and $100$~TeV.}
\end{center}
\end{figure}
\begin{figure}[t!]
\begin{center}
\includegraphics[scale=0.42]{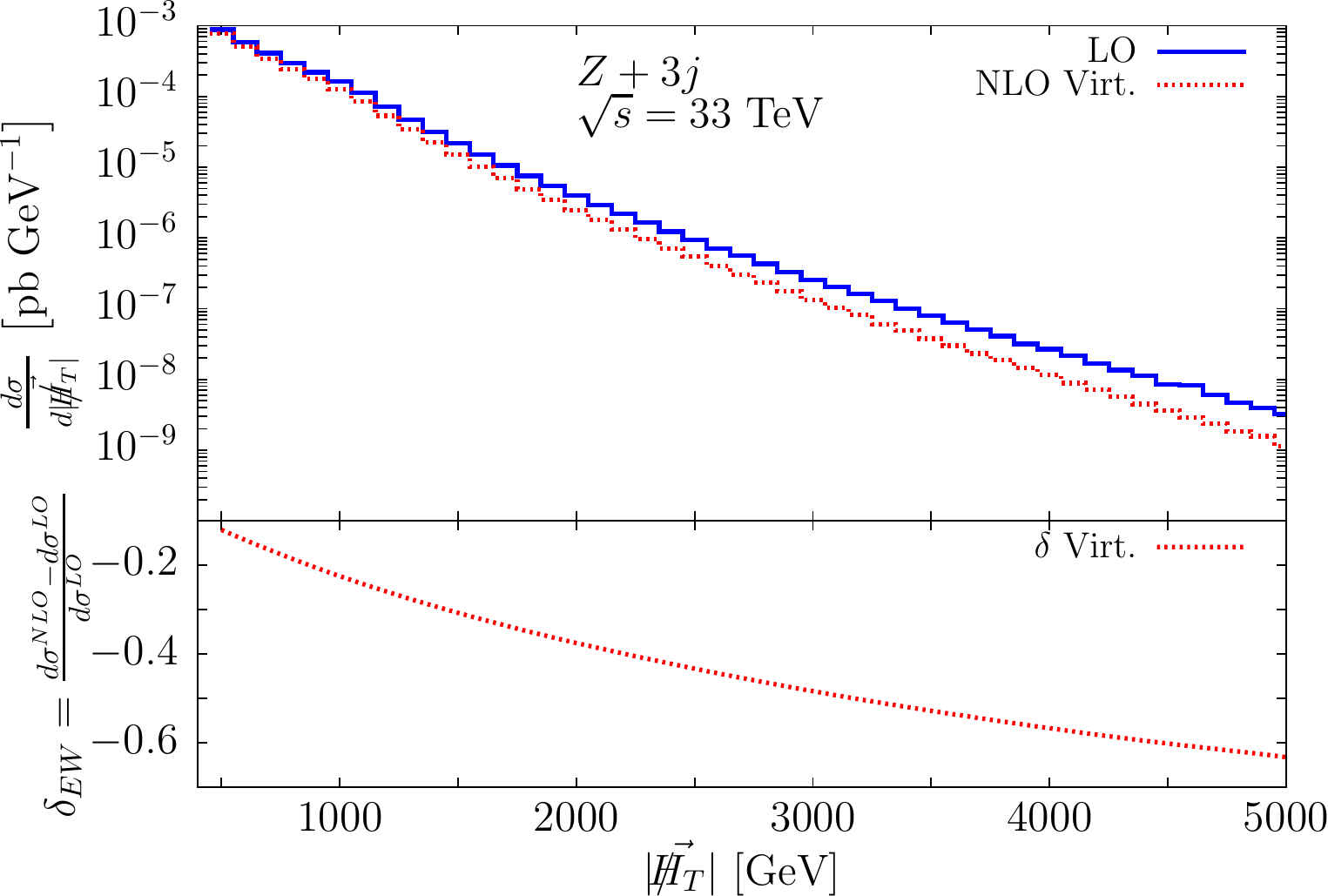}
\includegraphics[scale=0.42]{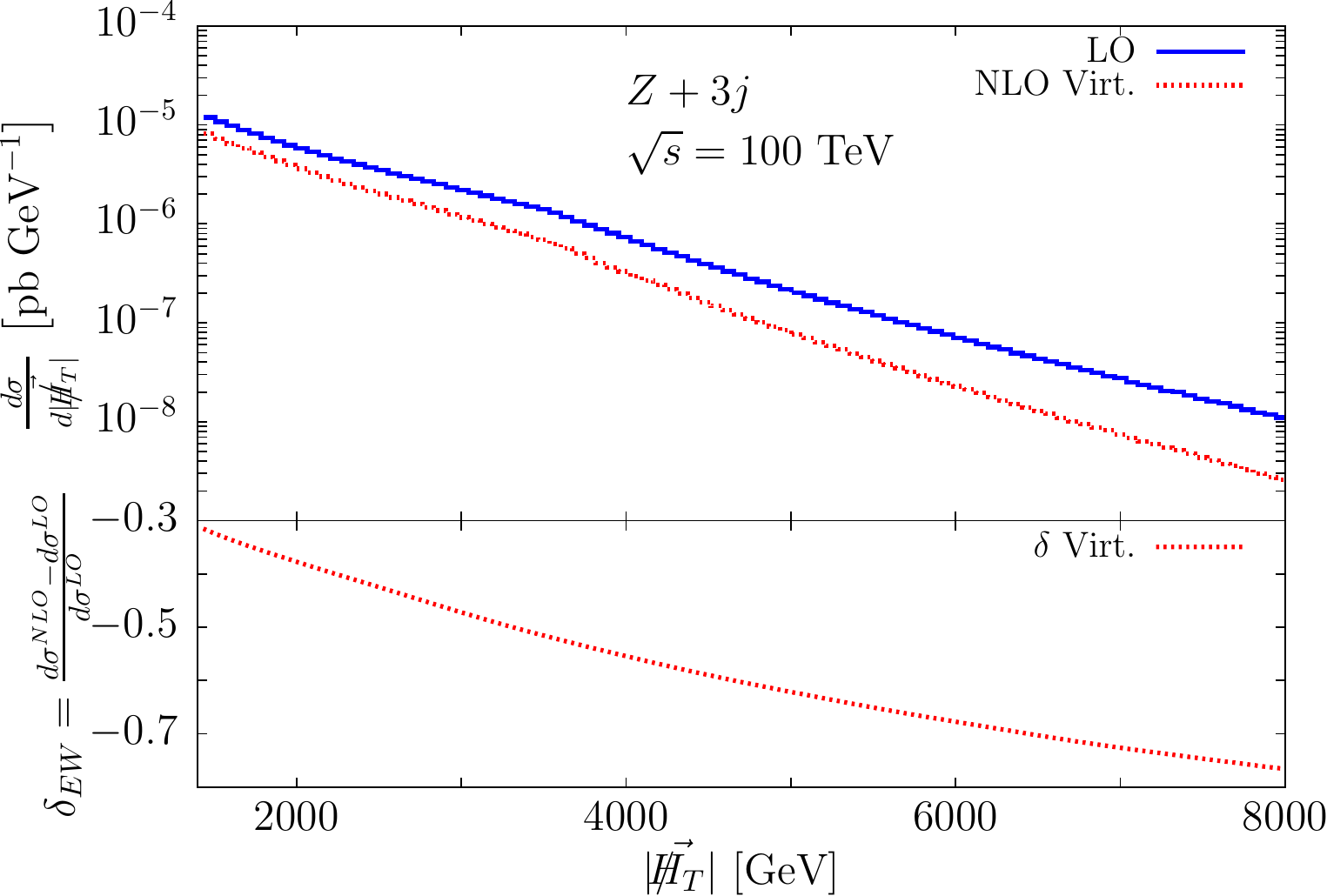}%
\caption{\label{fig:Z3j} The same as Fig.~\ref{fig:Z3j_lhc} with rescaled cuts at $33$ and $100$~TeV, 
as described in the text.}
\end{center}
\end{figure}

Fig.~\ref{fig:Z3j_lhc}
shows the effect of the Sudakov logs on the observable $|\rlap\slash{\!{\vec H}_T}|$  
in the process $Z+3$~jets according to the event selection of Eq.~(\ref{eq:cmscut}). 
All plots start from  $|\rlap\slash{\!{\vec H}_T}|= 0.2$~TeV and have different upper limits: 
$2.2$~TeV for $\sqrt{s} = 14$~TeV, $4$~TeV for $\sqrt{s} = 33$~TeV 
and $6$~TeV for $\sqrt{s} = 100$~TeV. 
As for the $Z + 2$~jets effective mass distributions, the effect of NLO weak corrections 
on $|\rlap\slash{\!{\vec H}_T}|$ is large and negative, 
raising to about 40\% (60\%, 70\%) in the extreme regions at 
$\sqrt{s} = 14$ ($33$, $100$)~TeV, respectively. For a chosen $|\rlap\slash{\!{\vec H}_T}|$ bin, 
the relative effects of the corrections are quite insensitive to the change of the 
collider energy and of the acceptance cuts. 
Figure~\ref{fig:Z3j} shows the $|\rlap\slash{\!{\vec H}_T}|$ with scaled acceptance 
cuts: 
\begin{eqnarray}
&& H_T > 1(3.5)~{\rm TeV} \, , \, \,  (\sqrt{s}= 33(100)~{\rm TeV)}\, \\
&& |\rlap\slash{\!\vec{H}_T}| > 0.4(1.4)~{\rm TeV} \, , \, \, (\sqrt{s}= 33(100)~{\rm TeV)}\, 
\nonumber\\
&& p_T^j > 100(350)~{\rm GeV} \, , \, \, (\sqrt{s}= 33(100)~{\rm TeV)}\, , \nonumber 
\label{eq:cmsrescaledcut}
\end{eqnarray}
with the rest of the cuts unchanged w.r.t. Eq.~(\ref{eq:cmscut}).
The changes in slope in the $|\rlap\slash{\!{\vec H}_T}|$ distributions are due to the 
presence of the cuts on the variable $H_T$. 
 
\paragraph{Summary} The NLO EW Sudakov corrections to $Z + n$~jets, $n = 2,3$ have been computed
using {\tt ALPGEN}, for 
two key observables, $m_{\rm eff}$ and $|\rlap\slash{\!{\vec H}_T}|$, that are
used in NP searches at the LHC.
The relative corrections do not show sensitivity to the collider energy for c.m. energies
up to $100$~TeV.  The corrections are negative 
and become very large (more than 50\% in absolute value) for extreme kinematically accessible 
values of the observables. With such large negative effects, also the possible 
compensation of real heavy gauge boson radiation and the higher-order contributions 
(beyond one-loop) requires further investigation.



\section{Jet vetoes and exclusive jet binning}

\begin{figure}[b!]
\begin{center}
\includegraphics[width=0.5\hsize]{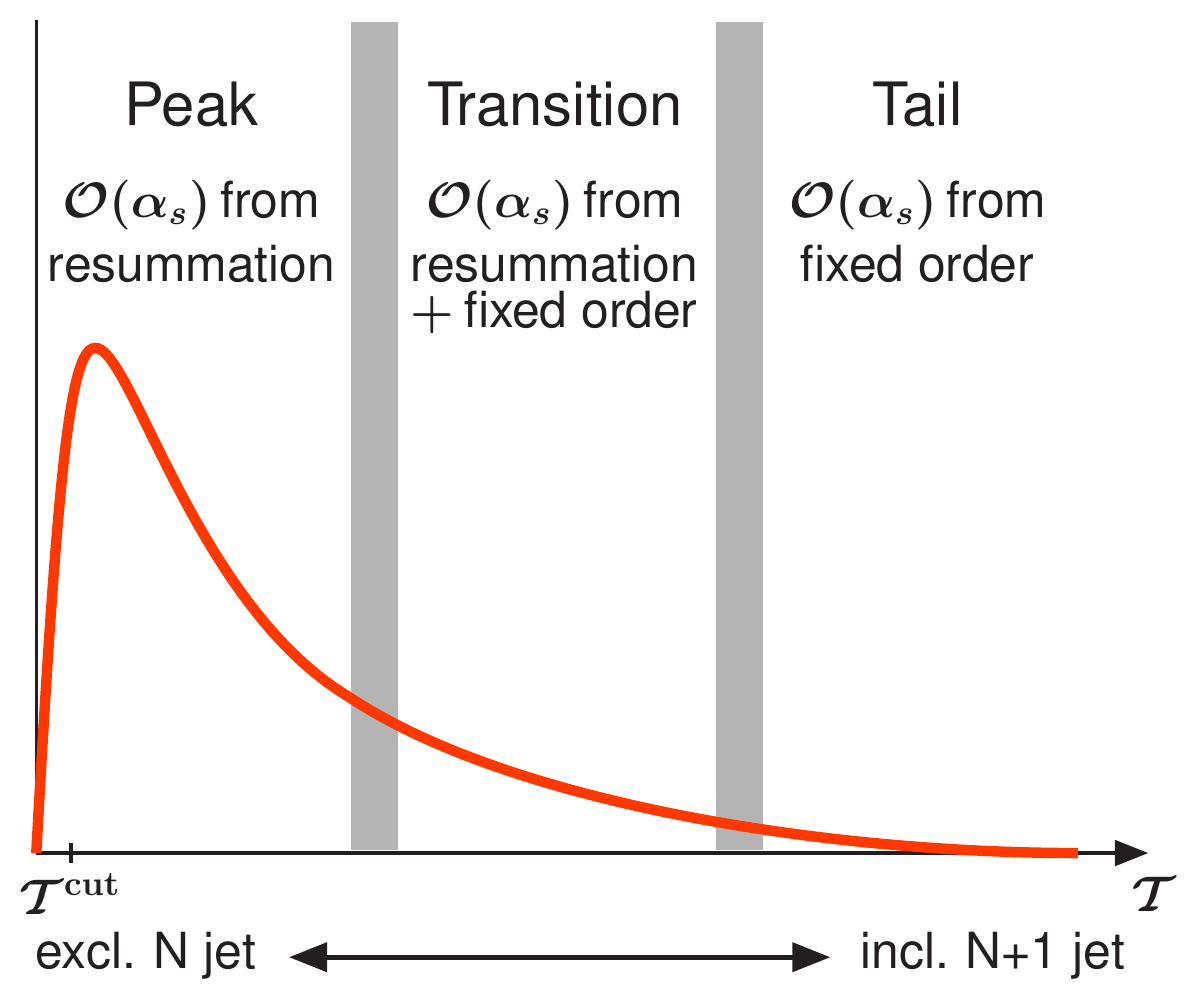}
\caption{Generic distribution for a variable $\tau$ obtained in QCD perturbation theory, taken from Ref.~\cite{Alioli:2012fc}.
}
\label{fig:qcd-transition}
\end{center}
\end{figure}

The prediction of cross sections in bins of exclusive jet multiplicity
poses an interesting theoretical challenge.  The distribution for an
observable $\tau$ in QCD perturbation theory has the structure shown in
Fig.~\ref{fig:qcd-transition}.  
This form of the distribution follows from the presence of Sudakov
double logarithms, $\alpha_s/\pi \times  {\rm ln}^2 (Q/\tau_{cut})$,
appearing in the perturbative expansion for the cross section, where $Q$
denotes the hard scale of the process such as the partonic
center-of-momentum scattering energy, and $\tau_{cut}$ denotes some
experimental constraint $\tau$.  For the cross section for the production of a fixed number of jets, 
$\tau_{cut}$ would denote the restriction on the transverse momentum of potential additional jets.  
When $\tau$ is relatively unconstrained
and $\tau_{cut} \sim Q$, as in the tail region of
Fig.~\ref{fig:qcd-transition}, these logarithms are of order unity, and
fixed-order QCD perturbation theory can be applied to predict the
distribution.  When $\tau_{cut} \ll Q$ in the peak region, the Sudakov
logarithms overwhelm the $\alpha_s$ suppression, and the perturbative
expansion must be resummed to all orders to ensure that the distribution
does not diverge as $\tau_{cut} \to 0$.  However, techniques exist to
perform this resummation, and the higher-order corrections in
fixed-order perturbation theory give only small corrections in the peak
region.  The transition region is the theoretically most intricate
region to predict.  The logarithms are large enough that resummation
must be performed, but not so large that the fixed-order corrections are
negligible.  Progress on both resummation and fixed-order calculations
are needed to accurately describe such observables.  Higgs production at
the LHC in exclusive jet bins is an example of an observable in the
transition region.  For Higgs plus jets observables, $Q \sim m_H$ and
$\tau_{cut} \sim p_{T}^{cut}=25-30 \, {\rm GeV}$, where $p_{T}^{cut}$
denotes the transverse momentum cut used to define a jet.  Both
higher-order corrections and resummation are needed to accurately
predict the rates and distributions for Higgs plus jets.  Since the
hard scale $Q$ grows with the partonic scattering energy, the high-$p_T$
jet region accessible at potential future 33 TeV and 100 TeV colliders
will potentially exhibit quite different behavior in QCD perturbation
theory.  This Section discusses predictions  for Higgs production in
both the gluon-fusion and associated production modes for 14 TeV, 33
TeV, and 100 TeV $pp$ collisions, focusing on issues and uncertainties
that arise because of the imposition of jet vetoes.

\draftnote{3-4 pages: F. Petriello et al \\
Resummation of jet veto logs, importance of jet veto logs as
a function of energy}


\subsection{Higgs production in gluon fusion}
\label{sec:qcd-higgs}

The discovery of the Higgs boson by the ATLAS and CMS collaborations has dominated the 
field of high energy physics during the past year.  A large component of the future worldwide effort in particle physics will be devoted to measuring the properties of this state in order to determine the underlying theory from which it arises.  Theoretical uncertainties from missing higher-order corrections are quickly becoming a limiting factor in this program.  In the $W^+W^-$ final state, the theoretical uncertainties are already a dominant systematic error~\cite{Aad:2012uub,Chatrchyan:2012ty}.  The reasons for this are two-fold.  The perturbative expansion for inclusive Higgs boson production is slowly convergent, and indeed even corrections beyond NNLO change the prediction in a significant way, as described in Section~\ref{sec:qcd-xsecs-approxn3lo} of this report.  In addition, and most importantly for this Section, significant cuts are imposed on the phase space of the hadronic radiation produced in association with the Higgs.  This is required because the background composition to this signal changes as a function of jet multiplicity.  In the zero-jet bin the background is dominated by continuum $WW$ production, while in the one-jet and two-jet bins, top-pair production becomes increasingly important.  The optimization of this search requires cuts dependent on the number of jets observed, and therefore also on theoretical predictions for exclusive jet multiplicities.

We present first a discussion of the NNLO QCD calculation for Higgs plus one-or-more jets, for which initial results for the gluon channel have recently been reported~\cite{Boughezal:2013uia}, and attempt to provide questions and guidance for phenomenological studies at future colliders when the full result is available.  The cross section for inclusive Higgs plus one-or-more jets enters the prediction for the exclusive one-jet bin through the relation $\sigma_1 = \sigma_{\geq 1}-\sigma_{\geq 2}$, where the subscript denote the number of final-state jets produced in addition to the Higgs.  We show the hadronic cross section for the production of the Higgs boson 
in association with one or more jets  at the 8 TeV LHC 
through NNLO in perturbative QCD. Jets are reconstructed using the $k_\perp$-algorithm
with $R = 0.5$ and $p_T^{cut}=30~{\rm GeV}$. 
The Higgs mass is taken to be $m_H=125$~GeV, and NNPDF parton distributions are used~\cite{Ball:2012cx}.  The central renormalization
and factorization scales are set to be $\mu_R=\mu_F=m_H$. 

\begin{figure}[!t]
\begin{center}
\includegraphics[width=0.45\textwidth,angle=270]{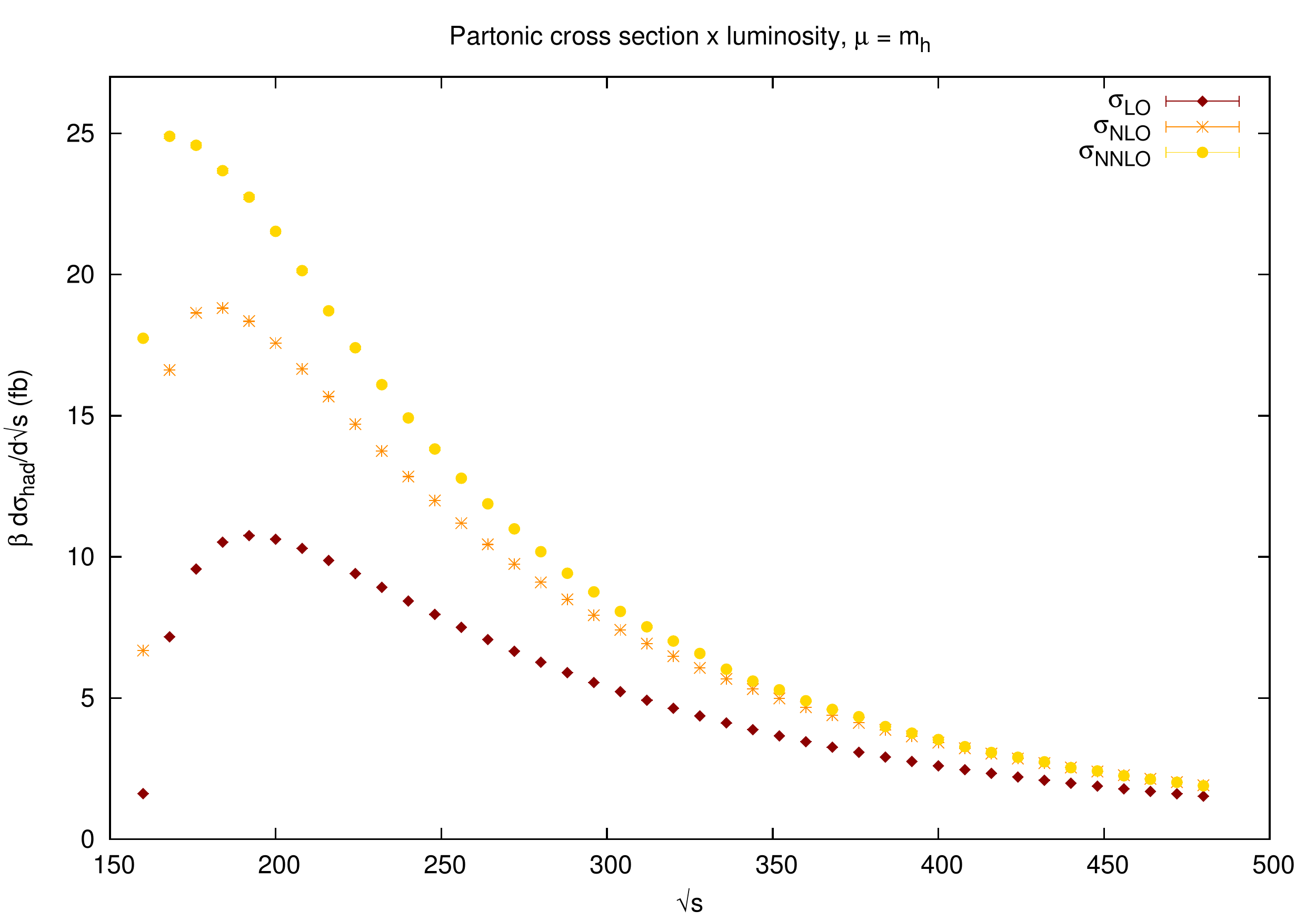}
\end{center}
\caption{Results for the product of partonic cross sections 
$gg \to H+{\rm jet}$ and parton luminosities in consecutive  orders in perturbative QCD
for an 8 TeV LHC at $\mu_R = \mu_F = m_h = 125~{\rm GeV}$.  The $x$-axis gives the value of partonic center-of-momentum frame energy.  See the text for 
further explanation. 
 }\label{fig:qcd-Hj-xsect}
\end{figure}

Fig.~\ref{fig:qcd-Hj-xsect} shows the partonic cross
section for $gg\to H+j$ multiplied by the gluon luminosity through NNLO in perturbative QCD:
\begin{equation}
\beta \frac{{\rm d}\sigma_{\rm had}}{{\rm d}\sqrt{s}} = \beta \frac{{\rm d}\sigma (s,\as,\mu_R,\mu_F)}{{\rm}d\sqrt{s}} \times
\mathcal L(\frac{s}{s_{\rm had}},\mu_F),
\end{equation}
where $\beta$ measures the distance from the partonic threshold,
\begin{equation}
\beta = \sqrt{1-\frac{E_{th}^2}{s}},~~~~~~~ E_{th} = \sqrt{m_H^2 + p_{\perp,j}^2} + p_{\perp,j}\approx 158.55~{\rm GeV}.
\end{equation}
It follows from Fig.~\ref{fig:qcd-Hj-xsect} that NNLO QCD corrections are significant in the 
region $\sqrt{s} < 500$~GeV. In particular, 
close to  partonic threshold $ \sqrt{s} \sim E_{th}$, radiative corrections are enhanced 
by threshold logarithms $\ln \beta$ that originate from the incomplete 
cancellation of virtual and real corrections.  There seems to be no significant enhancement 
of these corrections at higher  energies, where  the  NNLO QCD prediction for 
the partonic cross section becomes almost indistinguishable from 
the NLO QCD one.  This suggests that QCD corrections to inclusive Higgs plus one-jet production will be milder at potential future 33 TeV and 100 TeV $pp$ colliders.  Since more phase space for harder gluon emission will be available, the threshold region will contribute a smaller fraction of the cross section at these higher-energy machines, reducing the effect of $\ln \beta$ terms.  This is consistent with the pattern for inclusive Higgs production reported in Section~\ref{sec:qcd-xsecs-approxn3lo}.  It would be interesting to study ${\rm d}\sigma_{\rm had}/{\rm d}\sqrt{s}$ at higher-energy $pp$ machines upon completion of the full calculation.

Discussed next is the resummation of jet-veto induced logarithms in exclusive jet bins.  This has been the subject of intense discussion during the 7 TeV and 8 TeV LHC runs.  Results combining resummation and fixed-order are now available for the zero-jet~\cite{Banfi:2012yh,Becher:2012qa,Tackmann:2012bt,Banfi:2012jm,Becher:2013xia,Stewart:2013faa} and one-jet~\cite{Liu:2012sz,Liu:2013hba} bins.  The resummation of large logarithms significantly reduces the theoretical uncertainties for the signal cross sections, and has a moderate impact on their central values.  Given the continued importance of analyses utilizing exclusive jet bins during the 14 TeV run of the LHC and it potential future higher-energy colliders, it is important to study how predictions scale with increasing collider energy.  For the exclusive Higgs plus one-jet bin in the gluon-fusion channel, the parametric form of the large logarithmic corrections is ${\rm ln}(\sqrt{s}/p_T^{cut})$, where $\sqrt{s}$ denotes the partonic center-of-momentum energy.  As the collider energy increases, more events with large $\sqrt{s}$ contribute.  The impact of the large logs and the resummation is expected to increase at higher-energy machines, if the transverse momentum cut $p_T^{cut}$ remains the same.  

Shown below in Table~\ref{table:qcd-hj1} are cross sections for  the parameter choices $m_H=125$ GeV, $|\eta_J|<4.5$, and using MSTW 2008 PDFs~\cite{Martin:2009iq}.  A fixed transverse momentum cut $p_T^{cut}=30$ GeV is assumed, and the central scale choice $\mu=m_H/2$ is chosen for the fixed order predictions.  Uncertainties are computed as described in Ref.~\cite{Liu:2013hba}, where the central values for the resummation are also discussed.  Also shown is the $K$-factor describing the change in the prediction as the resummation is added.  Results are presented at the NLO in fixed-order perturbation theory, and at ${\rm NLL}^{\prime}+{\rm NLO}$ in resummed perturbation theory (the order-counting of resummed 
perturbation theory used here is described in Ref.~\cite{Berger:2010xi}).  The usefulness of the resummation in decreasing the theoretical uncertainties is clear.  The fixed-order uncertainties grow with collider energy, reaching nearly 100\% at 100 TeV.  In contrast, they remain between $20\%-25\%$ when the resummation is implemented.  In 14 TeV and 33 TeV $pp$ collisions, the effect of resummation is to slightly decrease the central value of the prediction, by up to 5\%.  This behavior changes significantly at 100 TeV, where instead a nearly 30\% increase in the cross section is found.  This is likely caused by the fixed-order perturbative expansion entirely breaking down and heading to negative values for such a large hierarchy between the hard scale and $p_T^{cut}$.  This is suggested by the large uncertainty present in the NLO result.  Resummation cures this behavior, and leads to an increase in the cross section.   However, it is likely that the minimum jet transverse momentum would have to be increased at future facilities, due to the increased energy deposited by the underlying soft physics.  We model this by increasing the minimum jet transverse momentum to 60 GeV in 33 TeV collisions, and to 80 GeV in 100 TeV collisions.  The results are shown in Table~\ref{table:qcd-hj2}.  The change in $K$-factor as the collider energy is increased is ameliorated; a less than 15\% increase in the cross section is found at 100 TeV.

\begin{table}
\begin{center}
\begin{tabular}{l || l | l | l |}
 & 14 TeV & 33 TeV & 100 TeV \\ \hline \hline
 ${\rm NLO}$ & $12.48^{+34\%}_{-46\%}$ & $40.17^{+54\%}_{-41\%}$  & $131.3^{+72\%}_{-98\%}$ \\ \hline
 ${\rm NLL}^{\prime}+{\rm NLO}$ &$11.73^{+27\%}_{-27\%}$ & $39.71^{+24\%}_{-24\%}$ & $166.9^{+22\%}_{-22\%}$ \\ \hline
 $K_{({\rm NLL}^{\prime}+{\rm NLO}){\rm NLO}}$ & 0.940 & 0.989 & 1.27 \\ \hline
\end{tabular}
\end{center}
\caption{Cross section central values and uncertainties for the exclusive Higgs plus one-jet bin for a fixed transverse momentum cut $p_T^{cut}=30$ GeV.  The results are shown in picobarns.}
\label{table:qcd-hj1}
\end{table}

\begin{table}
\begin{center}
\begin{tabular}{l || l | l | l |}
 & 14 TeV & 33 TeV & 100 TeV \\ \hline \hline
 ${\rm NLO}$ & $12.48^{+34\%}_{-46\%}$ & $26.90^{+30\%}_{-39\%}$  & $91.23^{+38\%}_{-46\%}$ \\ \hline
 ${\rm NLL}^{\prime}+{\rm NLO}$ &$11.73^{+27\%}_{-27\%}$ & $27.44^{+24\%}_{-24\%}$ & $103.0^{+24\%}_{-24\%}$ \\ \hline
 $K_{({\rm NLL}^{\prime}+{\rm NLO}){\rm NLO}}$ & 0.940 & 1.02 & 1.13 \\ \hline
\end{tabular}
\end{center}
\caption{Cross section central values and uncertainties for the exclusive Higgs plus one-jet bin, for the following transverse momentum cuts: $p_T^{cut}=30$ GeV at 14 TeV, $p_T^{cut}=60$ GeV at 33 TeV, and $p_T^{cut}=80$~GeV at 100 TeV .  The results are shown in picobarns.}
\label{table:qcd-hj2}
\end{table}

Studied next is the cross section for Higgs production via gluon-fusion in the exclusive zero-jet bin.  This cross section has been studied through ${\rm NNLL}^{\prime}+{\rm NNLO}$ in resummed perturbation theory in Ref.~\cite{Stewart:2013faa}.  A careful study of clustering contributions of the form ${\rm ln}\,{\rm R}$, where R denotes the anti-$k_T$ jet-radius parameter, was also performed in this reference.  Numerical results for the Higgs plus zero-jet cross section at the ${\rm NNLL}^{\prime}+{\rm NNLO}$ order are presented in Table~\ref{table:qcd-h0exc} for $pp$ collisions at 14, 33, and 100 TeV.  Also shown is $\epsilon_0$, the fraction of events which fall into the zero-jet bin.  A fixed transverse momentum cut $p_T^{cut}=30$ GeV is assumed.  The most notable effect upon increasing collider energy is the significant reduction of the fraction of events in the zero-jet bin, from 60\% at 14 TeV to 44\% at 100 TeV.  The range of Bjorken-$x$ becomes larger as the collider energy is increased, leading to a larger probability for additional radiation and consequently reducing the number of zero-jet events.  A small reduction of scale uncertainty is found when going from 14 TeV collisions to higher energies, similar to what was found for the one-jet cross section in Table~\ref{table:qcd-hj1}.
\begin{table}[htbp]
\begin{center}
\begin{tabular}{l || l | l | l |}
 & 14 TeV & 33 TeV & 100 TeV \\ \hline \hline
 $\sigma_{{\rm NNLL}^{\prime}+{\rm NNLO}}$ & $33.25^{+5.5\%}_{-5.5\%}$ & $104.2^{+3.9\%}_{-3.9\%}$  & $364.2^{+4.4\%}_{-4.4\%}$ \\ \hline
 $\epsilon_0^{{\rm NNLL}^{\prime}+{\rm NNLO}}$ &$0.596^{+4.4\%}_{-4.4\%}$ & $0.522^{+4.9\%}_{-4.9\%}$ & $0.438^{+4.4\%}_{-4.4\%}$ \\ \hline
\end{tabular}
\end{center}
\caption{Central values and uncertainties for the exclusive Higgs plus zero-jet bin cross section and zero-jet event fraction, for a fixed transverse momentum cut $p_T^{cut}=30$ GeV.  The results are shown in picobarns.}
\label{table:qcd-h0exc}
\end{table}


\subsection{\texorpdfstring{$W\!H$}{WH} production at NNLO}
\label{sec:qcd-xsecs-convergence}

\draftnote{2 pages: based on figures from M. Grazzini}

Sometimes it is experimentally necessary to require an exclusive final state,
for example by restricting the number of jets allowed to be present.  This is
true for some current LHC analyses and will no doubt continue to be true for
higher energies. This requirement of an exclusive final state can result in the
presence of large logarithms due to the unbalancing of the cancellation between
positive real emission terms and negative virtual corrections, thus affecting
the convergence and reliability of the perturbative series. As an example,
consider searches for Higgs boson production in the associated mode ($V\!H$). While this
has been the major Higgs boson search channel at the Tevatron, at the LHC it suffers
from large backgrounds. A precise measurement in this channel (with the Higgs
decaying into a $b\bar b$ pair) can still be very useful in the years
following the Higgs boson discovery in order to
understand the Higgs coupling to $b$-quarks. A relative reduction in the
background at the LHC can be achieved by requiring the vector boson and the
Higgs boson to be at large transverse momentum, with no additional jets present
greater than some threshold. While the impact of higher order QCD corrections is
mild for the inclusive measurement, such restrictions can greatly affect the
size of the higher order corrections and thus the convergence of the
perturbative series.  In this context, the issue was first raised in 
the original exclusive NNLO calculation of $W\!H$ production~\cite{Ferrera:2011bk}.

\begin{figure}[htbp]
\begin{center}
\includegraphics[width=0.43\hsize]{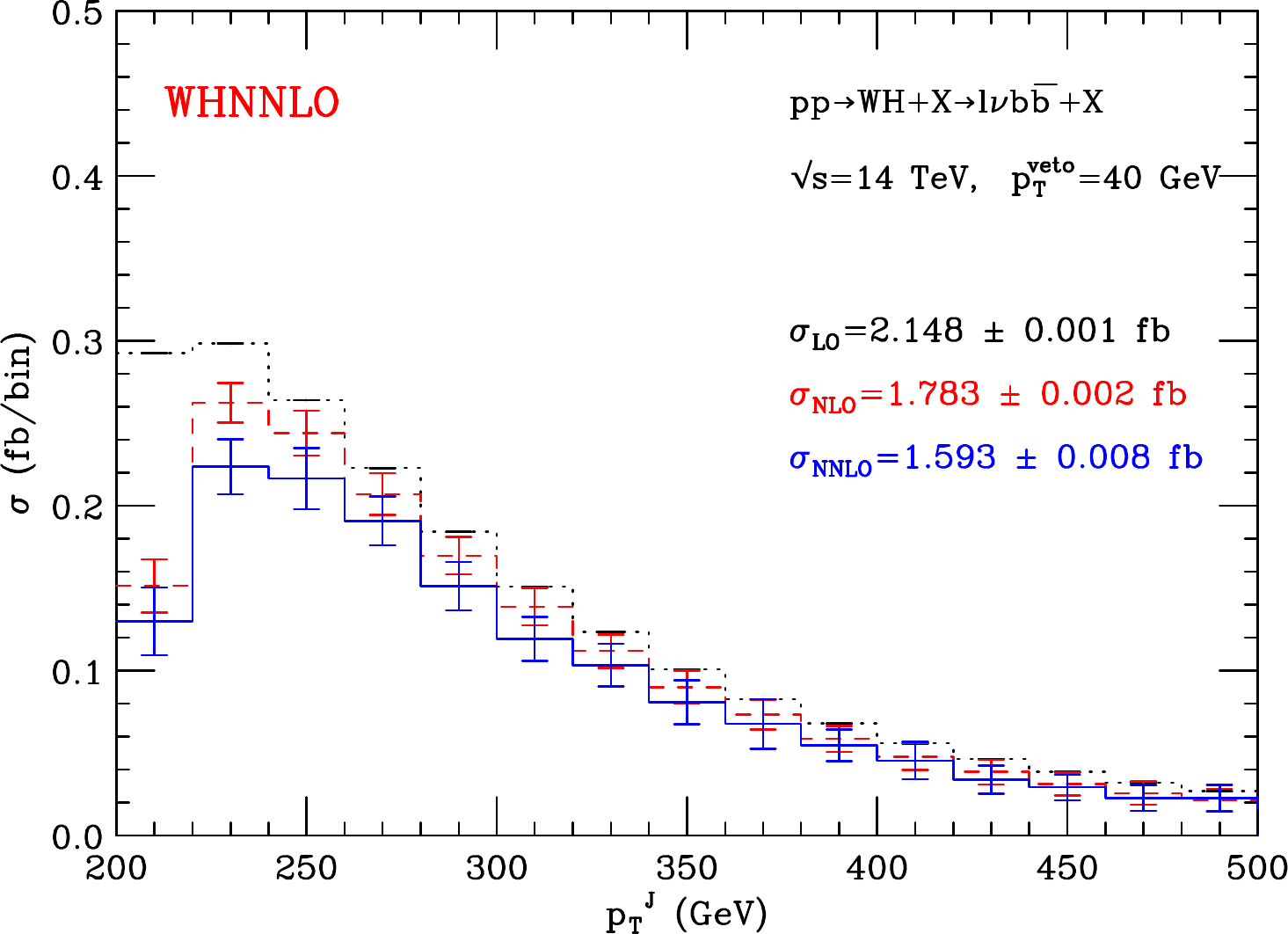}
\hspace*{0.43\hsize} \\
\includegraphics[width=0.43\hsize,bb=40 280 460 560]{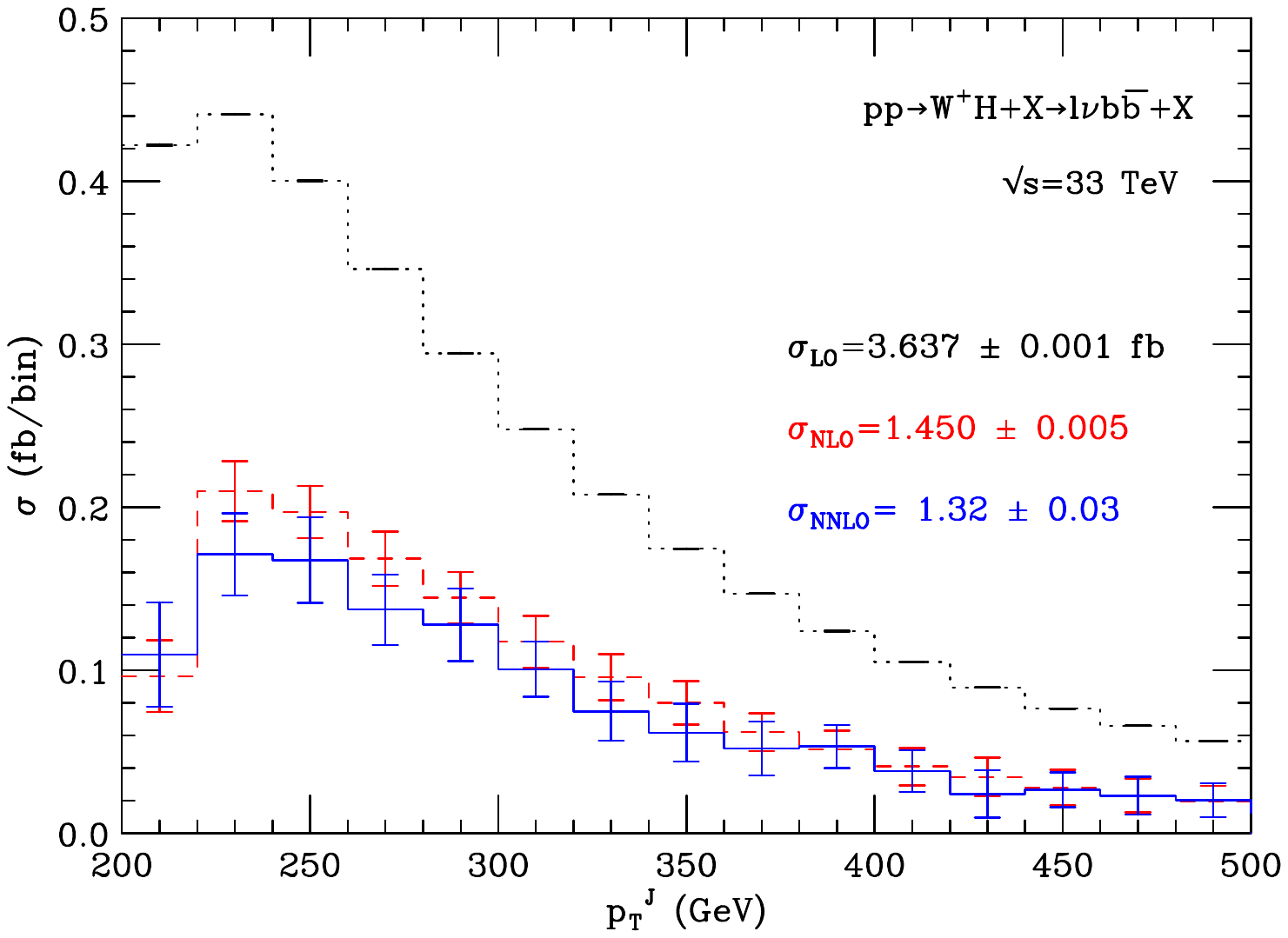}
\includegraphics[width=0.43\hsize]{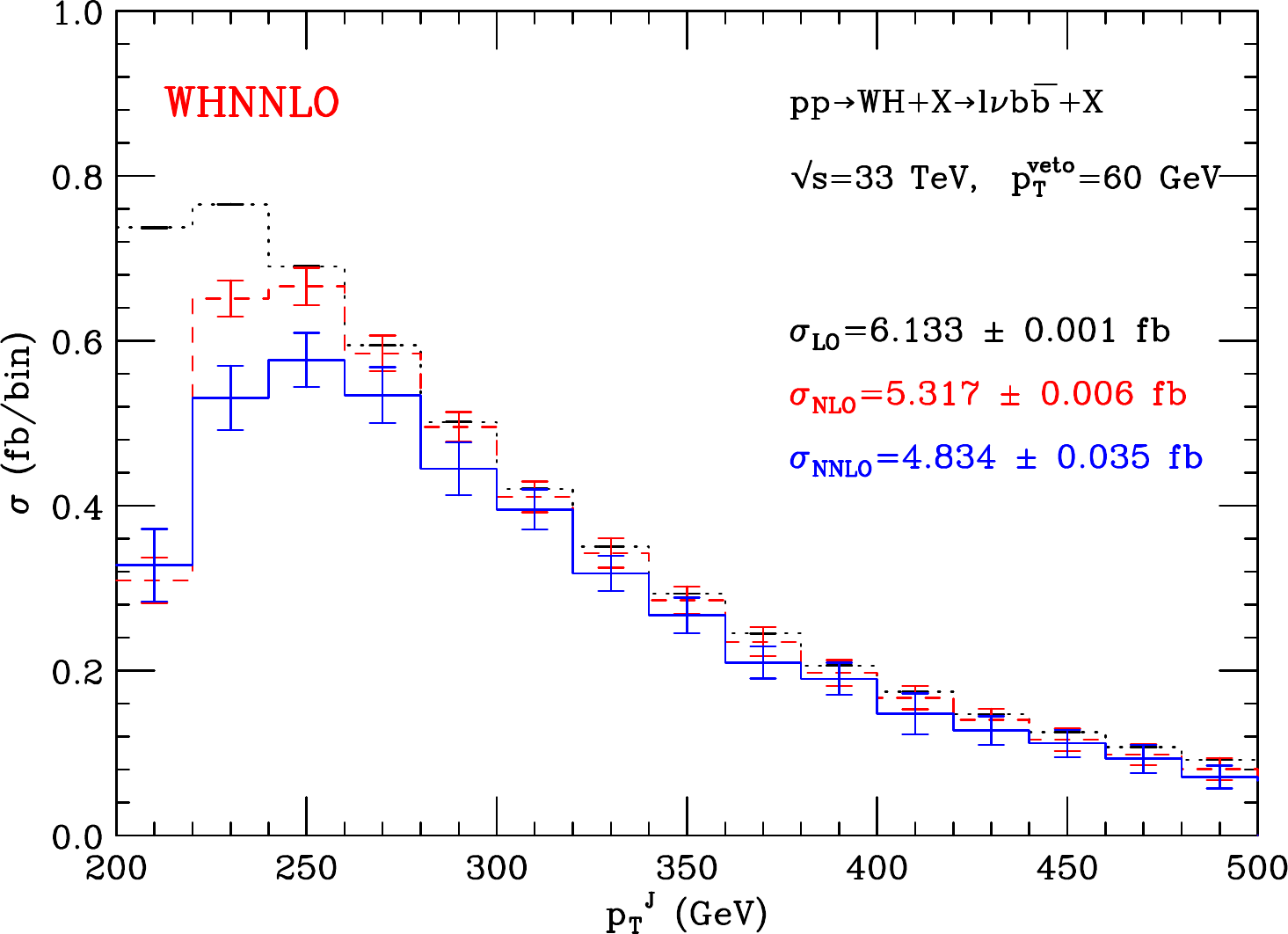} \\
\includegraphics[width=0.43\hsize,bb=40 280 460 560]{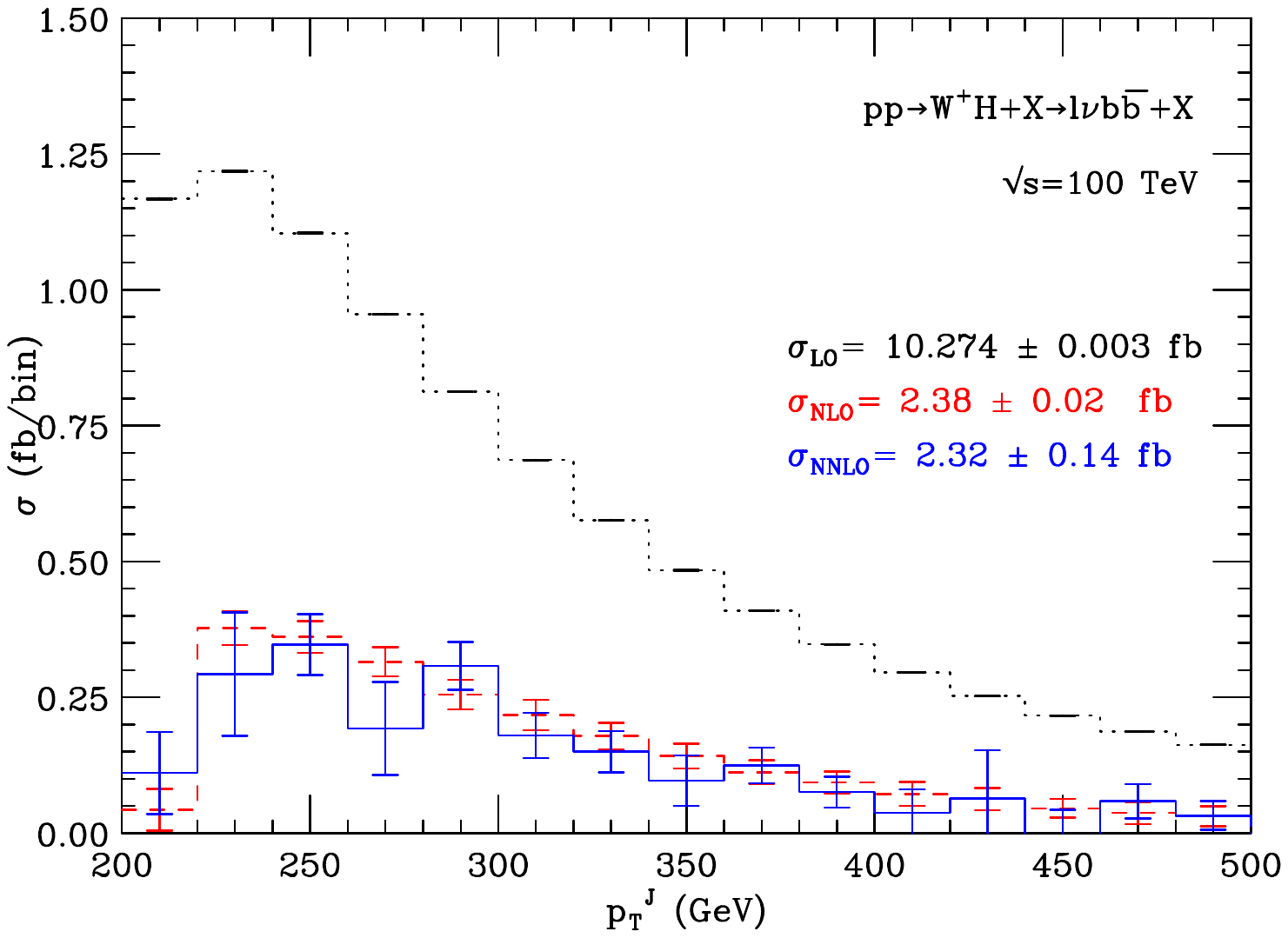}
\includegraphics[width=0.43\hsize]{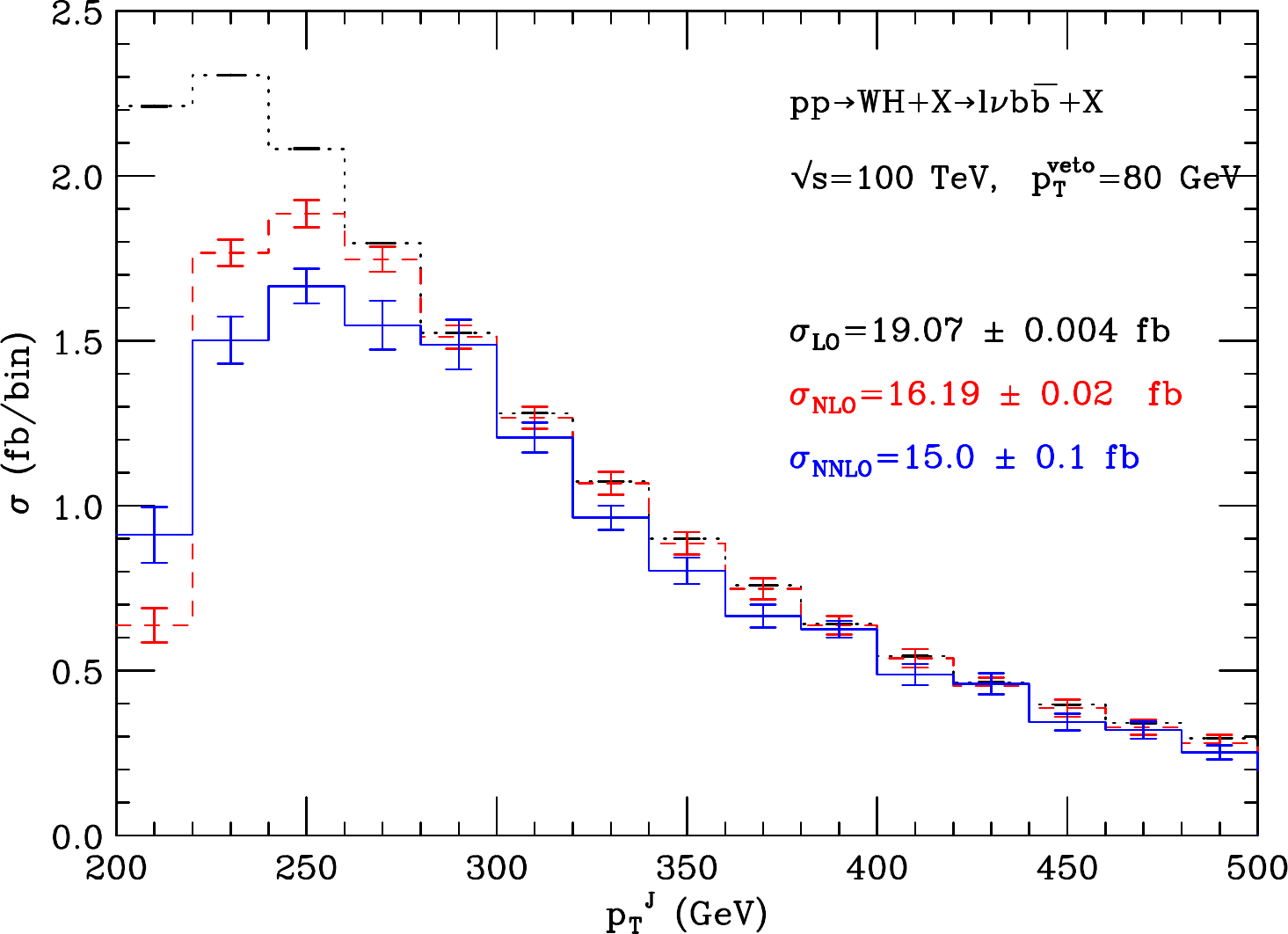}
\caption{The transverse momentum of the fat jet in $W\!H$ production, at $14$~TeV (top), $33$~TeV (middle)
and $100$~TeV (bottom). In the left-hand column additional jets with transverse momenta above
$40$~GeV are vetoed, while in the right this veto is raised to $60$~GeV at $33$~TeV and $80$~GeV
at $100$~TeV.
}
\label{fig:qcd-xsecs-whnnlo}
\end{center}
\end{figure}

In this report we address the extent to which this issue is exacerbated in
predictions for higher energy proton-proton colliders.  For the following study
the transverse momentum of the $W$ boson is required to be
greater than $200$~GeV. The Higgs boson decays into a $b\bar b$ pair that is
reconstructed as a ``fat jet'' using the Cambridge-Aachen algorithm with $R=1.2$. 
This fat jet must also be at large transverse momentum, greater than $200$~GeV.
In addition to these cuts, any additional jets were required
have transverse momenta below $40$~GeV.  The study was repeated at $14$, $33$
and $100$~TeV, with the results shown in Fig.~\ref{fig:qcd-xsecs-whnnlo} (left).
The cross section is plotted as a function of the transverse momentum of the
fat (Higgs) jet.   As the center-of-mass energy increases, the effect of such
stringent cuts increases.  There are large negative corrections when
going from LO to NLO, resulting in a NLO $K$-factor (NLO/LO) as low as $0.3$ (for $100$~TeV).
The corrections from NLO to NNLO by comparison are modest, but the strong
reduction from LO to NLO indicates that the fixed order prediction may be
unreliable, and a resummed cross section is necessary to achieve a reliable
prediction.  As alternative scenarios at higher energies, Fig.~\ref{fig:qcd-xsecs-whnnlo}
(right) shows the effect of raising the jet threshold to $60$~GeV (at $33$~TeV) and $80$~GeV
(at $100$~TeV). In these cases, the effect of allowing additional phase space
results in a better-behaved perturbative series, and cross
sections that behave order-by-order in a manner that is more similar to the pattern
observed for a $40$~GeV veto at $14$~TeV. 


\section{Quarkonium physics at future colliders}
\label{sec:qcd-quarkonium}

\draftnote{1 page: E. Braaten, G. Bodwin, J. Russ}

A detailed account of issues in quarkonium production at the Energy
Frontier can be found in ``Quarkonium at the Frontiers of High Energy
Physics: A Snowmass White Paper'' \cite{Bodwin:2013nua} and references
therein.

The production of heavy quarkonia at large transverse momentum $p_T$ is
an aspect of quarkonium physics in which dramatic progress can be
expected through the interaction of theory with experiments at the
energy frontier. Quarkonium production is an aspect of QCD that is not
yet fully understood and it can serve as a theoretical laboratory for
QCD in which powerful effective-field theory tools can be brought into
play. Quarkonium production can be used to test and to extend our
understanding of factorization theorems, which are the theoretical
foundation for all perturbative calculations in QCD, and to develop new
theoretical concepts for controlling large higher-order corrections that
could have wider applicability in the calculation of high-energy cross
sections. Quarkonium production is also useful as a laboratory for
exploring experimental techniques within a setting in which experimental
signatures are very clean and very high statistics can be accumulated.
Quarkonium production processes can be used to measure Higgs couplings,
and so to probe for physics beyond the standard model. If new physics
involves nonrelativistic bound states, then techniques that have been
developed for understanding quarkonium production will be directly
applicable.

The current standard method for calculating quarkonium production rates
is the nonrelativistic QCD (NRQCD) factorization approach, which is
based on the effective field theory NRQCD. In this approach, production
rates are expressed as perturbatively-calculable partonic cross sections
multiplied by nonperturbative constants called NRQCD matrix elements.
Some of the NRQCD matrix elements must be determined through fits of
NRQCD factorization predictions to experimental data. The universality
feature of NRQCD factorization then allows one to make predictions for
other quarkonium production processes---predictions that should be
tested for as many processes as possible. The NRQCD factorization
approach is a conjecture that has not been proven to all orders in
$\alpha_s$. Therefore, experimental tests take on an added importance.
Standard methods for proving factorization suggest that, if NRQCD
factorization is correct, then it holds only for $p_T$ much larger than the
quarkonium mass. Hence, high-$p_T$, high-statistics measurements of 
quarkonium production are key tests of factorization. 

NRQCD factorization predictions have now been computed at
next-to-leading order in $\alpha_s$ for many production processes.
In general, these predictions agree with the experimental data from
$pp$, $p\bar p$, $ep$, and $e^+e^-$ colliders for the production of
quarkonia at large $p_T$. The most notable exception is the polarization
of the $J/\psi$, $\psi(2S)$, and $\Upsilon(nS)$ ($n=1,2,3$) at the
Tevatron and at the LHC. This serious discrepancy between theory and
experiment deserves further investigation from both the theoretical and
experimental sides. Feeddown from quarkonium states of higher mass to
the $J/\psi$ and $\Upsilon(1S)$ states blurs the comparisons between
theory and experiment. Future experiments could sharpen these
comparisons by measuring direct-production rates of quarkonia.

At the frontiers of high-energy physics, there are many opportunities
for further interesting work on quarkonium production. In Run~2 of the
LHC, the extension of the energy frontier to $13$~TeV and the increase
in luminosity in comparison with Run~I will make it possible to extend
the $p_T$ reach of quarkonium studies. Such high-$p_T$ measurement are
crucial tests of NRQCD factorization. Measurements of production and
polarization for the $\chi_{cJ}$ and $\chi_{bJ}$ states and for new
processes, such as associated production with $W$ or $Z$ bosons, would
provide valuable additional tests of the theory.

The LHC luminosity upgrade would afford the opportunity to push studies of $b\bar
b$ states to still higher values of $p_T$. The $b\bar b$ systems are
particularly important tests of the validity of NRQCD because they are
more nonrelativistic than the $c\bar c$ systems. However, tests of
factorization for these systems will require that challenging
measurements be made at values of $p_T$ that are much larger than the
bottomonium mass.

A future high-energy $e^+e^-$ collider will allow studies of quarkonium
production in two-photon collisions. In a Higgs factory mode, it would
afford new opportunities to make precision measurements of the Higgs
couplings to Standard Model particles. The decay modes $H\to
J/\psi+\gamma$, $H\to J/\psi+Z$, and $H\to \Upsilon+Z$ might be
particularly useful in this regard.


\section{Jet and missing $E_T$ performance at HL-LHC }
\label{sec:}

\draftnote{A. Schwartzman}

Both the precision and the discovery physics programs at the 14 TeV LHC and at future colliders require
high instantaneous luminosities. This will result in the presence of a large number of additional
interactions for every crossing, and the subsequent difficulties in triggering on interesting physics
and then in removing the effects of the pileup to recover the original physics. Pileup is one of the
main challenges for jets and missing transverse energy, with both the presence of additional energy
(offset) as well as the creation of additional fake jets due to local fluctuations in the energy
density.   Thus, it is important to develop the tools needed for removing the effects of pileup in
order to insure that precise measurements can still be undertaken at high luminosities. 

A study was carried out within the ATLAS full simulation framework, using topological clustering and
local hadron calibration, where samples with up to 200 pileup events per crossing (with both $50$ns and
$25$ns bunch structure) were used. 
Topoclusters are an attempt to reconstruct three-dimensional energy deposits in the calorimeter and are
built using a nearest-neighbor algorithm that clusters calorimeter cells with energy significance
($|E_{cell}|/\sigma$)~$>4$~for the seed, $>2$~for neighbors, and $>0$~at the boundary. $E_{cell}$ is
calibrated using information derived from test beam and detailed GEANT4 simulations (EM scale) and
$\sigma$ is the sum in quadrature of the electronic and expected pileup noise, as described in
Sec. 10.5.2 of Ref.~\cite{Aad:2008zzm}. Topoclusters can be further calibrated using local
information (LCW) as described in Ref.~\cite{Cojocaru:2004jk}.  

\begin{figure}[t!]
\begin{center}
\includegraphics[width=0.4\linewidth]{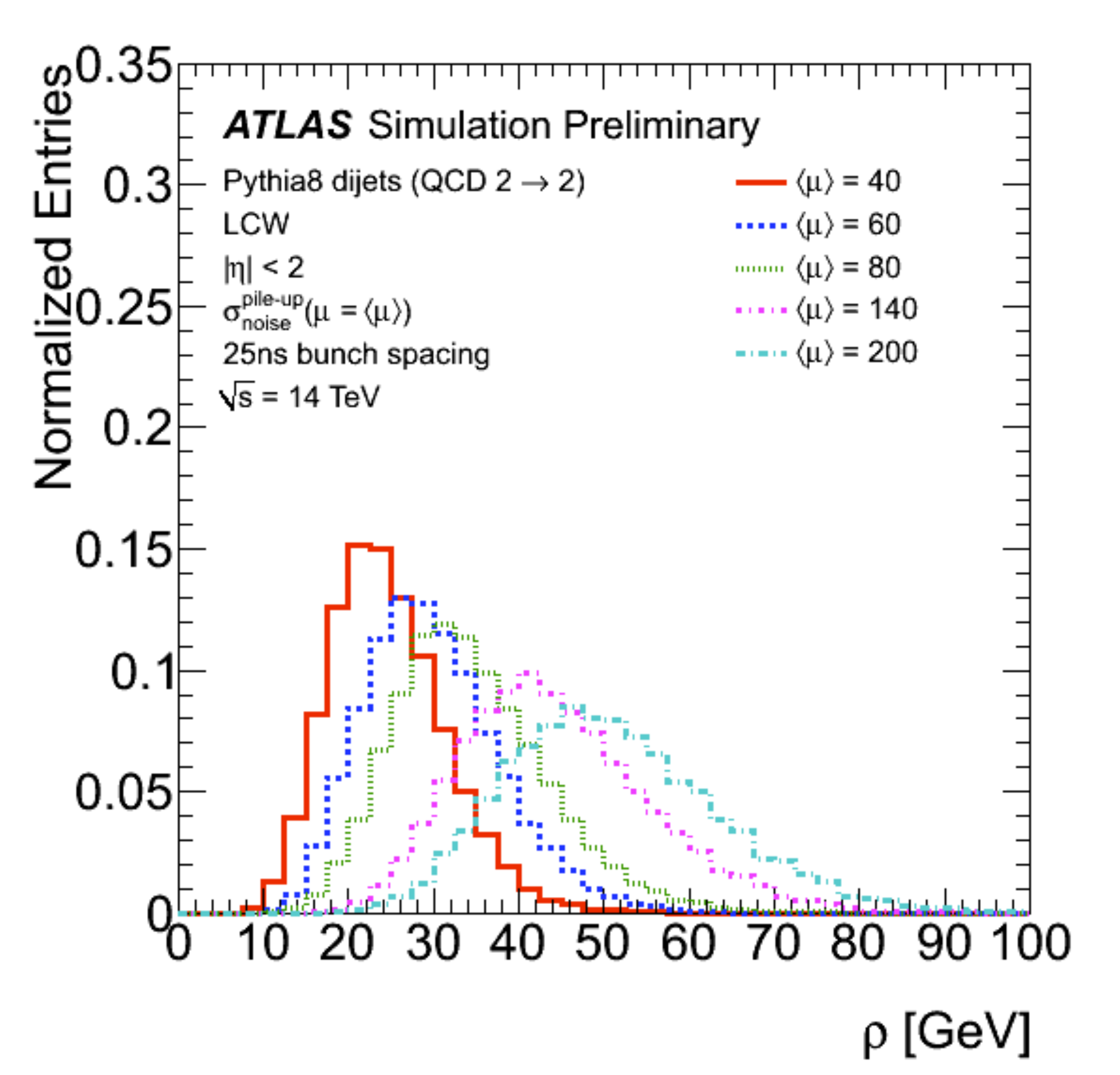}
\includegraphics[width=0.4\linewidth]{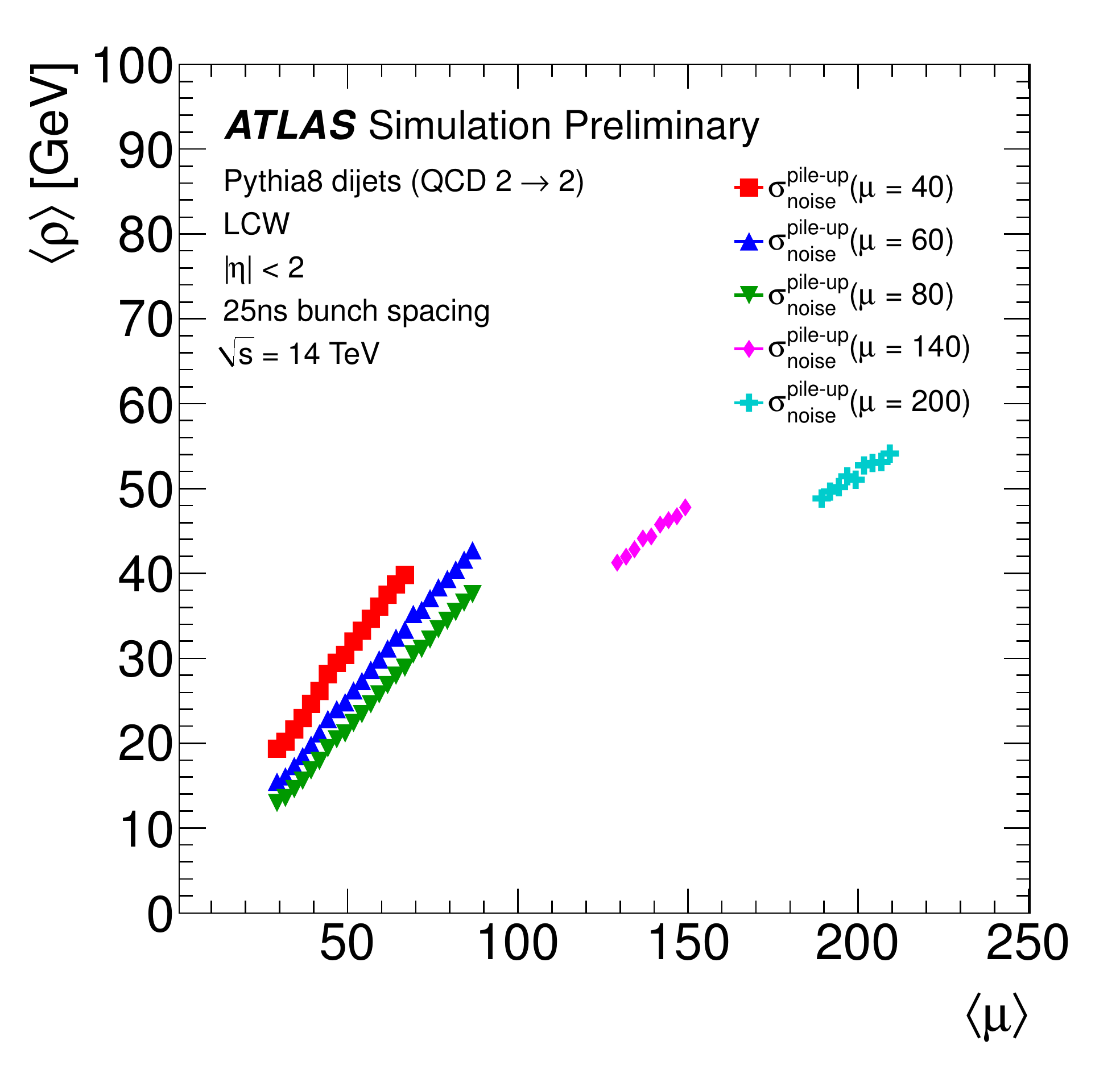}
\end{center}
\caption{Distributions of the event-by-event median $p_{T}$ density ($\rho$) for different values of the mean number of interactions per bunch crossing
$\langle\mu\rangle$ (left) and the dependence of $\rho$ on the mean number of interactions per bunch crossing (right).}
\label{fig:rho}
\end{figure}
Figure~\ref{fig:rho} shows an example of an event where hard jets are embedded in a background of soft
energy resulting from multiple minimum bias events. An effective density $\rho$ can be determined on an
event-by-event basis~\cite{Cacciari:2007fd} based on energy depositions outside hard jets.  The
corrected jet transverse momentum can be calculated by subtracting the contribution due to pileup
$\rho A^{jet}$, where $A^{jet}$ is the active area of the jet. The technique works since the pileup density depends
linearly on the number of interactions up to very high values of the average number of pileup events.
As can be seen in Fig.~\ref{fig:rho}, the optimization of $\sigma$ pileup noise during topocluster formation is
critical to suppress pileup contributions at high luminosity. When the pileup noise used in
topoclustering is chosen to match the pileup conditions, a significant reduction on the event $p_T$
density is achieved. 
Although the event-by-event pileup subtraction accounts for global pileup fluctuations from one event to
another, it is not sensitive to local fluctuations. A residual correction is necessary to account for
the higher occupancy inside jets and for out-of-time pileup effects. 

\begin{figure}[t!]
\begin{center}
\includegraphics[width=0.5\linewidth]{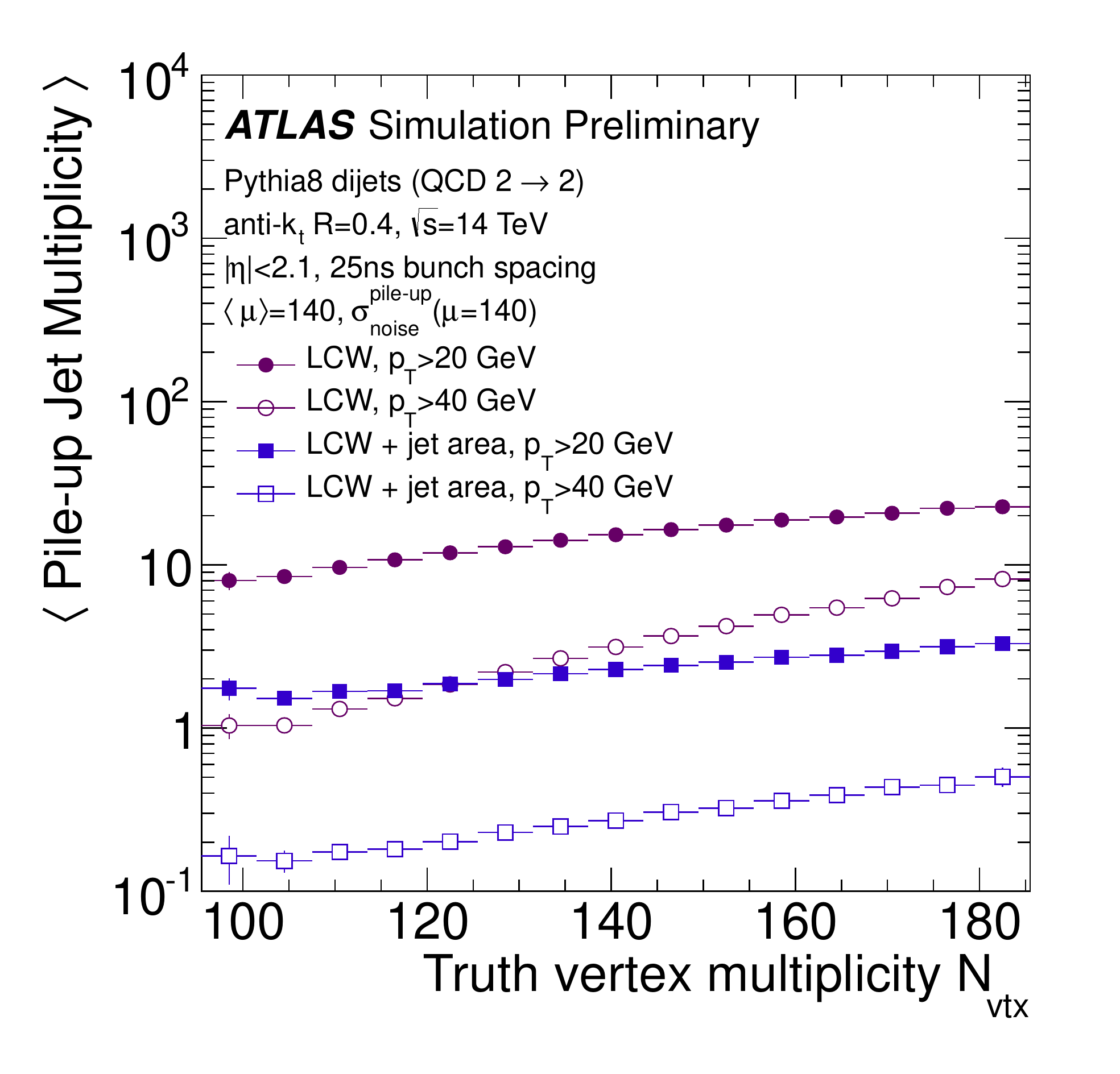}
\end{center}
\caption{The mean pileup jet multiplicity as a function of the number of truth vertices $N_{vtx}$ in events with an average number
of interactions per bunch crossing, $\langle\mu\rangle=140$. A pileup-jet is defined as a reconstructed jet that is not closely matched in $\Delta R$ to a truth jet.}
\label{fig:fakes}
\end{figure}

The jet calibration scheme has been shown to work well up to very high luminosities. The pileup
subtraction technique restores the jet energy response to that of jets in the presence of no pileup. 
The pileup subtraction also significantly reduces the number of fake (pileup) events per jet, as shown
in Fig.~\ref{fig:fakes}. There are approximately $3$ ($0.5$) pileup jets with $p_T> 20$ ($40$) GeV per event,
with approximately $140$ additional interactions per crossing. Further improvements can be expected using
tracking and vertexing information. 

\begin{figure}[t!]
\begin{center}
\includegraphics[width=0.45\linewidth]{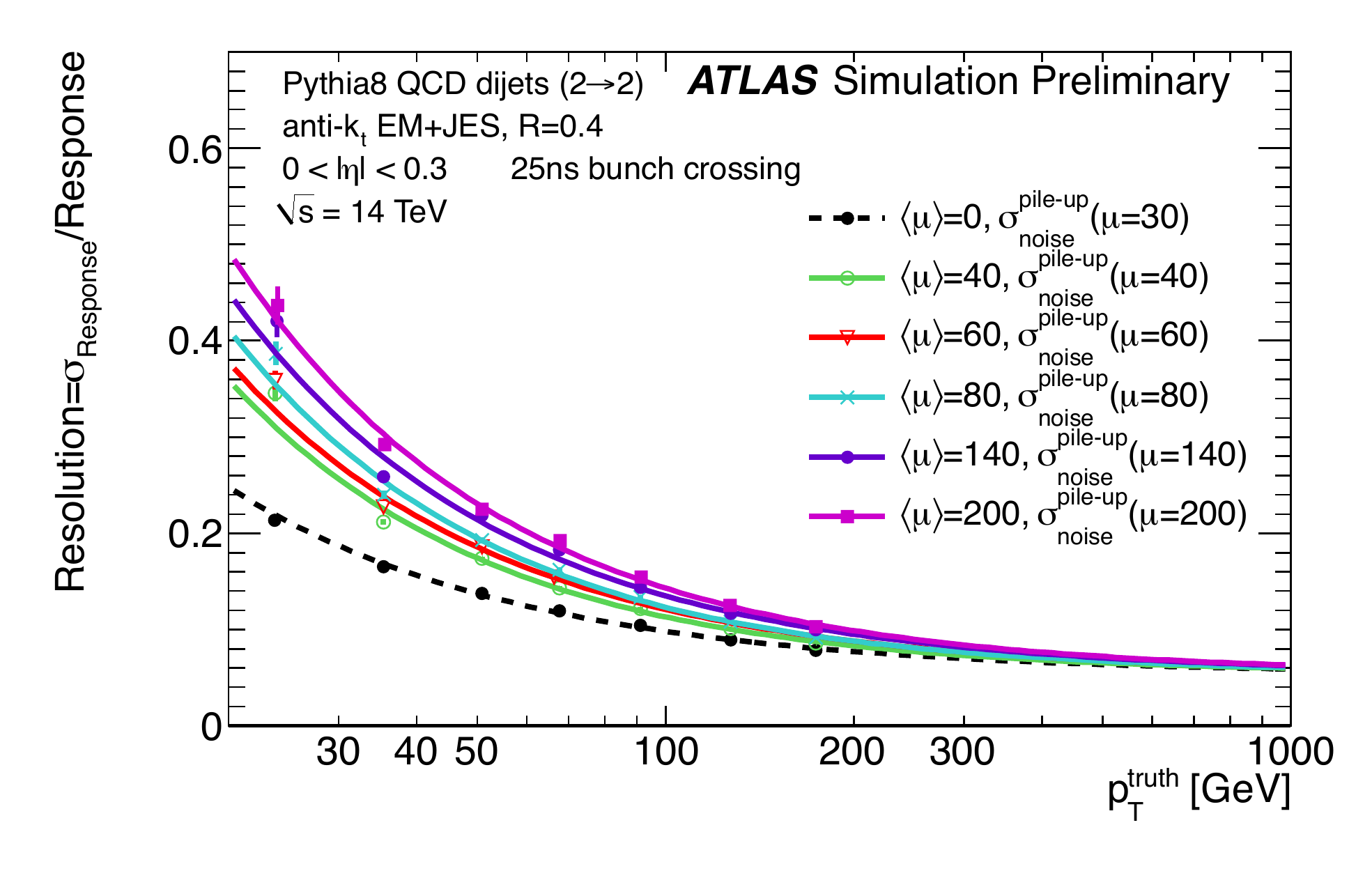}
\includegraphics[width=0.415\linewidth]{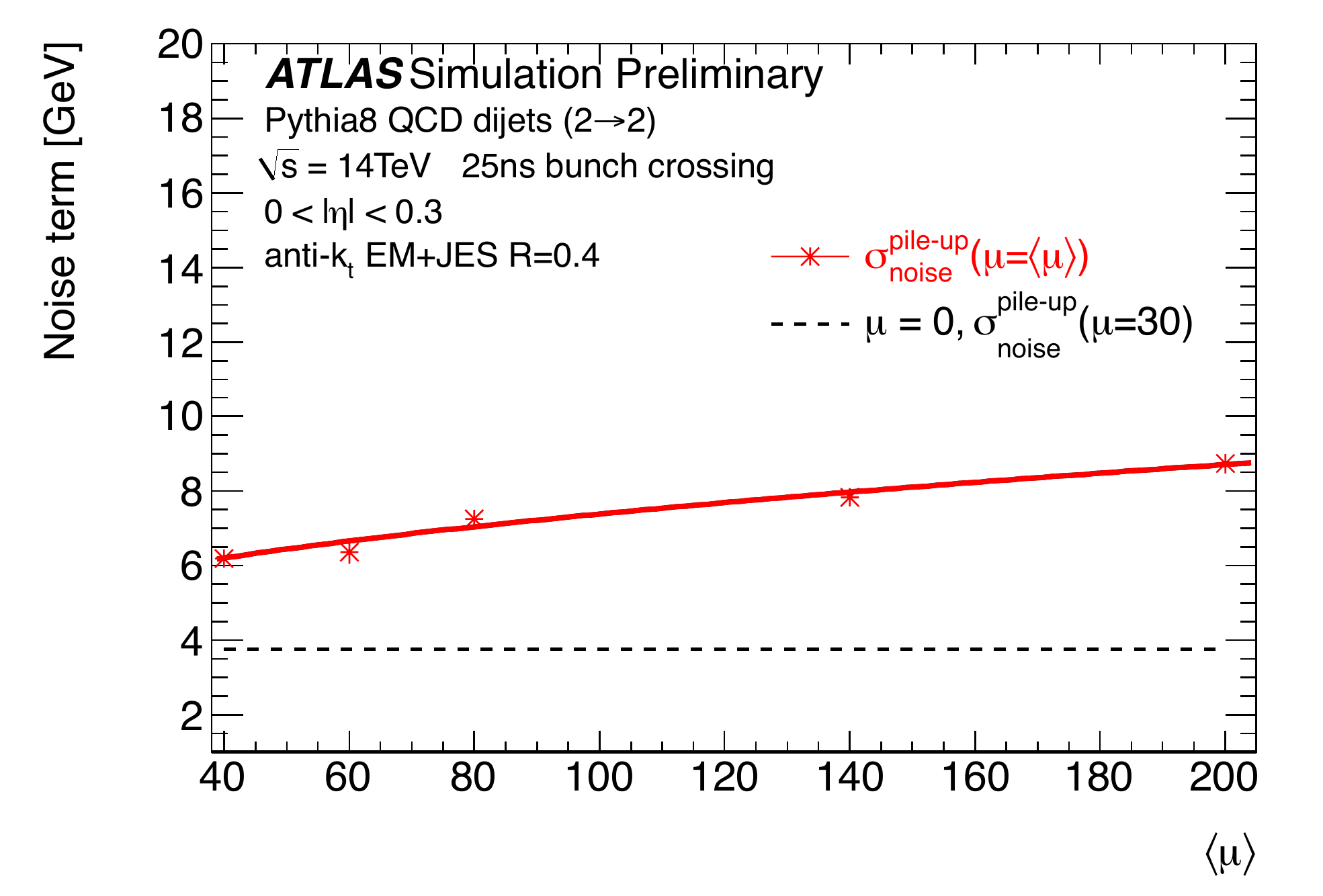}
\end{center}
\caption{The fractional jet energy resolution as a function of jet $p_{T}$ for different values of the mean number of interactions
per bunch crossing (left) and the noise term of the jet energy resolution as a function of $\langle\mu\rangle$ (right).}
\label{fig:resolution}
\end{figure}

The jet energy resolution at low jet transverse momentum degrades in the presence of pileup; there are
local pileup fluctuations within events, not captured by the global event-by-event energy density used
in the calibration. The result is an effective noise term whose magnitude increases as the square root
of the average number of additional interactions. A reduction of local pileup fluctuations may be
possible using tracking information and advanced subtraction techniques using more local information.
Figure~\ref{fig:resolution} shows the dependence of the fractional jet energy resolution and the noise term
on an average number of interactions per bunch crossing, $\langle\mu\rangle$. The square root dependence of the noise term indicates that the degradation in
resolution is due to pileup fluctuations and there are no additional contributions from no-linear
effects of topoclustering or local hadron calibration. 

\begin{figure}[!t]  
  \begin{center}
    \includegraphics[width=0.49\columnwidth]{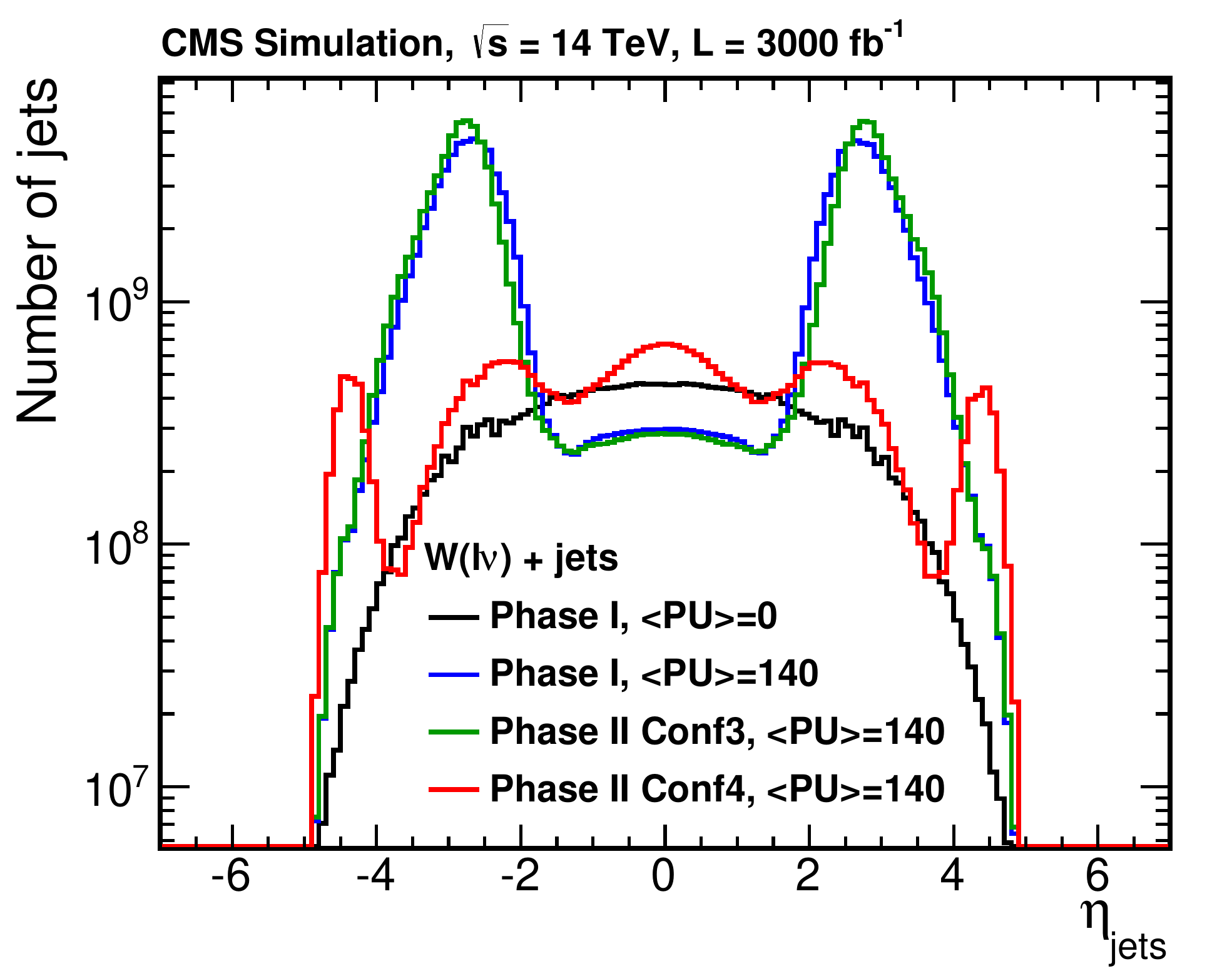}
    \includegraphics[width=0.49\columnwidth]{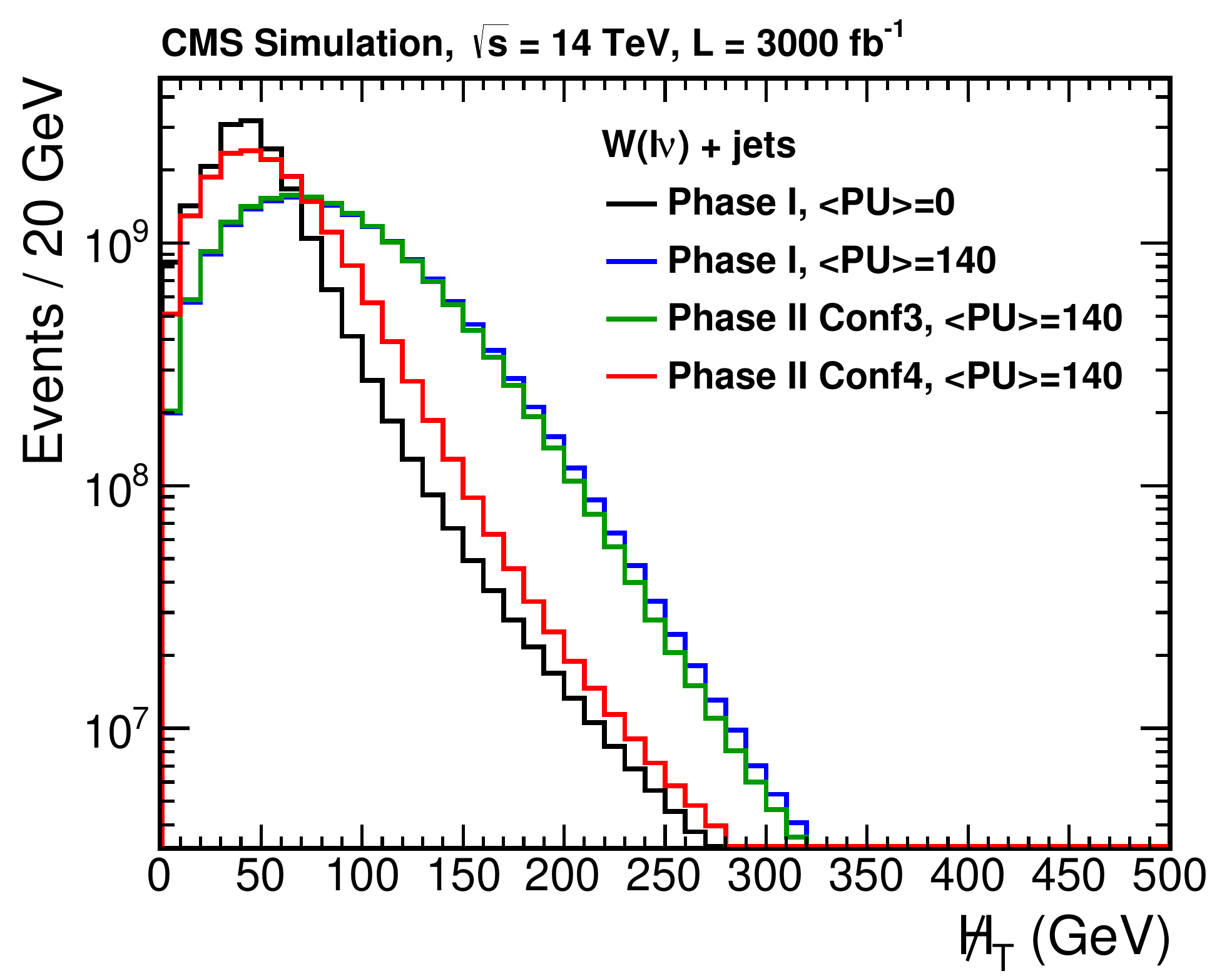} 
    \caption{\label{fig:Wlv_JetEta_MHT}
    Pseudorapidity distributions for jets with $\pt>30$ GeV (left) and \mht distributions (right)
    for various pileup and detector configurations in $W(\to\ell\nu)$ + jets events (left).
    The Phase~II configuration 4 (Conf4) corresponds to the detector design which includes a tracker
    extended up to $|\eta|=4$, while the Phase~I and Phase~II configuration 3 (Conf3) detector has a tracker
    covering only up to $|\eta|=2.5$.
    }
  \end{center}
\end{figure}

In addition to the optimization in event reconstruction, the proper detector upgrade
will be important to deal with high pileup environment.
CMS uses the particle-flow event reconstruction~\cite{CMS-PAS-PFT-09-001}, and pileup charged hadron subtraction can removes
a large fraction of pileup charged particles on a particle-by-particle basis in the region covered by a tracking system.
As shown in Fig.~\ref{fig:Wlv_JetEta_MHT}, an extension of the tracking detector to forward region is expected to help
reducing pileup jets substantially in the forward region and improve missing $E_T$ (\mht) resolution~\cite{CMS-PAS-FTR-13-014} in events
with $\langle\mu\rangle=140$ expected during high luminosity LHC (HL-LHC).

\begin{figure}[t!]
\begin{center}
\includegraphics[width=0.4\linewidth]{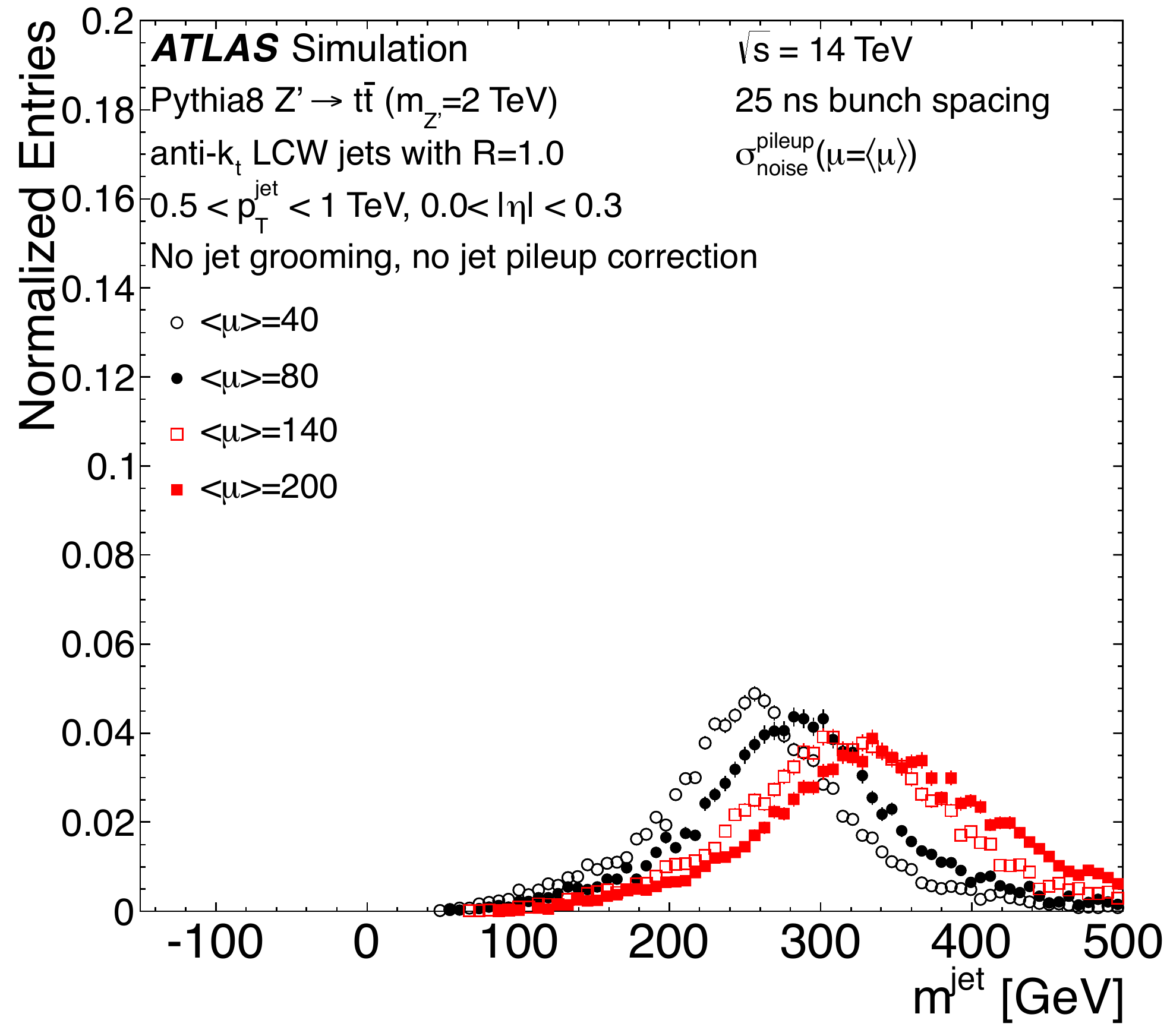}
\includegraphics[width=0.4\linewidth]{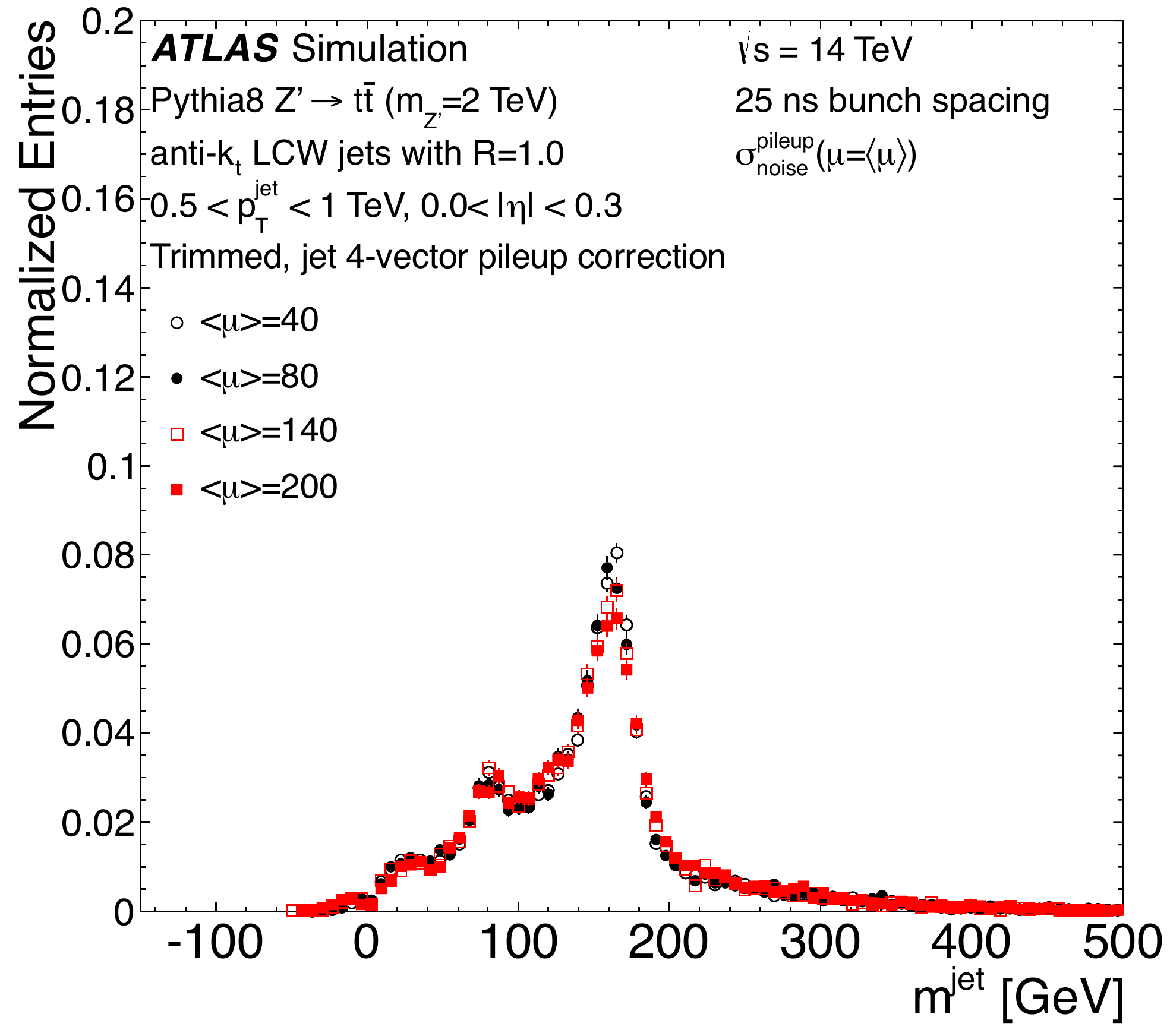}
\end{center}
\caption{The jet mass distribution before (left) and after (right) trimming in $Z^{\prime} \rightarrow t \bar{t}$ events,
for $m(Z^{\prime})=2$~TeV and anti-$k_T$ jets with $R=1.0$. The mean number of interactions per bunch crossing is 40 (open circles),
80 (closed circles), 140 (open squares) and 200 (closed squares). The value of the pileup noise ($\sigma$) used in the topoclustering is
optimized for each value of pileup, except in the case of $\langle\mu\rangle=40$ where the value of $\sigma$ is optimized
for $\langle\mu\rangle=30$, as in 2012 data.}
\label{fig:boost}
\end{figure}

Jet substructure is a key technique for the reconstruction of boosted objects. Grooming algorithms
significantly reduce the sensitivity to pileup, due to the reduced jet area. The original jet mass can
be recovered, with reasonably low degradation of mass resolution. Figure~\ref{fig:boost} shows the jet
mass distribution in $Z^{\prime} \rightarrow t \bar{t}$ events before and after the application of trimming,
for increasing values of $\langle\mu\rangle$. Trimming is implemented by re-clustering the jet into subjets of
radius $R=0.3$. The clusters belonging to subjets that carry at least $5\%$ of the original jet $p_T$ are
kept, and those clusters in subjets with less than $5\%$ of the original jet $p_T$ are discarded.

Finally, we conclude with a statement about the increased level of activity expected as the energy of a $pp$ collider increases. The
total inelastic cross section, $74.7\pm1.7$~mb at $8$~TeV~\cite{Antchev:2013paa} is expected to grow to $80$~mb at $13$~TeV, $90$~mb at $33$~TeV,
and $105$~mb at $100$~TeV. Inelastic events with at least one track in the central region have a central track density  ($\Delta\eta \times
\Delta \phi$) of $1.0$ at $8$~TeV and will have a central charged track density  of $1.1 \pm 0.1$ at $13$~TeV, growing to $1.33 \pm 0.14$
at $30$~TeV and  $1.8 \pm 0.4$ at $100$~TeV. The transverse energy for inelastic events deposited per unit $\Delta R^2$ in the central region of the
detector has been measured  at $\sim 0.8$~GeV at $8$~TeV, and will grow to $1.0 \pm 0.15$~GeV at $13$~TeV, $1.25 \pm 0.2$~GeV at $30$~TeV, and
$1.9 \pm 0.35$~GeV at $100$~TeV. The larger inelastic cross sections and the increased activity found in each inelastic event as the
center-of-mass energy increases will add to the difficulty for carrying out precision physics measurements.


\section{Conclusions}
\label{sec:conclusions}

\draftnote{J. Huston et al.}

We summarize here the main conclusions of our report, with an emphasis on where advances in our understanding of QCD are needed to fully explore the electroweak scale at run 2 of the LHC and at future machines.  We divide our conclusions below into four broad categories.

\begin{enumerate}

\item Parton distribution functions for precision and discovery physics:

\begin{itemize}

\item Improvements in our current understanding of PDFs are needed in the `precision region.'  For example, the differences between the CT, MSTW and NNPDF gluon PDFs are large exactly in the range relevant for Higgs studies. Efforts are underway to understand the source of the  differences between the three global PDF groups, and this will hopefully  result in a reduction of the total PDF uncertainty for Higgs production in 8 and 14 TeV proton-proton collisions. Collider data may also help to reduce the uncertainties in this precision region, but  a future electron-hadron collider, such as the LHeC, would be the ultimate machine to provide PDFs for precision HL-LHC physics.

\item Improvement in our current understanding of PDFs is needed in the `discovery region.'  The optimal use of current and future LHC data in global PDf fits is needed to reduce the uncertainties in order to facilitate high-mass discoveries.

\item Photon-induced reactions will become an increasingly large contribution to scattering processes at future high-energy colliders.  Further work is needed to constrain the photon distribution function from LHC data. Photon PDFs and their uncertainties need to be available from all PDF groups. 

\end{itemize}

\item The frontiers of perturbation theory:

\begin{itemize}

\item Higher precision calculations combining QCD at NNLO and beyond, together with EW corrections at NLO, are needed to fully realize the potential of future high energy pp collisions.  The capability to perform an NNLO calculation for any $2 \to 2$ process and selected $2 \to 3$ processes is desired, and seems within reach.

\item Should QCD and electroweak corrections be combined additively or multiplicatively?  Does either prescription correctly reproduce the result of an exact calculation?  Calculations of mixed corrections are needed to check how to perform this combination, and to develop an intuition about when each method is appropriate.

\item The battle for precision must be fought on several fronts.  Progress on both fixed-order calculations and on the resummation of large logarithms is crucial.  It will remain so at future colliders, where the larger phase space available will lead to increasingly larger ratios between the available scales.  The full realization of the potential of future proton-proton machines to unravel the identity of the Higgs boson requires advances in our QCD calculational abilities.

\item Achieving a theoretical precision at the 1\% level will require a detailed understanding of  possible corrections to the typically-assumed factorization picture, and a re-examination of assumptions such as the universality of PDFs and fragmentation functions.

\end{itemize}

\item The Sudakov zone:

\begin{itemize}

\item At high energies, electroweak corrections became as large as, or larger than, QCD corrections.  The inclusion of electroweak corrections into theoretical simulation programs is mandatory for physics studies at future high energy proton-proton colliders.

\item Electroweak corrections are technically challenging for high-multiplicity final states.  In many common kinematic situations these corrections are dominated by Sudakov logarithms.  Continued attention should be devoted to assessing frameworks for their approximate inclusion into Monte Carlo simulations.

\end{itemize}

\item The determination of fundamental constants:

\begin{itemize}

\item The current errors on $\as$, $m_b$ and $m_c$ induce sizable parametric uncertainties in predictions for Higgs boson decay rates.  Lattice and continuum extraction methods both feature smaller errors than used in coupling extractions, and have consistent central values.  The error assumptions in future Higgs coupling analyses should be revisited.

\item Improvements in lattice calculations, data from LHeC, and data from future $e^+e^-$ colliders could reduce the error on $\as$ to 0.1\%.

\end{itemize}

\end{enumerate}

The rapid progress in QCD and EW phenomenology that has been observed in the recent past, along with that expected in the near future, will, along with the data taken at the LHC, allow for full exploitation of the discovery and precision-measurement capability of the LHC and subsequent colliders.




\Acknowledgements

L. Barz\`e is supported by the ERC Grant 291377.

T.~Becher and X.Garcia~i~Tormo are supported by the Swiss National Science Foundation (SNF) under grant 200020-140978 and the Sinergia grant number CRSII2 141847 1.

G.~Bodwin and R.~Boughezal are supported by the U.S.\ Department of Energy under Contract No. DE-AC02-06CH11357.

J.~M.~Campbell, P.~Mackenzie and R.~Van de Water are supported by the U.S.\ Department of Energy under Contract No. DE-AC02-06CH11359.

S.~Carrazza and S.~Forte are supported by an Italian PRIN 2010 and by a European EIBURS grant. 

M. Chiesa, G. Montagna, M. Moretti, O. Nicrosini and F. Piccinini
are supported by the italian PRIN project 2010YJ2NYW and by the
Research Executive Agency (REA) of the European Union under the
Grant Agreement number PITN-GA-2010-264564 (LHCPhenoNet).

S.~Hoeche is supported by the U.S.\ Department of Energy under Contract No. DE-AC02-76SF00515.

F.~Petriello is supported by the U.S.\ Department of Energy under Grant No. DE-SC0010143.

J.~Rojo is supported by a Marie Curie  Intra--European Fellowshipa
of the European Community's 7th Framework Programme under contract
number PIEF-GA-2010-272515. 

F. Tramontano is supported by the italian PRIN project 2010YJ2NYW.

%
%
%
%
\bibliography{QCD/qcd-references}




\end{document}